\documentclass[11pt,a4paper,openright,twoside]{book}
\usepackage{color}
\DeclareMathAlphabet{\mathbbm}{U}{bbm}{m}{n}
\SetMathAlphabet\mathbbm{bold}{U}{bbm}{bx}{n} 
\def\1{\mbox{l\hspace{-0.53em}1}}
\usepackage{pstricks}
\usepackage[dvips,final]{graphicx}
\usepackage{multirow}
\newcommand{\fr}{\frac}
\usepackage{amsmath}
\usepackage{fancyheadings}
\usepackage{young}
\usepackage{float}
\usepackage{subfigure}
\usepackage{longtable}
\usepackage{caption}
\def\onedot{\makebox(0,0){$\scriptstyle 1$}}
\def\twodot{\makebox(0,0){$\scriptstyle 2$}}
\def\threedot{\makebox(0,0){$\scriptstyle 3$}}
\def\fourdot{\makebox(0,0){$\scriptstyle 4$}}

\newcommand{\lambdab}{\mbox{\boldmath$\lambda$}}
\newcommand{\rhob}{\mbox{\boldmath$\rho$}}

\usepackage{rotating}
\rotdriver{dvips}

\def\ssqr#1#2{{\vbox{\hrule height #2pt
      \hbox{\vrule width #2pt height#1pt \kern#1pt\vrule width #2pt}
      \hrule height #2pt}\kern- #2pt}}

\def\bsqr{\ssqr{10}{.2}}
\def\nbox{\vbox{\hbox{$\bsqr\bsqr\bsqr\bsqr\raise2.7pt\hbox{$\,\cdot\cdot
\cdot\cdot\cdot\,$}\bsqr\bsqr\bsqr$}\nointerlineskip
\kern-.2pt\hbox{$\phantom{\bsqr}$}}}

\def\ndots{\vbox{\hbox{$\phantom{\bsqr}\raise2.7pt\hbox{$\,\cdot\cdot\cdot
\,$}$}\nointerlineskip
\kern-.2pt\hbox{$\phantom{\bsqr}$}}}

\def\nboxA{\vbox{\hbox{$\bsqr\bsqr\bsqr\bsqr\raise2.7pt\hbox{$\,\cdot\cdot
\cdot\cdot\cdot\,$}\bsqr\bsqr\bsqr$}\nointerlineskip
\kern-.2pt\hbox{$\bsqr$}}}

\def\nboxE{\vbox{\hbox{$\bsqr\bsqr\bsqr\raise2.7pt\hbox{$\,\cdot\cdot
\cdot\cdot\cdot\,$}\bsqr\bsqr\bsqr\bsqr$}\nointerlineskip
\kern-.2pt\hbox{$\bsqr\bsqr\bsqr\raise2.7pt\hbox{$\,\cdot\cdot\cdot
\cdot\cdot\,$}\bsqr$}}}

\def\nboxF{\vbox{\hbox{$\bsqr\bsqr\bsqr\bsqr\raise2.7pt\hbox{$\,\cdot
\cdot\cdot\cdot\cdot\,$}\bsqr\bsqr$}\nointerlineskip
\kern-.2pt\hbox{$\bsqr\bsqr\bsqr\bsqr\raise2.7pt\hbox{$\,\cdot\cdot
\cdot\cdot\cdot\,$}\bsqr$}}}

\newcommand{\drawsquare}[2]{\hbox{%
\rule{#2pt}{#1pt}\hskip-#2pt
\rule{#1pt}{#2pt}\hskip-#1pt
\rule[#1pt]{#1pt}{#2pt}}\rule[#1pt]{#2pt}{#2pt}\hskip-#2pt
\rule{#2pt}{#1pt}}

\newcommand{\Yfund}{\raisebox{-.5pt}{\drawsquare{6.5}{0.4}}}
\newcommand{\Ythrees}{\raisebox{-.5pt}{\drawsquare{6.5}{0.4}}\hskip-0.4pt%
     \raisebox{-.5pt}{\drawsquare{6.5}{0.4}}\hskip-0.4pt%
     \raisebox{-.5pt}{\drawsquare{6.5}{0.4}}}

\newcommand{\Ythreea}{\raisebox{-6.75pt}{\drawsquare{6.5}{0.4}}\hskip-6.9pt%
    \raisebox{-0.25pt}{\drawsquare{6.5}{0.4}}\hskip-6.9pt
    \raisebox{6.25pt}{\drawsquare{6.5}{0.4}}}

\newcommand{\Yadjoint}{\raisebox{-3.5pt}{\drawsquare{6.5}{0.4}}\hskip-6.9pt%
    \raisebox{3pt}{\drawsquare{6.5}{0.4}}\hskip-0.4pt
    \raisebox{3pt}{\drawsquare{6.5}{0.4}}}
\newlength{\accohaut}

\begin{document}

%



\thispagestyle{empty}
\vspace*{-3cm}
\begin{center}

\centerline{\includegraphics[height=3.5cm]{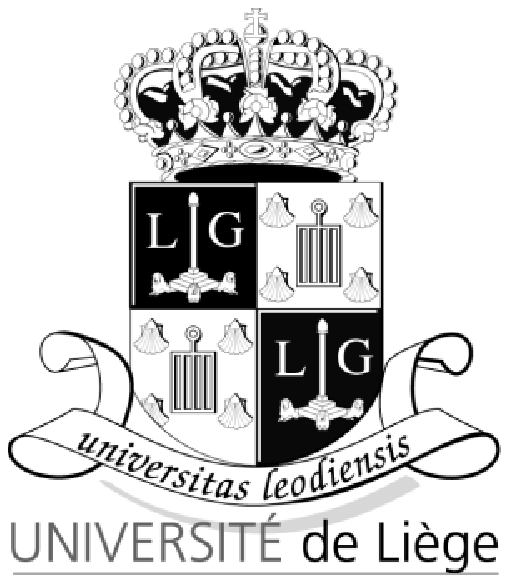}}
\sc \large Acad\'emie Universitaire Wallonie--Europe\\
\sc \normalsize Facult\'e des Sciences
\vspace*{4.5cm}

\sffamily

{\Huge  Baryon resonances \\ \vspace{0.5cm} in large $N_c$ QCD}
\vspace*{0.5cm}


\vspace*{4.5cm}

{\Large  \sc Nicolas Matagne} \\

\vspace*{.5cm}
\normalsize 

\vspace*{5.2cm}

{\normalsize
Th\`ese pr\'esent\'ee en vue de l'obtention du \\
Grade de Docteur en Sciences}

\vspace*{1.3cm}
\normalsize
D\'ecembre 2006
\end{center}

\newpage
\thispagestyle{empty} 
\ 
\newpage

\chapter*{Remerciements -- Acknowledgments}
\thispagestyle{empty}


Je tiens tout d'abord \`a exprimer ma profonde et sinc\`ere reconnaissance envers Mme  Fl. Stancu. Sans son soutien permanent durant ces quatre ans et deux mois de th\`ese, ce travail n'aurait jamais pu voir le jour. Au moment de l'\'ecriture de ces lignes, je me ressouviens de notre premi\`ere rencontre dans son bureau, un jour de juin 2002, la conversation ayant assez vite d\'evi\'e sur  les probl\`emes personnels qui m'inondaient \`a l'\'epoque. Tout de suite, j'ai pu compter sur sa confiance, sa gentillesse et sa compr\'ehension face \`a ces difficult\'es qui marqu\`erent de leur empreinte ind\'el\'ebile les deux premi\`eres ann\'ees de cette th\`ese. Progressivement le ciel se d\'ecouvra et l\`a encore, ce sont ses conseils judicieux et ses suggestions qui me permirent de faire mes premiers pas dans le monde de la recherche. Enfin, je ne peux laisser sous silence son gros travail de correction et de relecture de cette th\`ese, t\^ache souvent fastidieuse mais n\'ecessaire au vu des innombrables coquilles et approximations qui la parsemaient. 
\\

Ce travail n'existerait pas sans le soutien financier de l'Institut Interuniversitaire des Sciences Nucl\'eaires (I.I.S.N.). Je tiens \`a exprimer ma gratitude envers M. le Professeur J. Cugnon qui me confia il y a quatre ans et demi le poste de chercheur I.I.S.N. vacant \`a l'\'epoque. Je ne saurais oublier les nombreuses conf\'erences qu'il a  financ\'ees avec l'argent du groupe. Celles-ci me permirent de rencontrer d'autre physiciens de mon domaine ainsi que de d\'ecouvrir des contr\'ees qui m'\'etaient jusque l\`a inconnues. \\

Je tiens \`a remercier mes amis de ULg pour les bons moments partag\'es ensemble avec une pens\'ee particuli\`ere vers Lionel qui fut le plus fid\`ele et avec qui j'ai pass\'e une excellente semaine en R\'epublique Tch\`eque l'ann\'ee derni\`ere.   Je  pense \'egalement \`a Jean-Philippe  pour ces conseils strat\'egiques qui m'aid\`erent \`a mener \`a bien ce travail et pour les nombreuses fois qu'il m'invita \`a souper chez lui. \\

Enfin, j'adresse une derni\`ere pens\'ee vers ma m\`ere dont l'absence durant ces ann\'ees s'est fait \`a certains moments cruellement ressentir. 
\vspace{1cm}
\begin{flushright}
 Sart Tilman, le 15 novembre 2006  
\end{flushright}

\newpage
\thispagestyle{empty} 
\ 
\newpage

\tableofcontents
\chapter*{Introduction
           \markboth{%
           Introduction}{}}%
\addcontentsline{toc}{chapter}{Introduction}

\thispagestyle{empty}

Thirty-two years after its introduction by  Gerard 't Hooft, the $1/N_c$ expansion in QCD, where $N_c$ is the number of colors, has revealed itself to be a powerful tool which has played an important role in numerous theoretical aspects of QCD. Among these, one can cite the problems of color confinement and spontaneous chiral symmetry breaking. One has not to forget the large $N_c$   lattice calculations which can provide a quantitative control and information about the range of applicability of the theory. \\
  
  During the last fifteen years, the phenomenology of baryons has been another important application of the large $N_c$ QCD. At this energy regime, it is not possible to make a perturbative expansion of QCD, the coupling constant being too large. The traditional issue was to derive baryon properties with the help of effective theories or constituent quark models. The results are naturally model dependent. The $1/N_c$ expansion generates a new perturbative approach to  QCD, not in the coupling constant, but in the parameter $1/N_c$. This expansion can be applied to low energy regime and   provides a new theoretical method that is quantitative, systematic and predictive. \\

  This new theoretical framework was first applied to baryons by Witten  in 1979, who determined the so called  large $N_c$ counting rules which allow to find the $N_c$ order of Feynman diagrams. From the color confinement of quarks, he proved that baryons must be composed of $N_c$ valence quarks and that the baryon mass is of order $\mathcal{O}(N_c)$. He studied the meson--baryon scattering amplitude and found that this amplitude is of order $\mathcal{O}(N_c^0)$. \\
  
  This important result has led to the discovery by Gervais and Sakita in 1984 and independently by Dashen and Manohar in 1993 of  an exact contracted spin-flavor SU($2N_f)_c$ symmetry for ground-state baryons, $N_f$ being the number of light flavors.  This symmetry implies the existence of an infinite tower of degenerate ground-state baryons.  When $N_c \to \infty$ the SU($2N_f$) algebra, used in quark-shell models, is identical to  SU($2N_f)_c$ algebra. This property allows a systematic $1/N_c$ expansion of various static QCD operators like the baryon mass, the axial vector current  or the magnetic moment  in terms of operator products of the baryon spin-flavor generators SU($2N_f$). 
Not all the operator products are linearly independent. Operator identities were derived in 1995 by Dashen, Jenkins and Manohar to eliminate redundant operators  from the QCD operator in the $1/N_c$ expansion. The operator products, describing for example the spin-spin operator or SU($N_f$) breaking terms, appear in the $1/N_c$ expansion accompanied by unknown coefficients which represent reduced matrix elements embedding the QCD dynamics. These coefficients are obtained by a fit to experimental data. \\

The $1/N_c$ expansion of the ground-state baryon mass operator explains the remarkable accuracy of the well know Gell-Mann--Okubo or of the Coleman--Glashow mass relations. New mass relations, satisfied to a non trivial order in $1/N_c$  by the experimental data were also predicted. The application to magnetic moments gave new relations between baryon magnetic moments, also in agreement with the experimental data.\\

The baryons containing a heavy quark, like a charm or a bottom quark, were also studied in the $1/N_c$ expansion. In this case, the baryons are governed by a heavy quark spin-flavor symmetry as well as by the large $N_c$ light quark spin-flavor symmetry.  Then, for a baryon containing one heavy quark, there are two towers of infinite degenerate ground-state  baryons.  Mass relations implying baryons containing charm and bottom   quarks  allow predictions of the masses of unmeasured charm and bottom baryons. \\

A basic question is whether or not the $1/N_c$ expansion can be applied to excited baryons which traditionally have been the domain of the quark model, where the results are necessarily model dependent. In principle 't Hooft's suggestion would lead to an $1/N_c$ expansion which would hold in all QCD regimes. Thus the study of excited baryons seems quite natural in the $1/N_c$ expansion. The corresponding symmetry is now SU($2N_f$) $\times$ O(3). \\
 
The experimental observation supports a classification of excited baryons into groups, called bands, which can roughly be associated to $N=0,1,2, \ldots$ units of excitation energy, like in a harmonic oscillator well. Each band contains a number of SU($2N_f$) $\times$ O(3) multiplets. The $N=1$ band contains only one multiplet, the $[{\bf 70},1^-]$, the $N=2$ band has five multiplets, the $[{\bf 56'},0^+]$, the $[{\bf 56},2^+]$, the $[{\bf 70},0^+]$, the $[{\bf 70},2^+]$ and  the $[{\bf 20},1^+]$ but there are no physical resonances associated to the $[{\bf 20},1^+]$. The $N=3$ band has 10 multiplets, the lowest one being the $[{\bf 70'},1^-]$ and the $N=4$ band has 17 multiplets, the lowest one being the $[{\bf 56},4^+]$. \\

Until 2003 only the $[{\bf 70},1^-]$, the $[{\bf 56'},0^+]$ and the $[{\bf 56},2^+]$ multiplets were investigated. This was a natural choice because the $[{\bf 70},1^-]$ multiplet is the lowest in energy and quite well separated from the other multiplets. The $[{\bf 56'},0^+]$ and the $[{\bf 56},2^+]$ multiplets, from symmetries point of view, are the simplest in the $N=2$ band. \\

  Generally the study of excited multiplets is not free of difficulties.  The  traditional approach  suggests the decoupling of the wave function into   a symmetric ground-state core composed of $N_c-1$ quarks and  one excited quark. The method, based on a Hartree approximation, has the advantage  that the core can be treated  like a ground-state baryon. However, with this approach, the SU($2N_f$) symmetry is broken at   order $\mathcal{O}(N_c^0)$ instead of $\mathcal{O}(1/N_c)$ as for the ground state. Nevertheless, as the experiments show that the breaking must be very small, the $1/N_c$ expansion mass predictions for the non-strange and strange baryons belonging to the  $[{\bf 70}, 1^-]$ multiplet are consistent with the experimental data and in agreement with the constituent quark model results. Furthermore, because of the symmetry properties of the $[{\bf 56'},0^+]$ and the $[{\bf 56},2^+]$ multiplets, the decoupling of the wave function of baryons belonging to these multiplets, into a core and an excited quark, is not needed. In that case, the SU($2N_f$) symmetry is broken at order $\mathcal{O}(1/N_c)$, like for the ground-state baryons. \\

Baryon decays were analyzed too. One can cite the predictions for decays of radially excited baryons belonging to the $[{\bf 56'},0^+]$ multiplet or of non-strange negative parity baryons. \\

The aim of this thesis is to extend previous work and help in understanding the spin-flavor structure of  excited baryon states by searching for regularities in the $1/N_c$ expansion. We should recall that resonances have a typical width of order of 100 MeV but as in quark model calculations we treat the resonances as stable states. The first idea  was to complete the baryon mass analysis of  the $N=2$ band with the study non-strange and strange $[{\bf 70},\ell^+]$ ($\ell=0,2$) multiplets and then to explore  the $N=3$ and $N=4$ bands.   \\

However  we discovered that the task was harder than expected for mixed-symmetric multiplets $[{\bf 70}]$ and then we turned to a symmetric multiplet. So we started with the study of the $[{\bf 56},4^+]$ multiplet belonging to the $N=4$ band, which is quite easy because the treatment is very similar to the one of the  $[{\bf 56},2^+]$ multiplet. Next we analyzed the non-strange baryons in the $[{\bf 70},0^+]$ and $[{\bf 70},2^+]$ multiplets. The problems began when we decided to study the $[{\bf 70},\ell^+]$ strange baryons.  We had first to derive the matrix elements of the SU(6) spin-flavor group for a symmetric wave function. In addition, the wave functions used for the  $[{\bf 70},\ell^+]$ multiplets were based on the traditional decoupling picture described above. Progressively, we realized that this approach  violates the permutation symmetry S$_{N_c}$ due to an incomplete antisymmetrization of the wave functions.  \\

  Recently, we proposed a new approach in which we consider the baryon wave function in one block, without decoupling it into a symmetric core and an excited quark. The symmetrization of the wave function is now properly taken into account. The main outcome is  that the SU($2N_f$) symmetry is broken at order $\mathcal{O}(1/N_c)$, like for the ground state. One difficulty of this new picture is that it implies the knowledge of the matrix elements of the spin-flavor generators for a mixed-symmetric spin-flavor wave function composed of $N_c$ particles.  For the case of two flavors, the problem is solved.  We applied this new approach to the non-strange baryons belonging to the  $[{\bf 70},1^-]$ multiplet.\\
  
  This thesis is divided in six chapters and five appendices. Chapter 1 gives an introduction to large $N_c$ QCD. 't Hooft's double line notation is presented as well as Witten's large $N_c$ power counting rules which allow us to determine the order  $\mathcal{O}(N_c)$ of Feynman diagrams. We  discuss the main properties of mesons and baryons in the large $N_c$ limit. This chapter  finishes with the introduction to large $N_c$ meson-baryon coupling. \\
  
  The second chapter is devoted to the baryon structure. A simple quark-shell model picture is presented in order to introduce the main baryon quantum numbers like the spin $S$, the  isospin $I$ or the angular momentum $\ell$ and the general feature of the baryon spectrum. Even if this model is very simple, the picture remains general. Furthermore it can be easily generalized to baryon composed of $N_c$ quarks as it is shown in the second part of this chapter.  \\
  
The third chapter is dedicated to ground-state baryons. The exact SU($2N_f)_c$ symmetry is explained as well as its relations to the SU($2N_f$) symmetry introduced in Chapter 2 in the context of the quark-shell model. Then, the $1/N_c$  expansion of a baryon static operator is analyzed and the operator identities necessary to obtain an $1/N_c$ expansion composed of linearly independent operators are summarized.  This chapter ends with an explicit application to the ground-state baryon mass operator. \\

For deriving the mass operator of ground-state and excited state baryons, one needs to know the matrix elements of the SU(6) generators for a symmetric spin-flavor wave function composed of $N_c$ particles. Chapter 4 is devoted to the derivation of these matrix elements. A generalized Wigner-Eckart theorem is presented reducing the problem to the knowledge of SU(6) isoscalar factors. For a symmetric spin-flavor wave function one can calculate them by deriving explicitly the matrix elements of generators acting on one particle as they are identical for all particles of the wave function. In the last part of the chapter, the matrix elements of the generators of the SU(4) group for a symmetric spin-flavor wave function are recovered. \\

The decoupling picture of excited baryons is detailed in Chapter 5. This chapter begins with the Hartree approximation, exact in the $N_c\to \infty$ limit and serving as a starting point for the decoupling approach. The baryon wave function for symmetric and mixed-symmetric multiplets are written explicitly and the  $1/N_c$ expansion to excited states is presented. Applications to the $[{\bf 70},\ell^+]$ ($\ell=0,2$) and $[{\bf 56},4^+]$ multiplets constitute the second part of the chapter 5. \\

Chapter 6 presents the new approach of excited baryons without decoupling the baryon wave function into a symmetric core and an excited quark. It starts with a discussion of the baryon wave function in term of the permutation group S$_{N_c}$. An unified   treatment of  the spin-orbit operator for symmetric and mixed-symmetric orbital or spin-flavor wave function is proposed. The study of  the $[{\bf 70},1^-]$ multiplet in this new light  finishes this thesis. \\

Many of the analytical details related to the different chapters are given in the Appendices A--E. 
\chapter{Large $N_c$ QCD}

\thispagestyle{empty}

\section{Introduction}

Quantum Chromodynamics (QCD) is a non-Abelian gauge field theory based on the SU$_c$(3) group, where the subscript $c$ stands for color. It describes the strong interactions of colored quarks and gluons. As the quarks come in three colors, they belong to the fundamental representation of the group SU$_c$(3). One has eight gluons as they belong to  the adjoint representation of the color group.  Hadrons are  color singlet combinations of  quarks, antiquarks and gluons. \\

\begin{figure}[h]
\begin{center}
\includegraphics[width=7cm,keepaspectratio]{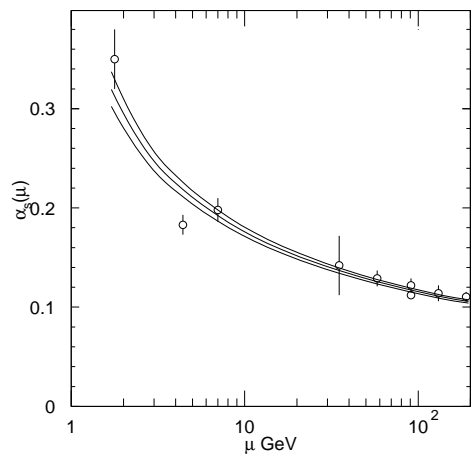}
\end{center}
\caption{Evolution of the effective coupling constant $\alpha_s=g^2/4\pi$ with the energy scale $\mu$ \cite{yao06}.}
\label{couplingconstant}
\end{figure}

It is impossible to solve QCD exactly because of the complexity of the phenomena described. Like for  Quantum Electrodynamics (QED), one can try to make a perturbative expansion of the theory in terms of the coupling constant $g$.  Figure \ref{couplingconstant} which represents the evolution of the QCD coupling constant with the energy scale suggests that QCD has  two energy distinct regimes. At high energy, where the coupling constant is small, leading to the asymptotic freedom, one can apply a perturbative approach and then explain the high energy behavior for the production and the interaction of hadrons. \\

At low energy regime, typical for baryon spectroscopy, such a perturbative method does not work. Some nonperturbative approaches have been developed to solve the problem. One can cite for example the  computational approach of lattice QCD. However, its application to baryon spectroscopy needs an improvement of the calculation methods and of the capacity of our computers. Another way to extend QCD to nucleons is  to construct  phenomenologic models inspired by QCD. It has already been  done with success in the past. For example, concerning nucleons, these studies gave a  description of baryon spectra generally in agreement with the experimental data. 
Here, we have chosen to use another promising approach, the large $N_c$ QCD. \\ 

Large $N_c$ QCD is a non-Abelian gauge theory based on the SU$_c(N_c)$ color group where $N_c$ is the number of colors. The generalization of the QCD Lagrangian to large $N_c$ is given by \cite{jenkins01} 
\begin{equation}
 \mathcal{L}=-\frac{1}{2} \mathrm{Tr}G^{\mu\nu}G_{\mu\nu}+\sum_{f=1}^{N_f}\bar{q}_f \left(i /\negthinspace\negthinspace \negthinspace\negthinspace D - m_f\right)q_f,
\end{equation}
where the gauge field strength and the covariant derivative, $D^{\mu}=\partial^\mu+igA^\mu$, are defined as in QCD and $N_f$ is the number of flavors.
This generalization of the classical SU$_c(3)$ QCD to an arbitrary large $N_c$ was introduced for the first time by 't Hooft in 1974  \cite{hooft74}, when $1/N_c$ was suggested as the expansion parameter of the theory. The idea was  considered in detail by Witten \cite{witten79} who introduced  large $N_c$ power counting rules for the Feynman diagrams. With this point of view, the number of quarks is $N_c$ and the number of gluons is $N_c^2-1=\mathcal{O}(N^2_c)$. As a result one obtains a new perturbative expansion not in the coupling constant, like in QED, but  in powers of $1/N_c$. \\

One may wonder about the meaning of an expansion in $1/N_c$ as one knows that in the real world one has $N_c=3$.  We can give only a pragmatic answer by comparing the predictions made by  such an expansion  with phenomenological models and experimental data. We have to stress  the fact that during the calculations, one considers the $1/N_c$ expansion rather than the $N_c\to \infty$ limit. One keeps only leading terms in $1/N_c$ and neglects subleading orders.\\
 
 To define the large $N_c$ limit, one  has first to find the $1/N_c$ order of quark-gluon diagrams. In a second stage, one can introduce mesons and baryons in large $N_c$ QCD and study the interactions between them.

\section{Feynman diagrams for large $N_c$}

In large $N_c$ QCD,  one can suppose that as the number of colors is very important,  a lot of new intermediate states, which do not exist for $N_c=3$, can appear during a process. They come form the fact that even if the color quantum numbers of the initial and final states are specified, one can have an indetermination on the color of the intermediate state. This introduces new intermediate states. The sum over these intermediate states  gives rise to large combinatoric factors. As the number of intermediate states will differ from one process to another, some Feynman diagrams will be suppressed. \\

\begin{figure}[h!]
\begin{center}
\includegraphics[width=4cm,keepaspectratio]{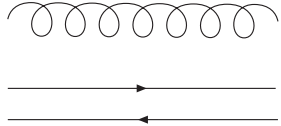}
\end{center}
\caption{A gluon in the traditional and in the double line notation.}
\label{double}
\end{figure}

To find the $1/N_c$ order of Feynman diagrams, it is very useful to use the double line notation introduced by 't Hooft \cite{hooft74}. In this notation one can replace the gluon gauge field $(A^\mu)^A$ in the adjoint representation by the tensor field $(A^\mu)^i_j$ with the indices $i$ and $j$ belonging to the fundamental representation of the SU$_c(N_c)$ group, as illustrated in Figure  \ref{double}. So, the gluon field has one upper index like the quark field $q^i$ and one lower index like the antiquark field $\bar{q}_j$. \\

\begin{figure}[h!]
\begin{center}
\includegraphics[width=15cm,keepaspectratio]{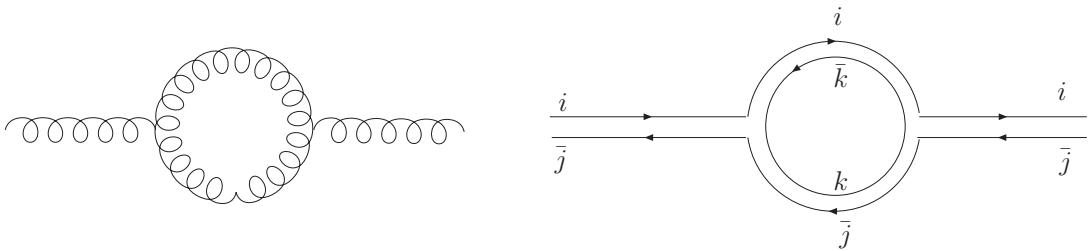}
\end{center}
\caption{The gluon vacuum polarization. The double line representation of the diagram is on the right part of the figure.}
\label{glueball}
\end{figure}

By using the double line notation, let us first consider the one-loop gluon vacuum polarization diagram (Figure \ref{glueball}). By looking at the right part of Figure \ref{glueball}, it is easy to determine the combinatoric factor of this diagram. Indeed, the color quantum numbers of the initial and final states are specified but not the central index $k$ which leads to a combinatoric factor of $N_c$ for this Feynman diagram. If $N_c \to \infty$ the contribution of this diagram is infinite. \\

To obtain a finite limit  for this process, one can renormalize the theory by introducing a new coupling constant $g/\sqrt{N_c}$ instead of $g$. Then 
\begin{equation}
\frac{g}{\sqrt{N_c}}\to 0\ \mathrm{when}\ N_c\to \infty,
 \label{renormalisation}
\end{equation}
where $g$ is fixed when $N_c$ becomes large. 
 In the one-loop gluon vacuum polarization we have two vertices and one combinatoric factor  $N_c$. With this renormalization, the order of the Feynman diagram in Figure \ref{glueball} becomes 
\begin{equation}
 \left(\frac{g}{\sqrt{N_c}}\right)^2 N_c = g^2
\end{equation}
independent of $N_c$, as expected. \\

\begin{figure}[h!]
\begin{center}
\includegraphics[width=10cm,keepaspectratio]{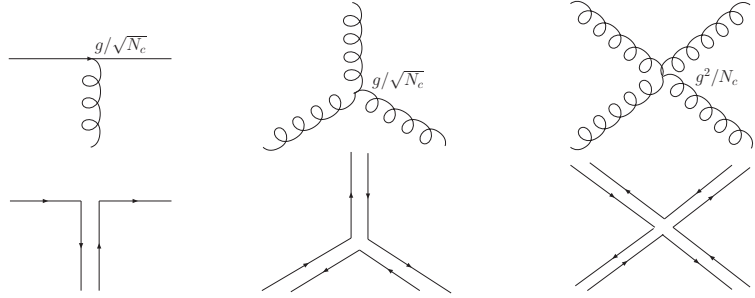}
\end{center}
\caption{Quark-gluon, three-gluons and four-gluons vertices in classical and double line notation.}
\label{vertex}
\end{figure}


With the renormalization, quark-gluon vertices and 3-gluons vertices are of order $1/\sqrt{N_c}$. However, 4-gluons vertices are of order $1/N_c$ because Feynman rules  give these vertices proportional to $g^2$. Figure \ref{vertex} summarizes these considerations.\\

\begin{figure}[h!]
\begin{center}
\includegraphics[width=12cm,keepaspectratio]{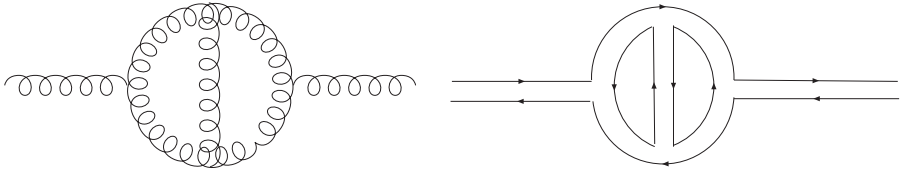}
\end{center}
\caption{A two-loop diagram of order 1.}
\label{twoloop}
\end{figure}

\begin{figure}[h!]
\begin{center}
\includegraphics[width=12cm,keepaspectratio]{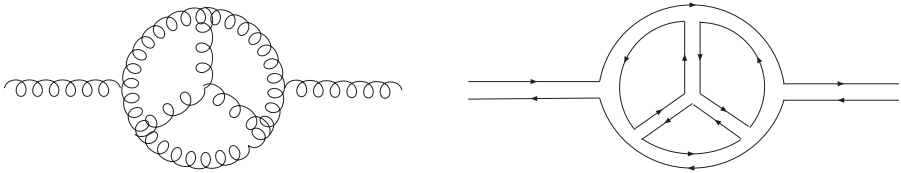}
\end{center}
\caption{A three-loop diagram of order 1.}
\label{threeloop}
\end{figure}

\begin{figure}[h!]
\begin{center}
\includegraphics[width=12cm,keepaspectratio]{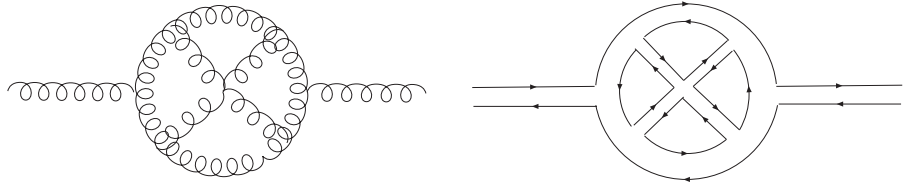}
\end{center}
\caption{A four-loop diagram of order 1.}
\label{fourloop}
\end{figure}

To extract the power counting rules for large $N_c$ Feynman diagrams, let us determine the order of some diagrams. Let us first have a look at the two-loop diagram drawn in Figure \ref{twoloop}. The order of this diagram is $(1/\sqrt{N_c})^4\times N_c^2=1$. Indeed, in the right-hand side of the figure, we have two internal loops so a combinatoric factor of $N_c^2$ appears and four vertices. Figure \ref{threeloop} is another example, with three-self-contracted loops and six interactions vertices, of a diagram of order 1, as well. Concerning the four-loop diagram (Figure \ref{fourloop}), the order is $(1/\sqrt{N_c})^8\times N_c^4=1$. All these planar diagrams -- they can be drawn in the plane -- survive in the large $N_c$ limit. \\

\begin{figure}[h!]
\begin{center}
\includegraphics[width=12cm,keepaspectratio]{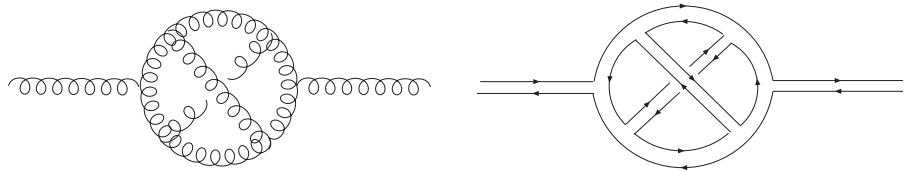}
\end{center}
\caption{A  non-planar diagram  of order $1/N^2_c$.}
\label{nonplanar}
\end{figure}

Let us examine a non-planar diagram (Figure \ref{nonplanar}).  This diagram has the particularity of being impossible to be drawn in a plane without line crossing (at points where there are no interaction vertices). It is suppressed in the large $N_c$ limit as it is of order $(1/\sqrt{N_c})^6\times N_c=1/N_c^2$. By considering some other examples, it is possible to show that all non-planar Feynman diagrams are of maximal order of $1/N_c^2$ so they vanish in the large $N_c$ limit. Contrary, all planar diagrams made with arbitrary number of gluon loops are of order 1.

\begin{figure}[h!]
\begin{center}
\includegraphics[width=10cm,keepaspectratio]{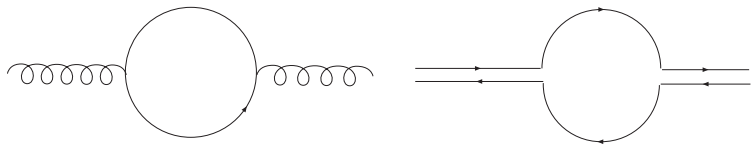}
\end{center}
\caption{Quark bubble of order $1/N_c$.}
\label{quarkloop}
\end{figure}

\begin{figure}[h!]
\begin{center}
\includegraphics[width=12cm,keepaspectratio]{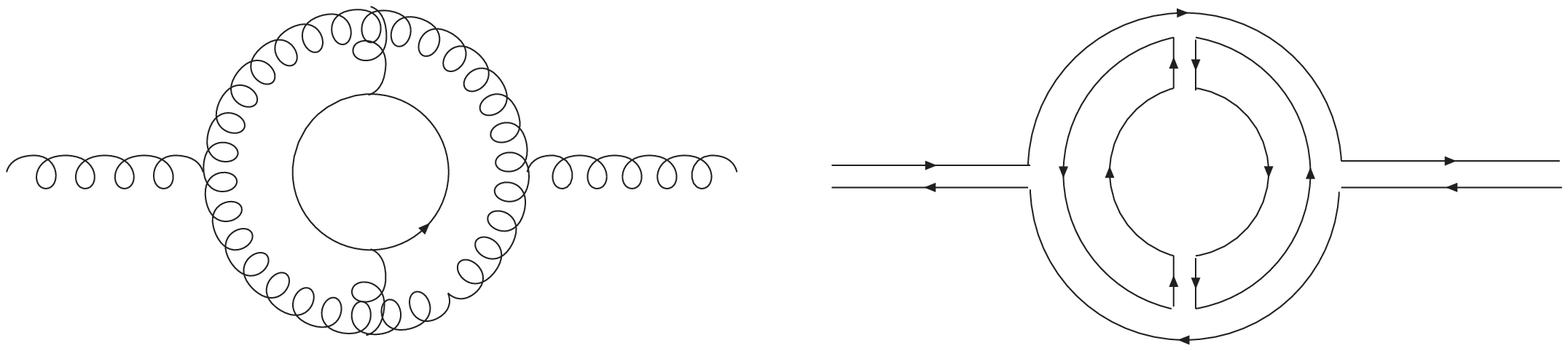}
\end{center}
\caption{A quark loop inside a gluon loop of order $1/N_c$.}
\label{quarkgluon}
\end{figure}

A quark loop is presented in Figure \ref{quarkloop}. We do not have a combinatoric factor as the color indices are fixed by the initial and final states. There are two vertices, so a factor $1/N_c$ appears. This diagram, of order $1/N_c$, is also suppressed at $N_c \to \infty$. Consider now Figure \ref{quarkgluon}, with one internal quark loop. This diagram also vanishes in the large $N_c$ limit as it is of order $(1/\sqrt{N_c})^6\times N_c^2 = 1/N_c$. \\

Form these considerations, one can establish the following large $N_c$ counting rules:
\begin{itemize}
 \item Planar gluon insertions do not change the order of the diagram.
 \item Each non-planar gluon line is suppressed by a factor of $1/N_c^2$.
 \item Internal quark loops are suppressed by factors of $1/N_c$.
\end{itemize}
{\it In conclusion, the leading Feynman diagrams are planar and contain a minimum number of quark loops}. \\

\begin{figure}[h!]
\begin{center}
\includegraphics[width=8cm,keepaspectratio]{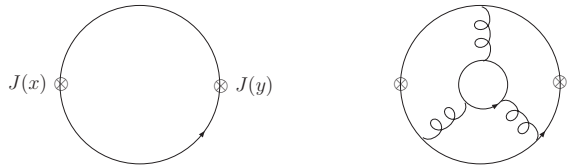}
\end{center}
\caption{On the left, a two point function diagram, depicting the propagation of a meson, of order $N_c$. On the right, the same diagram with an internal quark loop, of order 1.}
\label{quarkcorr}
\end{figure}

\begin{figure}[h!]
\begin{center}
\includegraphics[width=10cm,keepaspectratio]{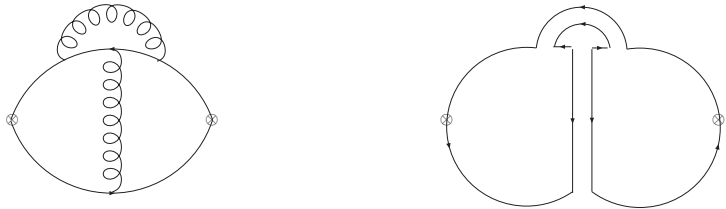}
\end{center}
\caption{A non-planar contribution of the current correlation function of order 1.}
\label{nonplanarquarkcorr}
\end{figure}

Before ending this section, we have to make few comments on the two-point function $\langle J(y)J(x)\rangle$ of a current correlation function $J$. This function creates a meson at the point $x$ and annihilates the meson at $y$.  In Figure \ref{quarkcorr} are drawn two examples of such diagrams. On the left part of the figure, a combinatoric factor of $N_c$ appears as there are no constraint on the color indices of the quark loop. On the right part, an internal quark loop is added. This diagram, with six vertices and three internal loops, of order $(1/\sqrt{N_c})^6\times N_c^3 = 1$, is non-dominant by comparison with the diagram drawn in left-hand side of the figure. One can directly show that diagrams with internal gluon loops are of order $N_c$. Figure \ref{nonplanarquarkcorr} shows that a non-planar gluon induces a suppression factor of $1/N_c$.\\

{\it The general conclusion is that dominant two-point function diagrams for large $N_c$ are planar diagrams with only a single quark loop which runs at the edge of the diagram.} \\

Leading order Feynman diagrams can be very complicated. For example, as we have seen, if we replace the internal quark loop of the diagram drawn in the right-hand side of Figure \ref{quarkcorr} by a gluon loop, we obtain a diagram of order $N_c$ like the left-hand side one. This kind of diagrams with a lot of gluons inside are impossible to evaluate. So we are  not able to sum the planar diagrams. However,  we assume that QCD is still a confining  theory at large $N_c$ \cite{witten79}. With this assumption we will deduce some properties interesting for  large $N_c$ mesons and baryons. We have just to keep planar diagrams and neglect other ones.

\section{Mesons in large $N_c$ QCD}

One might wonder about the interest to study mesons here, as we plan to apply large $N_c$ QCD to baryon spectroscopy. In fact, some important results used in the expansion of the mass operator of baryons depend on the baryon-meson scattering. Before studying this process, we need to learn some properties of large $N_c$ mesons. \\

Large $N_c$ mesons are colorless bound states composed of a quark and an antiquark. The color-singlet meson wave function is given by \cite{bhaduri88}
\begin{equation}
 |1\rangle_c = \frac{1}{\sqrt{N_c}} \underbrace{\left(\bar{l}l+\bar{m}m+\ldots+\bar{n}n\right)}_{N_c\ \mathrm{terms}},
 \label{largenmesonwavefunction}
\end{equation}
where $l,\ m$ and $n$ represent some color quantum numbers. One can notice that the sum over the $N_c$ possible color quantum numbers leads to a combinatoric factor of $\sqrt{N_c}$.\\

\begin{figure}[h!]
\begin{center}
\includegraphics[width=12cm,keepaspectratio]{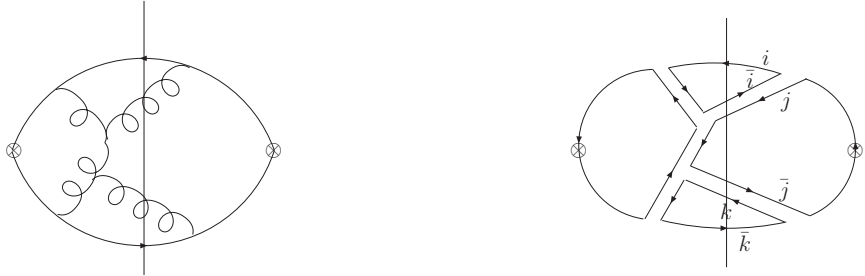}
\end{center}
\caption{An intermediate state contributing to $\langle J J \rangle$ \cite{witten79}.}
\label{mesonintermediate}
\end{figure}

Let us analyze meson creation and annihilation diagrams. 
Let us  prove  that the operator $J(x)$ which corresponds to a local quark bilinear such as $\bar{q}q$ or $\bar{q}\gamma^\mu q$ having good quantum number to create a meson, creates only one-meson states in the large $N_c$ limit. Let us consider Figure \ref{mesonintermediate} where a two-point function of a quark bilinear is represented. One has to prove that the intermediate state is a $\bar{q}q$ pair. In a confining theory, that means we have one meson. We have cut the diagram in a typical way. If we look at the right part of the diagram, we have an intermediate state with one quark, one antiquark and two gluons. To obtain only one meson state, this intermediate state must form only one colorless hadron. It is the case. Indeed, if we look at the color structure of the state we have
\begin{equation}
 \bar{q}_kA^k_jA^j_iq^i,
 \label{mesoncolor}
\end{equation}
which corresponds to a color singlet. In fact, if we analyse all possible planar diagrams, it is possible to show that we will obtain only this kind of combinations, a quark-antiquark pair at the end and gluon fields in the center. With non-planar diagrams like Figure \ref{nonplanarquarkcorr}, we can have a group  structure like 
\begin{equation}
 \bar{q}_kA^k_lq^lA^j_mA^m_j,
\end{equation}
which is a product of two color singlet operators, $\bar{q}_kA^k_lq^l$ and $A^j_mA^m_j$. These intermediate states can be interpreted as representing one meson and one color singlet glue state. As non-planar diagrams are suppressed in the large $N_c$ limit, only intermediate states like Eq. (\ref{mesoncolor}) exist in the large $N_c$ limit. \\

As the intermediate state of the two-point function of $J$ is a meson one can write
\begin{equation}
 \langle J(k)J(-k)\rangle = \sum_{n}\frac{\langle 0|J|n\rangle^2}{k^2-m_n^2} \sim N_c,
\end{equation}
 where we sum over physical intermediate states  without color degrees of freedom which means a sum over all planar Feynman diagrams contributing to the two-point function. The color combinatoric factor of each Feynman diagrams must be absorbed into $\langle 0|J|n\rangle^2$ which is the probability to create a meson from the vacuum. So, a factor $\sqrt{N_c}$ is attached to each meson annihilation or creation point. In other words the matrix element $\langle 0|J|n\rangle$ for a current to create a meson is of order $\sqrt{N_c}$. The meson decay constant $f_{m}$ is then of order,
 \begin{equation}
  f_m \sim \langle 0|J|n\rangle \sim \mathcal{O}(\sqrt{N_c}). 
 \end{equation}
   This result is not surprising if we look at Eq. (\ref{largenmesonwavefunction}). The sum over  $N_c$ colors in the equation, corresponding to  $N_c$ terms appearing from the indetermination of the color quantum number in the quark loop of Figure \ref{mesonintermediate}, leads to $\sqrt{N_c}$.\\
 
 One can summarize the result concerning the two-point function in large $N_c$: {\it $\langle J(x)J(y) \rangle$  is a sum of planar Feynman diagrams in which $J$ creates a meson with amplitude $\langle n|J|0\rangle$,  which propagates with a propagator $1/(k^2-m_n^2)$.} \\

\begin{figure}[h!]
\begin{center}
\includegraphics[width=14cm,keepaspectratio]{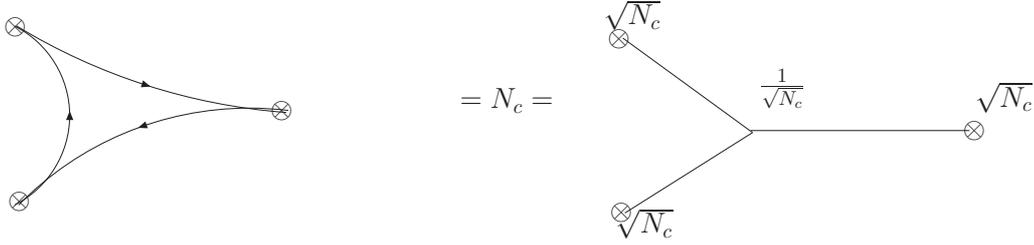}
\end{center}
\caption{Dominant three mesons diagram. On left part of the figure is drawn in double line notation the diagram of the right part. As this diagram is of order $N_c$, the three--meson vertex represented must be of order $1/\sqrt{N_c}$.}
\label{typicalmeson}
\end{figure}

 Consider now Figure \ref{typicalmeson} which represents  a typical three-point function. Each cross represents a meson creation or annihilation point. For each meson insertion on the quark loop, a factor of $\sqrt{N_c}$ is attached. As this diagram is of order $N_c$, the three-meson vertex is of order $1/\sqrt{N_c}$.  \\

Let us generalize and take a $k$ point function. Each current creates one meson and the quark loop planar diagram is of order $N_c$,
\begin{equation}
 \overbrace{N_c}^\mathrm{quark\ loop}=\underbrace{\left(\sqrt{N_c}\right)^{k}}_{k\ \mathrm{mesons}}\overbrace{\left(\frac{1}{\sqrt{N_c}}\right)^{k-2}}^{k-\mathrm{meson\ vertex}}.
 \label{mesonvertex}
\end{equation}
The $k-$meson vertex is then of order $(1/\sqrt{N_c})^{k-2}$. \\


 Let us summarize the properties of large $N_c$ mesons.
 \begin{itemize}
  \item The pion decay constant $f_{\pi}$ is of order $\sqrt{N_c}$.
  \item The mass of a meson is of order 1, as inferred by Eq. (\ref{largenmesonwavefunction}).
  \item The amplitude for a meson to decay into two mesons is of order $1/\sqrt{N_c}$.
  \item As the number of mesons entering in a vertex increases, the power of $1/N_c$ of the vertex increases (Eq. (\ref{mesonvertex})). For example, the self-coupling of three mesons is of order $1/\sqrt{N_c}$; the self-coupling of four mesons is of order $1/N_c$; etc. 
  \item As the number of planar diagrams describing two point functions is infinite,  the number of intermediate meson states is infinite.
 \end{itemize}
{\it In conclusion, large $N_c$ mesons are free, stable and non-interacting.}

\section{Large $N_c$ baryons}
Baryons in large $N_c$ were first studied in details by Witten \cite{witten79}. 
They  are colorless bound states composed of $N_c$ valence quarks antisymmetric in color indices. The color indices of the $N_c$ quarks are contracted with the SU($N_c$) $\varepsilon$-symbol to form the color-singlet state required by the color confinement
\begin{equation}
 \varepsilon_{i_1i_2i_3\ldots i_{N_c}}q^{i_1}q^{i_2}q^{i_3}\ldots q^{i_{N_c}}.
\end{equation}
 As the number of quarks grows with $N_c$, the baryon mass is of order $N_c$.  We have also to insist on the fact that to obtain a fermionic bound state, the number of colors $N_c$ must be odd. At leading order, a large $N_c$ baryon can be represented by quark-gluon diagrams consisting of $N_c$ valence quark lines with arbitrary planar gluons exchanged between the quark lines. As it follows from Section 1.2, the interaction between two quarks is of order $1/N_c$ as one gluon is exchanged between the two quark lines, with two quark-gluon vertices of order $1/\sqrt{N_c}$. In the following, we shall only consider ordinary baryons, \emph{i.e.}, baryons which are composed of $N_c$ valence quarks but not any valence antiquarks as the exotic hadrons named pentaquarks, formed of four quarks and an antiquark.\\

 
 From  Witten's large $N_c$ power counting rules, the following properties result for baryon in the $N_c \to \infty$ limit: 
 \begin{itemize}
 \item Baryon masses are of order $N_c$.
 \item Baryon size  and shape are of order 1. 
 \end{itemize} 
 
\section{Meson-baryon couplings}
 
\begin{figure}[h!]
\begin{center}
\includegraphics[width=15cm,keepaspectratio]{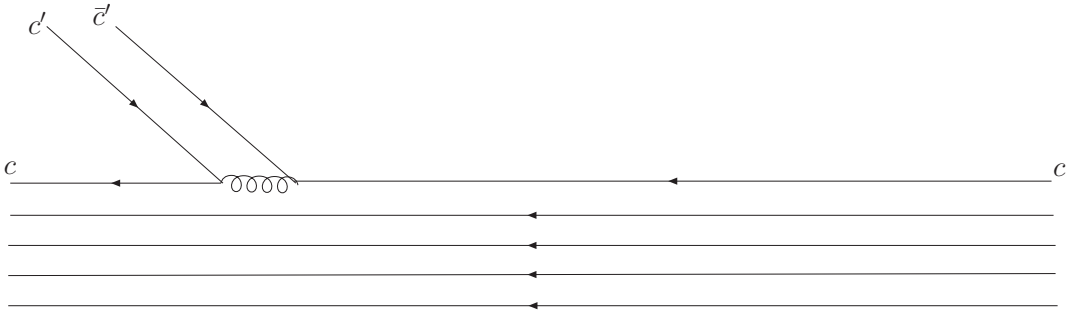} 
\end{center}
\caption{Coupling of a meson, denoted by colored  quarks $c'\bar{c}'$, to a baryon, containing colored quark $c$ \cite{muther87}, diagram of order $\sqrt{N_c}$.}
\label{onemesonbar}
\end{figure}

For the discussion given in Chapter 3, it is very important for us to study  meson-baryon couplings.
Let us first analyze Figure \ref{onemesonbar}. It represents the coupling of a meson to a baryon.
The two quark-gluon vertices  bring a factor $(1/\sqrt{N_c})^2$. The sum over colors $c$ in the nucleon gives a factor $N_c$. The normalisation of the  meson wave function Eq. (\ref{largenmesonwavefunction}) bring a factor $1/\sqrt{N_c}$ and the sum over the colors $c'$ in the meson gives a factor $N_c$ \footnote{It corresponds to the sum over the $N_c$ terms appearing in Eq.  (\ref{largenmesonwavefunction}).}. The net dependence on $N_c$ of this diagrams is then given by \cite{muther87}
\begin{equation}
 \left(\frac{1}{\sqrt{N_c}}\right)^2(N_c)^2\frac{1}{\sqrt{N_c}} = \sqrt{N_c}.
\end{equation}
So the baryon-meson coupling is of order $\sqrt{N_c}$. \\

\begin{figure}[h!]
\begin{center}
\includegraphics[width=15cm,keepaspectratio]{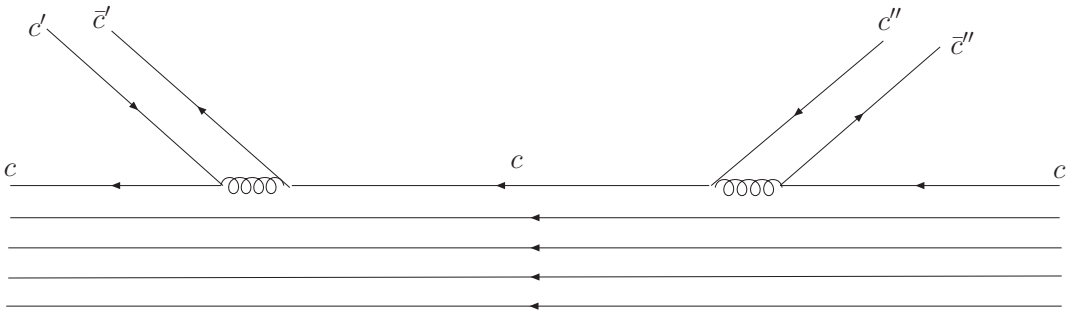} 
\end{center}
\caption{Baryon-meson scattering diagram of order 1. The two mesons are denoted by colored quarks $c'\bar{c}'$ and $c''\bar{c}''$.}
\label{twomesonbar}
\end{figure}

Let us now have a look  at Figure \ref{twomesonbar}. The diagram describes the baryon + meson $\to$ baryon + meson scattering. Here, we have four quark-gluons vertices giving a factor $(1/\sqrt{N_c})^4$. The sum over the colors $c$ in the nucleon brings a factor $N_c$. A factor $(1/\sqrt{N_c})^2$ appears from the normalisation constant of the two mesons. The sum over the colors $c'$ and $c''$ in each mesons gives a factor $(N_c)^2$. The order of this diagram is then given by
\begin{equation}
 \left(\frac{1}{\sqrt{N_c}}\right)^4N_c\left(\frac{1}{\sqrt{N_c}}\right)^2 \left(N_c\right)^2 = 1.
\end{equation}
The baryon-meson scattering diagram is then of order 1. \\

One can generalize the counting rules to multimeson-baryon scattering  amplitudes. Each additional meson in a scattering process reduces the scattering amplitude by a factor of $1/\sqrt{N_c}$. Then the process baryon + meson $\to$ baryon + ($n-1$) mesons is of order $N_c^{1-n/2}$ \cite{jenkins98}.

\chapter{The baryon structure}

\thispagestyle{empty}

\section{Introduction}

As we have seen in the first chapter, large $N_c$ QCD baryons are described by a number of $N_c$ valence quarks. Even if the formalism is developed for an arbitrary $N_c$, in applications one takes $N_c=3$. \\

In order to understand the structure of baryons formed of $N_c$ quarks it is useful first to recall the structure of baryons formed of three quarks and then to generalize the wave functions to $N_c$ quarks. This structure and the wave functions derived in this chapter will be used in the following chapters. \\

Our aim here is to write baryon wave functions and to introduce the quantum numbers of baryons like the spin $S$, the isospin $I$ or the parity $P$.  For this reason, we will use a simple \emph{quark-shell} model, similar to the shell model of nuclear physics, based on a Hartree-Fock approximation, the quarks being confined by an average central potential. With  that model, it is possible to write simple baryon wave functions. 
Although the presentation is restricted to a very simple model, the picture remains entirely general.
Furthermore, one can easily generalize the quark-shell model to $N_c$ quarks.  \\


In the quark-shell picture, the general form of a  baryon wave function  is 
\begin{equation}
 \Psi=\psi_{\ell m}\chi\phi C, 
 \label{dividedwavefunction}
\end{equation}
where $\psi_{\ell m}$ $\chi$, $\phi$ and $C$ are the orbital, spin, flavor and color parts respectively. As already noticed in the previous chapter, this wave function must be antisymmetric under the permutations of the quarks, as baryons are fermions. The nature requires $C$ to be an SU$_c(N_c)$ singlet. Because of the color confinement\footnote{As mentioned in the previous chapter, we assume  color confinement for $N_c>3$ as well.}, one obtains colorless bound states. As the color part must be an SU$_c(N_c)$ singlet, which is totally antisymmetric, the remaining part, \emph{i.e.} $\psi_{\ell m}\chi\phi$ is always symmetric. \\

The Hamiltonian $H$ of a quark-shell model  has the form
\begin{equation}
 H=\sum_{i=1}^{N_c}\left(T_i+V_i\right),
 \label{quarkmodelhamiltonian}
\end{equation}
where $T_i$ represents the kinetic energy of a quark $i$ and $V_i$ the average  confinement potential experienced by the quark $i$. For simplicity we take a toy model of a three dimensional harmonic oscillator confinement. This Hamiltonian is spin-flavor independent, provided the quarks $u,\ d$ and $s$ have identical masses. \\

 In this work, we consider only ``ordinary'' baryons made up of valence $u$, $d$ and $s$ quarks. The flavor-independence of the Hamiltonian implies an SU(3) symmetry, its spin-independence, an SU(2) symmetry and the rotational invariance, an O(3) symmetry. As a consequence, the symmetry properties of baryons can be described by SU(6) $\times$ O(3), where SU(6) represents the spin-flavor symmetry. Baryon states must  be SU(6) $\times$ O(3) symmetric. \\

This chapter is divided in two parts. We first present the  baryon wave functions for $N_c=3$. In the second part, a generalization for $N_c$ particles is proposed.

\section{The three quarks wave functions}

Let us  analyze separately the space part and the spin-flavor part of the wave function.

\subsection{The spin-flavor part $\chi\phi$}

If we combine the spin and the flavor degrees of freedom to an SU(6) spin-flavor symmetry, each quark can be in one of the six different states  $u\uparrow$, $u\downarrow$, $d\uparrow$, $d\downarrow$, $s\uparrow$, $s\downarrow$. As baryons are composed of three quarks, one can make the following direct product 
\begin{equation}
\stackrel{6}{\Yfund} \times\stackrel{6}{\Yfund} \times \stackrel{6}{\Yfund}\ =\
\stackrel{56}{\Ythrees} + 2 \stackrel{70}{\Yadjoint} +\stackrel{20}{\Ythreea}\, , \label{directprodtab}
\end{equation}
where the dimensions of the  irreducible representations (irreps) are indicated above each Young diagram. Baryons must belong to one of these SU(6) multiplets which are  symmetric, mixed-symmetric and antisymmetric respectively under the permutation of the quarks. It is possible to decompose the multiplets into theirs SU(2) $\times$ SU(3) content\footnote{See \cite{stancu96} p. 317.}:
\begin{eqnarray}
\stackrel{56}{\Ythrees} & = & \stackrel{2}{\Yadjoint}\times \stackrel{8}{\Yadjoint} + \stackrel{4}{\Ythrees} \times \stackrel{10}{\Ythrees}\, , \label{56rep} \label{su6symm3}\\
\stackrel{70}{\Yadjoint} & = & \stackrel{2}{\Yadjoint} \times \stackrel{1}{\Ythreea} + \stackrel{2}{\Yadjoint} \times \stackrel{10}{\Ythrees} +  \stackrel{2}{\Yadjoint} \times \stackrel{8}{\Yadjoint} + \stackrel{4}{\Ythrees} \times \stackrel{8}{\Yadjoint}\, , \label{su6smixed3} \\
\stackrel{20}{\Ythreea} & = & \stackrel{2}{\Yadjoint} \times \stackrel{8}{\Yadjoint} + \stackrel{4}{\Ythrees} \times \stackrel{1}{\Ythreea}.
\end{eqnarray}
The SU(3) flavor weight diagrams for the decuplet and the octet  are shown explicitly in Figures \ref{su3decup} and \ref{su3octet}. The SU(3) flavor wave functions are written in Tables \ref{symflstates}--\ref{antisymstate}. 

\pagebreak[4]

\begin{table}[h]
\begin{center}
{\scriptsize
\renewcommand{\arraystretch}{1.75}
\begin{tabular}{lc}
\hline \hline
Baryon  &   $\phi^S$ \\
\hline
$\Delta^{++}$ & $uuu$ \\
$\Delta^{+}$ & $\frac{1}{\sqrt{3}}\left( uud+udu+duu \right)$ \\
$\Delta^0$   & $\frac{1}{\sqrt{3}}\left( udd+dud+ddu \right)$ \\
$\Delta^-$   & $ddd$ \\
$\Sigma^+$ & $\frac{1}{\sqrt{3}}\left( uus+usu+suu \right)$ \\
$\Sigma^0$ & $\frac{1}{\sqrt{6}}\left( uds+dus+usd+sud+sdu+dsu\right)$\\
$\Sigma^-$ & $\frac{1}{\sqrt{3}}\left( sdd+dsd+dds\right)$\\
$\Xi^0$ & $\frac{1}{\sqrt{3}}\left( uss+sus+ssu\right)$\\
$\Xi^-$ & $\frac{1}{\sqrt{3}}\left( dss+sds+ssd\right)$\\
$\Omega^-$& $sss$\\
\hline \hline
\end{tabular}}
\caption{Symmetric flavor states of three quarks for the baryon decuplet.}
\label{symflstates}
\end{center}
\end{table} 

\begin{table}[h!]
\begin{center}
{\scriptsize
\renewcommand{\arraystretch}{1.75}
\begin{tabular}{lcc}
\hline \hline
Baryon         &   \hspace{0cm}  $\phi^{\lambda}$     & \hspace{0cm} $\phi^{\rho}$ \\
\hline
$p$              & \hspace{0cm}$-\frac{1}{\sqrt{6}}\left( udu+duu-2uud\right)$ & \hspace{0cm}$\frac{1}{\sqrt{2}}\left( udu-duu\right)$\\
$n$          & $\frac{1}{\sqrt{6}}\left( udd+dud-2ddu\right)$ & $\frac{1}{\sqrt{2}}\left( udd-dud\right)$\\
$\Sigma^+$ & $\frac{1}{\sqrt{6}}\left( usu+suu-2uus\right)$ & $-\frac{1}{\sqrt{2}}\left( usu-suu\right)$\\
\raisebox{-1.45ex}[0cm][0cm]{$\Sigma^0$} & {\renewcommand{\arraystretch}{0.4}\begin{tabular}[t]{c}
$-\frac{1}{\sqrt{12}}\left( 2uds+2dus\right.$ \\ $\left. -sdu-sud-usd-dsu\right)$\end{tabular}} & \hspace{0.5cm}\raisebox{-1.45ex}[0cm][0cm]{$-\frac{1}{2}\left( usd+dsu-sdu-sud\right)$}\\
$\Sigma^-$  & $\frac{1}{\sqrt{6}}\left( dsd+sdd-2dds \right)$ & $-\frac{1}{\sqrt{2}}\left( dsd-sdd\right)$ \\
\raisebox{-1.45ex}[0cm][0cm]{$\Lambda^0$} & \raisebox{-1.45ex}[0cm][0cm]{$\frac{1}{2}\left( sud-sdu+usd-dsu \right)$} & {\renewcommand{\arraystretch}{0.4}\begin{tabular}[t]{c} $\frac{1}{\sqrt{12}}\left( 2uds-2dus \right.$ \\ $\left. +sdu-sud+usd-dsu\right)$\end{tabular}}\\ 
$\Xi^0$  & $-\frac{1}{\sqrt{6}}\left( uss+sus-2ssu \right)$  & $-\frac{1}{\sqrt{2}}\left( uss-sus\right)$ \\
$\Xi^-$  & $-\frac{1}{\sqrt{6}}\left( dss+sds-2ssd \right)$ & $-\frac{1}{\sqrt{2}}\left( dss-sds\right)$\\
\hline \hline
\end{tabular}}
\caption{Mixed-symmetric flavor states of three quarks for the baryon octet.}
\label{msymflstates}
\end{center}
\end{table} 

\begin{table}[h!]
 \begin{center}
{\scriptsize
\renewcommand{\arraystretch}{1.75}
\begin{tabular}{lc}
\hline \hline
Baryon & $\phi^A$ \\
\hline
$\Lambda$ & $\frac{1}{\sqrt{6}}\left( sdu-sud+usd-dsu+dus-uds\right)$ \\
\hline \hline
\end{tabular}}
\caption{Antisymmetric flavor states of three quarks for the baryon singlet.}\label{antisymstate}
\end{center}
\end{table}

\begin{figure}[h]
\begin{center}
\setlength{\unitlength}{6mm}
\begin{picture}(6,5.1962)(-3,-2.5981)
\put(-3,2.5981){\makebox(0,0){$\Delta^-$}}
\put(-1,2.5981){\makebox(0,0){$\Delta^0$}}
\put(1,2.5981){\makebox(0,0){$\Delta^+$}}
\put(3,2.5981){\makebox(0,0){$\Delta^{++}$}}
\put(-2,0.866){\makebox(0,0){$\Sigma^{*-}$}}
\put(0,0.866){\makebox(0,0){$\Sigma^{*0}$}}
\put(2,0.866){\makebox(0,0){$\Sigma^{*+}$}}
\put(-1,-0.866){\makebox(0,0){$\Xi^{*-}$}}
\put(1,-0.866){\makebox(0,0){$\Xi^{*0}$}}
\put(0,-2.5981){\makebox(0,0){$\Omega^{-}$}}
\end{picture}
\caption{Flavor SU(3) weight diagram for the decuplet baryons. \label{su3decup}}
\end{center}
\end{figure}

\begin{figure}[h]
\begin{center}
\setlength{\unitlength}{6mm}
\begin{picture}(4,3.4642)(-2,-1.7321)
\put(-1,1.7321){\makebox(0,0){$n$}}
\put(1,1.7321){\makebox(0,0){$p$}}
\put(-2,0){\makebox(0,0){$\Sigma^-$}}
\put(0,0){\makebox(0,0){$\Lambda^0,\Sigma^0$}}
\put(2,0){\makebox(0,0){$\Sigma^+$}}
\put(-1,-1.7321){\makebox(0,0){$\Xi^-$}}
\put(1,-1.7321){\makebox(0,0){$\Xi^0$}}
\end{picture}
\end{center}
\caption{Flavor SU(3) weight diagram for the octet baryons. \label{su3octet}}
\end{figure}


\subsection{The orbital part $\psi_{\ell m}$}

\subsubsection{Single particle wave functions}
For finding the wave function of a particle moving in a central three dimensional harmonic oscillator potential, we have to solve the Schr\"odinger equation
\begin{equation}
 \left(\frac{p^2}{2m}+\frac{1}{2}m\omega^2r^2\right)\psi=E\psi.
\end{equation}
The eigenvalues of this equation are
\begin{equation}
 E=\left(N+\frac{3}{2}\right)\hbar\omega, n=0,1, \ldots
\end{equation}
where $N$ is the number of excited quanta which can be written as
 \begin{equation}
  N=2n+\ell,
 \end{equation}
where $\ell$ is the orbital angular momentum quantum number and $n=0, 1, 2, \ldots$ is associated with the number of nodes in the radial wave function. \\


The wave function can be written as \cite{faiman68}
\begin{equation}
 \psi_{n\ell m}=\mathcal{N}(\alpha r)^\ell L_n^{\ell+1/2}(\alpha^2r^2)e^{-\alpha^2r^2/2}Y^m
 _\ell(\theta\phi),
\end{equation}
where $\alpha^2=m\omega$, $L_n^{\ell+1/2}$ is a Laguerre polynomial and $\mathcal{N}$ the normalisation constant
\begin{equation}
 |\mathcal{N}|^2=\frac{2\alpha^3n!}{\sqrt{\pi}(n+\ell+\frac{1}{2})(n+\ell-\frac{1}{2})\cdots\frac{3}{2}\frac{1}{2}}.
\end{equation}
For each value of $N$ larger than one, several values for $n$ and $\ell$ are possible. Moreover, for each value of $\ell$, $m$ can vary from $-\ell$ to $\ell$.
The parity of the wave function is given by
\begin{equation}
 P=(-1)^\ell.
\end{equation} \\

Here are a few examples of wave functions
\begin{eqnarray}
 (0s) & = & \psi_{000}({\bf r})=\left(\frac{4\alpha^{3}}{\sqrt{\pi}}\right)^{1/2}\frac{1}{\sqrt{4\pi}}e^{-\alpha^2r^2/2}, \label{0swavefunction} \\
 (0p) & = & \psi_{11m}({\bf r})=\sqrt{\frac{2}{3}}\left(\frac{4\alpha^{3}}{\sqrt{\pi}}\right)^{1/2}\alpha r e^{-\alpha^2r^2/2} Y^m_1(\theta,\phi), \label{0pwavefunction} \\
 (0d) & = & \psi_{22m}({\bf r})=\sqrt{\frac{4}{15}}\left(\frac{4\alpha^{3}}{\sqrt{\pi}}\right)^{1/2}\alpha^2r^2e^{-\alpha^2r^2/2} Y^m_2(\theta,\phi), \\
 (1s) & = & \psi_{200}({\bf r})=\sqrt{\frac{2}{3}}\left(\frac{4\alpha^{3}}{\sqrt{\pi}}\right)^{1/2}\left(\frac{3}{2}-\alpha^2r^2\right)e^{-\alpha^2r^2/2}, \label{1swavefunction}
\end{eqnarray}
where on the left-hand side we introduce a compact notation to be used in the following.

\subsubsection{Baryon wave functions}

 The Hamiltonian $H$ introduced in Eq. (\ref{quarkmodelhamiltonian}) becomes
\begin{equation}
 H=\sum_{i=1}^3 \left(\frac{p^2_i}{2m}+\frac{1}{2}m\omega^2r^2_i\right),
\end{equation}
where it is understood that the color degree of freedom was integrated out. The orbital part of the wave function of a baryon is obtained as the product of three independent-quark wave functions. It  must have the correct permutation  symmetry to lead to a totally symmetric wave function when it is associated with the spin-flavor part $\chi\phi$. This means that the orbital  part must have the same symmetry as the spin-flavor part.  \\

To remove the center of mass motion one has to introduce the Jacobi coordinates. For equal quark masses they are
\begin{eqnarray}
 {\bf R} & = & \frac{1}{\sqrt{3}}({\bf r}_1+{\bf r}_2 + {\bf r}_3), \label{3centerofmass} \\ 
\lambdab & = & \frac{1}{\sqrt{6}}({\bf r}_1 + {\bf r}_2 - 2{\bf r}_3), \label{3rho} \\
\rhob & = & \frac{1}{\sqrt{2}}({\bf r}_1- {\bf r}_2) \label{3lambda}.
\end{eqnarray}
Note that these coordinates can be used as basis functions of the symmetric  [3] (Eq. (\ref{3centerofmass})) and mixed-symmetric [21] irreps (Eqs. (\ref{3rho})--(\ref{3lambda})) of the permutation group S$_3$.
 The Hamiltonian becomes \cite{faiman68}
\begin{equation}
 H=\frac{P^2}{6m}+\frac{3}{2}m\omega^2R^2+\left[\frac{1}{2m}(p^2_\lambda+ p^2_\rho)+\frac{1}{2}m\omega^2(\lambda^2+\rho^2)\right],
\end{equation}
where ${\bf P}$, ${\bf p}_\lambda$ and ${\bf p}_\rho$ are the conjugate momenta to ${\bf R}$, $\lambdab$ and $\rhob$ respectively.
An eigenfunction  of this Hamiltonian can be written in the form
\begin{equation}
 \chi({\bf R})\phi(\lambdab)\psi(\rhob),
 \label{quarkmodelharmonic}
\end{equation}
where $\chi$, $\phi$ and $\psi$ are one-body harmonic oscillator wave functions. The first part describes the center of mass motion and the $\phi$, $\psi$ parts the internal motion of the quarks.   To conserve the translation-invariance of the model, in the wave function (\ref{quarkmodelharmonic})   the center of mass motion $\chi({\bf R})$ is factored out and left in the ground state. States with excitation of the center of mass motion are referred to as being spurious. \\

Then the parity of the wave function is 
\begin{equation}
P=(-1)^{\underset{i}{\sum}\ell_i},
\end{equation}
with $\ell_i$, the angular momentum of the different quarks. \\


The energy of the state corresponds to the sum over the energies $E_i$ of the different quarks which compose the baryon minus the center of mass energy $3/4\hbar\omega$. To label the levels, the notation $[{\bf x},\ell^P]$ is used where ${\bf x}$ represents the dimension of an SU(6) irrep, $\ell$ is the angular momentum of the multiplet and $P$ the parity. Note that in a harmonic oscillator picture the levels belonging to a given shell (or band) $N$ are degenerate. This degeneracy is lifted in a model with a linear confinement. \\

The lowest level, with $N=0$ corresponds to the ground state. All the quarks are in a $(0s)$ state and so we obtain a $(0s)^3$ configuration. The orbital part of the baryon wave function is totally symmetric and therefore  the label is  $[{\bf 56},0^+]$. For this multiplet, one can directly conclude that decuplet baryons (Figure \ref{su3decup}) have spin $S=3/2$ and octet baryons (Figure \ref{su3octet}) have spin $S=1/2$. The flavor singlet baryon $\Lambda$ must be an excited state. In nature it is identified with $\Lambda (1450)$ which has $J^P = 1/2^-$.\\

The first excited state, $N=1$, must be a $(0s)^2(0p)$ configuration, with $\ell=1^-$. With this configuration, it is possible to form symmetric or mixed-symmetric wave functions. However, the symmetric wave function is a spurious state as it is proportional to the center of mass coordinate ${\bf R}$ \footnote{It corresponds to the internal quark motion being in its ground state but the center of mass moving in a $(0p)$ state \cite{faiman68}.}. Thus the first excited state of negative parity belongs to the  $[{\bf 70}, 1^-]$ multiplet.\\

The $N=2$ level is more complicated as three quarks configurations are allowed, \emph{i.e.} $(0s)^2(1s)$, $(0s)^2(0d)$ and $(0s)(0p)^2$ since they all correspond to the same energy. Some of these wave functions are spurious. One can make linear combinations of these wave functions to obtain non-spurious states. Table \ref{harmonicconfigurations} gives the proper linear combinations. All the other states are spurious.\\

\begin{figure}[h!]
\begin{center}
\includegraphics{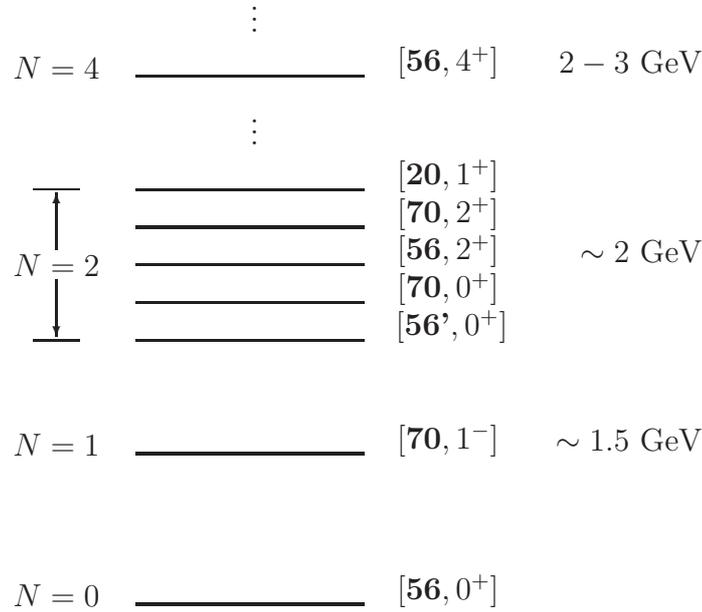}
\caption{The lowest levels of the baryon spectrum drawn schematically. An indication of the average mass of the baryons belonging to the $N=1$, 2 and 4 levels is written on the right part of the figure.}\label{baryonspectrum}
\end{center}
\end{figure}

Appendix A gives details about the orbital wave functions for the different states and the method used to derive the harmonic oscillator configurations written in Table  \ref{harmonicconfigurations}.
Table \ref{su(6)o(3)wave} details the necessary steps for the construction of symmetric SU(6) $\times$ O(3) wave functions from their component SU(3) flavor, SU(2) spin and O(3) for $N = 0,\ 1,\ 2$.  In Figure \ref{baryonspectrum} a schematic picture of the  baryon spectrum is sketched.

\begin{table}[h!]
\begin{center}
{\scriptsize
\renewcommand{\arraystretch}{1.75}
\begin{tabular}{lllll}
\hline
\hline
SU(6) multiplet  & N & L & P & Harmonic oscillator configuration \\
\hline
56 & 0 & 0 & + & $|[3](0s)^3\rangle$ \\
70 & 1 & 1 & $-$ & $|[21](0s)^20p\rangle$ \\
$56'$ & 2 & 0 & + & $\sqrt{\frac{2}{3}}|[3](0s)^2 (1s)\rangle - \sqrt{\frac{1}{3}}|[3](0s)(0p)^2\rangle$\\
70 & 2 & 0 & + & $\sqrt{\frac{1}{3}}|[21](0s)^2 (1s) \rangle + \sqrt{\frac{2}{3}}|[21](0s)(0p)^2\rangle$\\
56 & 2 & 2 & + & $\sqrt{\frac{2}{3}}|[3](0s)^2 (0d)\rangle - \sqrt{\frac{1}{3}}|[3](0s)(0p)^2\rangle$\\
70 & 2 & 2 & + & $\sqrt{\frac{1}{3}}|[21](0s)^2 (0d) \rangle + \sqrt{\frac{2}{3}}|[21](0s)(0p)^2\rangle$\\
20 & 2 & 1 & + & $|[1^3](0s)(0p)^2\rangle$ \\
\hline
\hline
\end{tabular}}
\caption{Quark-shell model configurations for a harmonic oscillator confinement and $N_c=3$.}
\label{harmonicconfigurations}
\end{center}
\end{table}

\begin{table}[h]
 \begin{center}
{\scriptsize
  \renewcommand{\arraystretch}{1.75}
\begin{tabular}{lllllll}
 \hline \hline
 SU(3) & & SU(2) & SU(6) & & O(3) & SU(6) $\times$ O(3) \\
 \hline
  &  &        &          & & $2^+_2$S   & $[{\bf 56},2^+]$ \\
  \raisebox{2.45ex}[0cm][0cm]{10}   &   \raisebox{2.45ex}[0cm][0cm]{$\times$}     &   \raisebox{2.45ex}[0cm][0cm]{$\frac{3}{2}$}            & 56    & $\times$ &  $0^+_2$S   & $[{\bf 56'},0^+]$ \\
  \raisebox{2.45ex}[0cm][0cm]{8} & \raisebox{2.45ex}[0cm][0cm]{$\times$} & \raisebox{2.45ex}[0cm][0cm]{$\frac{1}{2}$} &        &          &  $0_0^+$S   & $[{\bf 56},0^+]$  \\
  \hline
 10 & $\times$ & $\frac{1}{2}$ &        &          &   & \\
 8  & $\times$ & $\frac{3}{2}$ &           &       & \raisebox{2.45ex}[0cm][0cm]{$2_2^+$M} & \raisebox{2.45ex}[0cm][0cm]{$[{\bf 70},2^+]$}\\
 8  & $\times$ & $\frac{1}{2}$ & \raisebox{2.45ex}[0cm][0cm]{70} & \raisebox{2.45ex}[0cm][0cm]{$\times$}& \raisebox{2.45ex}[0cm][0cm]{$0^+_2$M}&  \raisebox{2.45ex}[0cm][0cm]{$[{\bf 70},0^+]$}\\
 1  & $\times$ & $\frac{1}{2}$ &        &          & \raisebox{2.45ex}[0cm][0cm]{$1^-_1$M}  &   \raisebox{2.45ex}[0cm][0cm]{$[{\bf 70},1^-]$} \\
 \hline
 8 & $\times$ & $\frac{1}{2}$ &          &        &             &            \\
 1 & $\times$ & $\frac{3}{2}$ & \raisebox{2.45ex}[0cm][0cm]{20} & \raisebox{2.45ex}[0cm][0cm]{$\times$}& \raisebox{2.45ex}[0cm][0cm]{$1_2^+$A} & \raisebox{2.45ex}[0cm][0cm]{$[{\bf 20},1^+]$}\\
 \hline
 \hline    
\end{tabular}}
\caption{Steps in forming SU(6) $\times$ O(3) symmetric wave functions for $N = 0,\ 1,\ 2$. The subscripts specify in each case the appropriate value of N, and the symmetry, S, M, A, is specified explicitly for each state. In SU(3) the irreps are labelled by their dimensions  10,  8 or  1. The irreps of SU(2) are labelled by the total spin 3/2 or 1/2 related to the eigenvalues of the Casimir operator of SU(2).  The labels S, M, A stand for symmetric, mixed-symmetric or antisymmetric \cite{horgan73}.\label{su(6)o(3)wave}}
 \end{center}
\end{table}

\section{Generalization to large $N_c$}



As already introduced, the large $N_c$ baryons are composed of $N_c$ valence quarks totally antisymmetric in color.
An important observation  is that the baryonic number of  large $N_c$ baryons becomes 
\begin{equation}
 \mathcal{B}=\frac{N_c}{3},
\end{equation}
inasmuch as the baryonic number of a quark is equal to 1/3.

\subsection{Ground-state baryons}
\begin{figure}[h!]
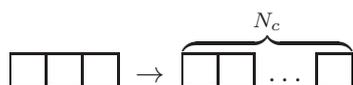

$$ \raisebox{-2pt}{\begin{Young}
      & &  \cr
      \end{Young}}\ \to \
      \raisebox{-2pt}{$\overbrace{\begin{Young}
       &  \cr
      \end{Young}\  \cdots \
      \begin{Young}
      \cr
      \end{Young}}^{N_c}$}  $$
\caption{Young diagrams  for the symmetric irrep of the permutation group S$_3$ and S$_{N_c}$ \cite{pirjol98a}.}
\label{sncsymm}
\end{figure}

Let us first analyze ground-state baryons. The orbital  and the spin-flavor parts  of the wave function must be separately  symmetric under the exchange of two quarks as all the $N_c$ quarks are in the ground state (0s). We label the symmetric irrep by the partition  $[N_c]$. Its Young diagram is shown in Figure \ref{sncsymm}.  For $N_c=3$ it corresponds to ${\bf 56}$ in SU(6). \\

The decomposition of the symmetric representation of SU(6)  into its  SU(2) $\times$ SU(3)  content can be written in terms of Young diagrams as
 \begin{eqnarray}\label{SU(6)symm}
\overbrace{\,\raisebox{-3.0pt}{\drawsquare{10.0}{0.4}}\hskip-0.4pt
        \raisebox{-3.0pt}{\drawsquare{10.0}{0.4}}\cdots
        \raisebox{-3.0pt}{\drawsquare{10.0}{0.4}}\,}^{N_c} &=&
\left(S=\frac12;\,
\raisebox{-8.0pt}{\drawsquare{10.0}{0.4}}\hskip-10.4pt
        \raisebox{2pt}{\drawsquare{10.0}{0.4}}\hskip-0.4pt
\raisebox{-8.0pt}{\drawsquare{10.0}{0.4}}\hskip-10.4pt
        \raisebox{2pt}{\drawsquare{10.0}{0.4}}\hskip-0.4pt
\cdots
\raisebox{-8.0pt}{\drawsquare{10.0}{0.4}}\hskip-10.4pt
        \raisebox{2pt}{\drawsquare{10.0}{0.4}}\hskip-0.4pt
        \raisebox{2pt}{\drawsquare{10.0}{0.4}}\hskip-0.4pt
\,\right) +
\left(S=\frac32;\,
\raisebox{-8.0pt}{\drawsquare{10.0}{0.4}}\hskip-10.4pt
        \raisebox{2pt}{\drawsquare{10.0}{0.4}}\hskip-0.4pt
\raisebox{-8.0pt}{\drawsquare{10.0}{0.4}}\hskip-10.4pt
        \raisebox{2pt}{\drawsquare{10.0}{0.4}}\hskip-0.4pt
\cdots
\raisebox{-8.0pt}{\drawsquare{10.0}{0.4}}\hskip-10.4pt
        \raisebox{2pt}{\drawsquare{10.0}{0.4}}\hskip-0.4pt
        \raisebox{2pt}{\drawsquare{10.0}{0.4}}\hskip-0.4pt
        \raisebox{2pt}{\drawsquare{10.0}{0.4}}\hskip-0.4pt
        \raisebox{2pt}{\drawsquare{10.0}{0.4}}\hskip-0.4pt
\,\right) \nonumber \\
& & \qquad\qquad\qquad +\,\cdots +
\left(S=\frac{N_c}{2};\,
\raisebox{-3.0pt}{\drawsquare{10.0}{0.4}}\hskip-0.4pt
        \raisebox{-3.0pt}{\drawsquare{10.0}{0.4}}\cdots
        \raisebox{-3.0pt}{\drawsquare{10.0}{0.4}}
\,\right)\,, \label{largentower}
\end{eqnarray}
where the SU(2) irreps are denoted by the corresponding value of the spin $S$ as in Table \ref{su(6)o(3)wave}.
 This decomposition is a natural generalization of Eq. (\ref{su6symm3}). 
  One can infer that large $N_c$ picture produces an infinite tower of baryon states when $N_c\to \infty$. 
  If we consider non-strange baryons, we obtain  a tower of baryon states with
\begin{equation}
 S=I=\frac{1}{2},\frac{3}{2},\ldots,\frac{N_c}{2}. \label{quarkconsttower}
\end{equation}
This tower is finite dimensional and reduces to $N$ and $\Delta$ for $N_c=3$.  \\

For baryons containing $n_s$ strange quarks, it is useful to  introduce the grand spin $\vec{K}$ defined as  
\begin{equation}
 \vec{K}=\vec{S}+\vec{I}, \label{grandspin}
\end{equation}
where $K=n_s/2$. The isospin content for each value of $K$ can be written as \cite{pirjol98b}
{\small
\begin{eqnarray}
\raisebox{-8.0pt}{\drawsquare{10.0}{0.4}}\hskip-10.4pt
        \raisebox{2pt}{\drawsquare{10.0}{0.4}}\hskip-0.4pt
\raisebox{-8.0pt}{\drawsquare{10.0}{0.4}}\hskip-10.4pt
        \raisebox{2pt}{\drawsquare{10.0}{0.4}}\hskip-0.4pt
\raisebox{-8.0pt}{\drawsquare{10.0}{0.4}}\hskip-10.4pt
        \raisebox{2pt}{\drawsquare{10.0}{0.4}}\hskip-0.4pt
        \raisebox{2pt}{\drawsquare{10.0}{0.4}}\hskip-0.4pt &\to&
(K=0\,, I=\frac12) + (K=\frac12\,, I=0,1) + (K=1\,, I=\frac12,\frac32)
 + \cdots\, , \label{octet} \\
\raisebox{-8.0pt}{\drawsquare{10.0}{0.4}}\hskip-10.4pt
        \raisebox{2pt}{\drawsquare{10.0}{0.4}}\hskip-0.4pt
\raisebox{-8.0pt}{\drawsquare{10.0}{0.4}}\hskip-10.4pt
        \raisebox{2pt}{\drawsquare{10.0}{0.4}}\hskip-0.4pt
\raisebox{-8.0pt}{\drawsquare{10.0}{0.4}}\hskip-10.4pt
        \raisebox{2pt}{\drawsquare{10.0}{0.4}}\hskip-0.4pt
        \raisebox{2pt}{\drawsquare{10.0}{0.4}}\hskip-0.4pt
        \raisebox{2pt}{\drawsquare{10.0}{0.4}}\hskip-0.4pt
        \raisebox{2pt}{\drawsquare{10.0}{0.4}}\hskip-0.4pt &\to&
(K=0\,, I=\frac32) + (K=\frac12\,, I=1,2) + (K=1\,, I=\frac12,\frac32,\frac52)
 + \cdots \, ,  \label{decuplet} 
\end{eqnarray}}

\noindent where one can see that Eq. (\ref{octet}) corresponds to $S=1/2$ baryons and Eq. (\ref{decuplet}) to $S=3/2$ baryons, consistent with Eq. (\ref{su6symm3}). Thus when $N_c=3$  the above representations correspond to the SU(3) octet and decuplet.\\

The dimension of an irrep of SU(3) is given by \cite{stancu96}
 \begin{equation}
 d^{\mathrm{SU(3)}}_{(\lambda\mu)}=\frac{1}{2}(\lambda+1)(\mu+1)(\lambda+\mu+2),
  \label{su3dimirrep}
  \end{equation}
  where $\lambda$ and $\mu$ correspond to

$$  \underbrace{\begin{Young}
    & & \cr
    & & \cr
   \end{Young}}_{\mu}\ \negthinspace \negthinspace\negthinspace\raisebox{12.5pt}{$\underbrace{\begin{Young} & & & \cr \end{Young}}_{\lambda}$}$$
One can  directly see that, using Eq. (\ref{su3dimirrep}), the dimension of each SU(3) irrep presented in the decomposition  (\ref{largentower}) depends on $N_c$.  \\

In Figures \ref{spin12} and \ref{spin32} are drawn large $N_c$ flavors weight diagrams which are an extension to $N_c$ of the octet and decuplet irreps of SU(3) when $N_c=3$. Contrary to the SU(2) case, the SU(3) weight diagrams vary with $N_c$. As a consequence the identification of $N_c=3$ baryons  is not unique. It is usual to identify $N_c=3$ baryons with the top of a large $N_c$ weight diagram corresponding to baryons with a small strangeness.
However, the hypercharge varies as
\begin{eqnarray}
 Y & = & \mathcal{S} + \mathcal{B} \nonumber \\
   & = & \mathcal{S} + \frac{N_c}{3},
\end{eqnarray}
where $\mathcal{S}$ is the strangeness. This means that the hypercharge of a large $N_c$ nucleon or $\Delta$ is equal to $N_c/3$.

\begin{figure}[h!]
\setlength{\unitlength}{3mm}
\centerline{\hbox{
\begin{picture}(20.79,18)(-10.395,-8)
\multiput(-1.155,10)(2.31,0){2}{\onedot}
\multiput(-2.31,8)(4.62,0){2}{\onedot}
\multiput(-3.465,6)(6.93,0){2}{\onedot}
\multiput(-4.62,4)(9.24,0){2}{\onedot}
\multiput(-5.775,2)(11.55,0){2}{\onedot}
\multiput(-6.93,0)(13.86,0){2}{\onedot}
\multiput(-8.085,-2)(16.17,0){2}{\onedot}
\multiput(-9.24,-4)(18.48,0){2}{\onedot}
\multiput(-10.395,-6)(20.79,0){2}{\onedot}
\multiput(-9.24,-8)(2.31,0){9}{\onedot}
\multiput(0,8)(2.31,0){1}{\twodot}
\multiput(-1.155,6)(2.31,0){2}{\twodot}
\multiput(-2.31,4)(2.31,0){3}{\twodot}
\multiput(-3.465,2)(2.31,0){4}{\twodot}
\multiput(-4.62,0)(2.31,0){5}{\twodot}
\multiput(-5.775,-2)(2.31,0){6}{\twodot}
\multiput(-6.93,-4)(2.31,0){7}{\twodot}
\multiput(-8.085,-6)(2.31,0){8}{\twodot}
\end{picture}
}}
\caption{SU(3) flavor weight diagram for the generalized octet baryons  (Eq. (\ref{octet})).  The numbers
denotes the multiplicity of the weights.  The long side of the weight diagram
contains ${1 \over 2}\left( N_c + 1 \right)$ weights \cite{jenkins98}.}
\label{spin12}
\end{figure}
\begin{figure}[h!]
\setlength{\unitlength}{3mm}
\centerline{\hbox{
\begin{picture}(20.79,18)(-8.085,-8)
\multiput(-1.155,10)(2.31,0){4}{\onedot}
\multiput(-2.31,8)(9.24,0){2}{\onedot}
\multiput(-3.465,6)(11.55,0){2}{\onedot}
\multiput(-4.62,4)(13.86,0){2}{\onedot}
\multiput(-5.775,2)(16.17,0){2}{\onedot}
\multiput(-6.93,0)(18.48,0){2}{\onedot}
\multiput(-8.085,-2)(20.79,0){2}{\onedot}
\multiput(-6.93,-4)(18.48,0){2}{\onedot}
\multiput(-5.775,-6)(16.17,0){2}{\onedot}
\multiput(-4.62,-8)(2.31,0){7}{\onedot}
\multiput(0,8)(2.31,0){3}{\twodot}
\multiput(-1.155,6)(6.93,0){2}{\twodot}
\multiput(-2.31,4)(9.24,0){2}{\twodot}
\multiput(-3.465,2)(11.55,0){2}{\twodot}
\multiput(-4.62,0)(13.86,0){2}{\twodot}
\multiput(-5.775,-2)(16.17,0){2}{\twodot}
\multiput(-4.62,-4)(13.86,0){2}{\twodot}
\multiput(-3.465,-6)(2.31,0){6}{\twodot}
\multiput(1.155,6)(2.31,0){2}{\threedot}
\multiput(0,4)(4.62,0){2}{\threedot}
\multiput(-1.155,2)(6.93,0){2}{\threedot}
\multiput(-2.31,0)(9.24,0){2}{\threedot}
\multiput(-3.465,-2)(11.55,0){2}{\threedot}
\multiput(-2.31,-4)(2.31,0){5}{\threedot}
\multiput(2.31,4)(2.31,0){1}{\fourdot}
\multiput(1.155,2)(2.31,0){2}{\fourdot}
\multiput(0,0)(2.31,0){3}{\fourdot}
\multiput(-1.155,-2)(2.31,0){4}{\fourdot}
\end{picture}
}}
\caption{SU(3) flavor weight diagram for the generalized decuplet baryons (Eq. (\ref{decuplet})). The numbers
denote the multiplicity of the weights.  The long side of the weight diagram
contains ${1 \over 2}\left( N_c - 1 \right)$ weights \cite{jenkins98}.}
\label{spin32}
\end{figure}

\subsection{Excited baryons}

For $N_c=3$, the spin-flavor  or orbital part of the baryon wave function can be symmetric, mixed-symmetric or antisymmetric under the exchange of two quarks. For large $N_c$ we present only the generalization of the symmetric and mixed-symmetric cases because we do not need the generalization of the {\bf 20} irrep of SU(6) in the following.  

\subsubsection{The orbital part}

For the generalization of the orbital part, we consider that the number of excited quarks of large $N_c$ excited baryons is the same as for $N_c=3$. This means the number of ground-state quarks is of order $N_c$ for low excitations. In this section we consider only $N=1$ and 2 excitations. \\

The generalization to arbitrary $N_c$ of a symmetric state is trivial because it is unique as shown in Figure \ref{sncsymm}. But the generalization of the mixed-symmetric state is not unique. 
So, one has  to generalize the mixed-symmetric representation [21] of S$_3$  to large $N_c$ in a way in which one can recover the $N_c=3$ states from $N_c>3$ states.   In Section 2.2.2 we have seen that the $N_c=3$ Jacobi coordinates can be used as basis functions for the symmetric [3] and mixed-symmetric [21]  irreps of S$_3$.  The Jacobi coordinates of an $N_c$ particle system are \cite{stancu96},
\begin{eqnarray}
 {\bf \dot{x}}^{N_c} & = & \frac{1}{\sqrt{N_c}}\sum_{t=1}^{N_c}{\bf x}^t, \label{ncsymmetrijacobi} \\
 {\bf \dot{x}}^s & = & \frac{1}{\sqrt{s(s+1)}}\left(\sum_{t=1}^{s}{\bf x}^t-s{\bf x}^{s+1}\right), \ \ \ \ \ 1\leq s\leq N_c-1. 
\label{ncmixedsymmetricjacobi}
\end{eqnarray}
One can see that taking $s=1$ and $2$  in Eq. (\ref{ncmixedsymmetricjacobi}) one obtains Eqs. (\ref{3rho}) and (\ref{3lambda}) respectively. \\

The center of mass coordinate (\ref{ncsymmetrijacobi}) is a symmetric function by construction and it can be used as basis  vector for the symmetric representation $[N_c]$. The internal coordinates (\ref{ncmixedsymmetricjacobi}) can form an invariant subspace for the mixed-symmetric representation $[N_c-1,1]$ \cite{stancu96}. The particular case of four particles with one unit of excitation has been presented in details in Ref. \cite{stancu98}. This observation suggests that  
the only possible solution is to choose the irrep $[N_c-1,1]$ of S$_{N_c}$ to describe excited states. Otherwise, one would obtain orbital mixed-symmetric  wave functions which do not reduce to [21] when one takes $N_c=3$.   

\begin{figure}[h!]
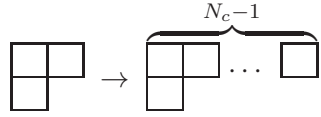

$$\raisebox{-9pt}{\begin{Young}
                 & \cr
		  \cr
                \end{Young}}\ \to \
\mbox{$\overbrace{\negthinspace\negthinspace\negthinspace\negthinspace\negthinspace\negthinspace\raisebox{-9pt}{\begin{Young}
                 & \cr
		  \cr
                \end{Young}}\  \raisebox{3.5pt}{$\cdots$}\negthinspace \negthinspace\negthinspace
\raisebox{3.5pt}{\begin{Young} \cr \end{Young}}}^{N_c-1}$}$$
\caption{Young diagram  for the mixed-symmetric irrep [21] of S$_3$ generalized to 
  $[N_c-1,1]$ of the permutation group S$_{N_c}$ \cite{pirjol98a}.}
\label{sncmixsym}
\end{figure}

Table \ref{harmonicconfigurations} indicates that one has one or two excited quarks depending on the configuration when $N_c=3$. The large $N_c$ generalization gives  the configurations $(0s)^{N_c-1}(0p)$ for $N=1$ and  $(0s)^{N_c-1}(1s)$, $(0s)^{N_c-1}(0d)$, $(0s)^{N_c-2}(0p)^2$ for $N=2$,  following the convention which keeps constant the number of excited quarks.  Like in the $N_c=3$ case, we need to remove the center of mass coordinate in the wave function for satisfying translational invariance and we obtain the configurations shown in Table \ref{harmonicconfigurationsnc}. These combinations were derived by recurrence. We applied  method used for  $N_c=3$ to    $N_c=4,5,6, \ldots$ and then generalized to $N_c$. The configurations of Table \ref{harmonicconfigurationsnc} will be used in Chapter 5 and 6 to analyze excited baryons.

\begin{table}[h]
\begin{center}
{\scriptsize
\renewcommand{\arraystretch}{1.75}
\begin{tabular}{lllll}
\hline
\hline
SU(6) multiplet  & N & L & P & Harmonic oscillator configuration \\
\hline
56 & 0 & 0 & + & $|[N_c](0s)^{N_c}\rangle$ \\
70 & 1 & 1 & $-$ & $|[N_c-1,1](0s)^{N_c-1}(0p)\rangle$ \\
$56'$ & 2 & 0 & + & $\sqrt{\frac{N_c-1}{N_c}}|[N_c](0s)^{N_c-1}(1s)\rangle - \sqrt{\frac{1}{N_c}}|[N_c](0s)^{N_c-2}(0p)^2\rangle$\\
70 & 2 & 0 & + & $\sqrt{\frac{1}{3}}|[N_c-1,1](0s)^{N_c-1}(1s)\rangle + \sqrt{\frac{2}{3}}|[N_c-1,1](0s)^{N_c-2}(0p)^2\rangle$\\
56 & 2 & 2 & + & $\sqrt{\frac{N_c-1}{N_c}}|[N_c](0s)^{N_c-1}(0d)\rangle - \sqrt{\frac{1}{N_c}}|[N_c](0s)^{N_c-2}(0p)^2\rangle$\\
70 & 2 & 2 & + & $\sqrt{\frac{1}{3}}|[N_c-1,1](0s)^{N_c-1}(0d) \rangle + \sqrt{\frac{2}{3}}|[N_c-1,1](0s)^{N_c-2}(0p)^2\rangle$\\
\hline
\hline
\end{tabular}}
\caption{Quark-shell model configurations for a harmonic oscillator confinement and $N_c$ particles.}
\label{harmonicconfigurationsnc}
\end{center}
\end{table}

\subsubsection{The spin-flavor part}
 
In Section 2.3.1 we  have already  discussed the decomposition of the symmetric irrep of SU(6) into its  SU(2) $\times$ SU(3) subgroup.
For the generalized {\bf 70} irrep of SU(6) this decomposition becomes \cite{pirjol98b}
{\scriptsize \begin{eqnarray}\label{SU(6)mixed}
\overbrace{\,
\raisebox{-8.0pt}{\drawsquare{10.0}{0.4}}\hskip-10.4pt
        \raisebox{2pt}{\drawsquare{10.0}{0.4}}\hskip-0.4pt
\raisebox{2pt}{\drawsquare{10.0}{0.4}}\,\cdots\,
\raisebox{2pt}{\drawsquare{10.0}{0.4}}\,}^{N_c-1} &=&
\left(S=\frac12;\, 
\raisebox{-13.0pt}{\drawsquare{10.0}{0.4}}\hskip-10.4pt
\raisebox{-3.0pt}{\drawsquare{10.0}{0.4}}\hskip-10.4pt
        \raisebox{7pt}{\drawsquare{10.0}{0.4}}\hskip-0.4pt
\raisebox{-3.0pt}{\drawsquare{10.0}{0.4}}\hskip-10.4pt
        \raisebox{7pt}{\drawsquare{10.0}{0.4}}\hskip-0.4pt
\raisebox{-3.0pt}{\drawsquare{10.0}{0.4}}\hskip-10.4pt
        \raisebox{7pt}{\drawsquare{10.0}{0.4}}\hskip-0.4pt
\cdots
\raisebox{-3.0pt}{\drawsquare{10.0}{0.4}}\hskip-10.4pt
        \raisebox{7pt}{\drawsquare{10.0}{0.4}}\hskip-0.4pt\,
\right) +
\left(S=\frac12,\frac32,\frac52;\,
\raisebox{-8.0pt}{\drawsquare{10.0}{0.4}}\hskip-10.4pt
        \raisebox{2pt}{\drawsquare{10.0}{0.4}}\hskip-0.4pt
\raisebox{-8.0pt}{\drawsquare{10.0}{0.4}}\hskip-10.4pt
        \raisebox{2pt}{\drawsquare{10.0}{0.4}}\hskip-0.4pt
\cdots
\raisebox{-8.0pt}{\drawsquare{10.0}{0.4}}\hskip-10.4pt
        \raisebox{2pt}{\drawsquare{10.0}{0.4}}\hskip-0.4pt
        \raisebox{2pt}{\drawsquare{10.0}{0.4}}\hskip-0.4pt
        \raisebox{2pt}{\drawsquare{10.0}{0.4}}\hskip-0.4pt
        \raisebox{2pt}{\drawsquare{10.0}{0.4}}\hskip-0.4pt\,
\right) \nonumber \\
&+&
\left(S=\frac12,\frac32;
\raisebox{-8.0pt}{\drawsquare{10.0}{0.4}}\hskip-10.4pt
        \raisebox{2pt}{\drawsquare{10.0}{0.4}}\hskip-0.4pt
\raisebox{-8.0pt}{\drawsquare{10.0}{0.4}}\hskip-10.4pt
        \raisebox{2pt}{\drawsquare{10.0}{0.4}}\hskip-0.4pt
\cdots
\raisebox{-8.0pt}{\drawsquare{10.0}{0.4}}\hskip-10.4pt
        \raisebox{2pt}{\drawsquare{10.0}{0.4}}\hskip-0.4pt
        \raisebox{2pt}{\drawsquare{10.0}{0.4}}\hskip-0.4pt\,
\right) +
\left(S=\frac12,\frac32;\,
\raisebox{-13.0pt}{\drawsquare{10.0}{0.4}}\hskip-10.4pt
\raisebox{-3.0pt}{\drawsquare{10.0}{0.4}}\hskip-10.4pt
        \raisebox{7pt}{\drawsquare{10.0}{0.4}}\hskip-0.4pt
\raisebox{-3.0pt}{\drawsquare{10.0}{0.4}}\hskip-10.4pt
        \raisebox{7pt}{\drawsquare{10.0}{0.4}}\hskip-0.4pt
\raisebox{-3.0pt}{\drawsquare{10.0}{0.4}}\hskip-10.4pt
        \raisebox{7pt}{\drawsquare{10.0}{0.4}}\hskip-0.4pt
\cdots
\raisebox{-3.0pt}{\drawsquare{10.0}{0.4}}\hskip-10.4pt
        \raisebox{7pt}{\drawsquare{10.0}{0.4}}\hskip-0.4pt
        \raisebox{7pt}{\drawsquare{10.0}{0.4}}\hskip-0.4pt
        \raisebox{7pt}{\drawsquare{10.0}{0.4}}\hskip-0.4pt\,
\right) \nonumber\\
&+& \cdots\, + \left(S=\frac{N_c}{2}-1;\,
\overbrace{\,
\raisebox{-3pt}{\drawsquare{10.0}{0.4}}\hskip-0.4pt
\raisebox{-3pt}{\drawsquare{10.0}{0.4}}\hskip-0.4pt\,\cdots\,
\raisebox{-3pt}{\drawsquare{10.0}{0.4}}\hskip-0.4pt\,}^{N_c}\,
\right) +
\left(S=\frac{N_c}{2}-1,\frac{N_c}{2};\,
\overbrace{\,
\raisebox{-8.0pt}{\drawsquare{10.0}{0.4}}\hskip-10.4pt
        \raisebox{2pt}{\drawsquare{10.0}{0.4}}\hskip-0.4pt
\raisebox{2pt}{\drawsquare{10.0}{0.4}}\,\cdots\,
\raisebox{2pt}{\drawsquare{10.0}{0.4}}\,}^{N_c-1}\,
\right)\, . \label{mixedsymmetricdecomposition}
\end{eqnarray}}

\noindent The first three  terms are identical to those of Eq. (\ref{su6smixed3}) where one can recognize the singlet, the decuplet and the octet  of SU(3) for large $N_c$. Indeed, the isospin content for each value of the grand spin $K$ of the first SU(3) irrep written on the right-hand side of Eq. (\ref{mixedsymmetricdecomposition}) can be written as
\begin{equation}
\raisebox{-13.0pt}{\drawsquare{10.0}{0.4}}\hskip-10.4pt
\raisebox{-3.0pt}{\drawsquare{10.0}{0.4}}\hskip-10.4pt
        \raisebox{7pt}{\drawsquare{10.0}{0.4}}\hskip-0.4pt
\raisebox{-3.0pt}{\drawsquare{10.0}{0.4}}\hskip-10.4pt
        \raisebox{7pt}{\drawsquare{10.0}{0.4}}\hskip-0.4pt
\raisebox{-3.0pt}{\drawsquare{10.0}{0.4}}\hskip-10.4pt
        \raisebox{7pt}{\drawsquare{10.0}{0.4}}\hskip-0.4pt
\raisebox{-3.0pt}{\drawsquare{10.0}{0.4}}\hskip-10.4pt
        \raisebox{7pt}{\drawsquare{10.0}{0.4}}\hskip-0.4pt  \to 
 (K=\frac12\,, I=0) + (K=1\,, I=\frac12) + \cdots\,,\label{singlet}
 \end{equation}
 where one can notice that for this irrep $K\neq 0$. For $N_c=3$, this irrep corresponds to the SU(3) singlet. The other terms of Eq. (\ref{mixedsymmetricdecomposition}) exist only for $N_c>3$. For each mixed-symmetric multiplet, this leads   to an infinite tower of excited states.  \\


\chapter{Ground-state baryons in the $1/N_c$ expansion}

\thispagestyle{empty}

\section{Introduction}
After 't Hooft's suggestion and Witten's conclusions, the success of the large $N_c$ QCD was due to the discovery in 1984, by Gervais and Sakita \cite{gervais84} and independently, 
during the ninetieths, by 
Dashen and Manohar \cite{dashen93a}, from the study of baryon-meson scattering process in 
the $N_c \to \infty$ limit, that ground-state baryons satisfy a contracted SU($2N_f)_c$ 
spin-flavor algebra where $N_f$ is the number of flavors.  The baryon contracted spin-flavor symmetry means that 
when $N_c \to \infty $, ground-state baryons form an infinite tower of degenerate states.
At $N_c \to \infty $, the SU($2N_f)$ algebra used in the  quark-shell model becomes 
the SU($2N_f)_c$ algebra. One can therefore use SU($2N_f$) to classify large $N_c$
baryons. The SU($2N_f)_c$ symmetry is broken at order $1/N_c$. Thus, the degenerate baryon states split at order $1/N_c$. \\

Static properties of baryons like mass, magnetic moment, axial currents  can be studied in an $1/N_c$ expansion.  The first term of this expansion must be SU($2N_f$) symmetric, the other terms break the symmetry. \\

In this chapter, we shall first study the emergence of the SU($2N_f)_c$ symmetry. In a second stage, we shall introduce the most general form of one-body static operators written in an $1/N_c$ expansion in terms of the generators of SU($2N_f$) which will serve to construct operators describing observables as linear combinations of  independent operators. 
The application of the $1/N_c$ expansion to the study of ground-state baryons will be considered next. An extension to the excited baryons  will be presented in Chapters 5 and 6.

\section{Large $N_c$ consistency conditions}

\begin{figure}[h!]
\begin{center}
\includegraphics[width=6cm,keepaspectratio]{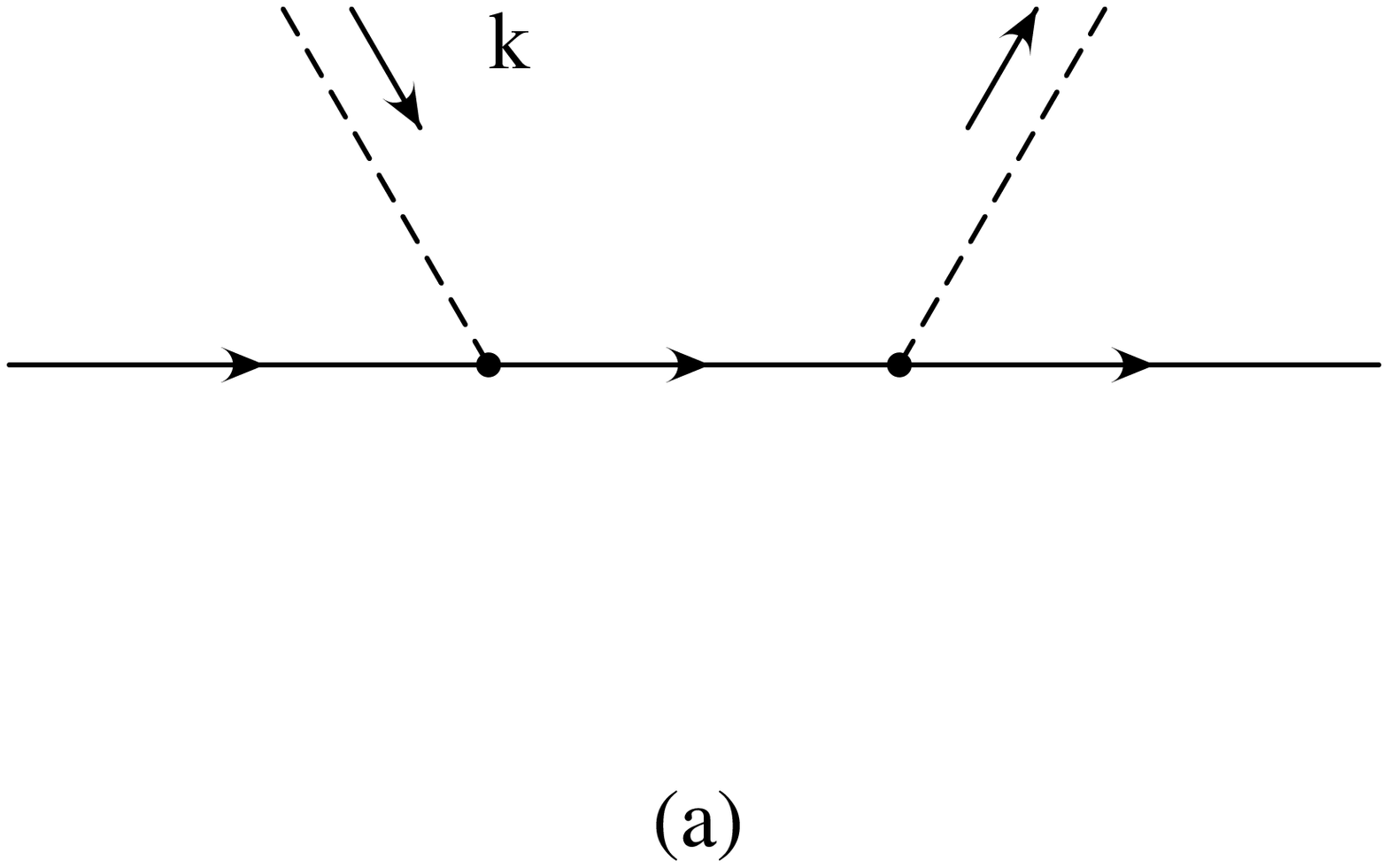} 
\hspace{2cm} 
\includegraphics[width=6cm,keepaspectratio]{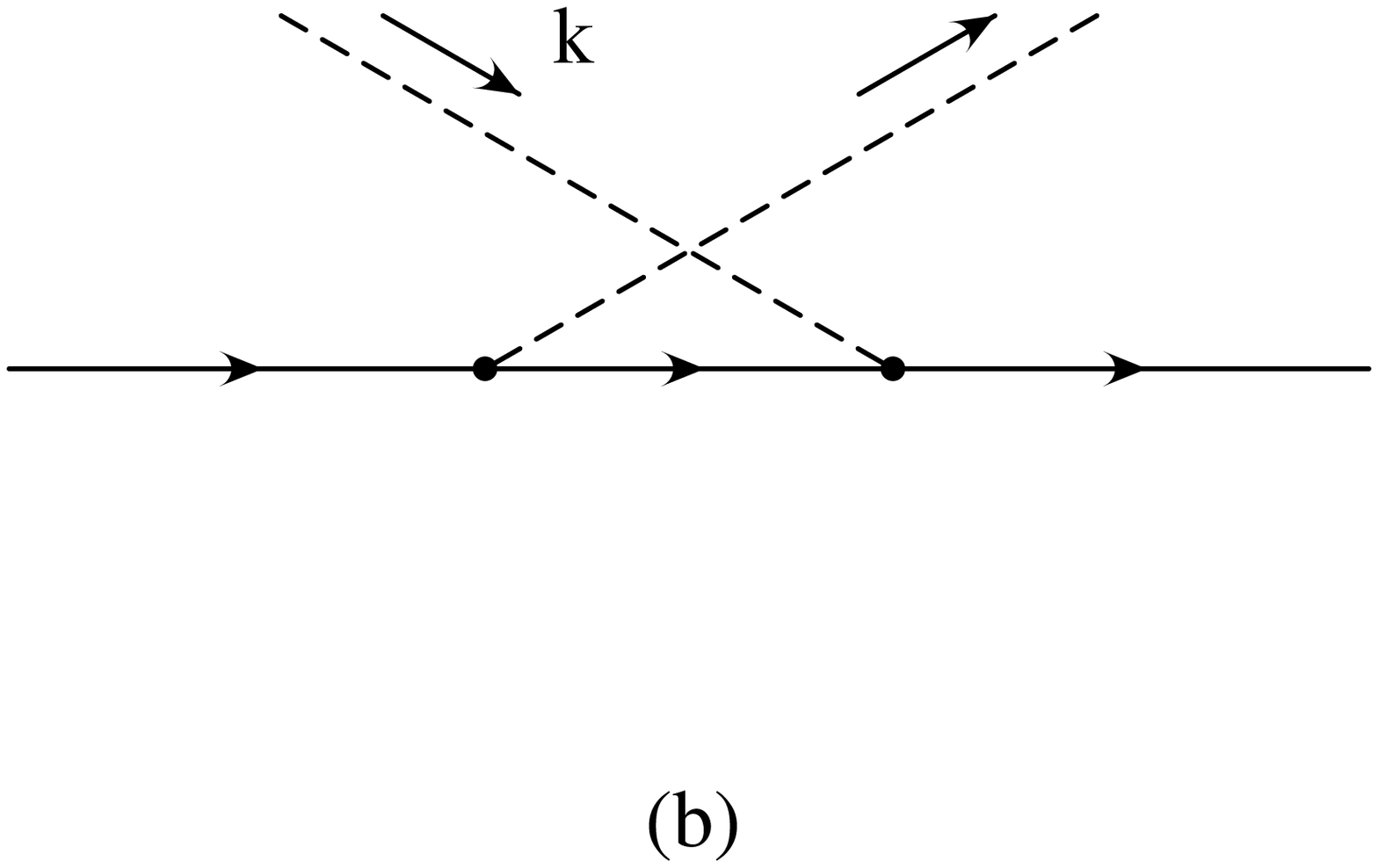}
\end{center}
\caption{Leading-order diagrams for the scattering $B+\pi \to B+\pi$.}
\label{pionbarver}
\end{figure}

In the first chapter  we have studied the baryon + meson $\to$ baryon + meson scattering amplitude in the large $N_c$ limit of QCD. Witten's large $N_c$ power counting rules \cite{witten79} gave amplitudes of order 1. We have also seen that the meson-baryon vertex is  of order $\sqrt{N_c}$  (Figure \ref{onemesonbar}). As drawn in Figure \ref{pionbarver}, the dominant diagrams representing the absorption and then the emission of a meson by a baryon contains two baryon-meson vertices and should grow as $N_c$ in contradiction with the counting rules and in violation of the unitarity. Large $N_c$ consistency conditions follow from this result. They imply that the large $N_c$ baryons satisfy an SU($2N_f)_c$ symmetry when $N_c \to \infty$. \\

Let us consider the elastic pion-baryon scattering. The pion-baryon vertex is given by
\begin{equation}
 \frac{\partial_\mu\pi}{f_\pi}\langle B |\bar{q}\gamma^\mu\gamma_5T^aq|B\rangle,
\end{equation}
 where $f_\pi$ is the pion decay constant of order $\sqrt{N_c}$ and  $\langle B |\bar{q}\gamma^\mu\gamma_5T^aq|B\rangle$ is the axial vector current matrix element of the nucleon, of order $N_c$, as there are $N_c$ quark lines inside the nucleon. As announced, the pion-baryon vertex is order $\sqrt{N_c}$. \\
 
 In the large $N_c$ limit, as the baryon is infinitely heavy compared with the pion, the baryon-pion coupling reduces to the static-baryon coupling. The axial vector current matrix element can be written as
 \begin{equation}
  \langle B |\bar{q}\gamma^i\gamma_5T^aq|B\rangle = g N_c \langle B|X^{ia}|B \rangle,
 \end{equation}
with $g$ and $\langle B|X^{ia}|B \rangle$   of order 1. If we consider only dominant diagrams (Figure \ref{pionbarver}), the pion-nucleon scattering amplitude is given by \cite{dashen93a}
 \begin{equation}
  \mathcal{A}= -i \frac{N_c^2g^2}{f_\pi^2}\frac{q^iq'^j}{q^0}\left[X^{ia},X^{jb}\right],
 \end{equation}
  if the initial and final baryons are on-shell. Here we sum over all the possible spins and isospins of the intermediate baryon.
This amplitude is of order $N_c$ and then violates unitarity and  Witten's large $N_c$ counting rules. To obtain a consistent theory, one must introduce other intermediate baryon  states, degenerate to leading order in $1/N_c$ that cancel the factor $N_c$ appearing in the scattering amplitude. We then obtain the following constraint
\begin{equation}
 N_c\left[X^{ia},X^{jb}\right]\leq \mathcal{O}(1).
\end{equation}
One can expand the operator $X^{ia}$ in terms of $1/N_c$ 
\begin{equation}
 X^{ia}=X^{ia}_0+\frac{1}{N_c}X^{ia}_1+\frac{1}{N_c^2}X^{ia}_2+ \cdots.
\end{equation}
The first order consistency condition  for large $N_c$ QCD is then given by
\begin{equation}
 \left[X^{ia}_0,X^{jb}_0\right]=0.
 \label{consist1}
\end{equation}

In the $N_c \to \infty$ limit, the spin operators $S^i$, the flavor operators $T^a$, and the amplitudes $X^{ia}$ can be identified with the generators of a contracted spin-flavor group  SU($2N_f)_c$, where $N_f$ is the number of flavors.
The algebra of the contracted SU($2N_f)_c$ group is given by
\begin{eqnarray}
 \left[S^i,T^a\right]& = & 0, \nonumber \\
 \left[S^i,S^j\right]& = & i\varepsilon^{ijk}S^k, \ \ \left[T^a,T^b\right] = if^{abc}T^c, \nonumber \\
 \left[S^i,X_0^{ja}\right] & = & i\varepsilon^{ijk}X_0^{ka}, \ \ \left[T^a,X_0^{ib}\right]=if^{abc}X_0^{ic}, \nonumber \\
 \left[X_0^{ia},X_0^{jb}\right]& = & 0. 
\end{eqnarray}
One can compare this algebra with the SU($2N_f$) algebra. The generators $S^i$, $T^a$ and the spin-flavor generators $G^{ia}$ satisfy the spin and flavor algebra \cite{dashen95}
\begin{eqnarray}
 \left[S^i,T^a\right]& = & 0, \nonumber \\
 \left[S^i,S^j\right]& = & i\varepsilon^{ijk}S^k, \ \ \left[T^a,T^b\right] = if^{abc}T^c, \nonumber \\
 \left[S^i,G^{ja}\right] & = & i\varepsilon^{ijk}G^{ka}, \ \ \left[T^a,G^{ib}\right]=if^{abc}G^{ic}, \nonumber \\
 \left[G^{ia},G^{jb}\right]& = & \frac{i}{4}\delta^{ij}f^{abc}T^c+\frac{i}{2N_f}\delta^{ab}\varepsilon^{ijk}S^k+\frac{i}{2}\varepsilon^{ijk}d^{abc}G^{kc}. 
 \label{su(2nf)algebra)}
\end{eqnarray}
The contracted algebra is obtained from the previous one by taking the limit
\begin{equation}
X_0^{ia}=\lim_{N_c\to \infty}\frac{G^{ia}}{N_c}.
\end{equation}
Only the commutation relation $[G^{ia},G^{jb}]$ is affected by this contraction. In dividing it by $N_c^2$  the left-hand side becomes  $[X_0^{ia},X_0^{jb}]$  and the right-hand side cancels out. This leads to the consistency condition (\ref{consist1}) in the large $N_c$ limit. \\

 Dashen et al \cite{dashen94a} have solved Eq. (\ref{consist1}) which means they found irreducible representations of SU($2N_f)_c$. The basis vectors form  infinite towers of degenerate $(S,I)$ baryon states. If we use the grand spin $\vec{K}$, Eq. (\ref{grandspin}), the first tower corresponds to the case with $K=0$, $S=I$ 
\begin{equation}
\left(\frac{1}{2},\frac{1}{2}\right),\left(\frac{3}{2},\frac{3}{2}\right),\left(\frac{5}{2},\frac{5}{2}\right), \ldots
\label{1/2tower}
\end{equation}
The case with $K=\frac{1}{2}$ leads to a second tower
\begin{equation}
\left(\frac{1}{2},0\right),\left(\frac{1}{2},1\right),\left(\frac{3}{2},1\right),\left(\frac{3}{2},2\right),\left(\frac{5}{2},2\right),\left(\frac{5}{2},3\right),\ldots,
\end{equation}
 for $K=1$ one has
\begin{equation}
\left(\frac{1}{2},\frac{1}{2}\right),\left(\frac{1}{2},\frac{3}{2}\right),\left(\frac{3}{2},\frac{1}{2}\right),\left(\frac{3}{2},\frac{3}{2}\right),\left(\frac{3}{2},\frac{5}{2}\right),\left(\frac{5}{2},\frac{3}{2}\right),\left(\frac{5}{2},\frac{5}{2}\right),\left(\frac{5}{2},\frac{7}{2}\right), \ldots,
\end{equation}
and with $K=\frac{3}{2}$ one has
\begin{equation}
\left(\frac{1}{2},1\right),\left(\frac{1}{2},2\right),\left(\frac{3}{2},0\right),\left(\frac{3}{2},1\right),\left(\frac{3}{2},2\right),\left(\frac{3}{2},3\right),\ldots.
\label{3/2tower}
\end{equation}
It is possible to do identifications with physical states when we assume, as in the previous chapter,  that $K=n_s/2$. For example, the $K=0$ tower contains strangeness zero baryons, like the nucleon state $(1/2,1/2)$ or the $\Delta$ state $(3/2,3/2)$. The $K=1/2$ tower contains strangeness $-1$ baryons $\Lambda(1/2,0)$, $\Sigma(1/2,1)$ and $\Sigma^*(3/2,1)$, the $K=1$ tower, baryons of strangeness $-2$ $\Xi(1/2,1/2)$ and $\Xi^*(3/2,1/2)$, the $K=3/2$ tower, the state $\Omega(3/2,0)$ of strangeness $-3$. The other states present inside the towers are spurious states, \emph{i.e.} baryons that exist for $N_c>3$ but not for $N_c=3$. \\


As a conclusion, one can notice a perfect connection between the infinite number of intermediate baryon states predicted by the planar Feynman diagrams and the results described here. The large $N_c$ power counting rules of  Witten imply a sum over all possible intermediate quark states, which is equivalent to summing over all intermediate baryon states. We have obtained an inconsistent large $N_c$ theory because,  in the first analysis of the baryon-pion scattering process, we took into account only one intermediate state. Large $N_c$ QCD requires the existence of infinite towers of degenerate states, solution of the consistency condition Eq. (\ref{consist1}). \\

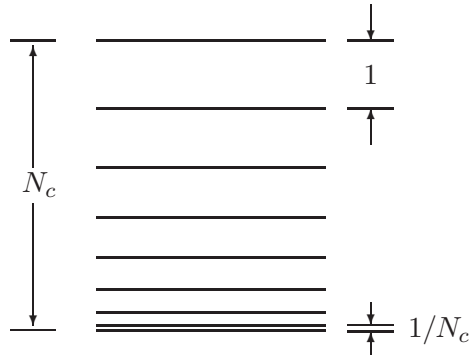
\begin{figure}[tbp]
\setlength{\unitlength}{6mm}
\centerline{\hbox{
\begin{picture}(10,7.725)(-1.9,-0.525)
\def\level{\line(1,0){5}}
\thicklines
\put(0,6.4){\level}
\put(0,4.9){\level}
\put(0,3.6){\level}
\put(0,2.5){\level}
\put(0,1.6){\level}
\put(0,0.9){\level}
\put(0,0.4){\level}
\put(0,0.1){\level}
\put(0,0){\level}
\thinlines
\put(5.5,6.4){\line(1,0){1}}
\put(5.5,4.9){\line(1,0){1}}
\put(5.5,0.125){\line(1,0){1}}
\put(5.5,-0.025){\line(1,0){1}}
\put(-1.9,0){\line(1,0){1}}
\put(-1.9,6.4){\line(1,0){1}}
\put(6,5.65){\makebox(0,0){$1$}}
\put(7.5,0){\makebox(0,0){$1/N_c$}}
\put(-1.65,3.05){$N_c$}
\put(-1.4,3.6){\vector(0,1){2.7}}
\put(-1.4,2.8){\vector(0,-1){2.7}}
\put(6,7.2){\vector(0,-1){0.8}}
\put(6,4.1){\vector(0,1){0.8}}
\put(6,-0.525){\vector(0,1){0.5}}
\put(6,0.625){\vector(0,-1){0.5}}
\end{picture}
}}
\caption{Mass splittings up to the $1/N_c$ order for the tower of large $N_c$ ground-state baryons  with $K=0, S=I=\frac{1}{2},\frac{3}{2},\ldots,\frac{N_c}{2}$. The $S^2/N_c$ operator leads to a mass splitting of $\mathcal{O}(1/N_c)$ between baryons with spins $S\sim \mathcal{O}(1)$ and to a mass splitting of $\mathcal{O}(1)$ between baryons with spins $S\sim \mathcal{O}(N_c)$. The mass splitting between the baryon states with $S=\frac{1}{2}$ and $S=\frac{N_c}{2}$ is $\mathcal{O}(N_c)$ \cite{jenkins98}.}
\label{barmasstower}
\end{figure}

It is possible to study $1/N_c$ corrections of static baryon matrix elements like the mass operator by deriving other consistency conditions. We will not give more details here about this method\footnote{The following section gives another way which will be used in Chapter 5 and 6 to expand baryon operators in powers of $1/N_c$.}. After calculation one can   obtain the following expansion for the mass operator \cite{jenkins98},
\begin{equation}
 M=m_0N_c\ \1 + m_2 \frac{1}{N_c}S^2+ \cdots,
 \label{su(2nf)cmass}
\end{equation}
where $m_i$ are unknown coefficients which contain large $N_c$ QCD dynamics and $S^2$ is the Casimir operator of SU(2). One can directly notice that the $1/N_c$ expansion is predictive only if $S$ is $\mathcal{O}(1)$, \emph{i.e.} for baryon situated at the beginning of the tower Eq. (\ref{1/2tower}). For baryons situated at the end of the tower, with $S$ of order $N_c$, the first and the second term of the expansion Eq. (\ref{su(2nf)cmass}) are of the same order. Thus, the mass splitting between two baryons with spin $S \sim \mathcal{O}(1)$ is of order $1/N_c$ and is of order $\mathcal{O}(1)$ between two baryons with spin $S \sim \mathcal{O}(N_c)$. Figure \ref{barmasstower} gives an illustration of the $1/N_c$ correction of the ground-state baryon mass operator Eq. (\ref{su(2nf)cmass}).\\

One can write a general form of the $1/N_c$ expansion for a baryon operator \cite{jenkins98}. For $N_f=2$, the $1/N_c$ expansion of a static baryon operator like the baryon mass operator has the form 
\begin{equation}
 N_c\mathcal{P}\left(X_0,\frac{S}{N_c},\frac{I}{N_c}\right),
 \label{polynomesu(2nf)c}
\end{equation}
where $\mathcal{P}$ is a polynomial.


\section{Operator expansion}
\label{groundoperatorexpansion}

In the previous section, we have shown that large $N_c$ non-strange baryons belong to an infinite tower of degenerate bound states. These states are identical to those obtained with the quark-shell model in the large $N_c$ limit satisfying an SU($4$) spin-flavor symmetry. Indeed, the tower (\ref{quarkconsttower}) is identical to the tower (\ref{1/2tower}) in the $N_c\to \infty$ limit. \\

\begin{figure}[h!]
\begin{center}
\includegraphics[width=6cm,keepaspectratio]{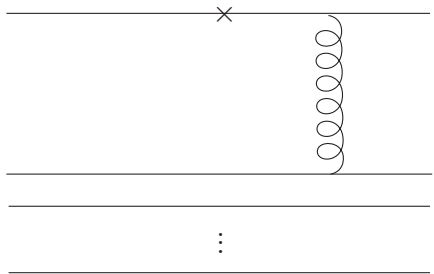}
\end{center}
\caption{An illustration of a two-body effective operator generated out of a one-body QCD operator, depicted by a cross. One can see directly that at least one gluon must be exchanged between the first quark line where the QCD one-body operator acts and a second quark line to generate an effective two-body operator.}
\label{twobody}
\end{figure}

Let us consider a QCD one-body operator such as $\bar{q}\gamma^\mu\gamma_5T^aq$. This operator is called one-body operator because it acts on one quark line. For example, in Figure \ref{twobody}, its action is depicted by a cross on the first quark line at the top of the figure. Its matrix element is obtained by inserting  the operator on any of the $N_c$ quark lines of the baryon. As each Feynman diagram depicting the insertion is of order 1 and as there are $N_c$ possible insertions, the QCD one-body has a matrix element which is at most of order $N_c$. There may be cancellations among the $N_c$ possible insertions, that is why the matrix element is not necessarily of order $N_c$.     \\

A QCD one-body  operator  transforming according to a given SU(2) $\times$ SU($N_f$) representation can be written as a series expansion of $n$-body operators, which involve $n$ quark lines. Due to the Wigner-Eckart theorem these operators can be decomposed into a product of unknown dynamical coefficients and effective operators which are expressed by  means of appropriate products of symmetry generators as \cite{dashen95}
\begin{equation}
 \mathcal{O}^{\mathrm{1-body}}_{\mathrm{QCD}}=\sum_{n}c^{(n)} \frac{1}{N_c^{n-1}}\mathcal{O}^{(n)},
 \label{qcdoperatorexpansion}
\end{equation}
where one sums over all possible $n$-body effective operators $\mathcal{O}^{(n)}$, $0\leq n \leq N_c$, with the same spin and flavor quantum number as $\mathcal{O}^{\mathrm{1-body}}_{\mathrm{QCD}}$ and where $c^{(n)}$ are  unknown dynamical coefficients. These effective operators are accompanied by a factor of $1/N_c^{n-1}$.  It comes from the fact that, as illustrated in Figure \ref{twobody}, one needs at least $n-1$ gluon exchanges at the quark level to generate $n$-body effective operators in the $1/N_c$ expansion out of a one-body QCD operator. \\

Two different $1/N_c$ expansions of a QCD operator are possible. The first one is based on the group SU($2N_f)_c$ and uses the spin-flavor generators $X_0^{ia}$ in operator products. This case was illustrated in Eq. (\ref{polynomesu(2nf)c}). The second one is based on the group SU($2N_f$) and where one uses the spin-flavor generators $G^{ia}$ with the large $N_c$ quark-shell model operator basis.  
In the following, we shall use the operator basis of the large $N_c$ quark-shell model which are easier to work with. \\


With this approach, on can interpret the effective $n$-body operators $\mathcal{O}^{(n)}$ as $n$-body quark operators acting on the spin and flavor indices of $n$ static quarks, 
\begin{equation}
 \mathcal{O}^{(n)}=\sum_{l,m}\left(S^i\right)^l \left(T^a\right)^m \left(G^{ia}\right)^{n-l-m},
 \label{productgenerator}
\end{equation}
and the expansion of a QCD one-body operator (\ref{qcdoperatorexpansion}) becomes
\begin{equation}
 \mathcal{O}^{\mathrm{1-body}}_{\mathrm{QCD}}=\sum_{l,m,n}c^{(n)} \frac{1}{N_c^{n-1}}\left(S^i\right)^l\left(T^a\right)^m \left(G^{ia}\right)^{n-l-m}.
 \label{qcdoperatorexpansion2}
\end{equation}
For ground state baryons, the orbital part of the wave function does not play any role. It will be different for excited baryons where we have to include SO(3) generators $\ell^j$ in the product of symmetry generators Eq. (\ref{productgenerator}) (see Section \ref{excitedstateoperatorexpansion}) and take into account the symmetry of the orbital part and the coupling of the angular momentum to the spin. \\

One can analyze the $N_c$ dependence of the $1/N_c$ expansion of the QCD one-body operator.  It depends on  $N_c$ in two different ways:
\begin{itemize}
 \item An explicit $1/N_c^{n-1}$ term which gives an operator of order $N_c$ as we can see when $n=0$. 
 \item An implicit dependence on $N_c$ which comes from the operators $\mathcal{O}^{(n)}$ themselves. Indeed, matrix elements of the generators of the group SU($2N_f$) have a nontrivial dependence on $N_c$ that varies for different states of the baryon spin-flavor multiplet. For example, if we consider the baryon tower Eq. (\ref{quarkconsttower}), states of the bottom of the tower have spin and isospin of order 1 but states at the top of the tower have spin and isospin of order $N_c$. It is possible to prove that the matrix elements of the SU(6) generators $T^8$ and $G^{ia}$ with $a=1,2,3$, when  applied on a symmetric SU(6) wave function, are of order $N_c$ between states with spin and strangeness of order $N_c^0$ (see Chapter 4). They are called \emph{coherent} generators.  This implies that operator $\mathcal{O}^{(n)}$ are of order $\leq \mathcal{O}(N_c^n)$.
\end{itemize}
\emph{The $1/N_c$ expansion of an operator can be not predictive if the explicit factor of $1/N_c^{n-1}$ is completely cancelled by the implicit factors resulting from the matrix elements of operators $\mathcal{O}^{(n)}$.}  \\

It is possible to generalise the $1/N_c$ expansion to a QCD $m$-body operator. As this operator acts on $m$ quarks inside the baryon, its matrix element is at most of order $N_c^m$.   In that case, one can expand an $m$-body QCD operator in terms of $n$-body effective operators with coefficients of order $N_c^{m-n}$

 \section{Quark operator classification}
 Let us introduce the creation and annihilation operators $q^\dagger_\alpha$ and $q^\alpha$ where $\alpha=1,\ldots,N_f$ represents the $N_f$ quark flavors with spin up, and $\alpha=N_f+1,\ldots,2N_f$, the $N_f$ quark flavors with spin down. As the color part of the baryon wave function is antisymmetric, one can omit the color quantum numbers of the quark operators. Thus, these operators satisfy the bosonic commutation relation
 \begin{equation}
  \left[q^\alpha,q^\dagger_\beta\right]=\delta^\alpha_\beta,
 \end{equation}
as the spin-flavor part of the wave function Eq. (\ref{dividedwavefunction}) is symmetric for ground-state baryons. 
One can write $n$-body quark operators in terms of these quark creation and annihilation operators and then classify the quark operators. \\

We have first the zero-body operator $\mathbbm{1}$,
 independent of $q$ or $q^\dagger$. It does not act on the quarks in the baryon. \\

 There are four one-body operators: the quark number operator 
 \begin{equation}
 q^\dagger q=\sum_{j=1}^{N_c}q^\dagger_j q_j = N_c \mathbbm{1},
 \label{quarknumber}
 \end{equation}
 and the SU($2N_f$) generators
 \begin{eqnarray}
  S^i & = & \sum_{j=1}^{N_c}q_j^\dagger\left(S^i \times \mathbbm{1}\right)q_j \ \ \ (3,1), \nonumber \\
  T^a & = & \sum_{j=1}^{N_c}q_j^\dagger\left(\mathbbm{1} \times T^a\right)q_j \ \ \ (1,adj), \label{quarkgen} \\
  G^{ia} & = & \sum_{j=1}^{N_c}q_j^\dagger\left(S^i \times T^a\right)q_j \ \ \ (3,adj). \nonumber
 \end{eqnarray}
 The brackets in the right-hand side denotes the (SU(2), SU($N_f$)) dimensions of the associated irreducible representations in SU(2) and SU($N_f$) respectively. The generators of a group correspond to the adjoint representation. For SU(2) its dimension is 3, as indicated, and    $adj$ denotes the dimension of the adjoint representation of SU($N_f$). The number 1 corresponds to the trivial representation in SU(2) or SU($N_f$). 
 In Eqs. (\ref{quarknumber}) and (\ref{quarkgen}), $j$ denotes the quark line number. Eq. (\ref{quarknumber}) shows that the quark number operator can be rewritten in terms of the  zero-body identity operator. In conclusion, there are only three independent one-body operators, $S^i$, $T^a$ and $G^{ia}$.  \\

Two-body operators act on two quarks and involve two creation and annihilation operators $q$ and $q^\dagger$. They can be written as operator products of baryon spin-flavor generators, for example,
\begin{equation}
 S^iT^a=\sum_{j,j'}\left(q_j^\dagger S^iq_j\right)\left(q^\dagger_{j'}T^a q_{j'}\right),
 \label{notnormalorder}
\end{equation}
where each one-body operator applies on one quark line independently. This operator is not a pure two-body operator as, in the case where $j=j'$, the two one-body operators apply in the same quark line and reduces to an one-body operator.  \\

It is possible to write two-body operators in a normal ordered form,
\begin{equation}
 q^\dagger q^\dagger S^iT^a q q =  \sum_{j\neq j'}\left(q_j^\dagger S^iq_j\right)\left(q^\dagger_{j'}T^a q_{j'}\right).
 \label{normalorder}
\end{equation}
As this operator has vanishing matrix elements on single-quark states, it is a pure  two-body operator.
One can generalize it to an $n$-body quark operator
\begin{equation}
 q^\dagger \ldots q^\dagger \mathcal{T} q \ldots q,
\end{equation}
where $\mathcal{T}$ is a traceless and completely symmetrized tensor.  One can permute the creation or annihilation operators since they are bosonic operators. The tensor $\mathcal{T}$ must be traceless to obtain a pure $n$-body operator and symmetric as the spin-flavor part of the ground-state baryon wave function is symmetric. Tensor operators antisymmetric or  mixed-symmetric vanish.  \\

 It is easier to derive matrix elements of operator products of the baryon spin-flavor generators  than of normal-order operators. However, only pure $n$-body operators appear in the $1/N_c$ expansion of QCD operator Eq. (\ref{qcdoperatorexpansion}). When operator products are rewritten in a normal-order form, the tensors $\mathcal{T}$ obtained are not traceless or completely symmetrized, so some linear combinations of the operator products are equivalent to lower-body operators or vanish identically on the completely symmetric ground-state baryon representation \cite{jenkins98}. Dashen \emph{et al.} \cite{dashen95} calculated operators identities which eliminate redundant operators appearing in the operator product combinations like the quark number operator Eq. (\ref{quarknumber}).  With these identities, it is not necessary to derive matrix elements of normal-order operators.

\section{Operator identities}
\label{groundoperatoridentities}
As we have seen in the previous section, some $n$-body operators can be reduced to linear combinations of $m$-body operators, with $m < n$ when we make an $1/N_c$ expansion of a QCD operator based on the group SU($2N_f$), Eq. (\ref{qcdoperatorexpansion}). With this reduction, all redundant operators are eliminated and we obtain pure $n$-body operators $\mathcal{O}^{(n)}$. Here, we  summarize the method used to compute  linear combinations of operators. Details can be found in Ref. \cite{dashen95}. \\

Eq. (\ref{quarknumber}) illustrates the first identity which reduces a one-body operator to a zero-body operator. \\ 


Let us have a look at two-body identities. One can write a two-body operator as a product of two one-body operators. We have seen that the only independent one-body operators are the  SU($2N_f$) generators Eq. (\ref{quarkgen}). They belong to the adjoint representation of SU($2N_f$). The product of two operators can be written as a symmetric product (anticommutator) or an antisymmetric product (commutator). The commutator can be eliminated by using the SU($2N_f$) algebra, Eqs. (\ref{su(2nf)algebra)}). The anticommutator transforms as the symmetric product of two SU($2N_f$) adjoint representations \cite{dashen95},
\begin{equation}
\left(adj\times adj\right)_S = 1+adj+\bar{a}a+\bar{s}s,
\label{1adj1adj}
\end{equation}
where $\bar{a}a=T^{[\alpha_1 \alpha_2]}_{[\beta_1 \beta_2]}$ is a traceless tensor which is antisymmetric in its upper and lower indices and $\bar{s}s=T^{(\alpha_1\alpha_2)}_{(\beta_1\beta_2)}$ transforms as a traceless tensor which is completely symmetric in its upper and lower indices. \\

From the analysis of Eq.  (\ref{1adj1adj}), two-body identities are divided into three different sets:
\begin{itemize}
 \item The first identity reduces the two-body operators to an SU($2N_f$) singlet. The SU($2N_f$) singlet in the symmetric product Eq. (\ref{1adj1adj}) is the Casimir operator. 
 \item The second set of identities corresponds to a linear combination of two-body operators which transforms as an SU($2N_f$) adjoint. 
 \item The result of a symmetric product of two generators of SU($2N_f$) can be an antisymmetric tensor operator, $\bar{a}a=T^{[\alpha_1 \alpha_2]}_{[\beta_1 \beta_2]}$. We have seen that this tensor operator  must vanish when acting on the ground-state baryons. This gives the third set of vanishing identities.
 \end{itemize}

 
 There are not new identities for $n$-body operators with $n>2$ because it is always possible to obtain $n$-body operators by recursively applying the two-body identities on all pairs of one-body identities appearing in a completely symmetric  product of $n$ operators \cite{dashen95}. \\ 
 
 Table \ref{tab:su2fiden} gives the three sets of two-body identities for the SU($2N_f$) group. The spin-flavor representation (SU(2), SU($N_f$)) of each operator identity  is given explicitly in the right-hand side column. This table is divided into three blocks  corresponding  to the three sets of two-body operators identities respectively. In the first block, one can recognize the SU($2N_f$) Casimir identity. In the second block, we have three identities transforming as linear combinations of two-body operators into SU($2N_f$) adjoints, the adjoints being decomposed under SU(2) $\times$ SU($N_f$) into the three representations $(3,1)$, $(1,adj)$ and $(3,adj)$ (see Eq. (\ref{quarkgen})). The third block of identities is obtained by combining the two adjoint one-body operators into the $\bar{a}a$ representation, and setting it equal to zero. More details are given in Ref. \cite{dashen95}.\\

\begin{table}[htbp]
\smallskip
\centerline{\vbox{ \tabskip=0pt \offinterlineskip
\def\tablerule{\noalign{\hrule}}
\def\space{height 4pt&\omit&&\omit&\cr}
\halign{
\vrule #&\strut\hfil\ $ # $\ \hfil&\vrule #&\strut\hfil\ $ # $\ \hfil&\vrule
#\cr
\tablerule\space
&2\ \left\{S^i,S^i\right\} + N_f\ \left\{T^a,T^a\right\} + 4 N_f\ \left\{G^{ia},
G^{ia}\right\} = N_c \left(N_c+2N_f\right) \left(2N_f-1\right)&&(1,1)&\cr
\space\tablerule\space
&d^{abc}\ \left\{G^{ia}, G^{ib}\right\} + {2\over N_f}\ \left\{S^i, G^{ic}
\right\} + {1\over4}\ d^{abc}\ \left\{T^a, T^b\right\} = \left(N_c+N_f\right)
\left(1-{1\over N_f}\right)\ T^c && (1,adj)&\cr
\space
&\left\{T^a,G^{ia}\right\} = \left(N_c+N_f\right) \left(1-{1\over N_f}\right)\
S^i
&&(3,1)&\cr
\space
&{1\over N_f}\ \left\{S^k,T^c\right\} +  d^{abc}\ \left\{T^a,G^{kb} \right\}
-\epsilon^{ijk} f^{abc} \left\{G^{ia}, G^{jb}\right\} = 2\left(N_c+N_f\right)
\left(1-{1\over N_f} \right)\ G^{kc} && (3,adj)&\cr
\space\tablerule\space
&4N_f\left(2-N_f\right)\ \left\{G^{ia},G^{ia}\right\} + 3 N_f^2\ \left\{T^a,
T^a\right\} + 4\left(1-N_f^2\right)\ \left\{S^i,S^i\right\}=0&& (1,1)&\cr
\space
&\left(4-N_f\right)d^{abc}\ \left\{G^{ia}, G^{ib}\right\} + {3\over 4} N_f
\ d^{abc}\ \left\{T^a, T^b\right\} - 2\left(N_f-{4\over N_f}\right)\
\left\{S^i,G^{ia}
\right\} = 0 && (1,adj) &\cr
\space
&4\ \left\{G^{ia},G^{ib}\right\} = -3\ \left\{T^a,T^b\right\}\qquad ({\bar a
a})
&& (1,{\bar a a})&\cr
\space
&4\ \left\{G^{ia},G^{ib}\right\} = \left\{T^a,T^b\right\}\qquad ({\bar s s})
&& (1,{\bar s s})&\cr
\space
&\epsilon^{ijk}\ \left\{ S^i,G^{jc}\right\} = f^{abc} \ \left\{T^a,G^{kb}
\right\}&&(3,adj)&\cr
\space
&d^{abc}\ \left\{T^a,G^{kb}\right\} = \left(1-{2\over N_f}\right) \left(
\left\{S^k,T^c\right\} -  \epsilon^{ijk} f^{abc}\ \left\{G^{ia}, G^{jb}
\right\}\right)&&(3,adj)&\cr
\space
&\epsilon^{ijk}\ \left\{G^{ia},G^{jb}\right\} = f^{acg} d^{bch}\
\left\{T^g,G^{kh}\right\}\qquad ({\bar a s}+{\bar s a})&&(3,{\bar a s}+{\bar s
a})&\cr
\space
&\left\{T^a,G^{ib}\right\} =0\qquad ({\bar a a})&&(3,{\bar a a})&\cr
\space
&\left\{G^{ia}, G^{ja}\right\} = {1\over2}\left(1-{1 \over N_f}\right)\
\left\{S^i, S^j \right\}
\qquad (S=2) && (5,1)&\cr
\space
&d^{abc}\ \left\{G^{ia}, G^{jb}\right\} = \left(1-{2\over N_f}\right) \
\left\{S^i,G^{jc}\right\}\qquad (S=2) && (5,adj)&\cr
\space
&\left\{G^{ia},G^{jb}\right\} = 0\qquad (S=2,{\bar a a})&& (5,{\bar a a})&\cr
\space\tablerule
}}}
\caption{SU($2N_f$) Identities: The second column gives the transformation
properties of the identities under SU(2)$\times$ SU($N_f$) \cite{dashen95}.}\label{tab:su2fiden}
\end{table}

\section{The mass operator}
\label{groundstatemassoperator}

As an illustration of the large $N_c$ operator expansion, we  analyze the baryon mass operator expansion for $N_f=3$ \cite{dashen95,jenkins95}. Let us divide this section into two part. In the first part, we suppose that the SU(3) symmetry is exact, \emph{i.e.} the up, down and strange quarks have the same mass. In the second part, we will consider  SU(3) breaking.

\subsection{Baryon masses with exact SU(3) symmetry}

We apply the $1/N_c$ expansion presented in Eq. (\ref{qcdoperatorexpansion2}).
As we assume that the SU(3) symmetry is exact, the mass operator transforms as a spin-flavor singlet $(1,1)$. As a consequence, all the operators $\mathcal{O}^{(n)}$ Eq. (\ref{productgenerator}) must be SU(2) $\times$ SU(3) scalars. \\

There is one zero-body operator transforming as $(1,1)$ under SU(2) $\times$ SU(3), the identity operator $\mathbbm{1}$,
\begin{equation}
\mathcal{O}^{(0)} = \mathbbm{1}.
\end{equation}
There are no one-body operators transforming as a spin-flavor singlet. Regarding two-body operators, if we look at Table \ref{tab:su2fiden}, taking $N_f=3$, one can notice that there are two $(1,1)$ operators identities. This means that we keep only the operator $S^2$ in the expansion,
\begin{equation}
\mathcal{O}^{(2)}=S^2,
\end{equation}
because $T^2$ and $G^2$ can be expressed in terms of $S^2$.
By applying the two-body identities, one can find that there are no three-body operators in the expansion. In general, there is a single $n$-body operator for each even $n$ \cite{jenkins95},
\begin{equation}
 \mathcal{O}^{(n)}=S^n,
\end{equation}
which transforms as $(1,1)$. \\

The $1/N_c$ expansion of the mass operator in the flavor symmetry limit becomes
\begin{equation}
 M^{(1,1)} = c_0N_c \mathbbm{1} + c_2 \frac{1}{N_c} S^2 + c_4 \frac{1}{N_c^3} S^4 + \cdots+ c_{N_c-1} \frac{1}{N_c^{N_c-2}}S^{N_c-1},\label{groundstatespin-flavor}
\end{equation}
the expansion being limited to the $N_c-2$ order. Indeed, as we have $N_c$ quark lines inside the baryon, the expansion goes until the order $1/N_c^{N_c-1}$. However, as $N_c$ is odd, the $N_c$-body operator is suppressed by operator identities. The expansion Eq. (\ref{groundstatespin-flavor}) corresponds to Eq. (\ref{su(2nf)cmass}) obtained by the resolution of large $N_c$ consistency conditions. 

\subsection{Baryon masses with SU(3) breaking operators}
\label{groundstatemasssu3breaking}
The SU(3) symmetry is not exact because the mass of the different light flavor quarks are different
\begin{equation}
 m_u\neq m_d \neq m_s.
\end{equation}
This means that the baryon mass operator can not transform as a spin-flavor singlet when the SU(3) symmetry is broken. The dominant perturbation transforms as $(1,8)$ \cite{jenkins95},
\begin{equation}
 M^{(1,8)} = \sum_{n=1}^{N_c}d_n\frac{1}{N_c^{n-1}}\mathcal{O}^a_n,
\end{equation}
where $d_n$ are unknown coefficients and $\mathcal{O}^a_n$ are products of SU(6) generators  with one free flavor index.
There are two different $(1,8)$ operators which are relevant for the analysis of the baryon mass splitting \cite{jenkins95}:  $\mathcal{O}^8$ which is isospin symmetric\footnote{The isospin symmetry assumes that $m_u=m_d$.} and $\mathcal{O}^3$ which breaks the isospin symmetry. Indeed, if we look at matrix elements of $T^8$ and $T^3$, one has (see Chapter 4):
 \begin{eqnarray}
  T^8 & = & \frac{1}{2\sqrt{3}}\left(N_c-3N_s\right), \\
  T^3 & = & \frac{1}{2}\left(N_u-N_d\right), 
 \end{eqnarray}
with $N_c=N_u+N_d+N_s$, $N_u,\ N_d,\ N_s$ correspond to the number of $u,\ d$ and $s$ inside the baryon.  
 By applying operator identities, one has
\begin{equation}
 M^{(1,8)}=\varepsilon d_1 T^8 + \varepsilon d_2 \frac{S^iG^{i8}}{N_c} + \varepsilon d_3 \frac{S^2T^8}{N_c^2} + \varepsilon d_4 \frac{S^2S^iG^{i8}}{N_c^3} + \varepsilon d_5 \frac{S^4T^8}{N_c^4}+\cdots,
\end{equation}
where  the isospin symmetry is conserved. The quantity $\varepsilon \sim 0.3$ measures the SU(3) breaking \cite{jenkins95}.
One can consider second order SU(3) breaking terms which transform as $(1,27)$ and which are proportional to $\varepsilon^2$. We will not discuss these terms here. \\

The $1/N_c$ expansion of the ground-state baryon mass operator becomes
\begin{eqnarray}
 M & = & M^{(1,1)}+M^{(1,8)} \nonumber \\
   & = & c_0N_c \mathbbm{1} + c_2 \frac{1}{N_c} S^2 +\varepsilon d_1 T^8 + \varepsilon d_2 \frac{S^iG^{i8}}{N_c} + \varepsilon d_3 \frac{S^2T^8}{N_c^2} +\mathcal{O}\left(\frac{1}{N_c^3}\right),
\end{eqnarray}
with SU(3) breaking to  first order.


\chapter{Matrix elements of SU(6) generators for a symmetric wave function}

\thispagestyle{empty}

\section{Introduction}

The study of the ground or excited state baryons requires the knowledge of matrix elements of SU($2N_f$) generators, as necessary ingredients in obtaining the mass formula. For the ground-state baryons the spin-flavor part of the wave function is symmetric. In a simplified version (see Chapter 5) the description of excited states also reduces to a symmetric state coupled to a single excited quark. If we consider the non-strange and strange baryons, we need to know the matrix elements of the SU(6) generators for a symmetric spin-flavor wave function and arbitrary $N_c$. The purpose of this chapter is to derive these matrix elements. So far, only the SU(4) case was solved analytically \cite{hecht69}. 
Here we present the approach we have developed in  Ref. \cite{matagne06a}. The results, summarized in Table \ref{su6isoscalar}, will be used in the following chapter. The SU(4) case was easily rederived within the same method. The results are given in Table \ref{genesu(4)sym}. \\

There are several ways to calculate the matrix elements of the SU(6) generators.
One is the standard group theory method. It is the way Hecht and Pang
\cite{hecht69} derived matrix elements of the SU(4) generators and it can straightforwardly 
be generalized
to SU(6). The difficulty in using this method is that it involves 
the knowledge of  isoscalar factors of SU(6). So far, the literature
provides a few examples of isoscalar factors:
${\bf 56} \times {\bf 35}  \rightarrow {\bf 56}$ \cite{carter65,schulke65},
${\bf 35} \times {\bf 35}  \rightarrow {\bf 35}$ \cite{schulke65} or
${\bf 35} \times {\bf 70} = {\bf 20}+{\bf 56}+2 \times {\bf 70}+{\bf 540}
+{\bf 560}+{\bf 1134}$ \cite{carter69}
which can be applied to 
baryons composed of three quarks or to pentaquarks. \\

Here we propose an alternative method, 
based on the decomposition of an SU(6) state into a product of
SU(3) and SU(2) states. It involves the knowledge of isoscalar factors of
the permutation group S$_n$ for a
baryon with an arbitrary number $N_c$ of quarks. As we shall see,
these isoscalar factors can easily be derived in various ways. \\

We recall that the group SU(6) has 35 generators ${S_i,T_a,G_{ia}}$
with $i = 1,2,3$ and $a = 1,2,\ldots,8$, where $S_i$ are the generators 
of the spin subgroup SU(2) and $T_a$ the generators of the flavor 
subgroup SU(3). The group algebra is
\begin{eqnarray}\label{ALGEBRA}
[S_i,S_j] & = & i \varepsilon_{ijk} S_k,
\ \ \ [T_a,T_b]  =  i f_{abc} T_c, \ \ \ [S_i,T_a]  =  0,\nonumber \\
\lbrack S_i,G_{ia}\rbrack & = & i\varepsilon_{ijk}G_{ka}, \ \ \ [T_a,G_{ib}]=if_{abc}G_{ic}, \nonumber \\
\lbrack G_{ia},G_{jb}\rbrack & = & \frac{i}{4} \delta_{ij} f_{abc} T_c
+\frac{i}{2} \varepsilon_{ijk}\left(\frac{1}{3}\delta_{ab} S_k +d_{abc} G_{kc}\right),
\end{eqnarray}
by which the normalization of the generators is fixed. It is a particular case of Eq. (\ref{su(2nf)algebra)}) for $N_f=3$. 

\section{SU(6) generators as tensor operators}

\label{SU(6) generators as tensor operators}
The SU(6) generators are the components of an irreducible tensor operator which transform
according to the adjoint representation $[21^4]$, equivalent to 
${\bf 35}$, in dimensional notation.   
The matrix elements of any irreducible tensor can be expressed in
terms of a generalized Wigner-Eckart theorem which is a factorization
theorem, involving the product between a reduced matrix element and 
a Clebsch-Gordan (CG) coefficient. 
The case SU(4) $\supset$ SU(2) $\times$ SU(2)
has been worked out by Hecht and Pang \cite{hecht69} and applied to
nuclear physics. \\

Let us consider that the tensor operator $[21^4]$
acts on an SU(6) state of symmetry $[f]$. The symmetry of the final state,
denoted by $[f']$, labels one of the irreps appearing in the
Clebsch-Gordan  series
\begin{equation}\label{CGS}
[f] \times [21^4] = \sum_{[f']} m_{[f']}[f'],
\end{equation}
where $m_{[f']}$ denotes the multiplicity of the irrep $[f']$.
The multiplicity problem arises if $[f'] = [f]$.
An extra label $\rho$ is then necessary. 
It is not the case here in connection with SU(6). Indeed, if $N_c = 3$,
for the symmetric state  ${\bf 56}$, one has
${\bf 56} \times {\bf 35} \rightarrow {\bf 56}$ with multiplicity 1.
For arbitrary $N_c$ and $[f] = [N_c]$ the reduction (\ref{CGS}) in terms of Young diagrams reads
\begin{eqnarray}
\overbrace{\,\raisebox{-3.0pt}{\drawsquare{10.0}{0.4}}\hskip-0.4pt
        \raisebox{-3.0pt}{\drawsquare{10.0}{0.4}}\hskip-0.4pt
        \raisebox{-3.0pt}{\drawsquare{10.0}{0.4}}\, \raisebox{-1pt}{\mbox{$\cdots$}}\,
        \raisebox{-3.0pt}{\drawsquare{10.0}{0.4}}\,}^{N_c}  \times \
\raisebox{-23.0pt}{\drawsquare{10.0}{0.4}}\hskip-10.4pt
\raisebox{-13.0pt}{\drawsquare{10.0}{0.4}}\hskip-10.4pt
\raisebox{-3.0pt}{\drawsquare{10.0}{0.4}}\hskip-10.4pt
\raisebox{7.0pt}{\drawsquare{10.0}{0.4}}\hskip-10.4pt
        \raisebox{17pt}{\drawsquare{10.0}{0.4}}\hskip-0.4pt
        \raisebox{17pt}{\drawsquare{10.0}{0.4}} & = &
\overbrace{\,\raisebox{-3.0pt}{\drawsquare{10.0}{0.4}}\hskip-0.4pt
        \raisebox{-3.0pt}{\drawsquare{10.0}{0.4}}\, \raisebox{-1pt}{\mbox{$\cdots$}}\,
	\raisebox{-3.0pt}{\drawsquare{10.0}{0.4}}\hskip-0.4pt
        \raisebox{-3.0pt}{\drawsquare{10.0}{0.4}}\,}^{N_c} + \
\overbrace{\,\raisebox{-8.0pt}{\drawsquare{10.0}{0.4}}\hskip-10.4pt
        \raisebox{2.0pt}{\drawsquare{10.0}{0.4}}\hskip-0.4pt
        \raisebox{2.0pt}{\drawsquare{10.0}{0.4}}\, \raisebox{4pt}{\mbox{$\cdots$}} \,
        \raisebox{2.0pt}{\drawsquare{10.0}{0.4}}\,}^{N_c-1}  \nonumber \\
	& & + \ \overbrace{\raisebox{-23.0pt}{\drawsquare{10.0}{0.4}}\hskip-10.4pt
\raisebox{-13.0pt}{\drawsquare{10.0}{0.4}}\hskip-10.4pt
\raisebox{-3.0pt}{\drawsquare{10.0}{0.4}}\hskip-10.4pt
\raisebox{7.0pt}{\drawsquare{10.0}{0.4}}\hskip-0.4pt
\raisebox{7.0pt}{\drawsquare{10.0}{0.4}}\hskip-20.4pt
        \raisebox{17pt}{\drawsquare{10.0}{0.4}}\hskip-0.4pt
        \raisebox{17pt}{\drawsquare{10.0}{0.4}}\hskip-0.4pt
        \raisebox{17pt}{\drawsquare{10.0}{0.4}}\, \raisebox{19pt}{\mbox{$\cdots$}} \,
	\raisebox{17pt}{\drawsquare{10.0}{0.4}}\hskip-0.4pt
        \raisebox{17pt}{\drawsquare{10.0}{0.4}}\,}^{N_c+1} + \
\overbrace{\raisebox{-23.0pt}{\drawsquare{10.0}{0.4}}\hskip-10.4pt
\raisebox{-13.0pt}{\drawsquare{10.0}{0.4}}\hskip-10.4pt
\raisebox{-3.0pt}{\drawsquare{10.0}{0.4}}\hskip-10.4pt
\raisebox{7.0pt}{\drawsquare{10.0}{0.4}}\hskip-10.4pt
        \raisebox{17pt}{\drawsquare{10.0}{0.4}}\hskip-0.4pt
        \raisebox{17pt}{\drawsquare{10.0}{0.4}}\hskip-0.4pt
        \raisebox{17pt}{\drawsquare{10.0}{0.4}}\, \raisebox{19pt}{\mbox{$\cdots$}} \,
	\raisebox{17pt}{\drawsquare{10.0}{0.4}}\hskip-0.4pt
        \raisebox{17pt}{\drawsquare{10.0}{0.4}}\hskip-0.4pt
        \raisebox{17pt}{\drawsquare{10.0}{0.4}}}^{N_c+2},
\end{eqnarray}
which gives $m_{[f']} = 1$ for all terms, including the case $[f'] = [f]$.
But the multiplicity problem arises at the level of the subgroup
SU(3). Introducing the  SU(3) $\times$ SU(2) content of the ${\bf 56}$ and
${\bf 35}$ irreps into their direct product one finds that the product
${\bf 8} \times {\bf 8}$ appears twice. For arbitrary $N_c$, this product is given by
\begin{eqnarray}
\overbrace{\raisebox{-8.0pt}{\drawsquare{10.0}{0.4}}\hskip-10.4pt
        \raisebox{2pt}{\drawsquare{10.0}{0.4}}\hskip-0.4pt
\raisebox{-8.0pt}{\drawsquare{10.0}{0.4}}\hskip-10.4pt
        \raisebox{2pt}{\drawsquare{10.0}{0.4}}\hskip-0.4pt
\raisebox{-8.0pt}{\drawsquare{10.0}{0.4}}\hskip-10.4pt
        \raisebox{2pt}{\drawsquare{10.0}{0.4}}
\, \raisebox{0pt}{\mbox{$\cdots$}} \,
\raisebox{-8.0pt}{\drawsquare{10.0}{0.4}}\hskip-10.4pt
        \raisebox{2pt}{\drawsquare{10.0}{0.4}}\hskip-0.4pt
        \raisebox{2pt}{\drawsquare{10.0}{0.4}}\hskip-0.4pt}^{\frac{N_c+1}{2}}\, \times \
\raisebox{-8.0pt}{\drawsquare{10.0}{0.4}}\hskip-10.4pt
        \raisebox{2pt}{\drawsquare{10.0}{0.4}}\hskip-0.4pt
        \raisebox{2pt}{\drawsquare{10.0}{0.4}}\hskip-0.4pt \, & = &
\overbrace{\raisebox{-8.0pt}{\drawsquare{10.0}{0.4}}\hskip-10.4pt
        \raisebox{2pt}{\drawsquare{10.0}{0.4}}\,
\raisebox{0pt}{\mbox{$\cdots$}} \,
\raisebox{-8.0pt}{\drawsquare{10.0}{0.4}}\hskip-10.4pt
        \raisebox{2pt}{\drawsquare{10.0}{0.4}}\hskip-0.4pt
        \raisebox{2pt}{\drawsquare{10.0}{0.4}}\hskip-0.4pt
        \raisebox{2pt}{\drawsquare{10.0}{0.4}}\hskip-0.4pt}^{\frac{N_c-1}{2}}\,
	+ \
\overbrace{\raisebox{-8.0pt}{\drawsquare{10.0}{0.4}}\hskip-10.4pt
        \raisebox{2pt}{\drawsquare{10.0}{0.4}}\hskip-0.4pt
\raisebox{-8.0pt}{\drawsquare{10.0}{0.4}}\hskip-10.4pt
        \raisebox{2pt}{\drawsquare{10.0}{0.4}}\hskip-0.4pt
\raisebox{-8.0pt}{\drawsquare{10.0}{0.4}}\hskip-10.4pt
        \raisebox{2pt}{\drawsquare{10.0}{0.4}}\,
\raisebox{0pt}{\mbox{$\cdots$}} \,
\raisebox{-8.0pt}{\drawsquare{10.0}{0.4}}\hskip-10.4pt
        \raisebox{2pt}{\drawsquare{10.0}{0.4}}\hskip-0.4pt
\raisebox{-8.0pt}{\drawsquare{10.0}{0.4}}\hskip-10.4pt
        \raisebox{2pt}{\drawsquare{10.0}{0.4}}\hskip-0.4pt
        \raisebox{2pt}{\drawsquare{10.0}{0.4}}\hskip-0.4pt
        \raisebox{2pt}{\drawsquare{10.0}{0.4}}\hskip-0.4pt}^{\frac{N_c+5}{2}}\,  \nonumber \\
 & & + \overbrace{\raisebox{-8.0pt}{\drawsquare{10.0}{0.4}}\hskip-10.4pt
        \raisebox{2pt}{\drawsquare{10.0}{0.4}}\,
\raisebox{0pt}{\mbox{$\cdots$}} \,
\raisebox{-8.0pt}{\drawsquare{10.0}{0.4}}\hskip-10.4pt
        \raisebox{2pt}{\drawsquare{10.0}{0.4}}\hskip-0.4pt
\raisebox{-8.0pt}{\drawsquare{10.0}{0.4}}\hskip-10.4pt
        \raisebox{2pt}{\drawsquare{10.0}{0.4}}}^{\frac{N_c-3}{2}}\, + \
\overbrace{\raisebox{-8.0pt}{\drawsquare{10.0}{0.4}}\hskip-10.4pt      \raisebox{2pt}{\drawsquare{10.0}{0.4}}\hskip-0.4pt
\raisebox{-8.0pt}{\drawsquare{10.0}{0.4}}\hskip-10.4pt
        \raisebox{2pt}{\drawsquare{10.0}{0.4}}\hskip-0.4pt
\raisebox{-8.0pt}{\drawsquare{10.0}{0.4}}\hskip-10.4pt
        \raisebox{2pt}{\drawsquare{10.0}{0.4}}\,
\raisebox{0pt}{\mbox{$\cdots$}} \,
\raisebox{-8.0pt}{\drawsquare{10.0}{0.4}}\hskip-10.4pt
        \raisebox{2pt}{\drawsquare{10.0}{0.4}}\hskip-0.4pt
\raisebox{-8.0pt}{\drawsquare{10.0}{0.4}}\hskip-10.4pt
        \raisebox{2pt}{\drawsquare{10.0}{0.4}}\hskip-0.4pt
\raisebox{-8.0pt}{\drawsquare{10.0}{0.4}}\hskip-10.4pt
        \raisebox{2pt}{\drawsquare{10.0}{0.4}}}^{\frac{N_c+3}{2}}\,  \nonumber \\
& & + \left(\,\,
\overbrace{\raisebox{-8.0pt}{\drawsquare{10.0}{0.4}}\hskip-10.4pt
        \raisebox{2pt}{\drawsquare{10.0}{0.4}}\hskip-0.4pt
\raisebox{-8.0pt}{\drawsquare{10.0}{0.4}}\hskip-10.4pt
        \raisebox{2pt}{\drawsquare{10.0}{0.4}}\,
\raisebox{0pt}{\mbox{$\cdots$}} \,
\raisebox{-8.0pt}{\drawsquare{10.0}{0.4}}\hskip-10.4pt
        \raisebox{2pt}{\drawsquare{10.0}{0.4}}\hskip-0.4pt
\raisebox{-8.0pt}{\drawsquare{10.0}{0.4}}\hskip-10.4pt
        \raisebox{2pt}{\drawsquare{10.0}{0.4}}\hskip-0.4pt
        \raisebox{2pt}{\drawsquare{10.0}{0.4}}\hskip-0.4pt}^{\frac{N_c+1}{2}}\,\right)_1\, + \
 \left(\,\,\overbrace{\raisebox{-8.0pt}{\drawsquare{10.0}{0.4}}\hskip-10.4pt
        \raisebox{2pt}{\drawsquare{10.0}{0.4}}\hskip-0.4pt
\raisebox{-8.0pt}{\drawsquare{10.0}{0.4}}\hskip-10.4pt
        \raisebox{2pt}{\drawsquare{10.0}{0.4}}\,
\raisebox{0pt}{\mbox{$\cdots$}} \,
\raisebox{-8.0pt}{\drawsquare{10.0}{0.4}}\hskip-10.4pt
        \raisebox{2pt}{\drawsquare{10.0}{0.4}}\hskip-0.4pt
\raisebox{-8.0pt}{\drawsquare{10.0}{0.4}}\hskip-10.4pt
        \raisebox{2pt}{\drawsquare{10.0}{0.4}}\hskip-0.4pt
        \raisebox{2pt}{\drawsquare{10.0}{0.4}}\hskip-0.4pt}^{\frac{N_c+1}{2}}\,\right)_2\, \nonumber \\
& & + \
\overbrace{\raisebox{-8.0pt}{\drawsquare{10.0}{0.4}}\hskip-10.4pt
        \raisebox{2pt}{\drawsquare{10.0}{0.4}}\hskip-0.4pt
\raisebox{-8.0pt}{\drawsquare{10.0}{0.4}}\hskip-10.4pt
        \raisebox{2pt}{\drawsquare{10.0}{0.4}} \,
\raisebox{0pt}{\mbox{$\cdots$}} \,
\raisebox{-8.0pt}{\drawsquare{10.0}{0.4}}\hskip-10.4pt
        \raisebox{2pt}{\drawsquare{10.0}{0.4}}\hskip-0.4pt
        \raisebox{2pt}{\drawsquare{10.0}{0.4}}\hskip-0.4pt
        \raisebox{2pt}{\drawsquare{10.0}{0.4}}\hskip-0.4pt
        \raisebox{2pt}{\drawsquare{10.0}{0.4}}}^{\frac{N_c+3}{2}}\,.
\end{eqnarray}
In the following, $\rho$ is used to distinguish between
various  ${\bf 8} \times {\bf 8}$ products. Then this label is carried
over by the isoscalar factors of SU(6) (see below). \\

In particle physics one uses the label  $s$ for the symmetric
and $a$ for the antisymmetric product (see \emph{e.g.} \cite{carter65}).
Here we shall use the notation $\rho$ of Hecht \cite{hecht65}. For $N_c$ = 3
the relation to other labels is: $\rho$ = 1 corresponds to $(8 \times 8)_a$
or to $(8 \times 8)_2$ of De Swart \cite{deswart63}; $\rho$ = 2 corresponds to
$(8 \times 8)_s$ or to  $(8 \times 8)_1$ of De Swart. Throughout the paper,
we shall use the SU(3) notations and phase conventions of Hecht \cite{hecht65}.
Accordingly, an irreducible representation of SU(3) carries the label $(\lambda \mu)$,
 introduced by Elliott \cite{elliot58}, who applied SU(3) for the first
time in physics, to describe rotational bands of deformed nuclei \cite{stancu96}. In particle physics the corresponding notation is $(p,q)$.
By analogy to SU(4)  \cite{hecht69} one can write
the  matrix elements of the SU(6) generators $E_{ia}$ as
{\small
\begin{eqnarray}\label{GEN}
\langle [N_c](\lambda' \mu') Y' I' I'_3 S' S'_3 | E_{ia} |
[N_c](\lambda \mu) Y I I_3 S S_3 \rangle = \sqrt{C^{[N_c]}(\mathrm{SU(6)})}  
  \left(\begin{array}{cc|c}
	S   &    S^i   & S'   \\
	S_3  &   S^i_3   & S'_3
  \end{array}\right)\nonumber \\
  \times     \left(\begin{array}{cc|c}
	I   &   I^a   & I'   \\
	I_3 &   I^a_{3}   & I'_3
   \end{array}\right)  
      \sum_{\rho = 1,2}
 \left(\begin{array}{cc||c}
	(\lambda \mu)    &  (\lambda^a\mu^a)   &   (\lambda' \mu')\\
	Y I   &  Y^a I^a  &  Y' I'
      \end{array}\right)_{\rho}
\left(\begin{array}{cc||c}
	[N_c]    &  [21^4]   & [N_c]   \\
	(\lambda \mu) S  &  (\lambda^a\mu^a) S^i  &  (\lambda' \mu') S'
      \end{array}\right)_{\rho} ,  
   \end{eqnarray}}
   
\noindent where $C^{[N_c]}(\mathrm{SU(6)}) = 5[N_c(N_c+6)]/12$ is the Casimir operator of SU(6), followed by CG
coefficients of SU(2)-spin and SU(2)-isospin. The sum over $\rho$ is over
terms containing products of isoscalar factors of SU(3) and SU(6) respectively.
We introduce $T_a$ as an SU(3) irreducible tensor operator of components
$T^{(11)}_{Y^aI^a}$. It is a scalar in
SU(2) so that the index $i$ is no more necessary. The generators $S_i$
form a rank 1 tensor in SU(2) which is a scalar in SU(3), so the index $i$ suffices.
Although we use the same symbol for the operator $S_i$ and its quantum numbers we hope that no confusion is created.
The relation with the algebra (\ref{ALGEBRA}) is
\begin{equation} \label{normes}
E_i =\frac{ S_i}{\sqrt{3}};~~~ E_a = \frac{T_a}{\sqrt{2}}; ~~~E_{ia} = \sqrt{2} G_{ia} .
\end{equation}
Thus, for the generators $S_i$ and $T_a$, which are elements of the $su$(2) and
$su$(3) subalgebras of (\ref{ALGEBRA}), the above expression
simplifies considerably. In particular, as  $S_i$  acts only on the
spin part of the wave function, we apply the usual
Wigner-Eckart theorem for SU(2) to get
\begin{eqnarray}\label{SPIN}
\langle [N_c](\lambda'\mu') Y' I' I'_3; S' S'_3 |S_i|
[N_c](\lambda \mu) Y I I_3; S S_3 \rangle =  \delta_{SS'}\delta_{\lambda \lambda'} \delta_{\mu\mu'} \delta_{YY'} \delta_{II'} \delta_{I_3I_3'}\nonumber \\
  \times \sqrt{C(\mathrm{SU(2)})} \left(\begin{array}{cc|c}
	S   &    1  &  S'   \\
	S_3 &    i  &  S'_3
      \end{array}\right),
   \end{eqnarray}
with $C(\mathrm{SU(2)}) = S(S+1)$.
Similarly, we use the Wigner-Eckart theorem for $T_a$ which is a generator of 
the subgroup SU(3)
\begin{eqnarray}\label{FLAVOR}
\langle [N_c](\lambda'\mu') Y' I' I'_3; S' S'_3 |T_a|
[N_c](\lambda \mu) Y I I_3; S S_3 \rangle =
\delta_{SS'} \delta_{S_3S'_3}\delta_{\lambda \lambda'} \delta_{\mu\mu'}
\nonumber \\
\times
\sum_{\rho = 1,2}
\langle (\lambda'\mu') || T^{(11)} || (\lambda \mu) \rangle_{\rho}
  \left(\begin{array}{cc|c}
	(\lambda \mu)    &  (11)   &   (\lambda'\mu')\\
	YII_3   &  Y^aI^aI^a_{3}  &  Y' I' I'_3
      \end{array}\right)_{\rho}, 
   \end{eqnarray}
where the reduced matrix element is defined as  \cite{hecht69} 
\begin{eqnarray}\label{REDUCED}
\langle (\lambda \mu) || T^{(11)} || (\lambda \mu) \rangle_{\rho} = \left\{
\begin{array}{cc}
\sqrt{C(\mathrm{SU(3)})}      & \mathrm{for}\ \rho = 1 \\
0 & \mathrm{for}\ \rho = 2 \\
\end{array}\right.
,\end{eqnarray}   
in terms of the eigenvalue of the Casimir operator       
$C(\mathrm{SU(3)}) = \frac{1}{3} g_{\lambda \mu}$ where
\begin{equation}\label{CSU3}
g_{\lambda\mu}= {\lambda}^2+{\mu}^2+\lambda\mu+3\lambda+3\mu.
\end{equation}
The SU(3) CG coefficient factorizes 
into an SU(2)-isospin CG coefficient and an SU(3) isoscalar factor
\cite{deswart63}
\begin{equation}\label{CGSU3}
\left(\begin{array}{cc|c}
	(\lambda \mu)    &  (11)   &   (\lambda'\mu')\\
	YII_3   &  Y^aI^aI^a_{3}  &  Y' I' I'_3
      \end{array}\right)_{\rho} =
\left(\begin{array}{cc|c}
	I   &    I^a  &  I'   \\
	I_3 &    I^a_3  &  I'_3
     \end{array}\right)
 \left(\begin{array}{cc||c}
	(\lambda \mu)    &  (11)   &   (\lambda'\mu')\\
 	 YI   &  Y^aI^a  &  Y' I'
      \end{array}\right)_{\rho}.
 \end{equation}   
The $\rho$ dependence is consistent with Eq. (\ref{GEN}) and reflects the 
multiplicity problem appearing in Eq. (\ref{PROD}) below. 
We shall return to this point in below. 

\section{SU(6) symmetric  wave functions}
\label{SU(6) symmetric  wave functions}
Here we consider a wave function which is symmetric in the spin-flavor space. To write
its decomposition into its SU(2)-spin and SU(3)-flavor  parts,
one can use the Kronecker or
inner product of the permutation group S$_n$. The advantage is that one
can treat the permutation symmetry separately in each degree of freedom
\cite{stancu96}.  A basis vector $|[f] Y \rangle $ of an irreducible representation of  S$_n$
is completely defined by the partition $[f]$, and by
a Young tableau $Y$ or its equivalent, an Yamanouchi symbol. In the following 
we do not need to specify the full Young tableau, we only need to know 
the position $p$ of the last particle in each tableau.
In this short-hand notation a symmetric state of $N_c$ quarks is
$|[N_c]1] \rangle$, because $p = 1$. A symmetric spin-flavor wave function can 
be obtained from the product $[f'] \times [f'']$  of spin and flavor states
of symmetries $[f']$ and $[f'']$ respectively, provided  $[f']$ = $[f'']$. \\

Let us consider a system of $N_c$ quarks having a total spin $S$.
The group SU(2) allows only partitions with maximum two rows, in this case
with  $N_c/2 + S$ boxes in the first row and $N_c/2 - S$
in the second row. So, one has
\begin{equation}\label{FPRIM}
[f'] = [\frac{N_c}{2} + S,\frac{N_c}{2} - S].
\end{equation}
By using the CG coefficients of S$_n$ and
their factorization property, described in the Appendix \ref{Isoscalar factors of the permutation group S$_n$}, one can write
a symmetric state 
of $N_c$ particles with spin $S$ as the linear combination
\begin{eqnarray}\label{FS}
|[N_c ] 1\rangle 
&=& c^{[N_c]}_{11}(S)| [f'] 1 \rangle | [f'] 1 \rangle
  + c^{[N_c]}_{22}(S)| [f'] 2 \rangle | [f'] 2 \rangle,
\end{eqnarray} 
where the coefficients $c^{\mathrm{[N_c]}}_{pp} (p = 1, 2)$ in the right-hand side are
isoscalar factors of the permutation group. 
Their meaning is the following.
The square of the first (second) coefficient is the fraction of Young tableaux
of symmetry $[f']$ having the last 
particle of both states $|[f']p\rangle$ in the first (second) row. That is why they carry the
double index $11$ and $22$ respectively, one index for each state.
Examples of such isoscalar factors can be found in Ref. \cite{stancu99}.  In the following, the first index refers to the spin part of the wave function and the second index to the flavor part.
The total number of Young
tableaux gives the dimension of the irrep
$[f']$,  so that the sum of squares of the two isoscalar
factors is equal to one.  \\

In the context of SU(6) $\supset$ SU(2) $\times$ SU(3) there are two alternative forms of each $c^{[N_c]}_{pp}$. They are
also derived in Appendix \ref{Isoscalar factors of the permutation group S$_n$}. The first form is
\begin{eqnarray}\label{SU2}
c^{[N_c]}_{11}(S) & = &  \sqrt{\frac{S[N_c+2(S + 1)]}{N_c(2 S + 1)}}, \nonumber \\
c^{[N_c]}_{22}(S) &  =  & \sqrt{\frac{(S + 1)(N_c - 2 S)}{N_c(2 S + 1)}}.
\end{eqnarray}
These expressions were
obtained by acting with $S_i$ on the spin part of the total wave function and by
calculating the matrix elements of  $S_i$ in two different ways, one
involving the Wigner-Eckart theorem and the other the linear combination
(\ref{FS}).
The coefficients (\ref{SU2}) are precisely the so called ``elements of orthogonal
basis rotation" of Refs. \cite{carlson98b} with the identification
$ c^{[N_c]}_{11} = c^{\mathrm{SYM}}_{0-}$ and
$ c^{[N_c]}_{22} = c^{\mathrm{SYM}}_{0+}$.
The other form of the same coefficients, obtained by acting
with $T_a$ on the flavor part of the total wave function is
\begin{eqnarray}\label{SU3}
c^{[N_c]}_{11} (\lambda \mu) & = & \sqrt{\frac{2g_{\lambda\mu}-N_c(\mu-\lambda+3)}{3N_c(\lambda +1)}},      \nonumber \\
c^{[N_c]}_{22} (\lambda \mu) & = & \sqrt{\frac{N_c(6+2\lambda+\mu)-2g_{\lambda\mu}}{3N_c(\lambda+1)}},
\end{eqnarray}
with $g_{\lambda\mu}$ given by Eq. (\ref{CSU3}).  One can use either form, (\ref{SU2}) or (\ref{SU3}),
depending on the  SU(2) or the SU(3) context of the quantity to calculate.
The best is to use the version which leads to simplifications.
One can easily see that the expressions (\ref{SU2}) and (\ref{SU3})
are equivalent to each other. 
By making the replacement $\lambda = 2 S$ and $\mu = N_c/2-S$ in
(\ref{SU3}) one obtains (\ref{SU2}).\\

The coefficients
$c^{\mathrm{MS}}_{0+}$ and $c^{\mathrm{MS}}_{0-}$ of Refs. \cite{carlson98b}, are
also isoscalar factors of the permutation group. They are 
needed to construct a mixed-symmetric state from the  inner product
of S$_n$ which generated the symmetric state as well. As shown in Appendix \ref{Isoscalar factors of the permutation group S$_n$}, they can be obtained from orthogonality relations. The identification is
$ c^{[N_c-1,1]}_{11} = c^{\mathrm{MS}}_{0-}$ and $ c^{[N_c-1,1]}_{22} = c^{\mathrm{MS}}_{0+}$.
There are also the coefficients  $ c^{[N_c-1,1]}_{12} = c^{\mathrm{MS}}_{++}$ and
$ c^{[N_c-1,1]}_{21} = c^{\mathrm{MS}}_{--}$.

\section{Matrix elements of the SU(6) generators}

Besides the standard group theory method of Sec. \ref{SU(6) generators as tensor operators}, another method to
calculate the matrix elements of the SU(6) generators
is based on the 
decoupling of the last particle from the 
rest, in each part of the wave function. This is easily done
inasmuch as the row $p$ of the last particle in a Young tableau
is specified.  \\

Let us first consider the spin part. The decoupling is
\begin{equation}\label{statessu(2)}
|S_1 1/2; S S_3; p\rangle  =
\sum_{m_1,m_2}
 \left(\begin{array}{cc|c}
	S_1  &    1/2  &  S   \\
	m_1  &    m_2  &  S_3 
      \end{array}\right)
|S_1 m_1 \rangle |1/2 m_2 \rangle,
\end{equation}
in terms of an SU(2)-spin CG coefficient
with $S_1 = S - 1/2$ for $p = 1$ and  $S_1 = S + 1/2$ for $p = 2$.

As for the flavor part, a wave function of
symmetry $(\lambda \mu)$ with the last particle in the row $p$
decouples to
\begin{eqnarray}\label{statessu(3)}
\lefteqn{| (\lambda_1 \mu_1) (10); (\lambda \mu)Y I I_{3}; p \rangle  = } 
\nonumber \\ \sum_{Y_1,I_1,I_{1_3},Y_2,I_2,I_{2_3}}
& \left(\begin{array}{cc|c}
	(\lambda_1 \mu_1)    &   (10)  & (\lambda \mu) \\
	 Y_1I_1I_{1_3} &   Y_2I_2I_{2_3} &  YII_{3}
      \end{array}\right) 
|(\lambda_1 \mu_1) Y_1I_1I_{1_3}\rangle
|(10) Y_2I_2I_{2_3} \rangle,
\end{eqnarray}
where $(\lambda_1 \mu_1) = (\lambda - 1, \mu)$ for $p = 1$ and
$(\lambda_1 \mu_1) = (\lambda + 1, \mu - 1)$ for $p = 2$. \\

Now we use the fact that  $S_i$, $T_a$ and $G_{ia}$ are one-body operators, \emph{i.e.} their general form is
$$O = \sum_{i = 1}^{N_c} O(i).$$
The expectation value of $O$ between symmetric states is equal to $N_c$ times the expectation value of any $O(i)$. Taking $i = N_c$
one has
\begin{equation}\label{ONEB}
\langle O \rangle = N_c \langle O(N_c) \rangle.
\end{equation}
This means that one can  reduce the calculation   of $\langle O \rangle$ to  the calculation of $\langle O(N_c)\rangle$. \\

To proceed, we recall that the flavor part of $G_{ia}$ is a $T^{(11)}$ tensor in
SU(3). To find out its matrix elements we have to consider the direct product
\begin{eqnarray}\label{PROD}
\lefteqn{(\lambda \mu) \times (11)  =  (\lambda+1, \mu+1)+ (\lambda+2, \mu-1) +
(\lambda \mu)_1 + (\lambda \mu)_2}  \nonumber \\
& & + \, (\lambda-1, \mu+2) + (\lambda-2, \mu+1)
+ (\lambda+1, \mu-2)+ (\lambda-1, \mu-1).
\end{eqnarray}
From the right-hand side, the only terms which give non-vanishing
matrix elements of $G_{ia}$ are $(\lambda' \mu') = (\lambda \mu),
(\lambda+2,\mu-1)$ and $(\lambda-2,\mu+1)$, \emph{i.e.} those with the same
number of boxes, equal to $\lambda + 2\mu$, as on the left-hand side.
In addition, as $G_{ia}$ is a rank 1 tensor in SU(2) it has non-vanishing matrix
elements only for $S'= S, S\pm 1$, \emph{i.e.} again only three distinct
possibilities.
The proper combinations of flavor and spin
parts to get $|[N_c]1
 \rangle$ will be seen in the following subsections.

\subsection{Diagonal matrix elements of $G_{ia}$}

The diagonal matrix element have $(\lambda' \mu') = (\lambda \mu)$
and $S' = S$. 
The first step is to use the relation (\ref{ONEB}) and the
factorization (\ref{FS}) of the spin-flavor wave function into
its spin and flavor parts. This gives
\begin{eqnarray}\label{GIAD}
\lefteqn{\langle [N_c](\lambda \mu) Y' I' I'_3 S S'_3 | G_{ia} |
[N_c](\lambda \mu) Y I I_3 S S_3 \rangle =N_c}  \nonumber \\
& & \times \  \sum_{p = 1,2}
\left({c^{[N_c]}_{pp}(S)}\right)^2 \langle  S_11/2;SS'_3; p|s_i(N_c)| S_11/2;SS_3; p \rangle \nonumber \\
& & \times \langle (\lambda_1 \mu_1) (10);(\lambda \mu) Y'I'I'_3; p |t_a(N_c)| (\lambda_1 \mu_1) (10);(\lambda \mu) YII_3; p \rangle,
\end{eqnarray}
where $s_i$ and $t_a$
are the SU(2) and SU(3) generators of the $N_c$-th particle respectively.
The matrix elements of $s_i(N_c)$ between the states (\ref{statessu(2)}) are
\begin{eqnarray}\label{SI}
\lefteqn{\langle S_11/2;SS'_3; p
|s_i(N_c)|S_11/2;SS_3; p \rangle =}
 \nonumber \\
& & \sqrt{\frac{3}{4}}
 \sum_{m_1 m_2 m'_2} 
\left(\begin{array}{cc|c}
	S_1  &    1/2  &  S   \\
	m_1  &    m_2  &  S_3 
      \end{array}\right)
 \left(\begin{array}{cc|c}
	S_1  &    1/2  &  S   \\
	m_1  &    m'_2  &  S'_3 
      \end{array}\right) 
  \left(\begin{array}{cc|c}
	1/2  &    1  &  1/2   \\
	m_2  &    i  &  m'_2   
      \end{array}\right)  \nonumber \\
& &= (-)^{S+S_1-1/2}\sqrt{\frac{3}{2}~(2S+1)}
   \left(\begin{array}{cc|c}
	S   &    1  &   S   \\
	S_3  &   i  &   S'_3   
      \end{array}\right) 
    \left\{\begin{array}{ccc}
	1   &    S  &   S   \\
	S_1  &   1/2  & 1/2
      \end{array}\right\}.
\end{eqnarray} \\

The matrix elements of the single particle operator $t_a$ between the 
states (\ref{statessu(3)}) are
\begin{eqnarray}\label{TA}
\lefteqn{\langle (\lambda_1 \mu_1) (10); (\lambda \mu) Y' I' I'_3; p
|t_a(N_c)|(\lambda_1 \mu_1) (10); (\lambda \mu) Y I I_3 ; p \rangle
= } \nonumber \\
& & \sqrt{\frac{4}{3}}\sum_{Y_1I_1I_{1_3} Y_2I_2I_{2_3}  Y'_2I'_2I'_{2_3}}\left(\begin{array}{cc|c}
	(\lambda_1 \mu_1)    &  (10)   &   (\lambda \mu)\\
	Y_1I_1I_{1_3}   &  Y_2I_2I_{2_3}  &  YII_3
      \end{array}\right)
     \left(\begin{array}{cc|c}
	(\lambda_1 \mu_1)    &  (10)   &   (\lambda \mu)\\
	Y_1I_1I_{1_3}   &  Y'_2I'_2I'_{2_3}  &  Y' I' I'_3
      \end{array}\right) \nonumber \\
  & &    \times \left(\begin{array}{cc|c}
	(10)    &  (11)   &   (10)\\
	Y_2I_2I_{2_3}   &  Y^aI^aI^a_{3}  &  Y'_2 I'_2 I'_{2_3}
      \end{array}\right)
\nonumber \\
& & = \sqrt{\frac{4}{3}}
  \left(\begin{array}{cc|c}
	I   &    I^a    &   I'   \\
        I_3 &    I^a_3  &   I'_3    
      \end{array}\right)
\sum_{\rho=1,2}\left(\begin{array}{cc||c}
        (\lambda \mu) & (11) &  (\lambda \mu)  \\
	 YI   &    Y^aI^a    &   Y'I'
      \end{array}\right)_ {\rho}\nonumber \\
 & & \times\
 U((\lambda_1\mu_1)(10)(\lambda \mu)(11);(\lambda \mu)(10))_{\rho}\, ,
\end{eqnarray}
where $U$ are SU(3) Racah coefficients. To obtain the last equality in (\ref{TA})  we used the identity  
\begin{eqnarray}
\lefteqn{\sum_{\rho=1,2}\langle (\lambda\mu) Y I;(11)Y^aI^a||(\lambda\mu)Y'I'\rangle_{\rho} U((\lambda_1\mu_1)(10)(\lambda \mu)(11);(\lambda \mu)(10))_{\rho}=} \nonumber \\
& & \sum_{Y_1I_1 Y_2I_2  Y'_2I'_2}\langle (\lambda_1\mu_1) Y_1 I_1;(10)Y_2I_2||(\lambda\mu)YI\rangle   \langle (10) Y_2I_2;(11)Y^aI^a||(10)Y'_2I'_2\rangle \nonumber \\ 
& & \times \langle (\lambda_1\mu_1)Y_1I_1;(10)Y'_2I'_2||(\lambda\mu)Y'I'\rangle U(I_1I_2I'I^a;II'_2), 
\end{eqnarray}
similar to the relation (12) of Ref. \cite{hecht65}.
Note that the sum over $\rho$ is consistent with Eq. (\ref{GEN}) and expresses the fact that the direct product $(\lambda \mu)
\times (11) \rightarrow (\lambda \mu)$ has multiplicity 2 in the
reduction from SU(6) to SU(3), as discussed in Sec. \ref{SU(6) generators as tensor operators}. \\

Introducing (\ref{SI}) and (\ref{TA}) into (\ref{GIAD}) one obtains
\begin{eqnarray}\label{GIADFINAL}
\lefteqn{\langle [N_c](\lambda \mu) Y' I' I'_3 ; S S'_3 | G_{ia} |
[N_c](\lambda \mu) Y I I_3 ; S S_3 \rangle = (-)^{2S} N_c \sqrt{2(2 S + 1)}}
\nonumber \\
  & & \times
  \left(\begin{array}{cc|c}
	S   &   1   & S   \\
	S_3 &   i   & S'_3
      \end{array}\right)           
   \left(\begin{array}{cc|c}
	I   &   I^a   & I'   \\
	I_3 &   I^a_3   & I'_3
      \end{array}\right)
  \sum_{\rho = 1,2}
  \left(\begin{array}{cc||c}
	(\lambda \mu)    &  (11)   &   (\lambda \mu)\\
	Y I   &  Y^a I^a  &  Y' I'
      \end{array}\right)_{\rho} 
      \nonumber \\ 
 & & \times \left[\left({c^{[N_c]}_{22}(S)}\right)^2
    \left\{\begin{array}{ccc}
	S + 1/2   &   1/2   & S   \\
	1       &   S     & 1/2 
      \end{array}\right\}
      U((\lambda+1,\mu-1)(10)(\lambda \mu)(11);(\lambda \mu)(10))_{\rho}\right.
      \nonumber \\
   & &  \left. - \left({c^{[N_c]}_{11}(S)}\right)^2
    \left\{\begin{array}{ccc}
	S - 1/2   &   1/2   & S   \\
	1       &   S     & 1/2 
      \end{array}\right\} 
      U((\lambda-1,\mu)(10)(\lambda \mu)(11);(\lambda \mu)(10))_{\rho}\right],
 \end{eqnarray}
where $c^{[N_c]}_{pp}$ are given by Eqs. (\ref{SU2}) or by the equivalent
form (\ref{SU3}).
 Using the definition of $U$ given in Appendix \ref{su3racahandisoscalarfactors}, Tables \ref{vergados}--\ref{hecht4}   we have
obtained the following expressions
\begin{equation}
U((\lambda+1,\mu-1)(10)(\lambda \mu)(11);(\lambda \mu)(10))_{\rho=1} =
\frac{\mu - \lambda +3}{4 \sqrt{g_{\lambda \mu}}},
\end{equation}
\begin{equation}
U((\lambda-1,\mu)(10)(\lambda \mu)(11);(\lambda \mu)(10))_{\rho=1} =
\frac{\mu + 2 \lambda +6}{4 \sqrt{g_{\lambda \mu}}},
\end{equation}
\begin{equation}
U((\lambda+1,\mu-1)(10)(\lambda \mu)(11);(\lambda \mu)(10))_{\rho=2} =
\frac{1}{4} \sqrt{
\frac{3 \lambda (\mu+2)(\lambda+\mu+1)(\lambda+\mu+3)}
{\mu (\lambda+2) g_{\lambda \mu}}},
\end{equation}
\begin{equation}
U((\lambda-1,\mu)(10)(\lambda \mu)(11);(\lambda \mu)(10))_{\rho=2} = 
- \frac{1}{4} \sqrt{
\frac{3 (\lambda+2) \mu (\mu+2) (\lambda+\mu+3)}
{{\lambda (\lambda+\mu+1) g_{\lambda \mu}}}},
\end{equation}
where $g_{\lambda \mu}$ is given by the relation (\ref{CSU3}).

\subsection{Off-diagonal matrix elements of $G_{ia}$}

We have applied the procedure of the previous subsection to obtain
the off-diagonal matrix elements of $G_{ia}$ as well. As mentioned above,
there are only two types of non-vanishing matrix elements, those with 
$(\lambda' \mu') = (\lambda+2, \mu-1)$, $S' = S+1$ and those with  
 $(\lambda' \mu')= (\lambda-2, \mu+1)$, $S' = S-1$.
We found that they are given by
\begin{eqnarray}\label{OFF1}
\lefteqn{\langle [N_c](\lambda+2, \mu-1) Y' I' I'_3 ; S+1, S'_3 | G_{ia} |
[N_c](\lambda \mu) Y I I_3 ; S S_3 \rangle =} \nonumber \\ & &  (-)^{2S+1} N_c \sqrt{2(2 S + 1)} 
  \left(\begin{array}{cc|c}
	S   &   1   & S+1   \\
	S_3 &   i   & S'_3 
      \end{array}\right)           
   \left(\begin{array}{cc|c}
	I   &   I^a   & I'   \\
	I_3 &   I_3^a   & I'_3
      \end{array}\right) \nonumber \\ 
    & & \times 
  \left(\begin{array}{cc||c}
	(\lambda \mu)    &  (11)   &   (\lambda+2, \mu-1)\\
	Y I   &  Y^a I^a  &  Y' I'
      \end{array}\right)
     c^{[N_c]}_{11}(S+1)  c^{[N_c]}_{22}(S)
    \left\{\begin{array}{ccc}
	S + 1/2   &   1/2   & S   \\
	1       &   S+1     & 1/2 
      \end{array}\right\}  \nonumber \\ 
    & & \times\
      U((\lambda+1,\mu-1)(10)(\lambda+2, \mu-1)(11);(\lambda \mu)(10)),
 \end{eqnarray}
and 
 \begin{eqnarray}\label{OFF2}
\lefteqn{\langle [N_c](\lambda-2, \mu+1) Y' I' I'_3 ; S-1,S'_3 | G_{ia} |
[N_c](\lambda \mu) Y I I_3 ; S S_3 \rangle =} \nonumber \\ & & (-)^{2S} N_c \sqrt{2(2 S + 1)} 
  \left(\begin{array}{cc|c}
	S   &   1   & S-1   \\
	S_3 &   i   & S'_3 
      \end{array}\right)           
   \left(\begin{array}{cc|c}
	I   &   I^a   & I'   \\
	I_3 &   I_3^a   & I'_3
      \end{array}\right)  \nonumber \\
 & &  \times     
  \left(\begin{array}{cc||c}
	(\lambda \mu)    &  (11)   &   (\lambda-2, \mu+1)\\
	Y I   &  Y^a I^a  &  Y' I'
      \end{array}\right) 
     c^{[N_c]}_{22}(S-1)  c^{[N_c]}_{11}(S)
    \left\{\begin{array}{ccc}
	S - 1/2   &   1/2   & S   \\
	1       &   S-1     & 1/2 
      \end{array}\right\} \nonumber \\ 
   & &  \times \
      U((\lambda-1,\mu)(10)(\lambda-2,\mu+1)(11);(\lambda\mu)(10)).
 \end{eqnarray}
Note that the off-diagonal matrix elements do not contain a summation over
$\rho$ because in the right-hand side of the SU(3) product (\ref{PROD}) the terms
$(\lambda+2, \mu-1)$ and $(\lambda-2, \mu+1)$ appear with multiplicity 1. \\

The above expression require one of the following $U$ coefficients
\begin{equation}
U((\lambda+1,\mu-1)(10)(\lambda+2, \mu-1)(11);(\lambda \mu)(10)) =
\frac{1}{2} \sqrt{\frac{3 (\lambda + \mu + \lambda\mu +1)}
{2 \mu (\lambda + 2)}},
\end{equation}
and 
\begin{equation}
U((\lambda-1,\mu)(10)(\lambda-2, \mu+1)(11);(\lambda \mu)(10)) =
- \frac{1}{2} \sqrt{\frac{3 (\lambda + 1)(\lambda + \mu + 2)}
{2 \lambda (\lambda + \mu + 1)}}.
\end{equation} \\

As a practical application for $N_c = 3$, the off-diagonal matrix element
are needed to couple $^{4}8$ and $^{2}8$ baryon states, for example.

\section{Isoscalar factors of SU(6) generators for arbitrary $N_c$}

Here we derive analytic formulas for isoscalar factors related to matrix elements 
of SU(6) generators between symmetric states by comparing the definition (\ref{GEN}) 
with results from Sec. 4. In doing this, we have to replace $E_{ia}$ by the 
corresponding generators $S_i$, $T_a$ or $G_{ia}$ according to the 
relations (\ref{normes}). Then from the result (\ref{GIADFINAL}) we obtain the 
isoscalar factor of $G_{ia}$   for $(\lambda' \mu') = (\lambda \mu)$,
$S' = S$ as
\begin{eqnarray}
\lefteqn{\left(\begin{array}{cc||c}
	[N_c]    &  [21^4]   & [N_c]   \\
	(\lambda \mu) S  &  (11) 1  &  (\lambda \mu) S
      \end{array}\right)_{\rho} = N_c (-1)^{2S}
      \sqrt{\frac{4(2S+1)}{C^{[N_c]}(SU(6))}}}
      \nonumber \\
     & & \times
     \left[ \left({c^{[N_c]}_{22}(S)}\right)^2
    \left\{\begin{array}{ccc}
	S + 1/2   &   1/2   & S   \\
	1       &   S     & 1/2
      \end{array}\right\}
      U((\lambda+1,\mu-1)(10)(\lambda \mu)(11);(\lambda \mu)(10))_{\rho}\right.
      \nonumber \\
    & & - \left.\left({c^{[N_c]}_{11}(S)}\right)^2
    \left\{\begin{array}{ccc}
	S - 1/2   &   1/2   & S   \\
	1       &   S     & 1/2
      \end{array}\right\}
      U((\lambda-1,\mu)(10)(\lambda \mu)(11);(\lambda \mu)(10))_{\rho} \right].
    \end{eqnarray}
Similarly, but using the formula (\ref{OFF2})
we obtain the isoscalar factors for
$(\lambda' \mu') \neq (\lambda \mu)$ $ S' \neq S$. These are
\begin{eqnarray}
\lefteqn{\left(\begin{array}{cc||c}
	[N_c]    &  [21^4]   & [N_c]   \\
	(\lambda \mu) S  &  (11) 1  &  (\lambda+2, \mu-1) S+1
      \end{array}\right) = N_c (-1)^{2S+1}
      \sqrt{\frac{4(2S+1)}{C^{[N_c]}(SU(6))}}}\nonumber \\
    & & \times \ c^{[N_c]}_{11}(S+1)~c^{[N_c]}_{22}(S)\left\{\begin{array}{ccc}
	S + 1/2   &   1/2   & S   \\
	1       &   S+1     & 1/2
      \end{array}\right\}\nonumber \\
 & & \times \ U((\lambda+1,\mu-1)(10)(\lambda+2,\mu-1)(11);(\lambda \mu)(10)),
\end{eqnarray}
and
\begin{eqnarray}
\lefteqn{\left(\begin{array}{cc||c}
	[N_c]    &  [21^4]   & [N_c]   \\
	(\lambda \mu) S  &  (11) 1  &  (\lambda-2, \mu+1)S-1
      \end{array}\right) = N_c (-1)^{2S}
      \sqrt{\frac{4(2S+1)}{C^{[N_c]}(SU(6))}}}
      \nonumber \\
      & & \times \ c^{[N_c]}_{11}(S) c^{[N_c]}_{22}(S-1)
    \left\{\begin{array}{ccc}
	S - 1/2   &   1/2   & S   \\
	1       &   S-1     & 1/2
      \end{array}\right\} \nonumber \\
 & & \times \ U((\lambda-1,\mu)(10)(\lambda-2, \mu+1)(11);(\lambda \mu)(10)).
\end{eqnarray}
Replacing $c^{[N_c]}_{pp}$ by definitions (\ref{SU2}) and the $U$ coefficients by their
expressions we have obtained the isoscalar factors for arbitrary  $N_c$ listed in
the first four rows of Table \ref{su6isoscalar}. \\

\begin{sidewaystable}
\begin{center}
\renewcommand{\arraystretch}{2.2}
{\scriptsize
\begin{tabular}{l|c|c|l}
\hline
\hline
$(\lambda_1\mu_1)S_1$ \hspace{0.5cm} & \hspace{0.5cm}$(\lambda_2\mu_2)S_2$ \hspace{0.5cm} &\hspace{0.5cm}$\rho$\hspace{0.5cm} & \hspace{0.5cm}$\left(\begin{array}{cc||c}                                         [N_c]  &  [21^4]  &  [N_c] \\
                           (\lambda_1\mu_1)S_1 & (\lambda_2\mu_2)S_2 & (\lambda\mu)S
                                      \end{array}\right)_\rho$  \\
\vspace{-0.5cm} &  &   & \\
\hline
$(\lambda + 2,\mu - 1)S+1$\hspace{0.5cm} & \hspace{0.0cm}$(11)1$ & $/$ &\hspace{0.5cm}$-\sqrt{\frac{3}{2}}\sqrt{\frac{2S+3}{2S+1}}\sqrt{\frac{(N_c-2S)(N_c+2S+6)}{5N_c(N_c+6)}}$ \\
$(\lambda\mu)S$  & \hspace{0.0cm}$(11)1$ & 1 & \hspace{0.5cm}$4(N_c+3)\sqrt{\frac{2S(S+1)}{5N_c(N_c+6)[N_c(N_c+6)+12S(S+1)]}}$ \\
$(\lambda\mu)S$  &\hspace{0.0cm}$(11)1$ & 2 & \hspace{0.5cm}$-\sqrt{\frac{3}{2}}\sqrt{\frac{(N_c-2S)(N_c+4-2S)(N_c+2+2S)(N_c+6+2S)}{5N_c(N_c+6)[N_c(N_c+6)+12S(S+1)]}}$ \\
$(\lambda - 2,\mu + 1)S-1$  & \hspace{0.0cm}$(11)1$  & $/$ & \hspace{0.5cm}$-\sqrt{\frac{3}{2}}\sqrt{\frac{2S-1}{2S+1}}\sqrt{\frac{(N_c+4-2S)(N_c+2+2S)}{5N_c(N_c+6)}}$ \\
$(\lambda\mu)S$  & \hspace{0.0cm}$(00)1$ & $/$  & \hspace{0.5cm}$\sqrt{\frac{4S(S+1)}{5N_c(N_c+6)}}$ \\
$(\lambda\mu)S$  & \hspace{0.0cm}$(11)0$ & $1$  & \hspace{0.5cm}$\sqrt{\frac{N_c(N_c+6)+12S(S+1)}{10N_c(N_c+6)}}$ \\
$(\lambda\mu)S$  & \hspace{0.0cm}$(11)0$ & $2$  & \hspace{0.5cm}0 \\
\hline
\hline
\end{tabular}}
\caption{Isoscalar factors SU(6) for $[N_c] \times [21^4] \rightarrow [N_c]$ defined by Eq. (\ref{GEN}).}\label{su6isoscalar}
\end{center}
\end{sidewaystable}

For completeness, we now return to the generators $S_i$ and $T_a$ which have
only diagonal matrix elements.  
For $S_i$  we use the equivalence between Eq. (\ref{GEN}) and the 
Wigner-Eckart theorem (\ref{SPIN}). This leads to row 5 of Table \ref{su6isoscalar}.
For $T_a$ we use the equivalence between Eq. (\ref{GEN}) and 
the Wigner-Eckart theorem (\ref{FLAVOR}). 
The calculation of the isoscalar factors for $\rho = 1$ and $\rho = 2$
gives the results shown in rows 6 and 7 of Table \ref{su6isoscalar}. \\

One can alternatively express the SU(6) isoscalar factors of Table \ref{su6isoscalar} in 
terms of $\lambda$ and $\mu$ by using the identities $\lambda=2S$, 
$\mu=N_c/2-S$ and  $g_{\lambda\mu} = [N_c(N_c+6)+12S(S+1)]/4$. \\

Before ending this section let us calculate, as an example,  the diagonal
matrix element of $G_{i8}$ by using Eq. (\ref{GEN}). We consider a system of 
$N_c$ quarks with spin $S$, isospin $I$ and strangeness $\mathcal{S}$ defined by 
$Y=N_c/3 + \mathcal{S}$. In this case Eq. (\ref{GEN}) becomes
\begin{eqnarray}\label{EXAMPLE}
\langle [N_c](\lambda\mu) Y'I'I'_3 S S'_3 | G_{i8} |
[N_c](\lambda \mu) Y I I_3 S S_3 \rangle =  \delta_{YY'}\delta_{II'}\delta_{I_3I'_3}\sqrt{\frac{C^{[N_c]}(\mathrm{SU(6)})}{2}}    \nonumber \\
 \times   \sum_{\rho = 1,2}
 \left(\begin{array}{cc||c}
	(\lambda \mu)    &  (11)   &   (\lambda \mu)\\
	Y I   &  00  &  Y' I'
      \end{array}\right)_{\rho}
\left(\begin{array}{cc||c}
	[N_c]    &  [21^4]   & [N_c]   \\
	(\lambda \mu) S  &  (11)1  &  (\lambda \mu) S
      \end{array}\right)_{\rho} 
      \nonumber \\
 \times
 \left(\begin{array}{cc|c}
	S   &    1   & S   \\
	S_3  &   i   & S'_3
  \end{array}\right)
     \left(\begin{array}{cc|c}
	I   &   0  & I'   \\
	I_3 &   0   & I'_3
   \end{array}\right), 
   \end{eqnarray}
Using Table 4 of Ref. \cite{hecht65} and our Table 1 for the isoscalar 
factors of SU(3) and SU(6) respectively we have obtained
\begin{eqnarray}
\langle [N_c](\lambda\mu) Y'I' I'_3 S S'_3 | G_{i8} |
[N_c](\lambda \mu) Y I I_3 S S_3 \rangle =
 \frac{\delta_{YY'}\delta_{II'}\delta_{I_3I'_3}}{ 4 \sqrt{3S(S+1)}}
  \left(\begin{array}{cc|c}
	S   &    1   & S   \\
	S_3  &   i   & S'_3
  \end{array}\right) \nonumber \\
    \times \,
   \left[3 I (I+1) - S (S+1) - \frac{3}{4} \mathcal{S} \left(\mathcal{S} - 2 \right)\right].
   \label{gi8matrixelements}
   \end{eqnarray}
With $\mathcal{S} = -n_s$, where $n_s$ is the number of strange quarks,
we can recover the relation 
\begin{equation}
S_i G_{i8} = \frac{1}{4 \sqrt{3}}
 \left[3 I (I+1) - S (S+1) -\frac{3}{4} n_s \left( n_s + 2 \right)\right],
 \label{sigi8matrixelement}
\end{equation}
used in Ref. \cite{matagne05a}. (Ref.  \cite{matagne05a}  contains a typographic
error. In the denominator of Eq. (13) one should read $\sqrt{3}$ instead of
$\sqrt{2}$.)

\section{Back to isoscalar factors of SU(4) generators}
\label{symmetricsu4generators}

In the case of SU(4) $\supset$ SU(2) $\times$ SU(2) the analogue of Eq. (\ref{GEN}) is \cite{hecht69}
\begin{eqnarray}\label{GENsu4}
\langle [N_c] I' I'_3 S' S'_3 | E_{ia} |
[N_c] I I_3 S S_3 \rangle = \sqrt{C^{[N_c]}(\mathrm{SU(4)})}   \nonumber \\
\times
\left(\begin{array}{cc||c}
	[N_c]    &  [21^2]   & [N_c]   \\
	I S  &   I^a S^i  &  I'S'
      \end{array}\right) 
  \left(\begin{array}{cc|c}
	S   &    S^i   & S'   \\
	S_3  &   S^i_3   & S'_3
  \end{array}\right)
     \left(\begin{array}{cc|c}
	I   &   I^a   & I'   \\
	I_3 &   I^a_3   & I'_3
   \end{array}\right),
   \end{eqnarray}
where $C^{[N_c]}(\mathrm{SU(4)})=[3N_c(N_c+4)]/8$ is the eigenvalue of the SU(4) Casimir operator. Note that for a symmetric state one has $I=S$.
We recall that the SU(4) algebra is
\begin{eqnarray}\label{ALGEBRASU4}
[S_i,S_j] & = & i \varepsilon_{ijk} S_k,
\ \ \ [T_a,T_b]  =  i \varepsilon_{abc} T_c, \ \ \ [S_i,T_a]  =  0, \nonumber \\
\lbrack S_i,G_{ia}\rbrack & = & i\varepsilon_{ijk}G_{ka}, \ \ \ [T_a,G_{ib}]=i\varepsilon_{abc}G_{ic}, \nonumber \\
\lbrack G_{ia},G_{jb}\rbrack & =  & \frac{i}{4} \delta_{ij} \varepsilon_{abc} T_c
+\frac{i}{2} \delta_{ab}\varepsilon_{ijk}S_k.
\end{eqnarray}
The tensor operators $E_{ia}$ are related to $S_i$, $T_a$ and $G_{ia}$ $(i=1,2,3;\ a=1,2,3)$ by
\begin{equation} \label{normes2}
E_i =\frac{S_i}{\sqrt{2}};~~~ E_a = \frac{T_a}{\sqrt{2}}; ~~~E_{ia} = \sqrt{2} G_{ia}.
\end{equation}
In Eq. (\ref{GENsu4}) they are identified by $I^aS^i=01,10$ and 11 respectively. Now we want to obtain the SU(4) isoscalar factors as particular cases of the SU(6) results with $Y^a=0$. In SU(4) the hypercharge of a system of $N_c$ quarks takes the value $Y=N_c/3$.
By comparing (\ref{GEN}) and (\ref{GENsu4}) we obtained the relation
\begin{eqnarray}\label{isosu4}
\lefteqn{\left(\begin{array}{cc||c}
	[N_c]    &  [21^2]   & [N_c]   \\
	I S  &   I^a S^i  &  I'S'
      \end{array}\right)
 = r^{I^aS^i} \sqrt{\frac{C^{[N_c]}(\mathrm{SU(6))}}{C^{[N_c]}(\mathrm{SU(4))}}}} \nonumber \\
 & & \times \sum_{\rho = 1,2}\left(\begin{array}{cc||c}
	(\lambda \mu)    &  (\lambda^a\mu^a)   &   (\lambda' \mu')\\
	\frac{N_c}{3} I   &  0 I^a  &  \frac{N_c}{3} I'
      \end{array}\right)_{\rho}
\left(\begin{array}{cc||c}
	[N_c]    &  [21^4]   & [N_c]   \\
	(\lambda \mu) S  &  (\lambda^a\mu^a) S^i  &  (\lambda' \mu') S'
      \end{array}\right)_{\rho}\, ,
\end{eqnarray}
where
\begin{eqnarray}
r^{I^aS^i}=\left\{
\begin{array}{cc} \sqrt{\frac{3}{2}} & \mathrm{for}\ I^aS^i=01 \\
1 &\mathrm{for}\ I^aS^i=10 \\
1 &\mathrm{for}\ I^aS^i=11
\end{array}\right.
,\end{eqnarray}
due to (\ref{normes}), (\ref{GENsu4}) and (\ref{normes2}). In Eq. (\ref{isosu4}) 
we have to make the replacement
\begin{equation}
\lambda = 2I,\ \mu=\frac{N_c}{2}-I;\
\lambda' = 2I',\  \mu'=\frac{N_c}{2}-I',
\end{equation}
and take
\begin{eqnarray}
(\lambda^a\mu^a)=\left\{
\begin{array}{cc}
(00) & \mathrm{for}\ I^a=0 \\
(11) & \mathrm{for}\ I^a=1 \\
\end{array}\right.
.\end{eqnarray}
In this way we obtain the SU(4) isoscalar factors presented  in Table \ref{genesu(4)sym} for a symmetric spin-flavor wave function. These isoscalar factors were originally derived by Hecht and Pang \cite{hecht69}. By introducing these isoscalar factors into the matrix elements (\ref{GENsu4}) we obtained the expressions given in Eqs. (A1--A3) of Ref. \cite{carlson98b}.  In Ref. \cite{pirjol98a} one can find another method to derive the matrix elements of $G^{ia}$.
\begin{table}[h!]
 \begin{center}\renewcommand{\arraystretch}{2.0}
 {\scriptsize
  \begin{tabular}{ll|c|c|l}
  \hline
\hline
\vspace{-0.5cm} &    &    &     & \\
$S_1$&$I_1$ \hspace{0.5cm} & \hspace{0.0cm}$S_2I_2$ \hspace{0cm}& \hspace{0.5cm}
$SI$ \hspace{0.5cm} & \hspace{0.5cm}$\left(\begin{array}{cc||c}
                                         [N_c]  &  [211]  &  [N_c] \\
                                          S_1I_1 & S_2I_2 & SI
                                      \end{array}\right)$  \\
\vspace{-0.5cm} &    &    &     & \\
\hline
\vspace{-0.5cm} &    &    &     & \\
$S+1$ & $S+1$ & $11$ & $SS$ & $-\sqrt{\frac{(2S+3)(N_c-2S)(N_c+4+2S)}{(2S+1)3N_c(N_c+4)}}$ \\
$S$   & $S$   & $11$ & $SS$ & $-\frac{N_c+2}{\sqrt{3N_c(N_c+4)}}$ \\
$S-1$ & $S-1$ & $11$ & $SS$ & $-\sqrt{\frac{(2S-1)(N_c+2-2S)(N_c+2+2S)}{(2S+1)3N_c(N_c+4)}}$ \\
$S$   & $S$   & $10$ & $SS$ & \\
$S$   & $S$   & $01$ & $SS$ & \raisebox{2.55ex}[0cm][0cm]{$\sqrt{\frac{4S(S+1)}{3N_c(N_c+4)}}$} \\
\hline\hline
\end{tabular}}
\caption{Isoscalar factors of SU(4) for $[N_c] \times [211] \to [N_c]$ defined by Eq. (\ref{GENsu4}). This table is identical, up to a phase factor, to Table A4.2 of Ref.  \cite{hecht69}.}\label{genesu(4)sym}
 \end{center}
\end{table}

\chapter{Excited baryons in large $N_c$ QCD, the decoupling picture}

\thispagestyle{empty}

\section{Introduction}

Few years after the pioneering work on  ground-state baryons in the $1/N_c$ expansion, excited baryons have begun   to be studied. As we have seen in Chapter 3, the $1/N_c$ expansion of QCD operators for ground-state baryons is based on the large $N_c$ consistency conditions. These conditions imply an SU($2N_f)_c$ contracted symmetry for ground-state baryons which can be related  to the usual SU($2N_f$) symmetry used in the  quark  models when $N_c \to \infty$. \\
 
The  approach used so far for  large $N_c$ excited baryons is based on a kind of quark-shell model picture where  the baryon  is splitted into a symmetric core and an excited quark.  However, this approach leads to a  conceptual problem. Indeed, for excited baryons it turns out that the SU($2N_f$) symmetry is broken to  first order \cite{goity97} for baryons belonging to multiplets where the spin-flavor part of the wave function is mixed-symmetric. Nevertheless, the experimental facts suggest a small breaking. \\

One can still wonder if excited baryons satisfy consistency conditions like their ground-state cousins. Pirjol and Yang   showed that a contracted SU($2N_f$) symmetry arises also for excited states \cite{pirjol98a}. However, they supposed that excited baryons are narrow. But as Witten's large $N_c$ power counting rules predict that the decay widths of  excited baryons are of order $N_c^0$, it means that they  do not become stable in the large $N_c$ limit. This leads to a second conceptual problem.   \\

Despite these two problems, the $1/N_c$ expansion method  has been applied with success to some excited multiplets, the $[{\bf 70},1^-]$ \cite{carlson98b,schat02b} for the $N=1$ band, the $[{\bf 56'},0^+]$ \cite{carlson00}, the $[{\bf 56},2^+]$ \cite{goity03} and the $[{\bf 70},\ell^+]$ ($\ell=0,2$) \cite{matagne05b,matagne06b} for the $N=2$ band and   the $[{\bf 56},4^+]$ \cite{matagne05a} belonging to the $N=4$ band\footnote{For the definition of bands, see Chapter 2.}. \\
 
Recently a new approach to excited baryons has been proposed \cite{matagne06c} where the baryon wave function is considered in one  block, without  decoupling it into a symmetric  core and an excited quark.  This approach  predicts that the $1/N_c$ expansion starts at the order $1/N_c$ for mixed-symmetric multiplets, as for the ground state. The following chapter is devoted to this new picture. \\

Recent studies predict  a multiplet structure of large $N_c$ baryons identical to the  quark-shell predictions at large $N_c$. They describe baryons as resonances in meson-baryon scattering \cite{cohen03,cohen05b}. \\

A summary of the results obtained for the baryons belonging to the $[{\bf 70},\ell^+]$ ($\ell=0,2$) multiplets can be found in Ref. \cite{matagne06e} for the non-strange case and in Refs. \cite{matagne06b1,matagne06d} for strange baryons.  Refs. \cite{matagne05c,matagne05d} give a general discussion of highly excited baryons in large $N_c$ QCD.  The drawbacks of the decoupling method are also discussed in Ref. \cite{matagne06e}\\

This chapter is divided into two parts. First we shall make a review of the decoupling picture of large $N_c$ excited baryons. In the second part, we shall illustrate this picture with our results for the multiplets,  $[{\bf 70},\ell^+]$ ($\ell=0,2$) and  $[{\bf 56},4^+]$.


\section{The Hartree approximation}
\label{chap5hartreeapprox}
The origin of the decoupling approach lies in Hartree work \cite{witten79}. Witten  suggested that baryons can be described by a Hartree wave function when one neglects spin-flavor dependent interactions.  \\ 

Let us first  have a look at  ground-state baryons in this simple picture.
As the space part of the wave function is symmetric, the spin-flavor part is also symmetric. Then we have to study the problem of $N_c$ bosons in an attractive central potential. If we apply Witten's large $N_c$ power  counting rules, the interaction between any given pair of quarks\footnote{This interaction implies a gluon exchange between two quark lines inside the baryon. A factor $1/\sqrt{N_c}$ appears at each quark-gluon vertex (see Chapter 1).} is of order $1/N_c$. Then, the total potential experienced by each quark is of order one, since every quark interacts with $N_c$ other quarks. In the large $N_c$ limit, one can assume that each quark moves in an average potential  generated by the  other $N_c-1$ quarks.  The Hartree wave function can then be written as \cite{goity97}
\begin{equation}
 \Phi(x_1,\xi_1;\ldots;x_{N_c},\xi_{N_c})=\prod^{N_c}_{i=1}\psi(x_i)\chi_S(\xi_1,\ldots,\xi_{N_c}),
\end{equation}
where $\psi(x_i)$ are $\ell=0$ wave functions, $x_i$ is the position of the $i^{th}$ quark and $\xi_i$ its spin-flavor quantum numbers. $\chi_S(\xi_1,\ldots,\xi_{N_c})$ is a totally symmetric tensor of rank $N_c$ in the spin-flavor space. \\

Let us now consider  excited baryons with an excitation energy of order one, which means that the number $n$ of excited quarks is small, \emph{i.e.} $n<<N_c$. These baryons are composed of $\mathcal{O}(N_c)$ ground-state quarks (the ``core'') and $\mathcal{O}(1)$ excited quarks. In this approach, the $N_c-n$ core quarks are described by a symmetric wave function in both orbital and spin-flavor parts, as the ground-state baryons \cite{goity97}. This means that the space and the spin-flavor parts of the core wave function are symmetric under the interchange of  any two quarks. \\

Let us take the example where only one quark is excited. The states with one excited quark can belong either to a symmetric $[N_c]$ or to a mixed symmetric $[N_c-1,1]$ irrep of O(3). To form a symmetric orbital-spin-flavor state from an orbital state of a given symmetry one has to combine it with a spin-flavor state of the same symmetry. 
The wave function formed of a symmetric  orbital and a symmetric spin-flavor parts takes the form 
\begin{equation}
 \Phi_S(x_1,\xi_1;\ldots;x_{N_c},\xi_{N_c})=\chi_S(\xi_1,\ldots,\xi_{N_c})\frac{1}{\sqrt{N_c}}\sum_{j=1}^{N_c} \left(\prod_{i\neq j} \psi(x_i)\right)\phi(x_j),
 \label{hartreesymm}
\end{equation}
where $\chi_S$ is defined as above and $\phi(x_j)$ is the wave function of the excited quark. The normalization constant is consistent with the sum over $j$. In term of Young tableaux, the wave function (\ref{hartreesymm}) becomes
 \begin{equation}
\raisebox{-2.0pt}{$\overbrace{\begin{Young}
       &  \cr
      \end{Young}\  \cdots \
      \begin{Young}
        $\times$ \cr
      \end{Young}}^{N_c}$}\ \ = \ \ 
      \raisebox{-2.0pt}{$\overbrace{\begin{Young}
       &  \cr
      \end{Young}\  \cdots \
      \begin{Young}
        $\times$ \cr
      \end{Young}}^{N_c}$}\ \ \times \ \
 \raisebox{-2.0pt}{$\overbrace{\begin{Young}
       &  \cr
      \end{Young}\  \cdots \
      \begin{Young}
        $\times$ \cr
      \end{Young}}^{N_c}$}\ ,     
\end{equation}
where the $N_c$-th quark, supposed to be the excited quark is marked by a cross. \\

To form a totally symmetric orbital-spin-flavor state starting from an orbital state of mixed symmetry $[N_c-1,1]$ the procedure is slightly more complicated. In this case a totally symmetric state is given by \cite{stancu96}
\begin{equation}
 \Phi'_S=\frac{1}{\sqrt{N_c-1}} \sum_Y |[N_c-1,1] Y\rangle_O |[N_c-1,1] Y\rangle_{SF},
 \label{nondecouplingchap5}
\end{equation}
where the indices designate the orbital ($O$) and the spin-flavor ($SF$) basis vectors spanning the invariant subspace of the irrep $[N_c-1,1]$. Each of these vectors carries a label $Y$ which corresponds to a given Young tableau. For illustration, let us consider the case of $N_c=5$. Explicitly the sum is 
\begin{eqnarray}
\label{newlookyoungproduct}
\raisebox{-2.0pt}{\mbox{\begin{Young}
      1 & 2 & 3 & 4 & 5 \cr
      \end{Young}}}
 & = &  \frac{1}{\sqrt{4}} \left(\negthinspace \negthinspace\negthinspace
\raisebox{-9.0pt}{\mbox{\begin{Young}
1 & 2 & 3 & 4 \cr
5 \cr
\end{Young}}}\
\raisebox{-9.0pt}{\mbox{\begin{Young}
1 & 2 & 3 & 4\cr
5 \cr
\end{Young}}}\ \
+ \negthinspace \negthinspace\negthinspace\raisebox{-9.0pt}{\mbox{\begin{Young}
1 & 2 & 3 & 5 \cr
4 \cr
\end{Young}}}\
\raisebox{-9.0pt}{\mbox{\begin{Young}
1 & 2 & 3 & 5 \cr
 4 \cr
\end{Young}}}\
\right. \nonumber \\
& & + \negthinspace \negthinspace\negthinspace \left. \raisebox{-9.0pt}{\mbox{\begin{Young}
1 & 2 & 4 & 5 \cr
3 \cr
\end{Young}}}\
\raisebox{-9.0pt}{\mbox{\begin{Young}
1 & 2 & 4 & 5\cr
3 \cr
\end{Young}}}\ \
+ \negthinspace \negthinspace\negthinspace\raisebox{-9.0pt}{\mbox{\begin{Young}
1 & 3 & 4 & 5 \cr
2 \cr
\end{Young}}}\
\raisebox{-9.0pt}{\mbox{\begin{Young}
1 & 3 & 4 & 5 \cr
 2 \cr
\end{Young}}}\
\right). \nonumber \\
\end{eqnarray}
The first term contains the 5-th quark in the second row and is given by the product of two normal Young tableaux. The other three terms have the 5-th quark in the first row. Explicit forms of the orbital wave functions are given in the next chapter in Table  \ref{nc=5basisfunctions}. For $N_c$ quarks there will be one term with normal Young tableaux and $N_c-2$ terms with the $N_c$-th particle in the second row. \\

In Ref. \cite{goity97}, the wave function constructed from orbital and spin-flavor parts of symmetry $[N_c-1,1]$ has the following form 
\begin{eqnarray}
  \lefteqn{\Phi_{MS}(x_1,\xi_1;\ldots;x_{N_c},\xi_{N_c})=} \nonumber \\
  & &\chi_{MS}(\xi_1,\ldots,\xi_{N_c}) \frac{1}{\sqrt{N_c(N_c-1)}}\sum_{i\neq j}\left[\left(\prod_{k\neq i,j} \psi(x_k)\right)\psi(x_i)\phi(x_j)-i\leftrightarrow j\right].
  \label{hartreems}
\end{eqnarray}
One can see that the orbital part contains all possible terms obtained from the permutation of the excited quark $j$ with the other quarks. This form strictly corresponds to a normal Young tableau with the $N_c$-th particle in the second row. The irreducible rank $N_c$ mixed-symmetric tensor $\chi_{MS}$ formed of single-particle spin-flavor states must have the same permutation properties. Examples are given in Eqs. (\ref{decupletch52})--(\ref{singletch52}). \\

It follows that the expression (\ref{hartreems}) is a truncated form of Eq. (\ref{nondecouplingchap5}). It contains only the term associated to the normal Young tableau, the other $N_c-2$ terms being omitted. The normalization constant $1/\sqrt{N_c-1}$ has been suppressed, although it plays a crucial role in the $N_c$ counting. Thus, although this function is symmetric under the permutation of the $N_c-1$ quarks of the core, it breaks S$_{N_c}$. The approximation (\ref{hartreems}) can thus be  illustrated as
\begin{equation}
 \raisebox{-2.0pt}{$\overbrace{\begin{Young}
       &  \cr
      \end{Young}\  \cdots \
      \begin{Young}
        $\times$ \cr
      \end{Young}}^{N_c}$}\ \ \approx \ \ \mbox{$\overbrace{\negthinspace\negthinspace\negthinspace\negthinspace\negthinspace\negthinspace\raisebox{-9pt}{\begin{Young}
                 & \cr
		 $\times$ \cr
                \end{Young}}\  \raisebox{3.5pt}{$\cdots$}\negthinspace \negthinspace\negthinspace
\raisebox{3.5pt}{\begin{Young} \cr \end{Young}}}^{N_c-1}$}\ \ \times \ \
\mbox{$\overbrace{\negthinspace\negthinspace\negthinspace\negthinspace\negthinspace\negthinspace\raisebox{-9pt}{\begin{Young}
                 & \cr
		 $\times$ \cr
                \end{Young}}\  \raisebox{3.5pt}{$\cdots$}\negthinspace \negthinspace\negthinspace
\raisebox{3.5pt}{\begin{Young} \cr \end{Young}}}^{N_c-1}$}\ ,
\label{hartreeyoungtabms}
\end{equation}
where the $N_c$-th quark, supposed to be the excited quark is marked by a cross, as above.

\section{Wave functions}
\label{excitedwavefunctions}

\label{excitedwavefunctionsintroduction}
The standard approach to excited baryon states is based on the wave functions (\ref{hartreesymm}) and (\ref{hartreems}), written in the spirit of the Hartree approximation. Indeed it was believed to be the only way to make the problem tractable \cite{carlson98b}. The wave function (\ref{hartreesymm}) suggests that an excited state described by a symmetric SU($2N_f$) representation $[N_c]$ as \emph{e.g.} the resonances belonging to the $[{\bf 56},2^+]$ or to the $[{\bf 56},4^+]$ multiplets, can be treated in a way similar to that of the ground state, the difference being that the total wave function contains an angular momentum part $|\ell m_\ell\rangle$ with $\ell \neq 0$. Then, as for the ground state, the needed ingredients are the matrix elements of the generators of SU(4) for $N_f=2$ or of SU(6) for $N_f=3$ between $[N_c]$ states. These are known and can be found, for example,  in Ref. \cite{carlson98b} for SU(4) and in Ref. \cite{matagne06a} for SU(6). \\

The difficulty arises for multiplets described by  the mixed-symmetric representation $[N_c-1,1]$, reduced to $[{\bf 70}]$ for $N_c=3$. One needs to know the matrix elements of the SU($2N_f$) generators for arbitrary $N_c$. So far only the matrix elements of the SU(4) generators between states of symmetry $[N_c-1,1]$ are known \cite{hecht69}. Their recent use to the study of the $[{\bf 70},1^-]$ multiplet is described in detail in the following chapter.  \\

In the present chapter we describe an approximate method, the origin of which is the Hartree-type wave function (\ref{hartreems}) that allows to reduce the treatment of mixed-symmetric states to a form similar to the ground-state description. This method has been  applied to  the $[{\bf 70},1^-]$ multiplet \cite{carlson98b} and to the $[{\bf 70},0^+]$ and $[{\bf 70},2^+]$ multiplets \cite{matagne05b,matagne06b}. The latter are presented in details in this chapter.  \\

This approximate method consists in describing an excited baryon state as a single excited quark coupled to a ``core'' for which both the orbital and the spin-flavor parts of the wave function are symmetric.  The excited quark is coupled to the core by its angular momentum. The core can be in the ground state or excited. If the core is  excited, it brings the corresponding angular momentum contribution to the total angular momentum of the system of $N_c$ quarks (see below). \\


 For generality we shall consider three flavors ($N_f=3$).

\subsection{Symmetric  wave functions}

The symmetric wave function in both SU(6) and O(3), used for example for the analysis of the $[{\bf 56}, 2^+]$ or $[{\bf 56}, 4^+]$ multiplets can be written as
\begin{equation}
|\ell S;JJ_3;(\lambda \mu) Y I I_3\rangle_S = \sum_{m_\ell,S_3}\left(\begin{array}{cc|c}
                                                                       \ell & S & J \\
								       m_\ell & S_3 & J_3 
                                                                     \end{array}\right)
|SS_3\rangle |(\lambda\mu) Y I I_3\rangle |\ell m_\ell \rangle,
\label{largencsymmetricexcited}
\end{equation}
where, as introduced in Chapter 4, $(\lambda\mu)$ labels the SU(3) irreducible representations and $Y$ stands for the hypercharge.

\subsection{Mixed-symmetric  wave functions}

\label{Mixed symmetric  wave functions}

As explained above, one needs to discuss separately the case with one excited quark and the case with $n>1$ excited quarks.

  \subsubsection{The case of one excited quark}
  First let us consider the case with one excited quark and $N_c-1$ core quarks. If we look at Table \ref{harmonicconfigurationsnc}, this configuration is allowed for the $[{\bf 70}, 1^-]$ and $[{\bf 70},\ell^+]$ ($\ell=0,2$)  multiplets. The symmetric core state is denoted by
\begin{equation}
 |S_cm_1 \rangle|(\lambda_c \mu_c) Y_cI_cI_{c_3}\rangle,
\end{equation}
the subscript $c$ referring to the core. The excited quark state is denoted by
\begin{equation}
 |1/2m_2 \rangle|(10) y i i_{3} \rangle|\ell m_\ell\rangle.
\end{equation}\\

 To form  states describing the whole system, we have to couple the core wave function to the excited quark wave function. Then, in SU(6) $\times$ SO(3), the most general form of the wave 
function is
\begin{eqnarray}\label{groundcore}
\lefteqn{|\ell S;JJ_3;(\lambda \mu) Y I I_3\rangle  =
\sum_{m_\ell,S_3}
   \left(\begin{array}{cc|c}
	\ell    &    S   & J   \\
	m_\ell  &    S_3  & J_3 
      \end{array}\right)}
\nonumber \\
& \times &
\sum_{p p'}   c^{[N_c-1,1]}_{p p'}(S)
|S S_3; p \rangle
|(\lambda \mu)Y I I_{3}; p'  \rangle
|\ell m_\ell\rangle,
\end{eqnarray}
where $p$ and $p'$ denote the row number where the $N_c$-th quark is located in a Young tableau. Then the spin wave function is 
\begin{equation}
|S S_3; p \rangle = \sum_{m_1,m_2}
 \left(\begin{array}{cc|c}
	S_c    &    \frac{1}{2}   & S   \\
	m_1  &         m_2        & S_3
      \end{array}\right)
      |S_cm_1 \rangle |1/2m_2 \rangle,
\end{equation}
with $p = 1$ if  $S_c = S - 1/2$ and $p = 2$ if $S_c = S + 1/2$ and the flavor wave function is
\begin{eqnarray}\label{statessu(3)ch5}
|(\lambda \mu)Y I I_{3}; p' \rangle  =
\sum_{Y_c,I_c,I_{c_3},y,i,i_{3}}
\left(\begin{array}{cc|c}
	(\lambda_c \mu_c)    &   (10)  & (\lambda \mu) \\
	 Y_cI_cI_{c_3} &   y i i_{3} &  YII_{3}
      \end{array}\right) 
|(\lambda_c \mu_c) Y_cI_cI_{c_3}\rangle
|(10) y i i_{3} \rangle,
\end{eqnarray}
where $p' = 1$ if $(\lambda_c \mu_c) = (\lambda - 1, \mu)$,
$p' = 2$ if $(\lambda_c \mu_c) = (\lambda + 1, \mu - 1)$ and $p' = 3$ if $(\lambda_c \mu_c) = (\lambda, \mu + 1)$.
The spin-flavor  part of the wave function (\ref{groundcore})
of symmetry $[N_c-1,1]$ results from the inner product of the 
spin and flavor wave functions.
 The  
coefficients  $c^{[N_c-1,1]}_{p p'}(S)$ are  isoscalar factors 
\cite{stancu96,stancu99} of the 
permutation group of ${N_c}$ particles. They have already been introduced in Section \ref{SU(6) symmetric  wave functions}. For the mixed-symmetric representation $[N_c-1,1]$,  we have
\begin{eqnarray}\label{isoscalarms}
c^{[N_c-1,1]}_{11}(S) & =  & - \sqrt{\frac{(S + 1)(N_c - 2 S)}{N_c(2 S + 1)}},
\nonumber \\
c^{[N_c-1,1]}_{22}(S) &  = & \sqrt{\frac{S[N_c+2(S + 1)]}{N_c(2 S + 1)}},
\nonumber \\
c^{[N_c-1,1]}_{12}(S) &  = & c^{[N_c-1,1]}_{21}(S) = 1,
\nonumber \\
c^{[N_c-1,1]}_{13}(S) &  = & 1.
\end{eqnarray} 
In Eqs. (\ref{decupletch52})-(\ref{singletch52})  below, we illustrate their
application for $N_c$ = 7. In each inner product 
the first Young diagram corresponds to spin and the second to flavor. Accordingly,
one can see that Eq. (\ref{decupletch52}) stands for $^210$, Eq. (\ref{octetch54}) 
for $^48$, Eq. (\ref{octetch52}) for $^28$ and Eq. (\ref{singletch52}) for $^21$.
Each inner product contains the corresponding isoscalar factors and 
the position of the last particle is marked with a cross. In the right-hand side, from the location of the cross one can read off the values of $p$ and of $p'$.
The equations are 
\begin{eqnarray}
\label{decupletch52}
\raisebox{-9.0pt}{\mbox{\begin{Young}
 & & & & & \cr
$\times$ \cr
\end{Young}}}\
& = &
c^{[6,1]}_{21}\! \! \!
\raisebox{-9.0pt}{\mbox{
\begin{Young}
& & & \cr
& & $\times$\cr
\end{Young}}} \ \times \! \! \! \! \!
\raisebox{-9.0pt}{\mbox{
\begin{Young}
& & & & $\times$\cr
& \cr
\end{Young}}}\ ,
\\ \nonumber
\\
\label{octetch54}
\raisebox{-9.0pt}{\mbox{\begin{Young}
 & & & & & \cr
$\times$ \cr
\end{Young}}}\
& = &
c^{[6,1]}_{12}\! \! \!
\raisebox{-9.0pt}{\mbox{
\begin{Young}
& & & & $\times$\cr
& \cr
\end{Young}}} \ \times \! \! \! \! \!
\raisebox{-9.0pt}{\mbox{
\begin{Young}
& & & \cr
& & $\times$\cr
\end{Young}}}\ ,
\\ \nonumber
\\
\label{octetch52}
\raisebox{-9.0pt}{\mbox{\begin{Young}
 & & & & & \cr
$\times$ \cr
\end{Young}}}\
&=& c^{[6,1]}_{11}\! \! \!
\raisebox{-9.0pt}{\mbox{
\begin{Young}
& & & $\times$\cr
& & \cr
\end{Young}}} \ \times \! \! \! \! \!
\raisebox{-9.0pt}{\mbox{
\begin{Young}
& & & $\times$\cr
& & \cr
\end{Young}}} \nonumber \\
& & + \ c^{[6,1]}_{22}\! \! \!
\raisebox{-9.0pt}{\mbox{
\begin{Young}
& & & \cr
& & $\times$\cr
\end{Young}}} \ \times \! \! \! \! \!
\raisebox{-9.0pt}{\mbox{
\begin{Young}
& & & \cr
& & $\times$\cr
\end{Young}}}\ ,
\\ \nonumber
\\
\label{singletch52}
\raisebox{-9.0pt}{\mbox{\begin{Young}
 & & & & & \cr
$\times$ \cr
\end{Young}}}\
&=& c^{[6,1]}_{13}\! \! \!
\raisebox{-9.0pt}{\mbox{
\begin{Young}
& & & $\times$\cr
& & \cr
\end{Young}}} \ \times \! \! \! \! \!
\raisebox{-15pt}{\mbox{
\begin{Young}
& & \cr
& & \cr
$\times$ \cr
\end{Young}}}\ .
\end{eqnarray}

\subsubsection{The case of $n>1$ excited quarks}

When more than one quark is excited, the Hartree approximation suggests that the baryon is composed $N_c-n$ core quarks in the ground state and $n$ excited quarks \cite{witten79}. Here, to generalize the approach developed for one excited quark, we suppose that    the core is still symmetric but not in the ground state. The core acquires an angular momentum $\ell_c$ and its wave function becomes 
\begin{equation}
 |S_cm_1 \rangle|(\lambda_c \mu_c) Y_cI_cI_{c_3}\rangle|\ell_c m_c\rangle,
\end{equation}
and the excited quark wave function is
\begin{equation}
|1/2m_2 \rangle|(10) y i i_{3} \rangle|\ell_q m_q\rangle,
\end{equation}
where the indices $c$ and $q$ stand for  the symmetric core and the excited quark respectively.
The SU(6) $\times$ O(3) symmetric wave function becomes
\begin{eqnarray}\label{EXCORE}
\lefteqn{|\ell S;JJ_3;(\lambda \mu) Y I I_3\rangle  =
\sum_{m_c,m_q,m_\ell,S_3}
\left(\begin{array}{cc|c}
	\ell_c    &  \ell_q   & \ell   \\
	m_c  &    m_q    & m_\ell 
      \end{array}\right) 
   \left(\begin{array}{cc|c}
	\ell    &    S   & J   \\
	m_\ell  &    S_3  & J_3 
      \end{array}\right)}
\nonumber \\
& \times &
\sum_{p p'}   c^{[N_c-1,1]}_{p p'}(S)
|S S_3; p \rangle
|(\lambda \mu)Y I I_{3}; p'  \rangle
|\ell_qm_q\rangle   |\ell_cm_c\rangle,
\end{eqnarray}
where the coupling between the angular momentum of the core and of the excited quark has been included through an extra Clebsch-Gordan coefficient.

\subsection{Operator expansion}
\label{excitedstateoperatorexpansion}
For excited states one has to generalize the operator expansion Eq. (\ref{qcdoperatorexpansion2}) by including the generators $\ell^i$ of the orthogonal group SO(3). The expansion of a one-body QCD operator becomes
\begin{equation}
\mathcal{O}^{\mathrm{1-body}}_{\mathrm{QCD}}=\sum_n c_n O^{(n)},
\label{excitedoperatorexpansion}
\end{equation}
with 
\begin{equation}
 O^{(n)}=\frac{1}{N_c^{n-1}}O^{(k)}_\ell\cdot O^{(k)}_{SF},
\end{equation}
where $O^{(k)}_\ell$ and $O^{(k)}_{SF}$ are respectively $k$-rank tensors in SO(3) and SU(2). They can be expressed in terms of products of generators $\ell^i,\ S^i,\ T^a$ and $G^{ia}$.   \\

For the mixed-symmetric multiplets,  the baryon wave function is splitted into a symmetric core and an excited quark, each  generators must be written as the sum of two terms, one acting on the core and the other on the excited quark. One has
\begin{equation}
 \ell^i = \ell^i_c + \ell^i_q;\ \ \ \ \ S^i=S_c^i+s^i;\ \ \ \ \ T^a=T^a_c+t^a;\ \ \ \ \ G^{ia}=G_c^{ia}+g^{ia},
\end{equation}
where $\ell^i_c,\ S_c^i,\ T_c^a$ and $G^{ia}$ are the core operators and $\ell^i_q,\ s^i,\ t^a$ and $g^{ia}$ the corresponding excited quark operators. \\

One may ask if there are SU($2N_f$) operators identities for mixed-symmetric multiplets like for the symmetric multiplets (see Section \ref{groundoperatoridentities}). As proved in \cite{carlson98b}, there is only  one identity which reduces the two-body operator to an SU($2N_f$) singlet, the Casimir operator. In terms of core and excited quark generators, this identity implies that we must eliminate the redundant operator $gG_c$ from the operator expansion. \\

One important consequence of the truncation of the wave function implied by the splitting is that the SU($2N_f$) symmetry is broken at order $\mathcal{O}(N_c^0)$ for mixed-symmetric multiplets. 
  This statement is best illustrated by considering the spin-orbit one-body operator $\ell^i_q s^i$. 
 The $\ell^i_q s^i$ operator acts on the excited quark only. As already mentioned, the excited quark is the $N_c$-th particle of the wave function and is located in the second row of the Young tableaux, see Eq. (\ref{hartreeyoungtabms}).    Here, we rewrite it  by using fractional parentage coefficients.  For one-body operators, we need one-body fractional parentage coefficients. In this way, one can decouple the last particle from the rest. In the simple case where the spatial wave function contains only one excited quark, for example, having the structure $(0s)^{N_c-1}(0p)$, and symmetry $[N_c-1,1]$ one can show that \cite{matagne05b}
\begin{eqnarray}
 \lefteqn{|[N_c-1,1](0s)^{N_c-1}(0p),1^-\rangle =}  \nonumber \\ & & \sqrt{\frac{N_c-1}{N_c}}\psi_{[N_c-1]} (0s)^{N_c-1}\phi_{[1]}(0p) - \sqrt{\frac{1}{N_c}} \psi_{[N_c-1]}\left((0s)^{N_c-2}(0p)\right)\phi_{[1]}(0s).
 \label{1exciteddecoupling}
\end{eqnarray}
where the first factor in each term in the right-hand side is a symmetric $(N_c-1)$-particle wave function and $\phi_{[1]}$ is a single particle wave function associated to the $N_c$-th particle. 
In the case of the spin-orbit operator one can see that only the first term contributes (the operator acts on the last particle only). Its matrix element is proportional to the square of the coefficient of the first term, \emph{i.e.} with $\frac{N_c-1}{N_c}$ which for large $N_c$ gives to the spin-orbit the order $\mathcal{O}(1)$. \\

For spin-flavor symmetric states, the corresponding expression in term of fractional parentage coefficients is \cite{matagne05b}
\begin{eqnarray}
 \lefteqn{|[N_c](0s)^{N_c-1}(0d),2^+\rangle =}  \nonumber \\ & & 
 \sqrt{\frac{1}{N_c}} \psi_{[N_c-1]}(0s)^{N_c-1}\phi_{[1]}(0d) + \sqrt{\frac{N_c-1}{N_c}} \psi_{[N_c-1]}\left((0s)^{N_c-2}(0d)\right)\phi_{[1]}(0s),
\end{eqnarray}
where we have taken the configuration $(0s)^{N_c-1}(0d)$ as an example. It corresponds to the $[{\bf 56},2^+]$ multiplet. As the spin-orbit operator $\ell^i_q s^i$ acts only on the last particle, it is of order $1/N_c$ for symmetric multiplets.  \\

Let us rewrite the operator $\vec{\ell}_q\cdot\vec{s}$ as $\vec{\ell}(N_c)\cdot\vec{s}(N_c)$ to point out  the fact that this operator acts on the $N_c$-th quark. This   notation will be used in Chapter 6.
The Hartree wave functions (\ref{hartreesymm}) and (\ref{hartreems}) can be used to obtain the expectation values of the operator $\vec{\ell}(N_c)\cdot\vec{s}(N_c)$ \cite{goity04,goity04b}. The result is 
\begin{eqnarray}
 \langle \Psi | \vec{\ell}(N_c)\cdot\vec{s}(N_c)| \Psi \rangle = 
 \renewcommand{\arraystretch}{1.75}\left\{\begin{array}{ll}
         \mathcal{O}\left(\frac{1}{N_c}\right) & \mbox{with $\Psi =\Phi_S$ (Eq. (\ref{hartreesymm}))} \\
	 \mathcal{O}(N_c^0)  & \mbox{with $\Psi=\Phi_{MS}$ (Eq. (\ref{hartreems}))}
        \end{array} \right. ,
\label{summarydecoupling}
\end{eqnarray}
in agreement with the discussion carried above.

\section{The $[{\bf 70},\ell^+]$ ($\ell=0,2$) baryon multiplets}

As we have seen in Figure \ref{baryonspectrum}, the $[{\bf 70},0^+]$ and the $[{\bf 70},2^+]$ multiplets belong to the $N = 2$ band, with baryon masses of $\sim 2$ GeV. \\
 
 For these multiplets, Table \ref{harmonicconfigurationsnc} suggests that there are two possible configurations. For the $[{\bf 70},0^+]$ one can write the orbital part of the wave function as
\begin{equation}
|[{\bf N_c-1,1}], 0^+\rangle = 
\sqrt{\frac{1}{3}}|[N_c-1,1](0s)^{N_c-1}(1s)\rangle + \sqrt{\frac{2}{3}}|[N_c-1,1](0s)^{N_c-2}(0p)^2\rangle.
\label{70,0+largenconfiguations}
\end{equation}
In the first term, $(1s)$ is the first single particle
radially excited state with $n=1$, $\ell = 0$ 
($N =2n+\ell$). In the second term     
the two quarks are excited to the $p$-shell to get $N = 2$. They are coupled to $\ell = 0$.
By analogy, for the $[{\bf 70}, 2^+]$ one has,
\begin{equation}
 |[{\bf N_c-1,1}], 2^+\rangle = \sqrt{\frac{1}{3}}|[N_c-1,1](0s)^{N_c-1}(0d) \rangle + \sqrt{\frac{2}{3}}|[N_c-1,1](0s)^{N_c-2}(0p)^2\rangle,
 \label{70,2+largenconfiguations}
\end{equation}
 where the two quarks in the $p$-shell are coupled to  $\ell = 2$. \\
 
 One can see that the coefficients of the linear combinations (\ref{70,0+largenconfiguations}) and (\ref{70,2+largenconfiguations}) are independent of $N_c$ which means that both terms have to be considered in the large $N_c$ limit. As discussed in Section \ref{Mixed symmetric  wave functions}, the first term of Eqs. (\ref{70,0+largenconfiguations}) or (\ref{70,2+largenconfiguations}) can be treated as an excited quark coupled to a symmetric ground-state core (Eq. (\ref{1exciteddecoupling})). The second term is treated as an excited quark coupled to an excited core. Indeed, by using the fractional parentage technique, one can write
 \begin{eqnarray}
\lefteqn{|[N_c-1,1] (0s)^{N_c-2}(0p)^2,\ell^+ \rangle  =} \nonumber \\ 
& & \sqrt{\frac{N_c-2}{N_c}} \Psi_{[N_c-1]}\left((0s)^{N_c-2}~(0p)\right) \phi_{[1]}(0p)
 -\sqrt{\frac{2}{N_c}} \Psi_{[N_c-1]}\left((0s)^{N_c-3}~(0p)^2\right)\phi_{[1]}(0s).\nonumber \\ 
\end{eqnarray}
 One can see that  the  coefficient of the first term is $\mathcal{O}(1)$ and  the  coefficient of the second term $\mathcal{O}(N_c^{-1/2})$.
Then, in the large  $N_c$ limit, one can  neglect the second term and take into account
only  the first term in the wave function, 
where the $N_c$-th particle has an $\ell =1$ excitation. \\

In the calculations, one has to add the two contributions coming from the two possible configurations of Eqs. (\ref{70,0+largenconfiguations}) and (\ref{70,2+largenconfiguations}). For this purpose, we use the two wave functions Eq. (\ref{groundcore}) and Eq. (\ref{EXCORE}) describing the two cases, with  one and two excited quarks respectively. \\
 
 Historically, we have first studied  the $[{\bf 70},\ell^+]$ multiplets for non-strange baryons \cite{matagne05b}. In a second stage, we have analyzed the strange baryons \cite{matagne06b}. The reason of this  sequence comes from the fact that meanwhile we obtained analytic expression for  the matrix elements of the SU(6) generators for a symmetric spin-flavor wave function. This was the subject of Chapter 3. Here we present separately the non-strange and the strange baryons belonging the $[{\bf 70},\ell^+]$ multiplets.

\subsection{The non-strange baryons}
\label{70,ell+nonstrange}

\subsubsection{The mass operator}
One can use the QCD one-body operator expansion Eq. (\ref{excitedoperatorexpansion}) to the mass operator to write
\begin{equation}
 M_{[{\bf 70},\ell^+]} = \sum_i c_i O_i,
\end{equation}
where $c_i$ are the unknown dynamical coefficients  (see Section \ref{groundoperatorexpansion}) and the operators $O_i$ are SU(4) $\times$ O(3) scalars. The building blocks of these operators are, as already explained, the core and the excited quark generators. The unknown coefficients are obtained from fitting the experimentally known masses (see Table \ref{70,ell+nonstrangemultiplet}).
 We also introduce the tensor 
operator\footnote{The irreducible spherical tensors are defined according to  
Ref. \cite{brink68}.}
\begin{equation}\label{70,ell+nonstrangeTENSOR}
\ell^{(2)ij}_{ab}=\frac{1}{2}\left\{\ell^i_a,\ell^j_b\right\}
-\frac{1}{3}\delta_{i,-j}\vec{\ell}_a\cdot\vec{\ell}_b~,
\end{equation}
with $a=c$, $b=q$ or vice versa or $a=b=c$ or $a=b=q$. For simplicity 
when $a=b$, we shall use a single index $c$, for the core, and $q$ for the 
excited quark so that the operators are $\ell^{(2)ij}_c$ and $\ell^{(2)ij}_q$ 
respectively. The latter case represents the tensor 
operator used in the analysis of the $[{\bf 70},1^-]$ multiplet (see \emph{e.g.} 
Ref. \cite{carlson98b}). \\

\begin{table}[h!]
\begin{center}
{\scriptsize
\renewcommand{\arraystretch}{1.75} 
\begin{tabular}{llrrl}
\hline
\hline
Operator & \multicolumn{4}{c}{Fitted coef. (MeV)}\\
\hline
\hline
$O_1 = N_c  \mathbbm{1} $                                   & \ \ \ $c_1 =  $  & 555 & $\pm$ & 11       \\
$O_2 = \ell_q^i s^i$                                & \ \ \ $c_2 =  $  &   47 & $\pm$ & 100    \\
$O_3 = \frac{3}{N_c}\ell^{(2)ij}_{q}g^{ia}G_c^{ja}$ & \ \ \ $c_3 =  $   & -191 & $\pm$ & 132  \\
$O_4 = \frac{1}{N_c}(S_c^iS_c^i+s^iS_c^i)$          & \ \ \ $c_4 =  $  &  261 & $\pm$ &  47  \\
\hline \hline
\end{tabular}}
\caption{List of operators and the coefficients resulting from the fit with $\chi^2_{\rm dof}  \simeq 0.83$ to resonances belonging to the $[{\bf 70},\ell^+]$ multiplets \cite{matagne05b}.}\label{70,ell+nonstrangeoperators}
\end{center}
\end{table}

We apply the $1/N_c$ 
counting rules presented in  Refs. \cite{carlson98b} and \cite{pirjol03a} and use their 
conclusions in selecting the most dominant operators in the practical
analysis. For non-strange baryons, 
Table I of Ref. \cite{carlson98b} gives a list of 18 linearly independent operators.
If the core is excited the number of operators appearing in the mass formula
is much larger.  However 
due to lack of experimental data, here we have to consider a restricted list. 
The selection is suggested by the conclusion  
of Ref. \cite{carlson98b}, $(N_f=2)$  and of Ref. \cite{schat02b} 
$(N_f = 3)$, that only 
 a few operators, of some specific structure, bring
a dominant
contribution to the mass. Following the notation of Ref. \cite{schat02b} these are
$O_1, O_2, O_3$ and $O_4$ exhibited  in Table \ref{70,ell+nonstrangeoperators}. The first is the trivial
operator of order $\mathcal{O}(N_c)$. The second is the one-body part of the spin-orbit 
operator of order $\mathcal{O}(1)$ which acts on the excited quark.
The third is a composite two-body operator formally of order $\mathcal{O}(1)$ as well. 
It involves 
the tensor operator (\ref{70,ell+nonstrangeTENSOR}) acting on the excited quark and the SU(4)
generators $g^{ia}$ acting on the excited quark and $G^{ja}_c$ acting on the
core. The latter is a coherent operator which introduces an extra power $N_c$
so that the order of $O_3$ is $\mathcal{O}(1)$.
In order to take into 
account its contribution we have applied the rescaling introduced 
in Ref. \cite{schat02b} which consists in introducing a multiplicative factor of 3. 
Without this factor the coefficient $c_3$ becomes too large,
as noticed in Ref. \cite{schat02b}.\footnote{
Alternatively the factor 3 could be included in the definition (\ref{70,ell+nonstrangeTENSOR})
of the tensor operator, as sometimes done in the literature. In practice
what it matters is the product $c_i O_i$.} 
The dynamics of the operator $O_3$ is less
understood. Previous studies \cite{carlson98b,schat02b}
speculate about its connection to a flavor exchange
mechanism \cite{glozman96,collins99} related to long distance meson exchange 
interactions. Finally, the last operator is the spin-spin interaction, the only one of
 order $\mathcal{O}(1/N_c)$ which we consider here. Higher order operators are neglected. \\

\begin{table}[h!]
\begin{center}
{\scriptsize
\renewcommand{\arraystretch}{1.75}
\begin{tabular}{lcccc}
\hline 
\hline
   &  \hspace{ .3 cm} $O_1$ \hspace{ .3 cm}  & \hspace{ .3 cm} $O_2$  \hspace{ .3 cm} & \hspace{ .3 cm} $O_3$  \hspace{ .3 cm}&  \hspace{ .3 cm} $O_4$  \hspace{ .3 cm}  \\
\hline
$^4N[{\bf 70},2^+]\frac{7}{2}^+$  &  $N_c$   &  $\frac{2}{3}$    & $-\frac{1}{6N_c}(N_c+1)$ & $\frac{5}{2N_c}$ \\
$^2N[{\bf 70},2^+]\frac{5}{2}^+$  &  $N_c$   &  $\frac{2}{9N_c}(2N_c-3)$ & 0  &   $\frac{1}{4N_c^2}(N_c+3)$  \\
$^4N[{\bf 70},2^+]\frac{5}{2}^+$  &  $N_c$   &  $-\frac{1}{9}$  & $\frac{5}{12N_c}(N_c+1)$  &   $\frac{5}{2N_c}$ \\
$^4N[{\bf 70},0^+]\frac{3}{2}^+$  &  $N_c$   &  0    & 0 &   $\frac{5}{2N_c}$ \\
$^2N[{\bf 70},2^+]\frac{3}{2}^+$  &  $N_c$   &   $-\frac{1}{3N_c}(2N_c-3)$   & 0 &  $\frac{1}{4N_c^2}(N_c+3)$ \\
$^4N[{\bf 70},2^+]\frac{3}{2}^+$  &  $N_c$   &   $-\frac{2}{3}$   & 0 &   $\frac{5}{2N_c}$ \\
$^2N[{\bf 70},0^+]\frac{1}{2}^+$  &  $N_c$   &   0   & 0 &  $\frac{1}{4N_c^2}(N_c+3)$ \\
$^4N[{\bf 70},2^+]\frac{1}{2}^+$  &  $N_c$   &    $-1$   &  $-\frac{7}{12N_c}(N_c+1)$ &  $\frac{5}{2N_c}$ \\
\hline
\hline
\end{tabular}}
\caption{Matrix elements of $N$ used to obtain the fit of Table \ref{70,ell+nonstrangeoperators} \cite{matagne05b}.}
\label{70,ell+nonstrangeNUCLEON}
\end{center}
\end{table}

\begin{table}[h!]
\begin{center}
{\scriptsize
\renewcommand{\arraystretch}{1.75}
\begin{tabular}{lcccc}
\hline 
\hline
   &  \hspace{ .3 cm} $O_1$ \hspace{ .3 cm}  & \hspace{ .3 cm} $O_2$  \hspace{ .3 cm} &  \hspace{ .3 cm} $O_3$  \hspace{ .3 cm} &  \hspace{ .3 cm} $O_4$  \hspace{ .3 cm}  \\
\hline

$^2\Delta[{\bf 70},2^+]\frac{5}{2}^+$  &   $N_c$   &  $-\frac{2}{9}$ & 0  & $\frac{1}{N_c}$ \\
$^2\Delta[{\bf 70},2^+]\frac{3}{2}^+$  &   $N_c$   &  $\frac{1}{3}$  & 0  & $\frac{1}{N_c}$ \\
$^2\Delta[{\bf 70},0^+]\frac{1}{2}^+$  &   $N_c$   &  0 & 0 & $\frac{1}{N_c}$ \\
\hline
\hline
\end{tabular}}
\caption{Matrix elements of $\Delta$ used to obtain the fit of Table \ref{70,ell+nonstrangeoperators} \cite{matagne05b}.}
\label{70,ell+nonstrangeDELTA}
\end{center}
\end{table}

Matrix elements of the operators $O_i$ are presented in Tables \ref{70,ell+nonstrangeNUCLEON} and \ref{70,ell+nonstrangeDELTA}. General formulas used to derived these elements are presented in Appendix \ref{generalformylasmatrix elements}.

\subsubsection{The spectrum of non-strange baryons ($N_f=2$)}

In Table \ref{70,ell+nonstrangemultiplet} we present the masses of the resonances which we have
interpreted as belonging to the $[{\bf 70},0^+]$ or to the $[{\bf 70},2^+]$ multiplet.
For simplicity, mixing of multiplets is neglected in this first attempt.
The resonances shown in column 8, correspond to
either  three stars (``very likely") or to two stars (``fair") or to
one star (``poor") status,
according to Particle Data Group (PDG) \cite{yao06}.
Therefore we used the full listings  to determine a  
mass average in each case. The experimental error to the mass was calculated 
as the quadrature of two uncorrelated errors, one being the average 
error from the same references and the other was the difference
between the average mass and the farthest observed mass.  
For the $P_{11}(2100)$* resonance we report results from
fitting the experimental value
of Ref. \cite{ploetzke98}, as being more recent than the average over the
PDG values. Note that the observed mass of Ref. \cite{ploetzke98} is in
agreement with the recent coupled channel analysis of Manley and
Saleski \cite{manley92}. \\

Several remarks are in order. Due to its large error in the mass,
the resonance $F_{15}$(2000) could be either described by the
$|^2N[{\bf 70},2^+]5/2^+\rangle$ state or by  the $|^4N[{\bf 70},2^+]5/2^+\rangle$ state 
(inasmuch as they appear
separated by about $60-70$ MeV only, in quark model studies, see, \emph{e.g.}, \cite{isgur79,sartor86}).
Here we identified $F_{15}$(2000) with the $|^4N[{\bf 70},2^+]5/2^+\rangle$ state
because it gives a better fit. Regarding the $F_{35}$(1905) resonance
there is also some ambiguity. In Ref. \cite{goity03} it was identified 
as a $|^4\Delta[{\bf 56},2^+]5/2^+\rangle$ state following Ref. \cite{isgur79}, but in
Ref. \cite{matagne05a} the interpretation $|^4\Delta[{\bf 70},2^+]5/2^+\rangle$ was preferred 
due to a better $\chi^2$ fit and other considerations related to the decay 
width. Here we return to the identification made in Ref. \cite{goity03}
and assign the $|^4\Delta[{\bf 70},2^+]5/2^+\rangle$ state to the second 
resonance from this sector, namely $F_{35}$(2000), as indicated in Table \ref{70,ell+nonstrangemultiplet}.
We hope that an analysis based on configuration
mixing and improved data could better clarify the resonance assignment
in this sector in the future. Presently the resulting 
$\chi^2_{\mathrm{dof}}$ is about 0.83 and the fitted values of $c_i$
are given in Table \ref{70,ell+nonstrangeoperators}.
Besides the seven fitted masses Table \ref{70,ell+nonstrangemultiplet} also contains few 
predictions. \\

We found that the contributions of
$S^i_c S^i_c$ and $s^i S^i_c$ are nearly equal when treated as
independent operators. Therefore, for simplicity, we assumed that they
have the same coefficient in the mass operator. 
One can see that the spin-spin interaction given by $O_4$ is the  dominant  
interaction, as in the $[{\bf 56},2^+]$ multiplet \cite{goity03} or 
in the $[{\bf 56},4^+]$ multiplet \cite{matagne05a}. 
Thus the main contributions to the mass come from  $O_1$ and $O_4$.
It is remarkable that $c_1$ and $c_4$ of the multiplets 
$[{\bf 56},2^+]$ and $[{\bf 70},\ell^+]$, both located in the $N = 2$ band, are very close to each other. 
In terms of the present notation the result of Ref. \cite{goity03} for $[{\bf 56},2^+]$ is 
$c_1 = 541 \pm 4$ MeV and $c_4 = 241 \pm 14$ MeV as compared to $c_1 = 555 \pm 11$ MeV and 
$c_4 = 261 \pm 47$ MeV here. Such similarity  gives
confidence in the large $N_c$ approach and in the present fit. \\

Finally note that the contributions of  $O_2$ 
and $O_3$ lead to large errors in the coefficients $c_i$ obtained in the $\chi^2$ fit,
which could possibly be removed with better data.
The operator $O_3$ containing the tensor term plays an important
role in the reduction of $\chi^2_{\mathrm{dof}}$ and it should be further investigated.

\begin{table}[h!]

\begin{center}
{\scriptsize
\renewcommand{\arraystretch}{1.75}
\begin{tabular}{lccccccl}\hline \hline
                    &      \multicolumn{5}{c}{$1/N_c$ expansion results}        &    &                     \\ 
\cline{2-6}		    
                    &      \multicolumn{4}{c}{Partial contribution (MeV)} & \hspace{.0cm} Total (MeV)  \hspace{.0cm}  & \hspace{.0cm}  Empirical (MeV) \hspace{0cm}&   Name, status \hspace{.0cm} \\
\cline{2-5}
                    &   \hspace{.0cm}   $c_1O_1$  & \hspace{.0cm}  $c_2O_2$ & \hspace{.0cm}$c_3O_3$ &\hspace{.0cm}  $c_4O_4$   &    &        \\
\hline
$^4N[{\bf 70},2^+]\frac{7}{2}^+$        & 1665 & 31 & 42 & 217 &      $ 1956\pm95$  & $2016\pm104$ &  $F_{17}(1990)$**  \\
$^2N[{\bf 70},2^+]\frac{5}{2}^+$      & 1665 & 10   & 0 & 43 &      $1719\pm34 $  &    \\

$^4N[{\bf 70},2^+]\frac{5}{2}^+$   & 1665 & -5  & -106 &  217 &     $ 1771\pm88$  & $1981\pm200$ & $F_{15}(2000)$**  \\
$^4N[{\bf 70},0^+]\frac{3}{2}^+$   & 1665  & 0   & 0 & 217 &    $1883\pm17$  & $1879\pm17$ &  $P_{13}(1900)$** \\
$^2N[{\bf 70},2^+]\frac{3}{2}^+$  &   1665   &   -16  & 0 &  43   & $1693\pm42$  &                                \\
$^4N[{\bf 70},2^+]\frac{3}{2}^+$     &  1665    &   -31  & 0 & 217    & $1851\pm69$  &                                   \\
$^2N[{\bf 70},0^+]\frac{1}{2}^+$   &   1665   &   0  &  0 & 43  &   $1709\pm25$  &     $1710\pm30$          &    $P_{11}(1710)$***                 \\
$^4N[{\bf 70},2^+]\frac{1}{2}^+$   & 1665  & -47   & 149 & 217 &     $1985\pm26 $  &   $1986\pm26$ &  $P_{11}(2100)$*\\
\hline
$^2\Delta[{\bf 70},2^+]\frac{5}{2}^+$  &  1665    &   -10  &  0 & 87  &    $1742\pm29$  &  $1976\pm237$             &  $P_{35}(2000)$**                 \\
$^2\Delta[{\bf 70},2^+]\frac{3}{2}^+$     &   1665   &  16   &  0 & 87    &  $1768\pm38$  &                                   \\
$^2\Delta[{\bf 70},0^+]\frac{1}{2}^+$   &   1665   & 0    &  0 &  87 &   $1752\pm19$  &   $1744\pm36$            &   $P_{31}(1750)$*                  \\
\hline
\hline
\end{tabular}}
\caption{The partial contribution and the total mass (MeV) for the non-strange $[{\bf 70}, \ell^+]$ multiplets predicted by the $1/N_c$ expansion as compared with the empirically known masses \cite{matagne05b}.}
\label{70,ell+nonstrangemultiplet}
\end{center}
\end{table}


\subsection{The strange  baryons}
\label{70,ell+strange}
\subsubsection{The mass operator}

For the strange baryons,  the mass operator
can be written as the linear combination 
\begin{equation}
\label{70,ell+strangemassoperator}
M_{[{\bf 70},\ell^+]} = \sum_{i=1}^6 c_i O_i + d_1 B_1 + d_2 B_2 + d_4B_4,
\end{equation} 
where $B_i$ are additional
SU(6) breaking operators which are defined to have vanishing matrix elements for non-strange baryons. The operators $B_1$ and $B_2$ are introduced by analogy with the ground state (Section \ref{groundstatemasssu3breaking}). The operator $B_4$ is new. 
The values of the coefficients $c_i$ and $d_i$
which encode the QCD dynamics, are given in Table \ref{70,ell+strangeoperators}.
They were found by a numerical fit described in the next section. \\

\begin{table}[h!]
\begin{center}
{\scriptsize
\renewcommand{\arraystretch}{1.75} 
\begin{tabular}{llrrl}
\hline
\hline
Operator & \multicolumn{4}{c}{Fitted coef. (MeV)}\\
\hline
\hline
$O_1 = N_c \mathbbm{1} $                                   & \ \ \ $c_1 =  $  & 556 & $\pm$ & 11       \\
$O_2 = \ell_q^i s^i$                                & \ \ \ $c_2 =  $  & -43 & $\pm$ & 47    \\
$O_3 = \frac{3}{N_c}\ell^{(2)ij}_{q}g^{ia}G_c^{ja}$ & \ \ \ $c_3 =  $  & -85 & $\pm$ & 72  \\
$O_4 = \frac{4}{N_c+1} \ell_q^i t^a G_c^{ia}$         & \ \ \            &     &       &     \\
$O_5 = \frac{1}{N_c}(S_c^iS_c^i+s^iS_c^i)$          & \ \ \ $c_5 =  $  & 253 & $\pm$ & 57  \\
$O_6 = \frac{1}{N_c}t^aT_c^a$                       & \ \ \ $c_6 =  $  & -25 & $\pm$ & 86    \\ 
\hline
$B_1 = t^8-\frac{1}{2\sqrt{3}N_c}O_1$               & \ \ \ $d_1 =  $  & 365 & $\pm$ & 169 \\
$B_2 = T_c^8-\frac{N_c-1}{2\sqrt{3}N_c}O_1$         & \ \ \ $d_2 =  $  &-293 & $\pm$ & 54 \\
$B_4 = 3 \ell^i_q g^{i8}- \frac{\sqrt{3}}{2}O_2$    & \ \ \            &     &       & \vspace{0.2cm}\\
\hline \hline
\end{tabular}}
\caption{List of operators and the coefficients resulting from the fit with 
$\chi^2_{\rm dof}  \simeq 1.0$ to non-strange and strange resonances belonging to $[{\bf 70},\ell^+]$ ($\ell=0,2$) \cite{matagne06b}.}
\label{70,ell+strangeoperators}
\end{center}
\end{table}

There are many linearly independent operators $O_i$ and $B_i$ 
which can be constructed from the excited quark and the core operators.  
Here, due to lack of data,
we have considered again a restricted list containing the most dominant
operators in the mass formula. The selection was determined from the
previous experience of Refs. \cite{carlson98b} and \cite{matagne05b} (see Section \ref{70,ell+nonstrange}) for $N_f = 2$
and of Ref. \cite{schat02b} for  $N_f$ = 3. The 
operators $O_i$ entering Eq. (\ref{70,ell+strangemassoperator}) are listed
in Table \ref{70,ell+strangeoperators}. $O_1$ is linear in $N_c$ and is the
most dominant in the mass formula. At $N_c \rightarrow \infty $
is the only one which survives. $O_2$ is the dominant part 
of the spin-orbit operator. It acts on the excited quark and is of 
order $N^0_c$. The operator $O_3$ is a generalization for SU(6) of the $O_3$ operator presented in Section \ref{70,ell+nonstrange}. It is, of course, of order $N_c^0$ because $G_c^{ia}$ is a coherent operator (Section \ref{groundoperatorexpansion}). 
For the same reason the matrix elements of  
$O_4$ are also of order  $N^0_c$. 
As explained in the next section, we could not obtain its coefficient $c_4$,
because of  scarcity of data for the $[{\bf 70},\ell^+]$ multiplet.  
The spin-spin operator $O_5$ is of order $1/N_c$, but its
contribution dominates over all the other terms of the mass operator
containing spin. \\

Here we take into account the isospin-isospin operator, denoted by $O_6$,
having matrix elements of order $N^0_c$ due to the
presence of $T_c$ which sums coherently. Up to a subtracting constant,
it is one of the four independent operators of order $N^0_c$,
which, together with $O_1$, are needed to describe the 
submultiplet structure of $[{\bf 70},1^-]$ \cite{cohen05b}. Incidentally, this operator 
has been omitted in the analysis of Ref. \cite{schat02b}. Its coefficient $c_6$ is 
indicated in Table \ref{70,ell+strangeoperators}. \\
 
In Tables \ref{70,ell+strangeNUCLEON}, \ref{70,ell+strangeDELTA} and \ref{70,ell+strangeSINGLET}
we show the diagonal matrix elements
of  the operators $O_i$ for octet, decuplet and flavor singlet states 
respectively. From these tables one can obtain the large $N_c$ behaviour mentioned 
above. Details about $O_3$ are given in Appendix \ref{generalformylasmatrix elements}. 
Its matrix elements  change the analytic dependence on $N_c$ 
in going from SU(2) to SU(3). This happens for octet resonances 
and can be seen 
by comparing the column 3 of Table \ref{70,ell+strangeNUCLEON} with the corresponding
result from Table \ref{70,ell+nonstrangeNUCLEON}. The change is that the factor $N_c + 1$ in SU(2)
becomes $N_c + 1/3$ in SU(3). The same change takes place for all operators
$O_i$ containing $G^{ja}_c$ as for example the operator $O_4$ also
presented in Appendix \ref{generalformylasmatrix elements}. \\

The SU(6) breaking operators, $B_1$ and $B_2$ and $B_4$ in the notation 
of Ref. \cite{schat02b},
expected to contribute to the mass are listed in Table \ref{70,ell+strangeoperators}.
The operators $B_1$, $B_2$ are the standard breaking operators 
while $B_4$ is directly related to the spin-orbit splitting.
They break the SU(3)-flavor symmetry to first order in $\varepsilon \simeq 0.3$ 
where $\epsilon$ is proportional to the mass difference between the strange 
and $u, d$ quarks.
Table  \ref{70,ell+strangeT8} gives the matrix elements of the excited quark operator 
$t_8$ and of the core operator $T^c_8$ which are necessary to construct
the matrix elements of $B_1$ and $B_2$.
These expressions have been obtained as indicated in Appendix \ref{generalformylasmatrix elements}.
It is interesting to note that they are somewhat different from those  
of Ref. \cite{schat02b}. However for all cases with physical quantum numbers but any $N_c$, our values are
identical to those of  Ref. \cite{schat02b}, so that for $N_c = 3$ there is
no difference. \\

For completeness, Table \ref{70,ell+strangeb3} gives the matrix elements 
of $3\ell^ig^{i8}$ needed to construct $B_4$. 
They were obtained from the formula (\ref{70,ell+strangeB4}) derived in Appendix \ref{generalformylasmatrix elements}. As above, they are different from those of Ref. \cite{schat02b} except for cases where $I$ and $\mathcal{S}$  correspond to physical states.
Unfortunately none of  
the presently known resonances has non-vanishing matrix elements for $B_4$.
By definition all $B_i$ have zero matrix elements for non-strange resonances.
In addition, the matrix elements of $B_4$ for $\ell$ = 0 resonances also cancel
and for the two remaining experimentally known strange resonances they also
cancel out.  
For this reason the coefficient $d_4$ could not be determined. \\

\begin{table}[h!]
\begin{center}
{\scriptsize
\renewcommand{\arraystretch}{1.75}
\begin{tabular}{lcccccc}
\hline
\hline
   &  \hspace{ .3 cm} $O_1$ \hspace{ .3 cm}  & \hspace{ .3 cm} $O_2$  \hspace{ .3 cm} & \hspace{ .3 cm} $O_3$  \hspace{ .3 cm}&   \hspace{ .3 cm} $O_4$  \hspace{ .3 cm} & \hspace{ .3 cm} $O_5$  \hspace{ .3 cm} & \hspace{ .3 cm} $O_6$  \hspace{ .3 cm} \\
\hline
$^48[{\bf 70},2^+]\frac{7}{2}$  &  $N_c$   &  $\frac{2}{3}$    & $-\frac{3N_c+1}{18N_c}$ & $-\frac{2(3N_c+1)}{9(N_c+1)}$ &  $\frac{5}{2N_c}$ & $\frac{N_c-13}{12N_c}$ \\
$^28[{\bf 70},2^+]\frac{5}{2}$  &  $N_c$   &  $\frac{2(2N_c-3)}{9N_c}$ & 0  & $-\frac{4(N_c+3)(3N_c-2)}{27N_c(N_c+1)}$ &  $\frac{N_c+3}{4N_c^2}$ & $\frac{N_c^2-4N_c-9}{12N_c^2}$ \\
$^48[{\bf 70},2^+]\frac{5}{2}$  &  $N_c$   &  $-\frac{1}{9}$  & $\frac{5(3N_c+1)}{36N_c}$ & $\frac{3N_c+1}{27(N_c+1)}$ &   $\frac{5}{2N_c}$ & $\frac{N_c-13}{12N_c}$\\
$^48[{\bf 70},0^+]\frac{3}{2}$  &  $N_c$   &  0    & 0 &  0 & $\frac{5}{2N_c}$ & $\frac{N_c-13}{12N_c}$\\
$^28[{\bf 70},2^+]\frac{3}{2}$  &  $N_c$   &   $-\frac{2N_c-3}{3N_c}$   & 0 & $\frac{2(N_c+3)(3N_c-2)}{9N_c(N_c+1)}$ & $\frac{N_c+3}{4N_c^2}$ & $\frac{N_c^2-4N_c-9}{12N_c^2}$\\
$^48[{\bf 70},2^+]\frac{3}{2}$  &  $N_c$   &   $-\frac{2}{3}$   & 0 &  $\frac{2(3N_c+1)}{9(N_c+1)}$ & $\frac{5}{2N_c}$ & $\frac{N_c-13}{12N_c}$\\
$^28[{\bf 70},0^+]\frac{1}{2}$  &  $N_c$   &   0   & 0 &  0 & $\frac{N_c+3}{4N_c^2}$ & $\frac{N_c^2-4N_c-9}{12N_c^2}$\\
$^48[{\bf 70},2^+]\frac{1}{2}$  &  $N_c$   &    $-1$   &  $-\frac{7(3N_c+1)}{36N_c}$ & $\frac{3N_c+1}{3(N_c+1)}$&  $\frac{5}{2N_c}$ & $\frac{N_c-13}{12N_c}$\vspace{0.2cm} \\
\hline
\hline
\end{tabular}}
\caption{Matrix elements for octet resonances \cite{matagne06b}.}
\label{70,ell+strangeNUCLEON}
\end{center}
\end{table}

\begin{table}[h!]
\begin{center}
{\scriptsize
\renewcommand{\arraystretch}{1.75}
\begin{tabular}{lcccccc}
\hline
\hline
   &  \hspace{ .3 cm} $O_1$ \hspace{ .3 cm}  & \hspace{ .3 cm} $O_2$  \hspace{ .3 cm} &  \hspace{ .3 cm} $O_3$  \hspace{ .3 cm} & \hspace{ .3 cm} $O_4$  \hspace{ .3 cm} & \hspace{ .3 cm} $O_5$  \hspace{ .3 cm} & \hspace{ .3 cm} $O_6$  \hspace{ .3 cm} \\
\hline

$^210[{\bf 70},2^+]\frac{5}{2}$  &   $N_c$   &  $-\frac{2}{9}$ & 0  & $\frac{2(3N_c+7)}{27(N_c+1)}$ & $\frac{1}{N_c}$ & $\frac{N_c+5}{12N_c}$\\
$^210[{\bf 70},2^+]\frac{3}{2}$  &   $N_c$   &  $\frac{1}{3}$  & 0  & $-\frac{3N_c+7}{9(N_c+1)}$ & $\frac{1}{N_c}$ & $\frac{N_c+5}{12N_c}$\\
$^210[{\bf 70},0^+]\frac{1}{2}$  &   $N_c$   &  0 & 0 & 0 & $\frac{1}{N_c}$ & $\frac{N_c+5}{12N_c}$\vspace{0.2cm}\\
\hline
\hline
\end{tabular}}
\caption{Matrix elements for decuplet resonances \cite{matagne06b}.}
\label{70,ell+strangeDELTA}
\end{center}
\end{table}

\begin{table}[h!]
\begin{center}
{\scriptsize
\renewcommand{\arraystretch}{1.75}
\label{SINGLET}
\begin{tabular}{lcccccc}
\hline
\hline
   &  \hspace{ .3 cm} $O_1$ \hspace{ .3 cm}  & \hspace{ .3 cm} $O_2$  \hspace{ .3 cm} &  \hspace{ .3 cm} $O_3$  \hspace{ .3 cm} & \hspace{ .3 cm} $O_4$  \hspace{ .3 cm} & \hspace{ .3 cm} $O_5$  \hspace{ .3 cm} & \hspace{ .3 cm} $O_6$  \hspace{ .3 cm} \\
\hline
$^21[{\bf 70},2^+]\frac{5}{2}$  &  $N_c$   &   $\frac{2}{3}$   &   0 & 0  & 0 & $-\frac{N_c+5}{6N_c}$\\
$^21[{\bf 70},2^+]\frac{3}{2}$  &  $N_c$   &   $-1$   &   0  & 0 & 0 & $-\frac{N_c+5}{6N_c}$\\
$^21[{\bf 70},0^+]\frac{1}{2}$  &  $N_c$   &   0  &   0   & 0 & 0 & $-\frac{N_c+5}{6N_c}$\vspace{0.2cm}\\
\hline
\hline
\end{tabular}}
\caption{Matrix elements for singlet resonances \cite{matagne06b}.}
\label{70,ell+strangeSINGLET}
\end{center}
\end{table}

\begin{table}[h!]
\begin{center}
{\scriptsize
\renewcommand{\arraystretch}{1.75}
\begin{tabular}{cc}
 \hline
 \hline
  & $3\ell^ig^{i8}$ \\ \hline
 $^48[{\bf 70},2^+]\frac{7}{2}$ & $\frac{2N_c-4I(I+1)+\mathcal{S}(\mathcal{S}+4)+1}{2\sqrt{3}(N_c-1)}$ \\
 $^28[{\bf 70},2^+]\frac{5}{2}$  & $\frac{4N_c^3+4I(I+1)(9+N_c(7N_c-12))-9(\mathcal{S}-1)^2-N_c^2(\mathcal{S}-1)(7\mathcal{S}-19)+12N_c(\mathcal{S}(\mathcal{S}-5)+1)}{6\sqrt{3}N_c(N_c-1)(N_c+3)}$ \\
 $^48[{\bf 70},2^+]\frac{5}{2}$ & $-\frac{2N_c-4I(I+1)+\mathcal{S}(\mathcal{S}+4)+1}{12\sqrt{3}(N_c-1)}$ \\
 $^48[{\bf 70},0^+]\frac{3}{2}$ & 0 \\
 $^28[{\bf 70},2^+]\frac{3}{2}$ & $-\frac{4N_c^3+4I(I+1)(9+N_c(7N_c-12))-9(\mathcal{S}-1)^2-N_c^2(\mathcal{S}-1)(7\mathcal{S}-19)+12N_c(\mathcal{S}(\mathcal{S}-5)+1)}{4\sqrt{3}N_c(N_c-1)(N_c+3)}$ \\
 $^48[{\bf 70},2^+]\frac{3}{2}$ & $-\frac{2N_c-4I(I+1)+\mathcal{S}(\mathcal{S}+4)+1}{2\sqrt{3}(N_c-1)}$ \\
 $^28[{\bf 70},0^+]\frac{1}{2}$ & 0 \\ 
 $^48[{\bf 70},2^+]\frac{1}{2}$ &  $-\frac{\sqrt{3}(2N_c-4I(I+1)+\mathcal{S}(\mathcal{S}+4)+1)}{4(N_c-1)}$ \\
 \hline
 $^210[{\bf 70},2^+]\frac{5}{2}$  & $-\frac{2N_c+4I(I+1)-\mathcal{S}(\mathcal{S}-8)-5}{6\sqrt{3}(N_c+5)}$ \\
 $^210[{\bf 70},2^+]\frac{3}{2}$  & $\frac{2N_c+4I(I+1)-\mathcal{S}(\mathcal{S}-8)-5}{6\sqrt{3}(N_c+5)}$ \\
 $^210[{\bf 70},0^+]\frac{1}{2}$  & 0 \\
 \hline
 $^21[{\bf 70},2^+]\frac{5}{2}$  & $\frac{3+12I(I+1)-2N_c(N_c+1)-3\mathcal{S}(\mathcal{S}+2N_c+2)}{\sqrt{3}(N_c+1)(N_c+3)}$ \\
 $^21[{\bf 70},2^+]\frac{3}{2}$  & $-\frac{\sqrt{3}(3+12I(I+1)-2N_c(N_c+1)-3\mathcal{S}(\mathcal{S}+2N_c+2))}{2(N_c+1)(N_c+3)}$ \\ 
 $^21[{\bf 70},0^+]\frac{1}{2}$  & 0 \\
 \hline \hline
\end{tabular}}
\caption{Matrix elements of the term $3\ell^i g^{i8}$ of $B_4$ \cite{matagne06b}.}\label{70,ell+strangeb3}
\end{center}
\end{table}

\begin{sidewaystable}[h!]
\renewcommand{\arraystretch}{1.75}
\begin{center}
{\scriptsize
\begin{tabular}{ccc}
\hline \hline
             &     $t_8$      &        $T^c_8$ \\
\hline
$^28_J$ & $\frac{N_c^3+[\mathcal{S}(5-\mathcal{S})+4I(I+1)-1]N_c^2-3[\mathcal{S}(2-\mathcal{S})+4I(I+1)-2]N_c+9\mathcal{S}}{2\sqrt{3}N_c(N_c-1)(N_c+3)}$ & $\frac{N_c^4+(3\mathcal{S}+1)N_c^3+[(\mathcal{S}(\mathcal{S}+1)-4I(I+1)-2]N_c^2-3[\mathcal{S}(\mathcal{S}+1)-4I(I+1)+2]N_c-9\mathcal{S}}{2\sqrt{3}N_c(N_c-1)(N_c+3)}$ \\
$^48_J$ & $\frac{2N_c-4I(I+1)+\mathcal{S}(\mathcal{S}+4)+1}{4\sqrt{3}(N_c-1)}$ & $\frac{2N_c^2+2(3\mathcal{S}-2)N_c+4I(I+1)-\mathcal{S}(\mathcal{S}+10)-1}{4\sqrt{3}(N_c-1)}$ \\
$^210_J$ & $\frac{2N_c+4I(I+1)-\mathcal{S}(\mathcal{S}-8)-5}{4\sqrt{3}(N_c+5)}$ & $\frac{2N_c^2+2(3\mathcal{S}+4)N_c-4I(I+1)+\mathcal{S}(\mathcal{S}+22)+5}{4\sqrt{3}(N_c+5)}$ \\
$^21_J$ & $\frac{-2N_c^2-2(3\mathcal{S}+1)N_c+12I(I+1)-3\mathcal{S}(\mathcal{S}+2)+3}{2\sqrt{3}(N_c+1)(N_c+3)}$ & $\frac{N_c^3+3(\mathcal{S}+2)N_c^2+(18\mathcal{S}+5)N_c-12I(I+1)+3\mathcal{S}(\mathcal{S}+5)-3}{2\sqrt{3}(N_c+1)(N_c+3)}$\\
$^28_J- {}^210_J$  & $\sqrt{\frac{2}{3}}\sqrt{\frac{N_c+3}{N_c(N_c-1)(N_c+5)}}$ &  $-\sqrt{\frac{2}{3}}\sqrt{\frac{N_c+3}{N_c(N_c-1)(N_c+5)}}$  \\
$^28_J- {}^21_J$ & $\frac{3(N_c-1)}{2\sqrt{N_c}(N_c+3)}$ &$-\frac{3(N_c-1)}{2\sqrt{N_c}(N_c+3)}$
\vspace{0.2cm} \\
\hline
\hline
\end{tabular}}
\caption{Matrix elements of $t_8$ and $T^c_8$ as a function of $N_c$, the
isospin $I$ and the strangeness $\mathcal{S}$. The off-diagonal matrix elements have $(\mathcal{S}=-1,I=1)$ or $(\mathcal{S}=-2,I=1/2)$ for $^28_J- {}^210_J$ and $(\mathcal{S}=0,I=0)$ for $^28_J- {}^21_J$ \cite{matagne06b}.}
\label{70,ell+strangeT8}
\end{center}
\end{sidewaystable}

\clearpage


\subsubsection{The spectrum of non-strange and strange baryons ($N_f=3$)}
Comparing Table \ref{70,ell+strangeoperators} with our previous results for non-strange baryons Table \ref{70,ell+nonstrangeoperators}, one can see that the
addition of strange baryons in the fit 
has not much changed  the values of the coefficients $c_1$ and $c_5$
(previously $c_4$).
The spin-orbit coefficient $c_2$
had changed sign but in absolute value remains small.
The resonance $F_{05} (2100)$ is mostly responsible for this change.
But actually the crucial experimental input for the spin-orbit contribution
should come from $\Lambda$'s, as in the case of the 
$[{\bf 70},1^-]$ multiplet \cite{schat02b}. Unfortunately data for 
the two flavor singlets with $\ell \neq 0$,
$^2\Lambda'[{\bf 70},2^+]5/2^+$ and $^2\Lambda'[{\bf 70},2^+]3/2^+$, 
which are spin-orbit partners are missing (see Table \ref{70,ell+MASSES}).
If observed, they
 will help to fix the strength and sign of 
the spin-orbit terms unambiguously inasmuch as $O_3$, $O_4$ and $O_5$ do 
not contribute to their mass. \\

Presently,  due to the large uncertainty obtained from the fit of $c_2$,
there is still some overlap with the  value obtained from non-strange 
resonances. 
The coefficient $c_3$ is about twice smaller in absolute value now. \\

 Regarding the SU(3) breaking terms, the coefficient
$d_1$  has opposite sign as compared to that of Ref. \cite{schat02b} and
is about four times larger in absolute value.
 The coefficient $d_2$ has the same sign and about the same order 
 of magnitude. One can conclude that the 
SU(3)-flavor breaking is roughly similar in the $[{\bf 70},1^-]$ and the 
$[{\bf 70},\ell^+]$ multiplets. \\

The resonances belonging to the $[{\bf 70},\ell^+]$ together with their
calculated masses are presented in Table \ref{70,ell+MASSES}. 
The angular momentum coupling allows for 8 octets, with $J$ ranging from
7/2 to 1/2, three decuplets with $J$ from 5/2 to 1/2 and three flavor singlets
with $J$ = 5/2, 3/2 or 1/2. Ignoring isospin breaking, there are in all 47 
resonances
from which 12 are fitted and 35 are predictions. 
The best fit gave $\chi^2_{\rm dof} \simeq 1$.
Among the presently 12 resonances only five are new, the strange resonances.
This reflects the fact that the experimental situation is still
rather poor in this energy range. The known resonances are three-, two- and one-star. \\

For all masses the main contribution comes from the operator $O_1$.
In the context of a constituent quark model this corresponds to the
contribution of the spin-independent part of the Hamiltonian, namely
the free mass term plus the kinetic and the confinement energy. A difference 
is that,  this contribution is constant for all resonances here, 
while in quark models the mass difference between the strange and the $u, d$ 
quarks is taken into account explicitly in the free mass term. Here 
this difference is embedded into the flavor breaking terms $B_i$. \\
 
The spin-orbit operator $O_2$ naturally contributes to states with $\ell \neq 0$ only.
The operator $O_3$ contributes to states with $S = 3/2$ only. For $S=1/2$ states it gives no contribution either due to the cancellation of a 6-j coefficient or when the wave function has $S_c=0$, as for example for flavor singlet states. \\

We have analyzed the role of the operator $O_4$ described in 
Appendix \ref{generalformylasmatrix elements}. This is an operator of order $N^0_c$, like
$O_2$, $O_3$ and $O_6$. 
As in Refs. \cite{carlson98b} and \cite{schat02b}, 
the combination $O_2 + O_4$ is of order $1/N_c$ for octets and decuplets,
but this is no longer valid for flavor singlets. It means that the operators 
$O_2$ and $O_4$ are independent in SU(3) and both have to be included in the fit. 
However, the inclusion of $O_4$ considerably deteriorated the fit,
by abnormally increasing the spin-orbit contribution with one order of
magnitude. Therefore the contribution of  $O_4$ 
can not be constrained with the present data and we have to wait 
until more data will be available, especially on strange resonances. \\

To estimate the role of the isospin-isospin operator $O_6$ we have made 
a fit without the contribution of this operator. This fit gave 
$\chi^2_{\rm dof} \simeq 0.9$ and about the same values for
$c_i$ and $d_i$ as that with $O_6$ included. This means that the 
presence of $O_6$ is not essential at the present stage. \\


\begin{table}[htp!]
\begin{center}
\renewcommand{\arraystretch}{1.5}
{\tiny
\begin{tabular}{cccccccccccl}
\hline \hline
                    &      \multicolumn{7}{c}{Part. contrib. (MeV)}  & \hspace{.0cm} Total (MeV)   & \hspace{.0cm}  Exp. (MeV)\hspace{0.0cm}& &\hspace{0.cm}  Name, status \hspace{.0cm} \\

\cline{2-8}
                    &   \hspace{.0cm}   $c_1O_1$  & \hspace{.0cm}  $c_2O_2$ & \hspace{.0cm}$c_3O_3$ &\hspace{.0cm}  $c_5O_5$ &\hspace{.0cm}  $c_6O_6$ & $d_1B_1$& $d_2B_2$&    &        \\
\hline
$^4N[{\bf 70},2^+]\frac{7}{2}$        & 1667 & -29 & 16 & 211 & 7 & 0   & 0     &   $1872\pm 46$  & $2016\pm104$ & & $F_{17}(1990)$**  \\
$^4\Lambda[{\bf 70},2^+]\frac{7}{2}$  &      &     &    &     &   &  0   & 254  &   $2125\pm 72$  & $2094\pm78$  & & $F_{07}(2020)$* \\
$^4\Sigma[{\bf 70},2^+]\frac{7}{2}$   &      &     &    &     &   & -211 & 85  &   $1745\pm95$  &              & & \\
$^4\Xi[{\bf 70},2^+]\frac{7}{2}$      &      &     &    &     &   &  -105 & 423 &   $2189\pm81$   &              & & \vspace{0.2cm}\\
\hline
$^2N[{\bf 70},2^+]\frac{5}{2}$        & 1667 & -10 & 0  & 42  & 3 &  0  &  0    &   $1703\pm29$   &              & & \\
$^2\Lambda[{\bf 70},2^+]\frac{5}{2}$  &      &     &    &     &   & -105   &  169 & $1766\pm26$   &              & & \\
$^2\Sigma[{\bf 70},2^+]\frac{5}{2}$   &      &     &    &     &   & -105   &  169 & $1766\pm26$   &              & & \\
$^2\Xi[{\bf 70},2^+]\frac{5}{2}$      &      &     &    &     &   & -211   &  338 & $1830\pm58$  &              & & \vspace{0.2cm}\\
\hline
$^4N[{\bf 70},2^+]\frac{5}{2}$        & 1667 &  5  &-39 & 211 & 7 &  0  &  0    &   $1850\pm44$   & $1981\pm200$ & & $F_{15}(2000)$** \\
$^4\Lambda[{\bf 70},2^+]\frac{5}{2}$  &      &     &    &     &   &  0    & 254 &  $2104\pm39$   & $2112\pm40$  & & $F_{05}(2110)$*** \\
$^4\Sigma[{\bf 70},2^+]\frac{5}{2}$   &      &     &    &     &   &  -211   & 85 &  $1724\pm111$  &              & & \\
$^4\Xi[{\bf 70},2^+]\frac{5}{2}$      &      &     &    &     &   &  -105   &  423 & $2167\pm54$   &              & & \vspace{0.2cm}\\
\hline
$^4N[{\bf 70},0^+]\frac{3}{2}$        & 1667 &  0  &  0 & 211 & 7 &  0  &  0    &   $1885\pm17$   & $1879\pm17$  & & $P_{13}(1900)$** \\
$^4\Lambda[{\bf 70},0^+]\frac{3}{2}$  &      &     &    &     &   &  0     &  254 &  $2138\pm42$   &              & & \\
$^4\Sigma[{\bf 70},0^+]\frac{3}{2}$   &      &     &    &     &   & -211   &  85  & $1758\pm100$  &              & & \\
$^4\Xi[{\bf 70},0^+]\frac{3}{2}$      &      &     &    &     &   & -105   &  423 &
 $2202\pm56$   &              & & \vspace{0.2cm}\\
\hline
$^2N[{\bf 70},2^+]\frac{3}{2}$        & 1667 & 14  &  0 & 42  & 3 &  0  &  0    &   $1727\pm31$   &              & & \\
$^2\Lambda[{\bf 70},2^+]\frac{3}{2}$  &      &     &    &     &   &  -105   & 169 &   $1790\pm29$   &              & & \\
$^2\Sigma[{\bf 70},2^+]\frac{3}{2}$   &      &     &    &     &   & -105   &  169 & $1790\pm29$   &              & & \\
$^2\Xi[{\bf 70},2^+]\frac{3}{2}$      &      &     &    &     &   & -211   &  338 & $1854\pm59$  &              & & \vspace{0.2cm}\\
\hline
$^4N[{\bf 70},2^+]\frac{3}{2}$        & 1667 & 29  &  0 & 211 & 7 &  0  &  0    &   $1914\pm33$   &              & & \\
$^4\Lambda[{\bf 70},2^+]\frac{3}{2}$  &      &     &    &     &   &   0    &  254 & $2167\pm41$   &              & & \\
$^4\Sigma[{\bf 70},2^+]\frac{3}{2}$   &      &     &    &     &   & -211  &   85 & $1787\pm103$  &              & & \\
$^4\Xi[{\bf 70},2^+]\frac{3}{2}$      &      &     &    &     &   & -105   &  423 & $2231\pm56$   &              & &
\vspace{0.2cm} \\
\hline
$^2N[{\bf 70},0^+]\frac{1}{2}$        &  1667&  0  &  0 & 42  & 3  &  0  &  0    &   $1712\pm27$   &  $1710\pm30$ & & $P_{11}(1710)$*** \\
$^2\Lambda[{\bf 70},0^+]\frac{1}{2}$  &      &     &    &     &    & -105   & 169 &  $1776\pm24$   &            & & \\
$^2\Sigma[{\bf 70},0^+]\frac{1}{2}$   &      &     &    &     &    &  -105   & 169 &  $1776\pm24$   &  $1760\pm27$            & & $P_{11}(1770)$* \\ 
$^2\Xi[{\bf 70},0^+]\frac{1}{2}$      &      &     &    &     &    & -211   &  338 & $1839\pm57$  &              & & \vspace{0.2cm}\\
\hline
$^4N[{\bf 70},2^+]\frac{1}{2}$        & 1667 & 43  & 55 & 211 &  7 &  0  &  0    &   $1983\pm26$   &  $1986\pm26$ & & $P_{11}(2100)$* \\
$^4\Lambda[{\bf 70},2^+]\frac{1}{2}$  &      &     &    &     &    &  0  & 254   &   $2237\pm57$   &              & & \\
$^4\Sigma[{\bf 70},2^+]\frac{1}{2}$   &      &     &    &     &    & -211 & 85  &   $1857\pm90$  &              & & \\
$^4\Xi[{\bf 70},2^+]\frac{1}{2}$      &      &     &    &     &    & -105  & 423 &   $2301\pm68$   &              & & \vspace{0.2cm}\\
\hline
$^2\Delta[{\bf 70},2^+]\frac{5}{2}$   & 1667 & 10  & 0  & 84  & -6 &  0  &  0    &
$1756\pm32$   & $1976\pm237$ & & $F_{35}(2000)$**\\
$^2\Sigma'[{\bf 70},2^+]\frac{5}{2}$  &      &     &    &     &    & -105 & 169   &   $1819\pm46$   &              & & \\
$^2\Xi'[{\bf 70},2^+]\frac{5}{2}$     &      &     &    &     &    & -211 & 338  &   $1883\pm77$  &              & & \\
$^2\Omega[{\bf 70},2^+]\frac{5}{2}$   &      &     &    &     &    & -316 & 507 &   $1946\pm113$  &              & &\vspace{0.2cm} \\
\hline
$^2\Delta[{\bf 70},2^+]\frac{3}{2}$   & 1667 & -14 & 0  & 84  & -6 &  0  & 0   & $1731\pm35$ &                &  & \\
$^2\Sigma'[{\bf 70},2^+]\frac{3}{2}$  &      &     &    &     &    & -105 & 169  & $1795\pm48$ &                &  & \\
$^2\Xi'[{\bf 70},2^+]\frac{3}{2}$     &      &     &    &     &    & -211 & 338 & $1859\pm78$&                &  & \\
$^2\Omega[{\bf 70},2^+]\frac{3}{2}$   &      &     &    &     &    & -316  & 507 & $1922\pm113$&                &  & \vspace{0.2cm}\\
\hline
$^2\Delta[{\bf 70},0^+]\frac{1}{2}$   & 1667 &  0  &  0 & 84  & -6 & 0  &  0  & $1746\pm31$ &  $1744\pm36$   &  & $P_{31}(1750)$* \\
$^2\Sigma'[{\bf 70},0^+]\frac{1}{2}$  &      &     &    &     &    & -105  & 169 & $1810\pm45$ &  $1896\pm95$   &  & $P_{11}(1880)$** \\
$^2\Xi'[{\bf 70},0^+]\frac{1}{2}$     &      &     &    &     &    & 211 & 338 & $1873\pm77$                 &  & \\
$^2\Omega[{\bf 70},0^+]\frac{1}{2}$   &      &     &    &     &    & 316& 507 & $1937\pm112$&                &  & \vspace{0.2cm}\\
\hline
$^2\Lambda'[{\bf 70},2^+]\frac{5}{2}$ & 1667 & -29 &  0 &  0  & 11 & -105 & 169 & $1713\pm51$ &                &  &\vspace{0.2cm} \\
\hline
$^2\Lambda'[{\bf 70},2^+]\frac{3}{2}$ & 1667 &  43 &  0 &  0  & 11 & -105 & 169 & $1785\pm62$ &                &  & \vspace{0.2cm}\\
\hline
$^2\Lambda'[{\bf 70},0^+]\frac{1}{2}$ & 1667 &  0  &  0 &  0  & 11 & -105& 169 & $1742\pm40$ &  $1791\pm64$   &  & $P_{01}(1810)$*** \vspace{0.2cm}\\
\hline
\hline
\end{tabular}}
\caption{The partial contribution and the total mass (MeV) predicted by the $1/N_c$ expansion \cite{matagne06b}. The last two columns give  the empirically known masses \cite{yao06}.}\label{70,ell+MASSES}
\end{center}
\end{table}

The fitted value of the $N(1990) F_{17}$ resonance slightly deteriorates
with respect to the SU(4) case. The reason is the negative contribution 
of the spin-orbit term. Further analysis, based on more data, is 
needed in the future, to clarify the change of sign in the spin-orbit term. \\

Of special interest is the fact that the resonance $\Lambda(1810) P_{01}$
gives the best fit when interpreted as a flavor singlet. Such an interpretation
is in agreement with that of Refs. \cite{glozman96,melde06} where the baryon spectra were
derived from a spin-flavor hyperfine interaction, rooted in
pseudo-scalar meson (Goldstone boson) exchange. Thus the 
spin-flavor symmetry is common to both calculations. Moreover, the
dynamical origin of the operator $O_3$, which does not directly contribute
to $\Lambda(1810) P_{01}$, but plays an important role
in the total fit, is thought to be related to pseudo-scalar meson exchange
\cite{carlson98b}. Hopefully, this study may help 
in shedding some light on the QCD dynamics hidden in the coefficients
$c_i$.  \\

In conclusions we have found that the SU(3) breaking corrections are comparable in
size with the $1/N_c$ corrections, as for the 
$[{\bf 70},1^-]$ multiplet \cite{schat02b} which successfully explained the
$\Lambda(1520) - \Lambda(1405)$ splitting. \\

The analysis of the $[{\bf 70}, \ell^+]$ remains an open problem. It  depends
on future experimental data which may help to clarify the role of
 various terms contributing to the mass operator and in particular of
$O_2$ and $O_4$ of Table \ref{70,ell+strangeoperators}. The present approach provides the theoretical
framework to pursue this study. \\


\section{The strange $[{\bf 56},4^+]$ baryon multiplet}

In this Section we explore the applicability of the $1/N_c$ expansion to the $[\textbf{56},4^+]$ multiplet 
($N = 4$ band).
The number of experimentally known resonances in the 2--3 GeV region
\cite{yao06}, expected to belong to this 
multiplet is quite restricted. Among the five possible candidates
there are two four-star resonances, $N(2220) 9/2^+$ and 
$\Delta(2420) 11/2^+$, one three-star resonance 
$\Lambda(2350) 9/2^+$, one two-star resonance 
$\Delta(2300) 9/2^+$ and one one-star resonance
$\Delta(2390) 7/2^+$. This is an exploratory study 
which will allow us to make some predictions, as shown in Sec. \ref{spindepopexcenergy}, regarding members of multiplets of fixed $J$ or regarding the behaviour of $c_i$ at high excitation energy. \\

In constituent quark models the $N = 4$ band has been studied so far either in 
a large harmonic oscillator basis \cite{capstick86} or in a 
variational basis \cite{stassart97}. We shall show that the present approach reinforces 
the conclusion that the spin-orbit contribution to the hyperfine interaction 
can safely be neglected in constituent quark model calculations. \\

\subsection{The wave functions}
The $N = 4$ band contains 17 multiplets having symmetries (56), (70) or
(20)  and angular momenta ranging from 0 to 4 \cite{stassart97}. Among them, the  
$[\textbf{56},4^+]$ multiplet has a rather simple structure. It is symmetric both in
SU(6) and O(3), where O(3) is the group of spatial rotation. Together with
the color part which is always antisymmetric, it gives a totally antisymmetric
wave function. In our study of 
the $[\textbf{56},4^+]$ multiplet, we have to couple the symmetric orbital 
part $|4m_\ell\rangle$ with $\ell=4$ 
to a symmetric spin-flavor  wave function. Following Eq. (\ref{largencsymmetricexcited}) we have
\begin{equation}
|4S;JJ_3;(\lambda\mu)YII_3; \rangle = \sum_{m_\ell, S_3}\left(
                            \begin{array}{cc|c}
                                4 & S & J \\
                                m_\ell & S_3 & J_3
                            \end{array}
       \right)  |SS_3\rangle |(\lambda\mu)YII_3\rangle |4m_\ell\rangle,
\label{states4+}
\end{equation}
where $S, S_3$ are the spin and its projection,
$(\lambda\mu)$  labels an SU(3) representation (here 8 and 10), $Y, I, I_3$  stand for
the hypercharge, isospin and its projection and $J, J_z$ for the total angular 
momentum and its projection. 
Expressing the states (\ref{states4+}) in the usual notation
$^{2S+1}d_J$, they are as follows: two SU(3) octets
$^28_{\frac{7}{2}}$, $^28_{\frac{9}{2}}$ and four decuplets
$^410_{\frac{5}{2}}$, $^410_{\frac{7}{2}}$, $^410_{\frac{9}{2}}$,
$^410_{\frac{11}{2}}$ (see Table \ref{56,4+multiplet}). \\

In the following,  we need the explicit form of the wave
functions. They depend on  $J_z$ but the matrix elements of the
operators that we shall calculate in the next sections do not
depend on $J_z$ due to the Wigner-Eckart theorem. So, choosing
$J_z=\frac{1}{2}$, we have for 
the octet states
\begin{eqnarray}
|^28[\textbf{56},4^+]\frac{7}{2}^+\frac{1}{2}\rangle & = & \sqrt{\frac{5}{18}}\psi^S_{41}\left(\chi^{\rho}_-\phi^\rho+\chi^\lambda_-\phi^\lambda\right)-\sqrt{\frac{2}{9}}\psi^S_{40}\left(\chi^{\rho}_+\phi^\rho+\chi^\lambda_+\phi^\lambda\right), \label{wavefunc1}\\
|^28[\textbf{56},4^+]\frac{9}{2}^+\frac{1}{2}\rangle & = &\sqrt{\frac{2}{9}}\psi^S_{41}\left(\chi^{\rho}_-\phi^\rho+\chi^\lambda_-\phi^\lambda\right)+\sqrt{\frac{5}{18}}\psi^S_{40}\left(\chi^{\rho}_+\phi^\rho+\chi^\lambda_+\phi^\lambda\right),
\end{eqnarray}
and for the decuplet states
{\scriptsize 
\begin{eqnarray}
|^410[\textbf{56},4^+]\frac{5}{2}^+\frac{1}{2}\rangle  & = &  \left(\sqrt{\frac{5}{21}}\psi^S_{42}\chi_{\frac{3}{2}-\frac{3}{2}}-\sqrt{\frac{5}{14}}\psi^S_{41}\chi_{\frac{3}{2}-\frac{1}{2}}+\sqrt{\frac{2}{7}}\psi^S_{40}\chi_{\frac{3}{2}\frac{1}{2}}-\sqrt{\frac{5}{42}}\psi^S_{4-1}\chi_{\frac{3}{2}\frac{3}{2}}\right)\phi^S, \\
|^410[\textbf{56},4^+]\frac{7}{2}^+\frac{1}{2}\rangle & = & \left(\sqrt{\frac{3}{7}}\psi^S_{42}\chi_{\frac{3}{2}-\frac{3}{2}}-\sqrt{\frac{2}{63}}\psi^S_{41}\chi_{\frac{3}{2}-\frac{1}{2}}-\sqrt{\frac{10}{63}}\psi^S_{40}\chi_{\frac{3}{2}\frac{1}{2}}+\sqrt{\frac{8}{21}}\psi^S_{4-1}\chi_{\frac{3}{2}\frac{3}{2}}\right)\phi^S, \\
|^410[\textbf{56},4^+]\frac{9}{2}^+\frac{1}{2}\rangle & = & \left(\sqrt{\frac{3}{11}}\psi^S_{42}\chi_{\frac{3}{2}-\frac{3}{2}}+\sqrt{\frac{49}{198}}\psi^S_{41}\chi_{\frac{3}{2}-\frac{1}{2}}-\sqrt{\frac{10}{99}}\psi^S_{40}\chi_{\frac{3}{2}\frac{1}{2}}-\sqrt{\frac{25}{66}}\psi^S_{4-1}\chi_{\frac{3}{2}\frac{3}{2}}\right)\phi^S, \\
|^410[\textbf{56},4^+]\frac{11}{2}^+\frac{1}{2}\rangle & = & \left(\sqrt{\frac{2}{33}}\psi^S_{42}\chi_{\frac{3}{2}-\frac{3}{2}}+\sqrt{\frac{4}{11}}\psi^S_{41}\chi_{\frac{3}{2}-\frac{1}{2}}+\sqrt{\frac{5}{11}}\psi^S_{40}\chi_{\frac{3}{2}\frac{1}{2}}+\sqrt{\frac{4}{33}}\psi^S_{4-1}\chi_{\frac{3}{2}\frac{3}{2}}\right)\phi^S.
\label{wavefunc4}
\end{eqnarray}}

\noindent with
$\phi^\lambda$, $\phi^\rho$, $\phi^S$ and $\chi$  given in  Appendix D. The orbital wave functions $\psi^S_{4m}$ are defined in the same way as the ones belonging to the $N=2$ band. The explicit form in term of Jacobi coordinates are not  necessary  for the calculations of this Section. One can found then in Table 2 of Ref. \cite{stassart97}. \\

\subsection{The mass operator}
The study of the $[\textbf{56},4^+]$ multiplet is similar to that of $[\textbf{56},2^+]$
as analyzed in Ref. \cite{goity03}, where the mass
spectrum is studied in the $1/N_c$ expansion up to and including 
$\mathcal{O}(1/N_c)$ effects. As already discussed, the mass operator must be rotationally invariant, parity and time reversal
even. The isospin breaking is neglected.  
The SU(3) symmetry breaking is implemented 
to $\mathcal{O}(\varepsilon)$,
where $\varepsilon \sim 0.3$ gives a measure of this breaking (see Section \ref{groundstatemassoperator}).
As the $[\textbf{56},4^+]$ baryons are 
described by a symmetric representation of SU(6), 
it is not necessary to distinguish between
excited and core quarks for the construction of a basis of mass operators,  
as explained in Ref. \cite{goity03} and in Section \ref{excitedwavefunctions}.
 Then 
the mass operator of the $[{\bf 56},4^+]$ multiplet
has the following structure 
\begin{equation}
\label{56,4+MASS}
M_{[{\bf 56},4^+]} = \sum_i c_i O_i + \sum_i d_i B_i,
\end{equation}
given in terms of the linearly independent operators $O_i$ and $B_i$.
Here
$O_i$ ($i = 1,2,3$) are rotational invariants and SU(3)-flavor singlets
\cite{goity97}, $B_1$ is the strangeness quark number operator
with negative sign, 
and the operators $B_i$ ($i = 2, 3$)
are also rotational invariants but contain the SU(6) flavor-spin generators 
$G_{i8}$ as well. The operators
$B_i$ ($i = 1, 2, 3$) provide SU(3) breaking and are defined to
have vanishing matrix elements for non-strange baryons (see Table \ref{56,4+octets}).
The relation (\ref{56,4+MASS}) contains the effective 
coefficients $c_i$ and $b_i$ as parameters. 
The above operators and the values of the corresponding coefficients 
obtained from fitting the experimentally known masses (see Section \ref{56,4+Fit and discussion}) 
are given in Table \ref{56,4+operators}. \\

\begin{table}[h!]
\begin{center}
{\scriptsize
\renewcommand{\arraystretch}{1.75}
\begin{tabular}{llrrl}
\hline
\hline
Operator & \multicolumn{4}{c}{Fitted coef. (MeV)}\\
\hline
\hline
$O_1 = N_c  \mathbbm{1} $                                               & \ \ \ $c_1 =  $  & 736 & $\pm$ & 30      $\ $ \\
$O_2 =\frac{1}{N_c} \ell_i  S_i$                                  & \ \ \ $c_2 =  $  &  4 & $\pm$ & 40   $\ $ \\
$O_3 = \frac{1}{N_c}S_i S_i$                                    & \ \ \ $c_4 =  $  &  135 & $\pm$ & 90   $\ $ \\
\hline
$B_1 = -{\cal S} $                                         & \ \ \ $ d_1 = $  & 110 & $\pm$ & 67   $\ $ \\
$ B_2 = \frac{1}{N_c} \ell_i G_{i8}-\frac{1}{2 \sqrt{3}} O_2$  & \ \ \  \\

$ B_3 = \frac{1}{N_c} S_i G_{i8}-\frac{1}{2 \sqrt{3}} O_3$  & \ \ \ &  & &  \\

\hline \hline
\end{tabular}}
\caption{Operators of Eq.  (\ref{56,4+MASS}) and coefficients 
resulting from the fit with $\chi^2_{\rm dof}  \simeq 0.26$ \cite{matagne05a}.  }
\label{56,4+operators}
\end{center}
\end{table}

\begin{table}[h!]
\begin{center}
\[
{\scriptsize
\renewcommand{\arraystretch}{1.75}
\begin{array}{crrr}
\hline
\hline
           & \ \ \ \ \ \ \ \  O_1  & \ \ \ \ \ \ \ O_2  & \ \ \ \ \ \ \ O_3  \\
\hline
^28_{7/2}  & N_c  & - \fr{5}{2 N_c} & \fr{3}{4 N_c}  \\
^28_{9/2}  & N_c  &   \fr{2}{  N_c} & \fr{3}{4 N_c}  \\
^410_{5/2} & N_c  & - \fr{15}{2 N_c} & \fr{15}{4 N_c} \\
^410_{7/2} & N_c  & - \fr{4}{  N_c} & \fr{15}{4 N_c} \\
^410_{9/2} & N_c  &   \fr{1}{2 N_c} & \fr{15}{4 N_c} \\
^410_{11/2} & N_c  &   \fr{6}{  N_c} & \fr{15}{4 N_c} \\[0.5ex]
\hline
\hline
\end{array}}
\]
\caption{Matrix elements of  SU(3) singlet operators \cite{matagne05a}.}
\label{56,4+singlets}
\end{center}
\end{table}
The matrix elements of $O_1$, $O_2$ and $O_3$ are trivial 
to calculate. They are given in Table \ref{56,4+singlets} for the octet and
the decuplet states belonging to the $[{\bf 56},4^+]$ multiplet. The  
$B_1$ matrix elements are also trivial.  To calculate
the $B_2$ matrix elements 
we use the expression 
\begin{equation}\label{56,4+Gi8}
G_{i8} = G^{i8} = \frac{1}{2 \sqrt{3}}(S^i - 3 S^i_s), 
\end{equation}
where $S^i$ and $S^i_s$ are the components of the total spin 
and of the total strange-quark spin respectively \cite{jenkins95}\footnote{This equation is equivalent to Eq. (\ref{gi8matrixelements}).}.
Using (\ref{56,4+Gi8}) we rewrite the expression of $B_2$ as 
\begin{equation}\label{B2}
B_2 = - \frac{\sqrt{3}}{2 N_c} \vec{l}\cdot \vec{S}_s
\end{equation}
with the decomposition 
\begin{equation}\label{decomposition}
\vec{l}\cdot \vec{S}_s = l_0S_{s0}+\frac{1}{2}\left( l_+S_{s-}+l_-S_{s+}\right),
\end{equation} 
which we apply on the wave functions 
(\ref{wavefunc1})--(\ref{wavefunc4}) in order to obtain the diagonal 
and off-diagonal matrix elements.  For $B_3$, one can use the following 
relation (see Eq. (\ref{sigi8matrixelement})) 
\begin{equation}\label{B3}
S_i G_{i8} = \fr{1}{4 \sqrt{3}} \left[3I(I+1) -S(S+1) - \fr{3}{4} n_s(n_s+2)\right].
\end{equation}
 Here $I$ is the isospin, $S$ is the total
spin and  $n_s$ the number of strange quarks. Both the diagonal and
off-diagonal matrix elements of  $B_i$  are exhibited in 
Table \ref{56,4+octets}. Note that only  $B_2$ has non-vanishing off-diagonal
matrix elements. Their role is very important in the state mixing, as discussed
in the next section. We found that the diagonal matrix elements of $O_2$, $O_3$,
$B_2$ and $B_3$ of strange baryons satisfy the following relation
\begin{equation}\label{dependence}
\frac{B_2}{B_3} =\frac{O_2}{O_3},
\end{equation}
for any state, irrespective of the value of $J$ in both the octet and the
decuplet. This can be used as a check of the analytic expressions in Table 
\ref{56,4+octets}. Such a relation also holds for the multiplet $[\textbf{56},2^+]$
studied in Ref. \cite{goity03} and might possibly be a feature of all
$[\textbf{56},\ell^+]$ multiplets. In spite of the relation (\ref{dependence}) 
which holds for the diagonal matrix elements, the operators $O_i$ and $B_i$ 
are linearly independent, as it can be easily proved. As an immediate proof, 
the off-diagonal matrix elements of $B_2$ are entirely different
from those of $B_3$.

\begin{table}[h!]
\begin{center}
\[
{\scriptsize
\renewcommand{\arraystretch}{1.75}
\begin{array}{cccc}
\hline
\hline
  &  \hspace{ .6 cm}  { B}_1 \hspace{ .6 cm}  &
     \hspace{ .6 cm}  { B}_2 \hspace{ .6 cm}  &
     \hspace{ .6 cm}  { B}_3 \hspace{ .6 cm}   \\
\hline
N_{J}       & 0 &                      0  &                           0  \\
\Lambda_{J} & 1 &   \fr{  \sqrt{3}\ a_J}{2 N_c} &   - \fr{3 \sqrt{3}}{8 N_c}   \\
\Sigma_{J}  & 1 & - \fr{  \sqrt{3}\ a_J}{6 N_c} &     \fr{  \sqrt{3}}{8 N_c}   \\
\Xi_{J}     & 2 &   \fr{ 2\sqrt{3}\ a_J}{3 N_c} &   - \fr{  \sqrt{3}}{2 N_c}   \\ [0.5ex]
\hline
\Delta_{J}  & 0 &                      0  &                           0  \\
\Sigma_{J}  & 1 &  \fr{ \sqrt{3}\ b_J}{2 N_c} &   - \fr{ 5 \sqrt{3}}{8 N_c}  \\
\Xi_{J}     & 2 &  \fr{ \sqrt{3}\ b_J}{  N_c} &   - \fr{ 5 \sqrt{3}}{4 N_c}  \\
\Omega_{J}  & 3 &  \fr{3\sqrt{3}\ b_J}{2 N_c} &   - \fr{15 \sqrt{3}}{8 N_c}  \\ [0.5ex]
\hline
\Sigma_{7/2}^8 - \Sigma^{10}_{7/2}    &  0 &  -\fr{  \sqrt{35}}{2 \sqrt{3} N_c}   &   0   \\
\Sigma_{9/2}^8 - \Sigma^{10}_{9/2}    &  0 &  -\fr{  \sqrt{11}}{  \sqrt{3} N_c}   &   0   \\
\Xi_{7/2}^8    - \Xi^{10}_{7/2}       &  0 &  -\fr{  \sqrt{35}}{2 \sqrt{3} N_c}   &   0   \\
\Xi_{9/2}^8    - \Xi^{10}_{9/2}       &  0 &  -\fr{  \sqrt{11}}{  \sqrt{3} N_c}   &   0   \\[0.5ex]
\hline
\hline
\end{array}}
\]
\caption{Matrix elements of SU(3) breaking operators, 
with  $a_J = 5/2,-2$ for $J=7/2, 9/2$ respectively and
$b_J = 5/2, 4/3, -1/6, -2$ for $J=5/2, 7/2, 9/2, 11/2$, respectively \cite{matagne05a}.}
\label{56,4+octets}
\end{center}
\end{table}

\subsection{State mixing}

As mentioned above, only the operator $B_2$ has non-vanishing off-diagonal 
matrix elements, so $B_2$ is the only one which 
induces mixing between the octet and decuplet states of $[\textbf{56},4^+]$
with the same quantum numbers, as a consequence of the SU(3)-flavor breaking.
Thus this mixing affects the octet and the decuplet $\Sigma$ and 
$\Xi$ states. As there are four off-diagonal matrix elements
(Table \ref{56,4+octets}), there are also 
four mixing angles, namely,  $\theta_J^{\Sigma}$ and $\theta_J^{\Xi}$, each with
$J =7/2$ and 9/2.  In terms of these mixing angles, the physical $
\Sigma_J$ and $\Sigma_J'$ states are defined by the following basis states
\begin{eqnarray}
|\Sigma_J \rangle  & = & |\Sigma_J^{(8)}\rangle \cos\theta_J^{\Sigma} 
+ |\Sigma_J^{(10)}\rangle \sin\theta_J^{\Sigma}, \\
|\Sigma_J'\rangle & = & -|\Sigma_J^{(8)}\rangle \sin\theta_J^{\Sigma}
+|\Sigma_J^{(10)} \rangle \cos\theta_J^{\Sigma},
\end{eqnarray}
and similar relations hold for $\Xi$. 
The masses of the physical states become
\begin{eqnarray}
M(\Sigma_J) & = & M(\Sigma_J^{(8)}) 
+ d_2 \langle \Sigma_J^{(8)}|B_2|\Sigma_J^{(10)} \rangle \tan \theta^{\Sigma}_J, \label{masssigmaj} \\
M(\Sigma_J') & = & M(\Sigma_J^{(10)}) 
- d_2 \langle \Sigma_J^{(8)}|B_2|\Sigma_J^{(10)} \rangle \tan \theta^{\Sigma}_J \label{masssigma'j},
\end{eqnarray}
where $M(\Sigma_J^{(8)})$ and $M(\Sigma_J^{(10)})$ are the diagonal matrix 
of the mass operator (\ref{56,4+MASS}), 
here equal to $c_1O_1+c_2O_2+c_3O_3+d_1B_1$, 
for $\Sigma$ states and  similarly for $\Xi$ states 
(see Table \ref{56,4+multiplet}). 
If replaced in the mass operator (\ref{56,4+MASS}), the relations (\ref{masssigmaj}) 
and  (\ref{masssigma'j}) and their counterparts for $\Xi$, introduce 
four new parameters which should be included in the fit.
Actually the procedure of Ref. \cite{goity03} was simplified to fit the coefficients 
$c_i$ and $b_i$ directly to the physical masses
and then to calculate the mixing angle from
\begin{equation}
\theta_J = 
\frac{1}{2}\arcsin\left( 2~ \frac{d_2\langle \Sigma_J^{(8)}|B_2|\Sigma_J^{(10)} \rangle}
{ M(\Sigma_J) - M(\Sigma'_J)}\right).
\label{56,4+mixingangles}
\end{equation} 
for $\Sigma_J$ states and analogously for $\Xi$ states.  \\

 Due to the scarcity of
data in the 2--3 GeV mass region, even such a simplified procedure
is not possible at present in the $[\textbf{56},4^+]$ multiplet. 

\subsection{Mass relations}

In the isospin symmetric limit, there are twenty-four independent masses, 
as presented in the first column of Table \ref{56,4+multiplet}. Our operator basis 
contains six operators, so there are eighteen mass relations that hold 
irrespective of the values of the coefficients $c_i$ and $d_i$. These relations 
can be easily checked from the definition (\ref{56,4+MASS}) and are presented in 
Table  \ref{56,4+masrel}.  \\

One can identify the Gell-Mann--Okubo (GMO) mass formula for each octet (two such relations) 
and the equal spacing relations (EQS) for each decuplet (eight such relations). 
There are eight relations left, that involve states belonging to different SU(3)
multiplets as well as to different values of $J$. 
Presently one can not test the accuracy of these relations due to lack of data.
But they may be used in making some predictions. The theoretical masses
satisfy the Gell-Mann--Okubo mass formula for octets and the Equal Spacing
Rule for decuplets, providing another useful test of the $1/N_c$ expansion.

\begin{table}[h!]
\begin{center}
\[
{\scriptsize
\renewcommand{\arraystretch}{1.75}
\begin{array}{crcl}
\hline
\hline
(1) & 9(\Delta_{7/2} - \Delta_{5/2})     & = & 7(N_{9/2} - N_{7/2})  \\
(2) & 9 (\Delta_{9/2} - \Delta_{5/2}) & = & 16 (N_{9/2} - N_{7/2})  \\
(3) & 9(\Delta_{11/2} - \Delta_{9/2}) & = & 11 (N_{9/2} - N_{7/2})  \\
\hline
(4) & 8 (\Lambda_{7/2} - N_{7/2}) +14 (N_{9/2}-\Lambda_{9/2})
&=&   3 (\Lambda_{9/2}-\Sigma_{9/2}) +6 (\Delta_{11/2}-\Sigma_{11/2})  \\
(5) & \Lambda_{9/2} - \Lambda_{7/2} + 3(\Sigma_{9/2} - \Sigma_{7/2}) & = &  4 (N_{9/2} - N_{7/2}) \\
(6) & \Lambda_{9/2} - \Lambda_{7/2} + \Sigma_{9/2} - \Sigma_{7/2} & = &  2 (\Sigma'_{9/2} - \Sigma'_{7/2})  \\
(7) & 11 \; \Sigma'_{7/2} + 9 \; \Sigma_{11/2} &=& 20 \; \Sigma'_{9/2}   \\
(8) & 20 \; \Sigma_{5/2} + 7 \; \Sigma_{11/2} &=& 27 \; \Sigma'_{7/2} \\
\hline
{\rm (GMO)  }    &   2 (N + \Xi) &=& 3 \; \Lambda + \Sigma  \\
{\rm (EQS)  }       &  \Sigma - \Delta &=& \Xi - \Sigma = \Omega - \Xi  \\
\hline
\hline
\end{array}}
\]
\caption{The 18 independent mass relations including
the GMO relations for the octets and the EQS relations for the decuplets \cite{matagne05a}.}
\label{56,4+masrel}
\end{center}
\end{table}

\subsection{Fit and discussion}
\label{56,4+Fit and discussion}
The fit of the  masses derived from  Eq. (\ref{56,4+MASS})
and the available empirical values used in the fit,
together with the corresponding resonance status in the Particle Data Group
\cite{yao06} are listed in Table \ref{56,4+multiplet}.
The values of the coefficients $c_i$ and $b_1$ obtained from 
the fit are presented in Table \ref{56,4+operators}, as already mentioned.
For the four and three-star resonances we used the empirical masses given in 
the summary  table. For the others, namely the one-star resonance
$\Delta(2390)$ and the two-star resonance $\Delta(2300)$
we adopted the following procedure. We considered as ``experimental'' mass
the average of all masses quoted in the full listings. The experimental
error to the mass was defined, as the quadrature of two uncorrelated errors,
one being the average error obtained from the same references in the 
full listings  and the other was the difference 
between the average mass relative to the farthest off observed mass, like for the $[{\bf 70}, \ell^+]$ multiplets. The 
masses and errors thus obtained are indicated in the before last column
of Table \ref{56,4+multiplet}. \\

\begin{table}[h!]
\begin{center}
\renewcommand{\arraystretch}{1.75}
{\scriptsize
\begin{tabular}{cccccccl}\hline \hline
                    &      \multicolumn{6}{c}{1/$N_c$ expansion results}        &                     \\ 
\cline{1-6}		    
                    &      \multicolumn{4}{c}{Partial contribution (MeV)} & \hspace{.0cm} Total (MeV)  \hspace{.0cm}  & \hspace{.0cm}  Empirical \hspace{.0cm} &  Name, status \\
\cline{2-5}
                    &   \hspace{.0cm}   $c_1O_1$  & \hspace{.0cm}  $c_2O_2$ & \hspace{.0cm}$c_3O_3$ &\hspace{.0cm}  $d_1B_1$   & \hspace{.0cm}   &  (MeV)   &    \\
\hline
$N_{7/2}$        & 2209 & -3 &  34 &   0  &  \hspace{.0cm} $ 2240\pm97 $ \hspace{.0cm} &\hspace*{.0cm}  \hspace{.0cm} &  \\
$\Lambda_{7/2}$  &      &     &    & 110  &  $2350\pm118 $  & & \\
$\Sigma_{7/2}$   &      &     &    & 110  &  $2350\pm118 $  &               & \\
$\Xi_{7/2}$      &      &     &    & 220  &  $2460\pm 166$  &               &  \\
\hline
$N_{9/2}  $      & 2209 & 2   & 34 &   0  &   $2245\pm95 $  & $ 2245\pm65 $ & N(2220)**** \\
$\Lambda_{9/2}$  &  &     &    & 110  &  $ 2355\pm116 $  & $ 2355\pm15 $ &  $\Lambda$(2350)***\\
$\Sigma_{9/2}$   &      &     &    & 110  &  $ 2355\pm116 $  &               &                    \\
$\Xi_{9/2}$      &      &     &    & 220  &  $2465\pm164$  &               &                    \\
\hline
$\Delta_{5/2}$   & 2209 & -9  &168 &   0  &  $ 2368\pm175$  & &  \\
$\Sigma^{}_{5/2}$&      &     &    & 110  & $2478\pm187$  & &   \\
$\Xi^{}_{5/2}$   &      &     &    & 220  &  $2588\pm220$  &               &                   \\
$\Omega_{5/2}$   &      &     &    & 330  & $2698\pm266$  &               &                    \\
\hline
$\Delta_{7/2}$   &2209  &-5   &168 &  0   &  $2372\pm153$  & $2387\pm88$ &  $\Delta$(2390)* \\
$\Sigma'_{7/2}$  &      &     &    & 110  & $2482\pm167$  &               &                  \\
$\Xi'_{7/2}$     &      &     &    & 220  & $2592\pm203$  &               &                    \\
$\Omega_{7/2}$   &      &     &    & 330  &  $2702\pm252$  &               &                    \\
\hline
$\Delta_{9/2}$   &2209  & 1   &168 &  0   &   $2378\pm144 $  & $2318\pm132  $ &  $\Delta$(2300)**\\
$\Sigma'_{9/2}$  &      &     &    & 110  &   $2488\pm159$  &               &                   \\
$\Xi'_{9/2}$     &      &     &    & 220  &  $2598\pm197$  &               &                    \\
$\Omega_{9/2}$   &      &     &    & 330  &  $2708\pm247$  &               &                    \\
\hline
$\Delta_{11/2}$  &2209  &7    &168 &  0   &  $2385\pm164$  & $ 2400\pm100$ &   $\Delta$(2420)**** \\
$\Sigma^{}_{11/2}$ &    &     &    & 110  & $2495\pm177$  &               &                     \\
$\Xi^{}_{11/2}$  &      &     &    & 220  &  $2605\pm212$  &               &                     \\
$\Omega_{11/2}$  &      &     &    & 330  &  $2715\pm260$  &               &                     \\
\hline
\hline
\end{tabular}}
\caption{The partial contribution and the total mass (MeV)
predicted by the $1/N_c$ expansion as compared with the 
empirically known masses \cite{matagne05a}.}
\label{56,4+multiplet}
\end{center}
\end{table}

Due to the lack of experimental data in the strange sector it was 
not possible to include all the operators $B_i$ in the fit in order 
to obtain 
some reliable predictions.  As the breaking of SU(3) is dominated by 
$B_1$ we included only this operator in  Eq. (\ref{56,4+MASS})
and neglected the contribution of the operators $B_2$ and $B_3$.
At a later stage, when more data will hopefully be available, all analytical 
work performed here could be used to improve the fit. That is why
Table \ref{56,4+operators} contains results for  $c_i$ ($i$ = 1, 2 and 3) 
and $d_1$ only. The $\chi^2_{\mathrm{dof}}$ of the fit is 0.26, where 
the number of degrees of freedom (dof) is equal to one (five data and four 
coefficients). \\

The first column of Table  \ref{56,4+multiplet}
 contains the 56 states (each state having a $2 I + 1$ multiplicity
from assuming an exact SU(2)-isospin symmetry) \footnote{%
Note that the notation $\Sigma_J$, $\Sigma'_J$ is consistent with 
the relations (\ref{masssigmaj}),~(\ref{masssigma'j}) inasmuch as 
the contribution of $B_2$ is neglected (same remark 
for $\Xi_J$, $\Xi'_J$ and corresponding relations).}. 
The columns two to five show the 
partial contribution of each operator included in the fit, multiplied by
the corresponding coefficient $c_i$ or $d_1$.
The column six gives the total mass according to Eq. (\ref{56,4+MASS}). 
The errors shown in the predictions result from the errors on the 
coefficients $c_i$ and $d_1$ given in Table \ref{56,4+operators}.
As there are only five experimental data available, nineteen of these 
masses are predictions. The breaking of SU(3)-flavor due to the operator
$B_1$ is 110 MeV as compared to 200 MeV produced in the 
$[\textbf{56},2^+]$ multiplet \cite{goity03}. \\

The main question is, of course, how reliable is this fit. The answer
can be summarized as follows:
\begin{itemize}
\item The main part of the mass is provided by the spin-flavor singlet operator
$O_1$, which is $\mathcal{O}(N_c)$. 

\item The spin-orbit contribution given by $c_2O_2$ is small. This fact 
reinforces the practice used in constituent quark models where the spin-orbit
contribution is usually neglected. Our result is consistent with the expectation
that the spin-orbit term vanishes at large excitation energies \cite{glozman02}.

\item The breaking of the $SU(6)$ symmetry keeping the flavor symmetry exact
is mainly due to the spin-spin operator $O_3$. This hyperfine interaction 
produces a splitting between octet and decuplet states of approximately 130 MeV 
which is smaller than that obtained in the $[\textbf{56},2^+]$ 
case \cite{goity03}, which gives 240 MeV.

\item The contribution of $B_1$
per unit of strangeness, 110 MeV, is also smaller here than in the 
$[\textbf{56},2^+]$ multiplet \cite{goity03}, where it takes a value of about 
200 MeV. That may be quite natural, as one expects a shrinking of the spectrum
with the excitation energy.

\item As it was not possible to include the contribution of $B_2$ and 
$B_3$ in our fit, a degeneracy appears between $\Lambda$ and $\Sigma$. 
\end{itemize}

In conclusion we have studied the spectrum of highly excited resonances in the 2--3 GeV
mass region by describing them as belonging to the $[\textbf{56},4^+]$ multiplet.
This is the first study of such excited states based on the $1/N_c$ 
expansion of QCD. A better description should include multiplet 
mixing, following the lines developed, for example, in Ref. \cite{goity04}. \\

We support previous assertions that better experimental values 
for highly excited non-strange baryons as well as more data 
for the $\Sigma^*$ and $\Xi^*$ baryons are needed in order to understand
the role of the operator $B_2$  within a multiplet
and for the octet-decuplet mixing. With better data the analytic work 
performed here will help 
to make reliable predictions in the large $N_c$ limit formalism.


\section{The spin dependence of the mass operator coefficients with the excitation energy}
\label{spindepopexcenergy}
The properties of low energy hadrons are interpreted to be a consequence of the
spontaneous breaking of chiral symmetry
\cite{glozman96,manohar84}.
For highly excited hadrons, as the ones considered here, there are
phenomenological arguments 
to believe that the chiral symmetry is restored.
This would imply
a weakening (up to a cancellation) of the spin-orbit and tensor interactions  
\cite{glozman02}. Then   
the main contribution to the hyperfine interaction 
remains the spin-spin term. \\

Indeed we have seen that for all resonances,  the spin-spin contribution is dominant, like in constituent quark model studies. 
Thus the $1/N_c$ expansion can provide a deeper understanding of the successes 
of the quark models. \\

\begin{figure}[h!]
 \begin{center}
  \includegraphics[width=10cm,keepaspectratio]{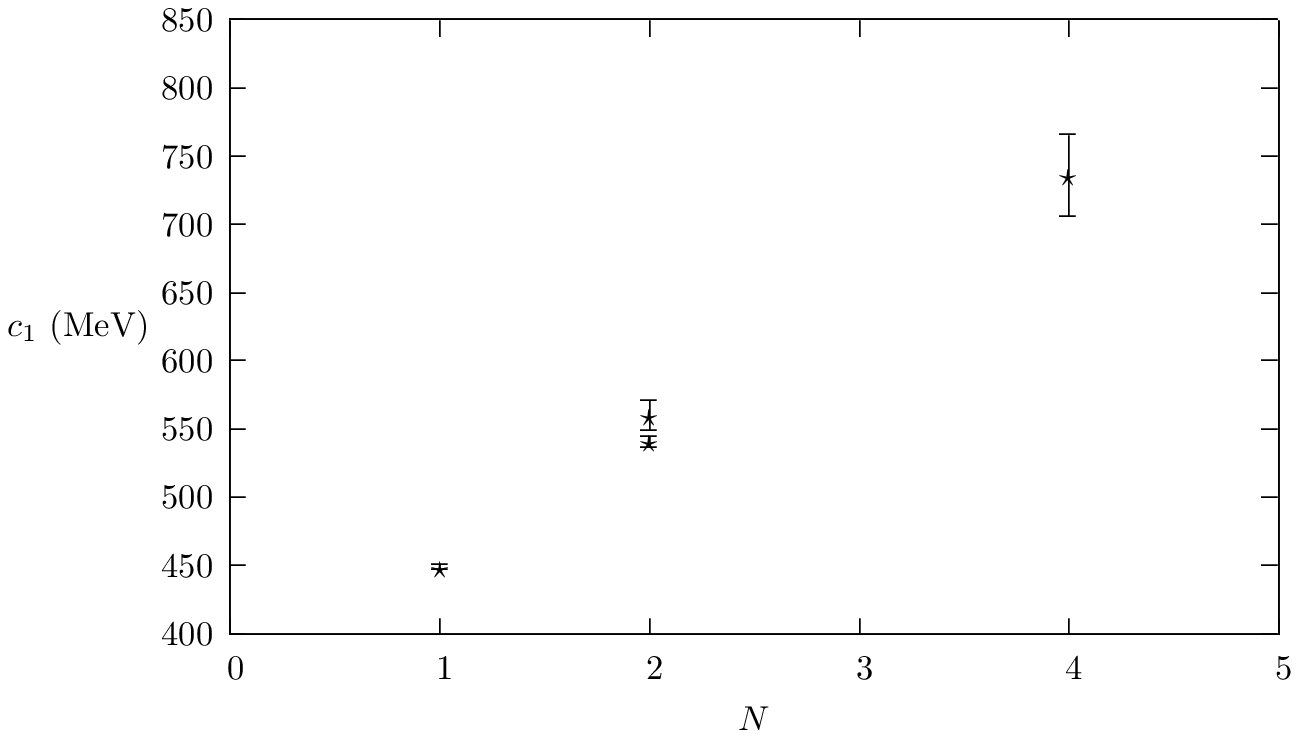} \\
  \includegraphics[width=10cm,keepaspectratio]{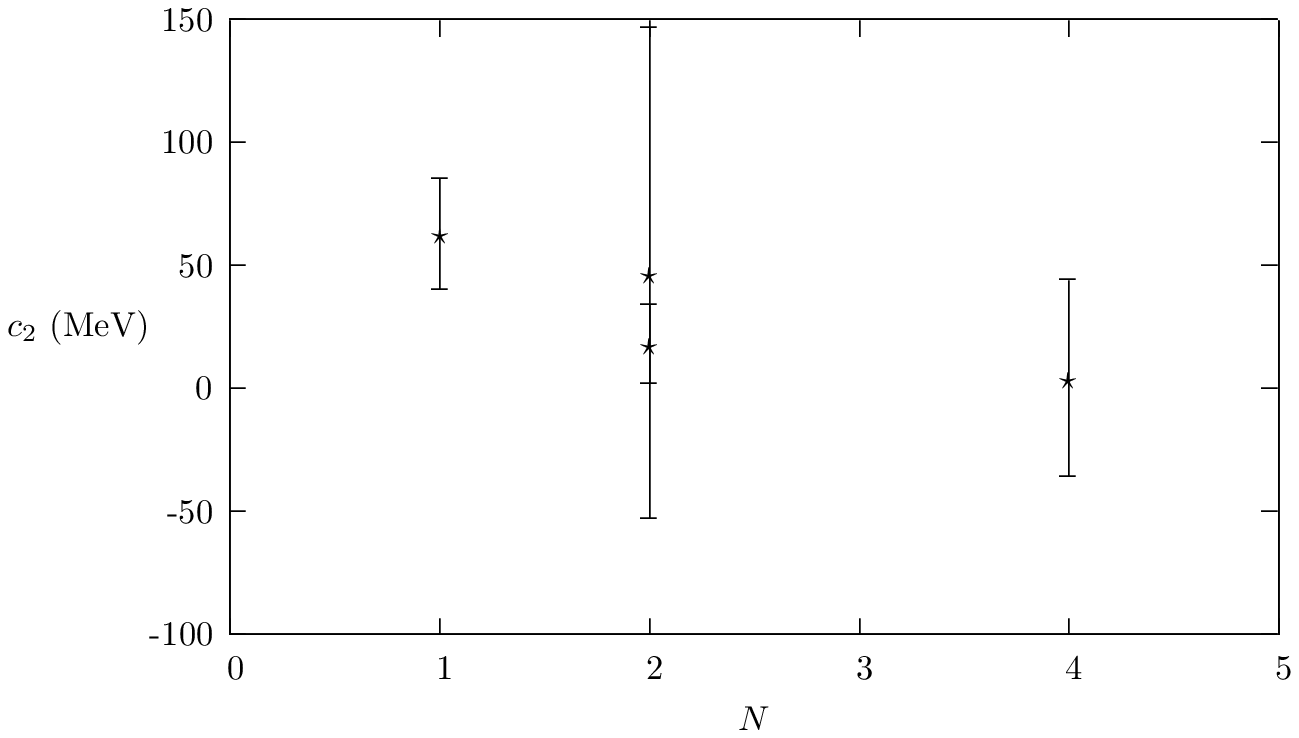} \\
  \includegraphics[width=10cm,keepaspectratio]{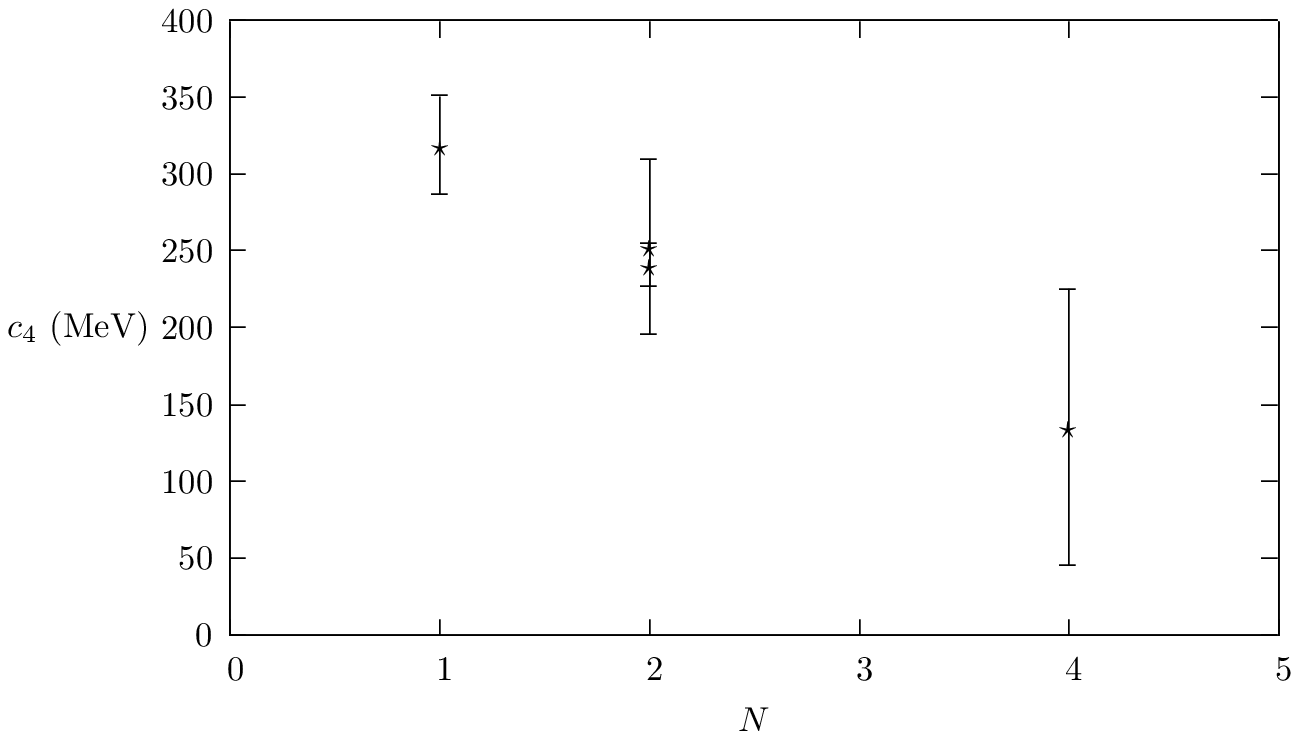}
  \caption{The coefficients $c_i$ vs. $N$ from various sources: $N = 1$ from Ref. \cite{schat02b}, for $N = 2$ from Ref. \cite{goity03} (lower values) and Ref. \cite{matagne05b} and for $N = 4$ from \cite{matagne05a}. The straight lines are to guide the eye. This figure suggests that the absolute value of $c_2$ and $c_4$ tend to zero when $N$ increase. Of course the coefficients $c_2$ and $c_4$ does not tend to $-\infty$ when $N$ increase.}\label{spindependencemassoperator}
 \end{center}
\end{figure}

It is interesting to see the evolution of some dynamical coefficients with the excitation energy. $c_1$ refers to the first order operator $N_c\mathbbm{1}$, $c_2$ to the spin-orbit term and $c_4$ to the spin-spin interaction\footnote{The denomination of the coefficients $c_i$ changes from one multiplet to another. Here, we use the one presented in Section \ref{70,ell+nonstrange} because the Figure \ref{spindependencemassoperator} was originally discussed in Ref. \cite{matagne05b}.}. \\

In Figure \ref{spindependencemassoperator} the presently known values of $c_1, c_2$ and $c_4$ with error bars 
are represented for the excited bands studied within the large $N_c$
expansion: $N = 1$ is from Ref. \cite{schat02b},  $N = 2$ from Ref. \cite{goity03} and
from the present work, $N = 4$ from Ref. \cite{matagne05a}. Extrapolation to higher 
energies, $N > 4$, suggests that the contribution of the spin dependent operators
would vanish, while the linear term in $N_c$,
which in a quark model picture would contain the free mass term, the kinetic 
and the confinement energy, would carry the entire
excitation.  Such a behaviour 
gives a deeper insight into the large $N_c$ mass operator
and is consistent with
the intuitive picture developed in Ref. \cite{glozman02} where at high energies the 
spin dependent interactions vanish as a consequence of the 
chiral symmetry restoration.

\chapter{New look at the $[{\bf 70},1^-]$ multiplet}

\thispagestyle{empty}

\section{Introduction}

The decoupling picture  presented in Chapter 5 describes the wave function of baryons belonging to mixed-symmetric multiplets as a symmetric core coupled to an excited quark. But as explained in Section \ref{chap5hartreeapprox}, this approach implies a truncation of the available basis vector space (see Eq. (\ref{hartreeyoungtabms})). An important consequence of this approximation is the breaking  to  first order of SU($2N_f$) symmetry for mixed-symmetric orbital wave functions (see Eq. (\ref{summarydecoupling})). This breaking to  first order  is an artifact of the decoupling picture as we shall see below. \\

Another consequence of the decoupling picture is that the number of independent operators appearing in the $1/N_c$ expansion of the mass operator for mixed-symmetry multiplets increases tremendously and the number of coefficients to be determined by a fit to experimental data becomes larger or much larger than the number of data. For example, for the $[{\bf 70},1^-]$ multiplet with $N_f = 2$ one has 12 linearly independent
operators up to order $1/N_c$  included \cite{carlson98b}, instead of 6 
(see below) when the 
splitting is not performed. We recall that there are only 7 non-strange 
resonances belonging to this band.
Consequently, in selecting the most dominant operators one has to make an
arbitrary choice  \cite{carlson98b} (see   Section \ref{70,ell+strange} for an explicit example of the $[{\bf 70},\ell^+]$ multiplets). \\

In this chapter, we propose a method to treat the $[{\bf 70},1^-]$ multiplet where the  decoupling of the system into a symmetric core and an excited quark is unnecessary. All one needs to know are the matrix
elements of the SU($2N_f$) generators between mixed-symmetric
states   $[N_c-1,1]$. For $N_f = 2$ these are provided by the 
work of Hecht and Pang  performed in the sixtieths in the context of nuclear physics where the SU(4) symmetry is very important \cite{hecht69}.
To our knowledge such matrix elements are yet 
unknown for  $N_f = 3$. Thus this analysis of the $[{\bf 70},1^-]$ multiplet  deals with non-strange 
baryon resonances only. \\

This chapter is divided into two parts. In the first part we shall show that when we consider  the baryon wave function  (\ref{nondecouplingchap5}) for the $[{\bf 70},1^-]$  multiplet and not the approximate form symbolized by Young tableaux in Eq. (\ref{hartreeyoungtabms}), the SU($2N_f$) symmetry is not broken to  first order but to order $1/N_c$, like for symmetric multiplets. In the second part we will illustrate this new approach with the study of the non-strange $[{\bf 70},1^-]$ baryon multiplet. The results presented here can be found in Ref. \cite{matagne06b}. A summary of the $1/N_c$ expansion method for excited baryons in the decoupling and the new  picture presented here can be found in Ref. \cite{matagne06f}

\section{The wave function of $[{\bf 70},1^-]$ excited states}

As presented in Table \ref{harmonicconfigurationsnc}, the baryons  belonging to the $[{\bf 70},1^-]$ multiplet are composed of $N_c-1$ ground-state quarks  and of one excited quark which give the configuration $(0s)^{N_c-1}(0p)$. The orbital part must have a mixed-symmetry  denoted by the partition $[N_c-1,1]$. The spin-flavor part
must have the same symmetry in order to obtain a totally symmetric
state in the orbital-spin-flavor space.  \\
 
 For the configuration $(0s)^{N_c-1}(0p)$ the basis vectors forming the invariant subspace of the mixed-symmetric $[N_c-1,1]$ irrep of the permutation group S$_{N_c}$  can be illustrated by the Weyl tableau represented in Figure  \ref{weyl[N_c-1,1]}. The dimension of this irrep is $N_c-1$. The basis vectors of this irrep can be described, for example,  by the generalized Jacobi coordinates Eq. (\ref{ncmixedsymmetricjacobi}), given in Chapter 2.   Table \ref{nc=5basisfunctions} shows  an example  for $N_c=5$ where the four independent basis vectors are written explicitly in terms of single particle states $s$ and $p$. The corresponding Young tableaux are also indicated. Note that in writing the content in $s$ and $p$ states the order is always the normal order $1,2,3,4,5$. Then one can see that the excited quark can be the $5$-th quark only for the normal Young tableau shown in the first row of Table \ref{nc=5basisfunctions}, but never for the other three basis vectors shown in the subsequent.

\begin{figure}[h!]
\begin{center}
\mbox{$\overbrace{\negthinspace\negthinspace\negthinspace\negthinspace\negthinspace\negthinspace\raisebox{-9pt}{\begin{Young}
                 $s$ & $s$ \cr
		 $p$ \cr
                \end{Young}}\  \raisebox{3.5pt}{$\cdots$}\negthinspace \negthinspace\negthinspace
\raisebox{3.5pt}{\begin{Young} $s$ \cr \end{Young}}}^{N_c-1}$}
\caption{Weyl tableau symbolizing the irreducible representation $[N_c-1,1]$ of the group $S_{N_c}$ for the configuration $(0s)^{N_c-1}(0p)$. The dimension of this irrep is $N_c-1$.}  \label{weyl[N_c-1,1]}
\end{center}
\end{figure}

\begin{table}[h!]
 \begin{center}
  \begin{tabular}{c|c}
   \hline \hline \\ \vspace{-0.5cm} &     \\
   Young tableau & Young-Yamanouchi basis vectors of [41] \\ \\ \vspace{-0.5cm} &       \\
   \hline \\
\raisebox{-10.5pt}{\mbox{\begin{Young}
1 & 2 & 3 & 4 \cr
5 \cr
\end{Young}}} & 
$\frac{1}{\sqrt{20}} \left(4ssssp-sssps-sspss-spsss-pssss\right)$ \\   \\
\raisebox{-10.5pt}{\mbox{\begin{Young}
1 & 2 & 3 & 5 \cr
4 \cr
\end{Young}}} &
$\frac{1}{\sqrt{12}}\left(3sssps-sspss-spsss-pssss\right)$ \\ \\
\raisebox{-10.5pt}{\mbox{\begin{Young}
1 & 2 & 4 & 5\cr
3 \cr
\end{Young}}} &
$\frac{1}{\sqrt{6}}\left(2sspss-spsss-pssss\right)$ \\ \\
\raisebox{-10.5pt}{\mbox{\begin{Young}
1 & 3 & 4 & 5 \cr
 2 \cr
\end{Young}}} &
$\frac{1}{\sqrt{2}} \left(spsss-pssss\right)$ \\ \\
\hline\hline
  \end{tabular}
  \caption{Young tableaux and the corresponding basis vectors of the irrep $[41]$  of S$_{5}$ for the configuration $(0s)^4(0p)$ \cite{stancu96}.}\label{nc=5basisfunctions}
 \end{center}
\end{table}

For $N_c$ quarks the generalization of the first row of Table \ref{nc=5basisfunctions} is
\begin{equation}
 \frac{1}{\sqrt{N_c(N_c-1)}}\left((N_c-1)\overbrace{sss\ldots s}^{N_c-1}p-\overbrace{sss\ldots s}^{N_c-2}ps-\overbrace{sss\ldots s}^{N_c-3}pss- \cdots -p\overbrace{sss\ldots s}^{N_c-1}\right). \label{n_cquarkbasis}
\end{equation}
This is the only basis vector taken into account in the decoupling picture, Eq. (\ref{hartreems}).
 However, in order to obtain a symmetric orbital-spin-flavor state one has to consider all  possibilities  presented in Eq. (\ref{nondecouplingchap5}). 
 If one does so the matrix elements of operators applied on the wave function (\ref{hartreems}) are identical for all $Y$'s, due to Weyl's duality between a linear group and a symmetric group in a given tensor space. The duality holds provided both the Lie and the permutation group are exact symmetries \cite{stancu96}. Thus there is no need to specify $Y$.  Then the explicit form of a  wave function 
of a total angular momentum $\vec{J} = \vec{\ell} + \vec{S}$ and an isospin $I$ can be written as
\begin{equation}\label{WFnewlook}
|\ell S I I_3;JJ_3 \rangle  = 
\sum_{m_\ell,S_3}
      \left(\begin{array}{cc|c}
	\ell    &    S   & J   \\
	m_\ell  &    S_3  & J_3 
      \end{array}\right)  |[N_c-1,1] S S_3 I I_3 \rangle 
|[N_c-1,1]\ell m_{\ell} \rangle.
\end{equation}
In the following there will be no need to specify the symmetry of the orbital part
$|\ell m_{\ell} \rangle $.

\section{Order of the spin-orbit operator}

As we have seen in Chapter 5, the decoupling picture predicts that the  SU($2N_f$) symmetry is broken to  first order by the operator  $\vec{\ell}(N_c) \cdot \vec{s}(N_c)$ acting on the $N_c$-th quark. Indeed, as $\langle\vec{\ell}(N_c)\cdot \vec{s}(N_c)\rangle$ is $\neq 0$ only when it acts on the first term of the orbital wave function  (\ref{n_cquarkbasis}), the operator is of order $\sqrt{\frac{N_c-1}{N_c}}\sim 1$. \\

But let us consider the total wave function  (\ref{nondecouplingchap5}) and accordingly take into account  the global factor $1/\sqrt{N_c-1}$ appearing in it. This factor is missing in the approximate form (\ref{hartreems}). By taking it into account  one finds immediately that the one-body spin-orbit operator $\vec{\ell}(N_c)\cdot \vec{s}(N_c)$ acting on the $N_c$-th quark is of order $\mathcal{O}(1/N_c)$. One can easily verify that  the same result holds for symmetric multiplets. Thus
\begin{eqnarray}
\langle \Psi |\vec{\ell}(N_c)\cdot \vec{s}(N_c) | \Psi \rangle = 
 \renewcommand{\arraystretch}{1.75}\left\{\begin{array}{ll}
         \mathcal{O}\left(\frac{1}{N_c}\right) & \mbox{with $\Psi=\Phi_s$ (Eq. (\ref{hartreesymm}))} \\
	 \mathcal{O}\left(\frac{1}{N_c}\right)  & \mbox{with $\Psi=\Phi'_s$ (Eq. (\ref{nondecouplingchap5}))}
        \end{array} \right. ,\label{newlooklsnc}
\end{eqnarray}
where one can notice the difference with Eq. (\ref{summarydecoupling}).
Therefore, when the antisymmetrization is properly taken into account the matrix elements of the one-body spin-orbit operator is of order $1/N_c$ both for symmetric and mixed-symmetric multiplets. \\

However, in Ref. \cite{carlson98b} the one-body spin-orbit $\vec{\ell}\cdot \vec{s}$ has the order $\mathcal{O}(N_c^0)$. It is a consequence of the wave function given by Eq. (3.4) of that work, where the coefficient $c_{\rho,\eta}$ are isoscalar factors of the permutation group \cite{matagne05b}. Their expressions (3.5) indicate, as explained in Section \ref{Mixed symmetric  wave functions}, that the $N_c$-th quark is located in the second row of the Young tableau describing the spin-flavor part of the wave function, which allows the decoupling into a symmetric core and a single quark in the spin-flavor space. However the position of the $N_c$-th quark  is out of control in the wave function (3.4) of Ref. \cite{carlson98b}.  If the position of the $N_c$-th quark was specified the orbital part $|\ell m_\ell\rangle$ would have carried the label $N_c$, for example $Y_\ell^m(\hat{x}_{N_c})$. But this is not the case. For this reason all  $N_c$ quarks can contribute to the one-body spin-orbit operator $\vec{\ell}\cdot \vec{s}$, this operator acting on every quark in the wave function. The whole contribution is equivalent with summing over all $N_c$ quarks
\begin{equation}
 \sum_{i=1}^{N_c} \langle \Psi |\vec{\ell}(i)\cdot \vec{s}(i) | \Psi \rangle = N_c \langle \Psi |\vec{\ell}(N_c)\cdot \vec{s}(N_c) | \Psi \rangle. \label{spinorbitonebody}
\end{equation}
That is why the one-body part of the spin-orbit operator is of order $\mathcal{O}(1)$. By doing explicit calculations for $N_c=3$ one can prove that matrix elements of the one-body operator $\vec{\ell}\cdot \vec{s}$ used by Carlson \emph{et al.} \cite{carlson98b} corresponds to $\sum_{j=1}^{N_c} \langle \Psi |\ell^i(j)s^i(j) | \Psi \rangle$ provided $\Psi$ is properly symmetrized (see Appendix E). \\

One can notice that Eq. (\ref{spinorbitonebody}) is valid for any symmetry of the orbital and the spin-flavor part of the wave function, \emph{i.e.} $\Psi$ can be identified either with $\Phi_S$,  Eq. (\ref{hartreesymm}) or $\Phi'_s$, Eq. (\ref{nondecouplingchap5}).
 Then  this operator must be of order $N_c^0$ for symmetric and mixed-symmetric multiplets. \\

 One can now introduce  a pure two-body operator $\vec{\ell}(i)\cdot \vec{s}(j),\ i\neq j$. The matrix elements of this operator are of course of order $1/N_c$.  As above, this operator  acts on whole  the orbital-spin-flavor baryon wave function. In that case, one has
 \begin{equation}
  \sum_{i=1}^{N_c} \sum_{j\neq i =1}^{N_c}\langle \Psi |\vec{\ell}(i)\cdot \vec{s}(j) | \Psi \rangle = N_c(N_c-1)\langle \Psi |\vec{\ell}(N_c)\cdot \vec{s}(N_c-1) | \Psi \rangle, \label{spinorbittwobody}
 \end{equation}
 which is of order $N_c$. 
  For $N_c=3$ (see appendix E) one can show that the matrix elements of the operator $\ell^i S^i_c$ used in Ref. \cite{carlson98b} are equal to those of  $\sum_{i=1}^{N_c} \sum_{j\neq i =1}^{N_c}\langle \Psi |\vec{\ell}(i)\cdot \vec{s}(j) | \Psi \rangle$. \\

To end this section, let us  consider  the total spin-orbit operator $\vec{\ell}\cdot \vec{S}$ written as
 \begin{eqnarray}
  \langle \Psi |\vec{\ell}\cdot \vec{S} | \Psi \rangle & = & \frac{1}{2}\left(J(J+1)-\ell(\ell+1)-S(S+1)\right) \nonumber \\
  						       & = & \sum_{i=1}^{N_c} \langle \Psi |\vec{\ell}(i)\cdot \vec{s}(i) | \Psi \rangle +  \sum_{i=1}^{N_c} \sum_{j\neq i =1}^{N_c}\langle \Psi |\vec{\ell}(i)\cdot \vec{s}(j) | \Psi \rangle.
 \end{eqnarray}
This operator is not a pure two-body operator. However, it was assumed as   a two-body operator in Ref. \cite{goity03,matagne05a}. Indeed, as explained above, the one-body part is of order 1 (it contributes as in the sum (\ref{spinorbitonebody})) but the two-body part is of order $N_c$. Therefore in this chapter, we shall consider the spin-orbit operator as a two-body operator.  \\

This is consistent with the fact that for large $N_c$ the  order of the  spin-orbit operator is $\leq \mathcal{O}(N_c)$.  In Chapter 2, we have seen that a two-body operator is $\leq \mathcal{O}(N_c^2)$. However we consider low excitations only. Let us take as for example the case with $S=\frac{N_c}{2}$, $\ell=1$ and $J=\frac{N_c}{2}+1$. One has
 \begin{eqnarray}
  \lefteqn{\langle (\frac{N_c}{2}+1)J_3;1\frac{N_c}{2}II_3|\ell^iS^i|1\frac{N_c}{2}II_3;(\frac{N_c}{2}+1)J_3\rangle  =} \nonumber \\ & & \frac{1}{2} \left[\left(\frac{N_c}{2}+1\right)\left(\frac{N_c}{2}+2\right)-2-\frac{N_c}{2}\left(\frac{N_c}{2}+1\right)\right] \nonumber \\ 
  								& & =  \frac{N_c}{2} \sim \mathcal{O}(N_c),
 \end{eqnarray}
which is the maximal order of the matrix elements of the spin-orbit operator for low excitations. 

\section{Matrix elements of the SU(4) generators for the $[N_c-1,1]$ spin-flavor irrep}

The matrix elements of the SU(4) generators $S^i,\ T^a$ and $G^{ia}$ for a symmetric irrep have been presented in Section  \ref{symmetricsu4generators}. According to Ref. \cite{hecht69} the matrix elements of every SU(4)  generator $E_{ia}$, Eq. (\ref{normes2}),  between states belonging to the representation $[N_c-1,1]$ 
can be expressed under the form of a generalized Wigner-Eckart theorem
which reads
\begin{eqnarray}\label{GENsu4newlook}
\langle [N_c-1,1] I' I'_3 S' S'_3 | E_{ia} |
[N_c-1,1] I I_3 S S_3 \rangle  = \sqrt{C^{[N_c-1,1]}(\mathrm{SU(4)})}   \nonumber \\
 \times
    \left(\begin{array}{cc||c}
	[N_c-1,1]    &  [211]   & [N_c-1,1]   \\
	S I  &   S^i I^a   &   S' I'
      \end{array}\right)_{\rho=1}
  \left(\begin{array}{cc|c}
	S   &    S^i   & S'   \\
	S_3  &   S^i_3   & S'_3
  \end{array}\right)
     \left(\begin{array}{cc|c}
	I   &   I^a   & I'   \\
	I_3 &   I^a_3   & I'_3
   \end{array}\right),
   \end{eqnarray}
where $C^{[N_c-1,1]}(\mathrm{SU(4)})=[N_c(3N_c+4)]/8$ is the eigenvalue of the 
SU(4) Casimir operator for the representation $[N_c-1,1]$. 
The other three factors in (\ref{GENsu4newlook}) are: an isoscalar factor of SU(4), 
a Clebsch-Gordan (CG)
coefficient of SU(2)-spin and a CG coefficient of SU(2)-isospin. 
Note that the isoscalar factor carries a lower index $\rho$ = 1 for
consistency with Ref. \cite{hecht69}. In general, this index is necessary to distinguish 
between irreducible representations, whenever the multiplicity  
in the inner product $[N_c-1,1] \times [211] \rightarrow [N_c-1,1]$ is larger
than one. In that case, following Ref. \cite{hecht69}, the matrix elements 
of the SU(4) generators for a fixed irreducible representation $[f]$ 
are defined such as the reduced matrix elements take the following values
\begin{equation}
\langle [f] || \mathrm{E} || [f] \rangle = 
\left\{
\begin{array}{ll}
\sqrt{C^{[N_c-1,1]}(\mathrm{SU(4)})} & \rm{for} \hspace{3mm} \rho = 1 \\
0 \hspace{3mm} & \rm{for} \hspace{3mm} \rho \neq 1 \\
\end{array}\right. .
\end{equation} \\

Thus the knowledge of the matrix elements of SU(4) generators amounts to the
knowledge of isoscalar factors. In Ref. \cite{hecht69} a variety of 
isoscalar factors were obtained in an analytic form. We need those 
for $[f] = [N_c-1,1]$. 
They are reproduced  in Table \ref{isoscalarsu4ms} in terms of our notation
and  typographical errors corrected.
They contain the phase factor introduced in Eq. (35) of Ref.  \cite{hecht69}. 
As compared to a symmetric spin-flavor state $N_c$, where  $I = S$ always, 
note that for a mixed representation,  one has  $ I = S $ (here 13 isoscalar factors) 
but also $I \neq S$ (here 10 isoscalar factors). The grouping in the table is
justified by the observation that the isoscalar factors obey the following 
orthogonality relation
\begin{equation}
\sum_{S_1 I_1 S_2 I_2}
 \left(\begin{array}{cc||c}
	[N_c-1,1]    &  [211]   & [N_c-1,1]   \\
	S_1 I_1  &   S_2 I_2   &   S I
      \end{array}\right)_{\rho}
   \left(\begin{array}{cc||c}
	[N_c-1,1]    &  [211]   & [N_c-1,1]   \\
	S_1 I_1  &   S_2 I_2   &   S' I'
      \end{array}\right)_{\rho} = \delta_{S S'} \delta_{I I'},   
 \end{equation}
 which can be easily checked. For example, by taking $S = S'$  and $I = I'$
 one can find that the squares of the first 13 coefficients sum up to one. \\
 
 One can now identify the isoscalar factors associated 
 to the generators of SU(4). In Table \ref{isoscalarsu4ms} one has $ S_2 I_2 $ = 10 corresponding to $S_i$,
 $S_2 I_2 $ = 01  to $T_{a}$ and  $ S_2 I_2 $ = 11 to $G_{ia}$,
 where 1 or 0 is the rank of the SU(2)-spin or SU(2)-isospin tensor
 contained in the generator. \\
 
For completeness also note that the isoscalar factors obey the following symmetry 
property
\begin{equation}
 \left(\begin{array}{cc||c}                                         [N_c-1,1]  &  [211]  &  [N_c-1,1] \\
                           I_1S_1 & I_2S_2 & (S-1)S
                                      \end{array}\right)_{\rho} =
\left(\begin{array}{cc||c}                                         [N_c-1,1]  &  [211]  &  [N_c-1,1] \\
                           S_1I_1 & S_2I_2 & S(S-1)
                                      \end{array}\right)_{\rho}.
\end{equation}


\begin{table}[h!]
{\small
\begin{center}
\renewcommand{\arraystretch}{2.0}
\begin{tabular}{ll|c|c|l}
\hline
\hline
\vspace{-0.5cm} &    &    &     & \\
$S_1$&$I_1$ \hspace{0.5cm} & \hspace{0.0cm}$S_2I_2$ \hspace{0cm}& \hspace{0.5cm}
$SI$ \hspace{0.5cm} & \hspace{0.5cm}$\left(\begin{array}{cc||c}                                         [N_c-1,1]  &  [211]  &  [N_c-1,1] \\
                           S_1I_1 & S_2I_2 & SI
                                      \end{array}\right)_{\rho=1}$  \\
\vspace{-0.5cm} &    &    &     & \\
\hline
\vspace{-0.5cm} &    &    &     & \\
$S+1$&$S+1$ & $11$ & $SS$ &  $\sqrt{\frac{S(S+2)(2S+3)(N_c-2-2S)(N_c+2+2S)}{(2S+1)(S+1)^2N_c(3N_c+4)}}$ \\
$S+1$&$S$ & $11$  & $SS$ & \\
$S$&$S+1$ & $11$  & $SS$ & \raisebox{2.45ex}[0cm][0cm]
{$\frac{1}{S+1}\sqrt{\frac{(2S+3)(N_c+2+2S)}{(2S+1)(3N_c+4)}}$} \\
$S$&$S$   & $11$  & $SS$ & $-\frac{N_c-(N_c+2)S(S+1)}{S(S+1)\sqrt{N_c(3N_c+4)}}$ \\
$S$&$S-1$ & $11$  & $SS$ & \\
$S-1$&$S$ & $11$  & $SS$ & \raisebox{2.45ex}[0cm][0cm]{$\frac{1}{S}\sqrt{\frac{(2S-1)(N_c-2S)}{(2S+1)(3N_c+4)}}$} \\
$S-1$&$S-1$ & $11$ & $SS$ & $\frac{1}{S}\sqrt{\frac{(S-1)(S+1)(2S-1)(N_c+2S)(N_c-2S)}{(2S+1)N_c(3N_c+4)}}$ \\
$S+1$&$S$   & $10$ & $SS$ & \\
$S$&$S+1$   & $01$ & $SS$ & \raisebox{2.45ex}[0cm][0cm]{0} \\
$S-1$&$S$   & $10$ & $SS$ & \\
$S$&$S-1$   & $01$ & $SS$ & \raisebox{2.45ex}[0cm][0cm]{0} \\
$S$&$S$     & $10$ & $SS$ & \\
$S$&$S$     & $01$ & $SS$ & \raisebox{2.45ex}[0cm][0cm]{$\sqrt{\frac{4S(S+1)}{N_c(3N_c+4)}}$} \\
$S+1$&$S$   & $11$ & $SS-1$ & $\sqrt{\frac{(2S+3)(N_c+2+2S)(N_c-2S)}{(2S+1)N_c(3N_c+4)}}$ \\
$S$&$S$     & $11$ & $SS-1$ & $-\frac{1}{S}\sqrt{\frac{N_c-2S}{3N_c+4}}$ \\
$S$&$S-1$   & $11$ & $SS-1$ & $\frac{1}{S}\sqrt{\frac{(S-1)(S+1)N_c}{3N_c+4}}$ \\
$S-1$&$S$   & $11$ & $SS-1$ & $-\frac{N_c+4S^2}{S \sqrt{(2S-1)(2S+1)N_c(3N_c+4)}}$ \\
$S-1$&$S-1$ & $11$ & $SS-1$ & $-\frac{1}{S}\sqrt{\frac{N_c+2S}{3N_c+4}}$ \\
$S-1$&$S-2$ & $11$ & $SS-1$ & $\sqrt{\frac{(2S-3)(N_c+2-2S)(N_c+2S)}{(2S-1)N_c(3N_c+4)}}$ \\
$S$&$S-1$   & $10$ & $SS-1$ & $\sqrt{\frac{4S(S+1)}{N_c(3N_c+4)}}$ \\
$S-1$&$S-1$ & $10$ & $SS-1$ & 0 \\
$S$&$S$     & $10$ & $SS-1$ & 0 \\
$S$&$S-1$   & $01$ & $SS-1$ & $\sqrt{\frac{4(S-1)S}{N_c(3N_c+4)}}$ \\
\hline
\hline
\end{tabular}
\caption{Isoscalar factors of SU(4) for $[N_c-1,1] \times [211] 
\rightarrow [N_c-1,1]$ defined by Eq. (\ref{GENsu4}).}\label{isoscalarsu4ms}
\end{center}}
\end{table}

\clearpage
The relations (\ref{GENsu4newlook}) are used to calculate the matrix elements
of the operators   entering to the $1/N_c$ expansion of the mass operator, as described
in the following section.

\section{The mass operator}

The $1/N_c$ expansion of the mass operator of the $[{\bf 70},1^-]$ multiplet is given by
\begin{equation}
\label{massoperatorsu4newlook}
M_{[{\bf 70},1^-]} = \sum_{i=1}^6 c_i O_i.
\end{equation} 
The operators contained in the mass operator (\ref{massoperatorsu4newlook})
are shown in Table \ref{operatorssu4newlook} together with the values of the dynamical coefficients $c_i$
found from the numerical fit described in the next section. 
The building  blocks of  $O_i$ 
are  now $S^i$, $T^a$, $G^{ia}$, $\ell^i$ and $\ell^{(2)ij}$. \\

Table   \ref{operatorssu4newlook} contains the six possible operators up to order $1/N_c$.
Here the spin-orbit operator $O_2$ is a two-body operator and its 
matrix elements are  of order  $1/N_c$, as  for
the ground state and the other symmetric states $[\bf{56}, \ell]$ 
with $\ell \neq 0$
\cite{goity03,matagne05a}. This is in
contrast to  the $[{\bf 70},1^-]$ multiplet analyzed in the 
decoupling procedure where the spin-orbit matrix elements are
of order $\mathcal{O}(1)$ (see Chapter 5).
The   spin-spin operator $O_3$ and 
the isospin-isospin operator $O_4$ are also two-body operators. 
They are linearly independent. However in the decoupling procedure the 
corresponding isospin-isospin operator 
$t^a T^a_c/N_c$ has not always been included in the analysis \cite{schat02b}.
The operators $O_5$
and $O_6$ are three-body operators but, as $G^{ia}$ sums coherently,
it introduces an extra factor $N_c$ (see Section \ref{groundoperatorexpansion}) and makes the matrix 
elements of $O_5$  and $O_6$ of order $1/N_c$ as well.


\begin{table}[h!]
{\scriptsize
\begin{center}
\renewcommand{\arraystretch}{1.8} 
\begin{tabular}{llrrrrr}
\hline
\hline
Operator &  & Fit 1 (MeV) &  Fit 2 (MeV) & Fit 3 (Mev) & Fit 4 (MeV) & Fit 5 (MeV) \\
\hline
$O_1 = N_c \ \1 $                                   &  $c_1 =  $  & $481 \pm5$  & $482\pm5$ &  $484\pm4$ &  $484\pm4$ & $498\pm3$\\
$O_2 = \frac{1}{N_c}\ell^i S^i$                     &  $c_2 =  $  & $-47 \pm39$ & $-30\pm34$ & $-31\pm20$ & $8\pm15$ & $38\pm34$\\
$O_3 = \frac{1}{N_c}S^iS^i$                         &  $c_3 =  $  & $161\pm 16$ & $149\pm11$ & $159\pm16$ & $149\pm11$ & $156\pm16$\\
$O_4 = \frac{1}{N_c}T^aT^a$                         &  $c_4 =  $  & $169\pm36$  & $170\pm36$ & $138\pm27$ & $142\pm27$ &\\
$O_5 = \frac{3}{N_c^2}\ell^{(2)ij}G^{ia}G^{ja}$     &  $c_5 =  $  & $-443\pm459$&            & $-371\pm456$&        & $-514\pm458$\\
$O_6 = \frac{1}{N_c^2}\ell^iT^aG^{ia}$              &  $c_6 =  $  & $473\pm355$ & $433\pm353$ &            &      & $-606\pm273$
\vspace{0.15cm} \\
\hline
$\chi_{\mathrm{dof}}^2$                             &             & $0.43$      & $0.68$ & $1.1$           & $0.96$ & $11.5$\\
\hline \hline
\end{tabular}
\caption{List of operators and the coefficients resulting from the fits \cite{matagne06c}.}
\label{operatorssu4newlook}
\end{center}}
\end{table}

The matrix elements of the operator $O_i$ are  presented in Table \ref{Matrixsu4newlook}.
They have been calculated for all available states of the multiplet
$[{\bf 70},1^-]$ starting from the wave function (\ref{WFnewlook}) and using the 
isoscalar factors of Table \ref{isoscalarsu4ms}.  One can see that all diagonal matrix
elements are of order $1/N_c$, of course except for $O_1$. For completeness
we also indicate the off-diagonal matrix elements.  

\begin{table}[h!]
{\scriptsize
\renewcommand{\arraystretch}{1.8} 
\begin{center}
\begin{tabular}{lcccccc}
\hline
\hline
  &  $O_1$ & $O_2$ & $O_3$ & $O_4$ & $O_5$ & $O_6$  \\
  \hline
 $^2N_{\frac{1}{2}}$ & $N_c$  & $-\frac{1}{N_c}$ & $\frac{3}{4N_c}$ & $\frac{3}{4N_c}$ & 0 & $\frac{N_c-6}{12N_c^2}$ \\
$^4N_{\frac{1}{2}}$ & $N_c$   &  $-\frac{5}{2N_c}$ & $\frac{15}{4N_c}$ & $\frac{3}{4Nc}$ & $\frac{5}{24N_c}$ & $-\frac{5}{24N_c}$ \\
$^2N_{\frac{3}{2}}$ & $N_c$  & $\frac{1}{2N_c}$ & $\frac{3}{4N_c}$  & $\frac{3}{4N_c}$ & 0 & $-\frac{N_c-6}{24N_c^2}$ \\
$^4N_{\frac{3}{2}}$ & $N_c$  &  $-\frac{1}{N_c}$ & $\frac{15}{4N_c}$ & $\frac{3}{4N_c}$ & $-\frac{1}{6N_c}$ & $-\frac{1}{12N_c}$ \\
$^4N_{\frac{5}{2}}$ & $N_c$ & $\frac{3}{2N_c}$ & $\frac{15}{4N_c}$ & $\frac{3}{4N_c}$ & $\frac{1}{24N_c}$ & $\frac{1}{8N_c}$ \\
$^2\Delta_{\frac{1}{2}}$ & $N_c$ & $-\frac{1}{N_c}$ & $\frac{3}{4N_c}$ & $\frac{15}{4N_c}$ & 0 & $-\frac{5}{12N_c}$ \\
$^2\Delta_{\frac{3}{2}}$ & $N_c$ & $\frac{1}{2N_c}$ & $\frac{3}{4N_c}$ & $\frac{15}{4N_c}$ & 0 & $\frac{5}{24N_c}$ \\
$^4N_{\frac{1}{2}}-$ $^2N_{\frac{1}{2}}$ & 0 & 0 & 0 & 0 & $-\frac{5}{12N_c^2}\sqrt{\frac{N_c(N_c+3)}{2}}$ & $-\frac{1}{6N_c^2}\sqrt{\frac{N_c(N_c+3)}{2}}$ \\
$^4N_{\frac{3}{2}}-$ $^2N_{\frac{3}{2}}$ & 0 & 0 & 0 & 0 & $\frac{1}{24N_c^2}\sqrt{5N_c(N_c+3)}$ & $-\frac{1}{12N_c^2}\sqrt{5N_c(N_c+3)}$\vspace{0.15cm} \\
\hline \hline
\end{tabular}
\caption{Matrix elements for all states belonging to the 
$[{\bf 70},1^-]$ multiplet \cite{matagne06c}.}
\label{Matrixsu4newlook}
\end{center}}
\end{table}


\begin{table}[h!]
{\scriptsize
\begin{center}
\renewcommand{\arraystretch}{1.5}
\begin{tabular}{crrrrrrcccl}\hline \hline
                    &      \multicolumn{6}{c}{Part. contrib. (MeV)}  & \hspace{.0cm} Total (MeV)   & \hspace{.0cm}  Exp. (MeV)\hspace{0.0cm}& &\hspace{0.cm}  Name, status \hspace{.0cm} \\

\cline{2-7}
                    &   \hspace{.0cm}   $c_1O_1$  & \hspace{.0cm}  $c_2O_2$ & \hspace{.0cm}$c_3O_3$ &\hspace{.0cm}  $c_4O_4$ &\hspace{.0cm}  $c_5O_5$ & $c_6O_6$   &        \\
\hline
$^2N_{\frac{1}{2}}$        & 1444 & -16 & 40 & 42 & 0 & -13  &   $1529\pm 11$  & $1538\pm18$ & & $S_{11}(1535)$****  \\
$^4N_{\frac{1}{2}}$        & 1444 &  39 & 201& 42 & -31& -33 &   $1663\pm 20$  & $1660\pm20$ & & $S_{11}(1650)$**** \\
$^2N_{\frac{3}{2}}$        & 1444 & -8  & 40 & 42 & 0  &  7  &   $1525\pm 8$   & $1523\pm8$  & & $D_{13}(1520)$****\\
$^4N_{\frac{3}{2}}$        & 1444 & 16  & 201& 42 & 25 & -13 &   $1714\pm45$   & $1700\pm50$ & & $D_{13}(1700)$***\\
$^4N_{\frac{5}{2}}$        & 1444 & -24 & 201& 42  & -6 & 20 &   $1677\pm8$    & $1678\pm8$  & & $D_{15}(1675)$****\\
\hline
$^2\Delta_{\frac{1}{2}}$  &  1444 & 16  & 40 & 211 & 0  & -66   & $1645\pm30$  & $1645\pm30$ & & $S_{31}(1620)$**** \\
$^2\Delta_{\frac{3}{2}}$  &  1444 & -8  & 40 & 211 & 0  & -33   & $1720\pm50$  & $1720\pm50$ & & $D_{33}(1700)$**** \vspace{0.15cm} \\
\hline \hline
\end{tabular}
\caption{The partial contribution and the total mass (MeV) predicted by the 
$1/N_c$ expansion. 
The last two columns give  the empirically known masses \cite{matagne06c}.}\label{MASSESsu4newlook}
\end{center}}
\end{table}


\section{Results}

In Table \ref{MASSESsu4newlook} we present the masses of the non-strange resonances
belonging to the $[{\bf 70},1^-]$ multiplet obtained from the fit to the experimental
values \cite{yao06}. 
We also indicate the partial contribution (without error bars) of each of the six terms
contributing to the total mass obtained from the values of $c_i$
of Table  \ref{operatorssu4newlook}, column Fit 1, \emph{i.e.} the case with all $O_i$
included.
The fit is indeed excellent, it gives $\chi_{\mathrm{dof}}^2 \simeq $ 0.43.
From Table  \ref{operatorssu4newlook} one can see that the values of the 
coefficients $c_3$ and $c_4$ are comparable to each other which shows the importance 
of including the isospin operator $O_4$, besides the usual spin
term $O_3$.  In addition,
looking at the partial contributions of Table \ref{MASSESsu4newlook},
 one can see that the spin term $O_3$
is dominant for the $^4N_{J}$ resonances while   
the isospin-isospin term is dominant for the $\Delta$ resonances. This brings a new
aspect into the description of excited states studied so far, where  
the dominant term was always the spin-spin term \cite{matagne05b}, the isospin term
being absent in the analysis. To get a better idea about the role of the 
operator $O_4$ we have also made a fit by removing it from the definition
of the mass operator (\ref{massoperatorsu4newlook}). The result is shown in 
Table \ref{operatorssu4newlook} column Fit 5. The $\chi_{\mathrm{dof}}^2 $ deteriorates
considerably becoming 11.5 instead of 0.43 obtained with all $O_i$ 
operators included. This clearly shows that $O_4$ is crucial in the fit. \\

As mentioned above, in studies based on the splitting of the system into an excited
 quark and a core another operator,  
 namely
 $\frac{1}{N_c}t^aT_c^a$, has been included in the mass analysis, instead of $O_4$.
 That operator is only a part of $O_4$. Its matrix elements are spin dependent.
 The sources of the spin dependence are the isoscalar factors 
 of the permutation group  (see Eq. (\ref{isoscalarms})), 
  which appear in the spin-flavor part of the wave function,
  as a consequence of the truncation of the spin-flavor space, 
  as explained in Section \ref{Mixed symmetric  wave functions}.
  We consider that the spin dependence  of an isospin operator matrix
  elements  \cite{carlson98b},   is not natural. 
 The present approach removes this anomaly. \\
 
The coefficient $c_2$ of the spin-orbit term is small and its magnitude and 
sign remains comparable to that of Ref. \cite{matagne06b} obtained in the
analysis of the $[{\bf 70},\ell^+]$ multiplet (see Table \ref{70,ell+strangeoperators}). The value of $c_2$ implies 
a small spin-orbit contribution to the total mass, in agreement
with the general pattern observed for the excited states \cite{matagne05b} (see Figure  \ref{spindependencemassoperator}) and in 
agreement with constituent quark models. \\

The coefficients $c_5$ and $c_6$ are large but they come out with large
error bars too. This suggests that other observables should be involved in
order to constrain the contribution of such operators,
as for example the mixing angles. 
This is the subject of further investigations.
However, the removal of $O_5$ and $O_6$ 
from the mass operator does not deteriorate the fit too badly, as shown
in  Table \ref{operatorssu4newlook} Fits 2--4, the  $\chi_{\mathrm{dof}}^2$
becoming at most 1.1. The partial contribution of $O_5$ or of
$O_6$ is generally comparable to that of the spin-orbit operator. Note that
the structure of $O_6$ is  related to that of the spin-orbit term, which makes
its small contribution plausible. 


\section{Conclusions}

The present chapter sheds an entirely new light into the description 
of the baryon multiplet  $[{\bf 70},1^-]$ in the  $1/N_c$ expansion. 
The main findings are:
\begin{itemize}
 \item In the mass formula the expansion starts at order $1/N_c$, as for the ground 
 state, instead of $N^0_c$ as previously concluded.
 \item The isospin operator  $\frac{1}{N_c} T^aT^a$ is crucial in the
 fit to the existing data and its contribution is as important as that 
 of the spin term $\frac{1}{N_c} S^iS^i$. 
 \end{itemize}
 It would be interesting to reconsider the study of higher excited 
  baryons, for example those belonging to $[{\bf 70},\ell^+]$ multiplets, 
  in the spirit of the present approach.
  Based on group theory  arguments it is expected that the mass 
  splitting starts at order $1/N_c$, as a general rule, 
  irrespective of the angular momentum and parity of the state and also 
  of the number of flavors.  In practical terms, the extension to
  three flavors would involve a considerable amount of algebraic
  group theory work on isoscalar factors of SU(6) generators for mixed-symmetric representations. \\
  
  Thus the conceptual problem related to the leading 
  correction to the mass of the excited baryon states, 
  thought to be of order $\mathcal{O}(1)$ 
  (see \emph{e.g.} Ref. \cite{cohen05}), 
  as originated from previous studies of the $[{\bf 70},1^-]$ multiplet, 
  has been solved. The leading correction is of order $1/N_c$, as for
  the ground state. 
\chapter*{Conclusions and perspectives
          \markboth{Conclusions and perspectives}{}}
	 
\addcontentsline{toc}{chapter}{Conclusions and perspectives}

\thispagestyle{empty}

At the time when we have started this work, the $1/N_c$ expansion to baryon spectroscopy was extensively applied to the ground state and partly to excited states. The only excited multiplets considered until 2003 were the $[{\bf 70},1^-]$, the $[{\bf 56'},0^+]$ and the $[{\bf 56},2^+]$.  The $[{\bf 56'},0^+]$ and the $[{\bf 56},2^+]$ multiplets were easy to treat as the spin-flavor and the orbital parts of the wave function are symmetric. In this case,   the SU($2N_f$) symmetry is broken at order $\mathcal{O}(1/N_c)$. The  approach  to the baryon wave function belonging to the  $[{\bf 70},1^-]$ multiplet  was to decouple it into a symmetric ground-state core  composed of $N_c-1$ quark and one excited quark. This point of view,  based on a Hartree approximation, has the advantage to treat the core like a ground-state baryon and  allows an easy generalization of the operator expansion in $1/N_c$ for the $N=1$ band. However this procedure  predicts a breaking of the SU($2N_f$) symmetry to  order $\mathcal{O}(N_c^0)$ in the mass instead of $\mathcal{O}(1/N_c)$, as for ground-state baryons. The conflict leads to a conceptual problem. It follows that excited states belonging to the $[{\bf 70},1^-]$ multiplet do not form  an infinite tower of degenerate states because of the breaking to  first order of the SU($2N_f$) symmetry. Nevertheless, experimental results suggest to treat the spin-flavor breaking terms as subleading. \\

Historically, our first study was devoted to resonances belonging to the $[{\bf 56},4^+]$ multiplet which is the lowest in the $N=4$ band. Although high in energy, $2.5\sim3$ GeV, it was appealing to look at this multiplet for its simplicity. Indeed, its  mass operator and the wave function are similar to those of $[{\bf 56},2^+]$. As mentioned above and  throughout this thesis, the resonances belonging to a symmetric SU($2N_f$) representation are easier to study than those belonging to mixed-symmetric representations. \\

 We next concentrated on the $[{\bf 70},\ell^+]$ ($\ell=0,2$) multiplets which present technical complications from group theory point of view. Here we adopted the traditional procedure described above \emph{i.e.} to split the system of $N_c$ quarks into an excited quark and a core described by a symmetric wave function both in the orbital and in the spin-flavor space. The main difference between the  $[{\bf 70},1^-]$ case and these multiplets comes from the fact that here the core is excited. \\

  In a first step, our study was devoted to non-strange baryons. To include strange baryons it was necessary to know the matrix elements of the SU(6) generators for a symmetric spin-flavor wave function composed of $N_c$ quarks.   Based on a rather simple procedure involving a generalized Wigner-Eckart theorem we have derived analytic expressions for these matrix elements and used them to complete the study of resonances belonging to $[{\bf 70},\ell^+]$ multiplets by including strange resonances into our previous studies on non-strange resonances. In this way, the study of the $N=2$ band was completed. As for the $[{\bf 20},1^+]$ multiplet, no candidate has been found so far experimentally. \\
  
  Our results on the baryon masses in the $N=2$ and $N=4$ bands, combined with results available in the literature for the $[{\bf 70},1^-]$ and the $[{\bf 56},2^+]$ multiplets, enabled us to understand the energy dependence of the dynamical coefficients entering  the mass operator of the $1/N_c$ expansion. We found out that the spin-spin contribution remains dominant over other contributions and that all the spin dependent terms tend to vanish at large excitation energy. We think this is an important conclusion that brings support to the phenomenology of baryons and to the constituent quark models where the spin-orbit interaction is usually neglected. \\
  
  The $1/N_c$ expansion is also a powerful method to classify baryon resonances into octets, decuplets and flavor singlets 
  which means to extend the Gell-Mann and Ne'eman classification to excited states. This is an   important task in general. The proposed classification can confirm or infirm quantum numbers as spin and parity which have been predicted by quark models. There are some resonances in the Particle Data Group Tables  where these quantum numbers have not been measured. This is generally the case for heavy baryons. Our Tables \ref{70,ell+nonstrangemultiplet}, \ref{70,ell+MASSES}, \ref{56,4+multiplet} and \ref {MASSESsu4newlook} give a classification of the presently known resonances and also make predictions for higher excited states to be discovered. \\
  
  While studying the positive parity {\bf 70}-plet resonances, we realized that it is  possible to treat the mixed-symmetric multiplets in the $1/N_c$ expansion of QCD by avoiding the cumbersome procedure of decoupling the system of $N_c$ quarks into two subsystems, one being a symmetric core and the other an excited quark. As described in the beginning of Chapter 6, this method implies a truncation of the available basis of vector space belonging to the irreducible representation $[N_c-1,1]$ of the permutation group S$_{N_c}$. By assuming that the $N_c$-th quark of the baryon wave function is always the excited quark, the decoupling picture predicts that the SU($2N_f$) symmetry is broken to  order $\mathcal{O}(N_c^0)$. But, when we take into account the complete basis vector space, this conceptual problem disappears. This is the case when we consider the wave function in one block, without splitting it. Moreover, such a procedure allows to have an equivalent treatment of symmetric and mixed-symmetric multiplets, predicting a breaking of the SU($2N_f$) symmetry to  order $\mathcal{O}(1/N_c)$ in both cases. In addition the anomaly created by  the decoupling picture which generates a dependence on the spin of the isospin-isospin operator matrix elements is removed. \\
  
  This new approach for mixed-symmetric multiplets requires the knowledge of  the matrix elements  of the SU($2N_f$) generators  for  mixed-symmetric wave functions of $N_c$ particles in the spin-flavor. The solution came from previous literature on group theory which was used in the sixtieths in nuclear physics. A system of $u,\ d$ quarks and a nucleus made out of protons and neutrons have in common the SU(4) symmetry. Thus we found that the matrix elements of the SU(4) generators for  mixed-symmetric states of $N_c$ quarks can be easily extracted from results relevant to nuclear physics. \\
  
  As a first application of the new method, we have revisited the $[{\bf 70},1^-]$ multiplet. It was the subject of the second part of Chapter 6. The two main conclusions were that the $1/N_c$ expansion starts at order $\mathcal{O}(1/N_c)$, as announced above and that the isospin operator plays a crucial role in the fit to existing data. Its  contribution is as important as that of the spin-spin term which brings support to models based on a flavor dependent interaction.  \\
  
  It would be interesting to extend the new approach to incorporate strange baryons into the  $[{\bf 70},1^-]$ multiplet. Unfortunately, to our knowledge, the isoscalar factors necessary to determine the generator matrix elements of the SU(6) spin-flavor group are not yet know for $N_c$ particles. Some studies were done during the sixtieths for the irreducible representation {\bf 70} but only for three particles. Considerable group theory work should be done to generalize this results to $N_c$ particles. The mixed-symmetric representation $[N_c-1,1]$ is more complex that the symmetric one. The method presented in Chapter 4 is not applicable to a mixed-symmetric representation because it is based on the fact that the matrix elements of the generators acting on one particle are identical for all the particles. The matrix elements of SU(6) generators are then easily obtained for a symmetric spin-flavor representation containing $N_c$ particles. This property  is not valid for a mixed-symmetric irreducible representation. \\
  
  Another interesting extension should be to reanalyze the $[{\bf 70},\ell^+]$ ($\ell=0,2$) multiplets in this new light. Of course, only non-strange baryons could be studied at the present time as explained above. Nevertheless, one could obtain a global vision of the $N=2$ band based on an  equivalent group theoretical description of the wave function and of the spin-orbit operator for the $[{\bf 56'},0^+]$, the $[{\bf 56},2^+]$ and the $[{\bf 70},\ell^+]$ multiplets. It would be interesting to point out the role of the isospin-isospin operator in the $[{\bf 70},\ell^+]$ multiplets. \\

  The study of multiplets belonging to the $N=3$ band remains entirely open. As in the $N=2$ band  we encounter symmetric and mixed-symmetric multiplets. The analysis of at least one of these multiplets would bring an additional point to Figure \ref{spindependencemassoperator} which could possibly  confirm the behaviour we found that the spin-dependent interactions decrease with the excitation energy.  \\
    
  As mentioned above, an important contribution of the $1/N_c$ expansion in QCD to baryon spectroscopy is that it allows mass predictions for all members of a multiplet. So far, whenever possible,  the best $\chi^2_{\mathrm{dof}}$ were obtained by identifying the multiplet members in the same way as in  constituent quark models. Otherwise, the identifications proposed become predictions. However, generally, a given resonance corresponds to a mixing of two or more states and are not pure states. It would be very interesting to consider mixing of states in the analysis and to predict  mixing angles. 
  The inclusion of mixing between the  $[{\bf 56},2^+]$ multiplet and the $[{\bf 70},\ell^+]$ ($\ell=0,2$) multiplets should be a very attractive future subject of investigation. \\
  
 In this thesis, we assumed  that excited baryons are bound states. However, excited baryons are resonances even at large $N_c$. Indeed, the Witten power counting rules predict a characteristic width of  excited baryons of $N_c^0$. Thus baryons do not become stable at large $N_c$. Some papers tried to take into account  this fact by analysing the baryon-pion scattering amplitude.  Nevertheless, we think that the quark model assumptions used in the calculations give satisfactory results and a good start. \\
  
  Finally, an important difficulty for the study of excited multiplets comes form the lack of experimental data. This can be seen in Tables \ref{70,ell+MASSES} and \ref{56,4+multiplet} which summarize the results for the strange baryon masses belonging to the  $[{\bf 70},\ell^+]$ ($\ell=0,2$) multiplets. Among the 47 possible theoretical states we have only 12 experimental know resonances. Furthermore, the quality of the data is not very good, the experimental error on the masses being quite high. The situation of the $[{\bf 56},4^+]$ multiplet is worse. With more data, it would be possible to include more operators in the fit. With an improvement of the quality, the error bars on the coefficients $c_i$ and $d_i$  appearing in the mass operator expansion  should decrease. \\
  
 
 This thesis aimed at completing the study of the $N=2$ band and at beginning the exploration of the $N=3$ and $N=4$ bands. The $N=2$ band was indeed completely covered and the lowest multiplet of the $N=4$ band was analysed. The $N=3$ band was not tackled but we proposed a new and elegant method to study baryons described by mixed-symmetric spin-flavor states which removed an important conceptual conflict in the $1/N_c$ expansion of the mass operator. The large $N_c$ approach to QCD is a new developing field, complementary to the traditional constituent quark model and we hope to have introduced a new perspective of research and new discussions in baryon spectroscopy.

\appendix
\chapter{Examples of orbital wave functions}

\thispagestyle{empty}

In this Appendix, we gives details about the orbital wave functions used in the quark-shell model presented in Chapter 2. The wave function of each state is derived by doing the direct product of the 1-particle wave functions Eqs. (\ref{0swavefunction})--(\ref{1swavefunction}) to obtain a state with a given angular momentum and a desired symmetry. For simplicity we use harmonic oscillator wave functions, but the angular momentum, parity and symmetry properties remain general irrespective of the form of the radial part. We give all the wave functions for  $N=0,1,2$ bands  for $N_c=3$ \cite{faiman68}. These wave functions may equivalently be written in terms of the Jacobi coordinates ${\bf R}$, $\lambdab$ and $\rhob$ defined in Eqs. (\ref{3centerofmass})--(\ref{3lambda}). There is a systematic procedure detailed in Ref. \cite{moshinsky69} which can be extended to obtain  states of analog symmetries  for large $N_c$.

\section{Harmonic polynomials and spherical harmonics}
Before writing down  harmonic oscillator wave functions describing a system of three particles, we give a Table containing of the  spherical harmonics used below.
\begin{equation}
\mathcal{Y}^m_\ell({\bf r})=r^\ell Y^m_\ell(\theta\phi)
\end{equation}

\begin{table}[h!]
\renewcommand{\arraystretch}{2}
 \begin{center}
\begin{tabular}{ll|c|c}
 \hline \hline
 $\ell$ & $m$ & $\mathcal{Y}^m_\ell({\bf r})$ & $Y^m_\ell(\theta\phi)$ \\
 \hline
 0      & 0   & $\frac{1}{\sqrt{4\pi}}$       & $\frac{1}{\sqrt{4\pi}}$ \\
 1      & 0   & $\sqrt{\frac{3}{4\pi}}z$      & $\sqrt{\frac{3}{4\pi}}\cos\theta$ \\
 1    &$\pm 1$& $\mp\sqrt{\frac{3}{8\pi}}(x\pm iy)$  & $\mp\sqrt{\frac{3}{8\pi}}\sin\theta e^{\pm i\phi}$ \\
 2    &   0   & $\sqrt{\frac{5}{16\pi}}(2z^2-x^2-y^2)$ & $\sqrt{\frac{5}{16\pi}}(2\cos^2 \theta-\sin^2\theta)$\\
 2  & $\pm 1$ & $\mp \sqrt{\frac{15}{8\pi}}z(x\pm iy)$    & $\mp \sqrt{\frac{15}{8\pi}} \cos\theta \sin\theta e^{\pm i \phi}$\\
 2  & $\pm 2$ & $\sqrt{\frac{15}{32\pi}}(x\pm iy)^2$      & $\sqrt{\frac{15}{32\pi}}\sin^2\theta e^{\pm 2i\phi}$ \\
 \hline \hline
\end{tabular}
\caption{Spherical harmonics for $\ell\leq 2$ \cite{edmonds57}.} \label{Spherical harmonics}
\end{center}
\end{table}

\section{Wave functions}

\subsubsection{$(0s)^3$: the symmetric representation, $\ell=0^+$}
\begin{equation}
\psi=\left(\frac{4\alpha^3}{\sqrt{\pi}}\right)^{3/2}\frac{1}{(4\pi)^{3/2}}e^{-\alpha^2(r^2_1+r^2_2+r^2_3)/2}.
\end{equation}
\subsubsection{$(0s)^2(0p)$: the symmetric representation, $\ell=1^-$}
\begin{equation}
 \psi = \frac{\sqrt{2}}{3}\left(\frac{4\alpha^3}{\sqrt{\pi}}\right)^{3/2}\frac{1}{4\pi}\alpha({\bf r}_1 + {\bf r}_2 + {\bf r}_3)e^{-\alpha^2(r^2_1+r^2_2+r^2_3)/2}.
\end{equation}
As one can directly notice, this wave function is a spurious state as it is proportional to the center of mass coordinate Eq. (\ref{3centerofmass}), \emph{i.e.} involves excitation of the center of mass. It is the orbital wave functions of baryons belonging to the $[{\bf 56},1^-]$ multiplet.

\subsubsection{$(0s)^2(0p)$: the mixed-symmetric representation, $\ell=1^-$}
\begin{eqnarray}
 \psi_{\lambda} & = & \frac{1}{3}\left(\frac{4\alpha^3}{\sqrt{\pi}}\right)^{3/2}\frac{1}{4\pi}\alpha({\bf r}_1 + {\bf r}_2 -2{\bf r}_3)e^{-\alpha^2(r^2_1+r^2_2+r^2_3)/2}, \\
 \psi_{\rho} & = & \frac{1}{\sqrt{3}}\left(\frac{4\alpha^3}{\sqrt{\pi}}\right)^{3/2}\frac{1}{4\pi}\alpha({\bf r}_1 - {\bf r}_2)e^{-\alpha^2(r^2_1+r^2_2+r^2_3)/2}.
\end{eqnarray}
These wave functions are not spurious as they are proportional of the coordinates $\lambdab$ and $\rhob$ Eqs. (\ref{3rho})--(\ref{3lambda}) and involve excitations of the internal motions of the quarks. They are orbital wave functions of baryons belonging to the $[{\bf 70},1^-]$ multiplet.

\subsubsection{$(0s)^2(1s)$: the symmetric representation, $\ell=0^+$}
\begin{equation}
 \psi = \frac{\sqrt{2}}{3}\left(\frac{4\alpha^3}{\sqrt{\pi}}\right)^{3/2}\frac{1}{(4\pi)^{3/2}}\left[\frac{9}{2}- \alpha^2(r^2_1 +r^2_2+r^2_3)\right]e^{-\alpha^2(r^2_1+r^2_2+r^2_3)/2}.
\end{equation}

\subsubsection{$(0s)^2(1s)$: the mixed-symmetric representation, $\ell=0^+$}
\begin{eqnarray}
 \psi_\lambda & = &  \frac{1}{3}\left(\frac{4\alpha^3}{\sqrt{\pi}}\right)^{3/2}\frac{1}{(4\pi)^{3/2}}\alpha^2(r_1^2+r_2^2-2r_3^2)e^{-\alpha^2(r^2_1+r^2_2+r^2_3)/2}, \\
 \psi_\rho & = & \frac{1}{\sqrt{3}}\left(\frac{4\alpha^3}{\sqrt{\pi}}\right)^{3/2}\frac{1}{(4\pi)^{3/2}}\alpha^2(r_1^2-r_2^2)e^{-\alpha^2(r^2_1+r^2_2+r^2_3)/2}. 
\end{eqnarray}

\subsubsection{$(0s)^2(0d)$: the symmetric representation, $\ell=2^+$}
\begin{equation}
 \psi=\frac{2}{\sqrt{45}}\left(\frac{4\alpha^3}{\sqrt{\pi}}\right)^{3/2}\frac{1}{4\pi}\alpha^2\left[r_1^2Y_2(\Omega_1)+r_2^2Y_2(\Omega_2)+r_3^2Y_2(\Omega_3)\right]e^{-\alpha^2(r^2_1+r^2_2+r^2_3)/2} \label{os2odsymm2}.
\end{equation}

\subsubsection{$(0s)^2(0d)$: the mixed-symmetric representation, $\ell=2^+$}
{\small \begin{eqnarray}
 \psi_\lambda & = & \sqrt{\frac{2}{45}} \left(\frac{4\alpha^3}{\sqrt{\pi}}\right)^{3/2}\frac{1}{4\pi}\alpha^2\left[r_1^2Y_2(\Omega_1)+r_2^2Y_2(\Omega_2)-2r_3^2Y_2(\Omega_3)\right]e^{-\alpha^2(r^2_1+r^2_2+r^2_3)/2}, \\
 \psi_\rho & = & \sqrt{\frac{2}{45}} \left(\frac{4\alpha^3}{\sqrt{\pi}}\right)^{3/2}\frac{1}{4\pi}\alpha^2\left[r_1^2Y_2(\Omega_1)-r_2^2Y_2(\Omega_2))\right]e^{-\alpha^2(r^2_1+r^2_2+r^2_3)/2}. 
\end{eqnarray}}

\subsubsection{$(0s)^2(0p)$: the symmetric representation, $\ell=0^+$}
\begin{equation}
 \psi=\frac{2}{3}\left(\frac{4\alpha^3}{\sqrt{\pi}}\right)^{3/2}\frac{1}{(4\pi)^{3/2}}\alpha^2({\bf r}_1\cdot{\bf r}_2+{\bf r}_2\cdot{\bf r}_3+{\bf r}_3\cdot{\bf r}_1)e^{-\alpha^2(r^2_1+r^2_2+r^2_3)/2}. 
\end{equation}

\subsubsection{$(0s)(0p)^2$: the mixed-symmetric representation, $\ell=0^+$}
\begin{eqnarray}
\psi_\lambda & = & \frac{\sqrt{2}}{3}\left(\frac{4\alpha^3}{\sqrt{\pi}}\right)^{3/2}\frac{1}{(4\pi)^{3/2}}\alpha^2({\bf r}_2\cdot{\bf r}_3+{\bf r}_3\cdot{\bf r}_1-2{\bf r}_1\cdot{\bf r}_2)e^{-\alpha^2(r^2_1+r^2_2+r^2_3)/2}, \\
\psi_\rho & = & \sqrt{\frac{2}{3}}\left(\frac{4\alpha^3}{\sqrt{\pi}}\right)^{3/2}\frac{1}{(4\pi)^{3/2}}\alpha^2({\bf r}_2\cdot{\bf r}_3-{\bf r}_3\cdot{\bf r}_1)e^{-\alpha^2(r^2_1+r^2_2+r^2_3)/2}. 
\end{eqnarray}

\subsubsection{$(0s)(0p)^2$: the mixed-symmetric representation, $\ell=1^+$}
 {\small 
\begin{eqnarray}
\psi_\lambda & = & \frac{1}{3} \left(\frac{4\alpha^3}{\sqrt{\pi}}\right)^{3/2}\frac{1}{(4\pi)^{3/2}}\alpha^2\left[({\bf r}_3\times {\bf r}_1)+({\bf r}_2\times {\bf r}_3)-2({\bf r}_1\times {\bf r}_2)\right]e^{-\alpha^2(r^2_1+r^2_2+r^2_3)/2}, \\
 \psi_\rho & = & \frac{1}{\sqrt{3}} \left(\frac{4\alpha^3}{\sqrt{\pi}}\right)^{3/2}\frac{1}{(4\pi)^{3/2}}\alpha^2\left[({\bf r}_3\times {\bf r}_1)-({\bf r}_2\times {\bf r}_3)\right]e^{-\alpha^2(r^2_1+r^2_2+r^2_3)/2}.
\end{eqnarray}}

\subsubsection{$(0s)(0p)^2$: the antisymmetric representation, $\ell=1^+$}
{\small
\begin{equation}
 \psi=\frac{\sqrt{2}}{3}\left(\frac{4\alpha^3}{\sqrt{\pi}}\right)^{3/2}\frac{1}{(4\pi)^{3/2}}\alpha^2\left[({\bf r}_3\times {\bf r}_1)+({\bf r}_2\times {\bf r}_3)+({\bf r}_1\times {\bf r}_2)\right]e^{-\alpha^2(r^2_1+r^2_2+r^2_3)/2}.
\end{equation}}

\subsubsection{$(0s)(0p)^2$: the symmetric representation, $\ell=2^+$}
{\small
\begin{equation}
 \psi=\frac{2}{\sqrt{27}}\left(\frac{4\alpha^3}{\sqrt{\pi}}\right)^{3/2}\frac{1}{(4\pi)^{1/2}}\alpha^2\left[r_1r_2Y_2(\Omega_1\Omega_2)+r_3r_2Y_2(\Omega_3\Omega_2)+r_3r_2Y_1(\Omega_3\Omega_1)\right]e^{-\alpha^2(r^2_1+r^2_2+r^2_3)/2} \label{0s0p2symm2}.
\end{equation}}

\noindent where $Y_2(\Omega_i\Omega_j)$ is the angular part obtained by combining ${\bf r}_i$, ${\bf r}_j$ to form $\ell = 2$, and normalized to unity over $\Omega_i$, $\Omega_j$, \emph{e.g.} 
\begin{equation}
Y_2(\Omega_i\Omega_j)= \sum_{m_im_j} \langle 11m_im_j|2M\rangle Y^{m_i}_1(\theta_i\phi_i)Y^{m_j}_1(\theta_j\phi_j)
\end{equation}
where $\langle 11m_im_j|2M\rangle$ are R$_3$ Clebsch-Gordan coefficients.

\subsubsection{$(0s)(0p)^2$: the mixed-symmetric representation, $\ell=2^+$}
{\small
\begin{eqnarray}
 \psi_\lambda & = & \sqrt{\frac{2}{27}}\left(\frac{4\alpha^3}{\sqrt{\pi}}\right)^{3/2}\frac{1}{(4\pi)^{1/2}}\alpha^2\left[r_2r_3Y_2(\Omega_2\Omega_3)+r_3r_1Y_2(\Omega_3\Omega_1)-2r_1r_2Y_2(\Omega_1\Omega_2)\right]e^{-\alpha^2(r^2_1+r^2_2+r^2_3)/2},\nonumber \\ \\
 \psi_\rho & = &  \frac{\sqrt{2}}{3}\left(\frac{4\alpha^3}{\sqrt{\pi}}\right)^{3/2}\frac{1}{(4\pi)^{1/2}}\alpha^2\left[r_2r_3Y_2(\Omega_2\Omega_3)-r_3r_1Y_2(\Omega_3\Omega_1)\right]e^{-\alpha^2(r^2_1+r^2_2+r^2_3)/2}.
\end{eqnarray}}

In each case one can check the symmetry properties of the wave functions by applying transformation of the group S$_3$. \\

We need now to make combinations of these wave functions to remove the center of mass coordinate dependence. Let us show an example for the $[{\bf 56}, 2^+]$ multiplet.  We need to rewrite the $(0s)^20d$ and $(0s)(0p)^2$ symmetric   wave functions (Eqs. (\ref{os2odsymm2}) and (\ref{0s0p2symm2})) in terms of {\bf R}, $\lambdab$ and $\rhob$.  Let us choose $M=0$.
Using Table \ref{Spherical harmonics}, one can rewrite Eq. (\ref{os2odsymm2}) as
{\small
\begin{eqnarray}
 \psi_{(0s)^2(0d)} & = & \frac{2}{\sqrt{45}}\left(\frac{4\alpha^3}{\sqrt{\pi}}\right)^{3/2}\frac{1}{4\pi}\sqrt{\frac{5}{16\pi}}\alpha^2\left[2(z_1^2+z_2^2+z_3^2)-(x_1^2+x^2_2+x_3^2)-(y_1^2+y_2^2+y_3^2)\right]\nonumber \\ & & \times e^{-\alpha^2(r^2_1+r^2_2+r^2_3)/2} \nonumber \\ 
 & = & \frac{1}{3}\left(\frac{4\alpha^3}{\sqrt{\pi}}\right)^{3/2}\frac{1}{(4\pi)^{3/2}}\alpha^2\left[3(z_1^2+z_2^2+z_3^2)-(r_1^2+r_2^2+r_3^2)\right] e^{-\alpha^2(r^2_1+r^2_2+r^2_3)/2},  
 \end{eqnarray}}
and Eq. (\ref{0s0p2symm2}) as
\begin{eqnarray}
 \psi_{(0s)(0p)^2} & = &  \frac{\sqrt{2}}{3}\left(\frac{4\alpha^3}{\sqrt{\pi}}\right)^{3/2}\frac{1}{(4\pi)^{3/2}}\alpha^2\left[3(z_1z_2+z_3z_2+z_1z_3)-(r_1r_2+r_3r_2+r_1r_2)\right]\nonumber \\ & & \times e^{-\alpha^2(r^2_1+r^2_2+r^2_3)/2}.
\end{eqnarray}
We use 
\begin{eqnarray}
 r_1^2+r_2^2+r_3^2 & = & R^2+\rho^2+\lambda^2,  \\
 r_1r_2+r_1r_3+r_2r_3 & = & R^2-\frac{1}{2}\left(\rho^2+\lambda^2\right), \\
 z_1^2+z_2^2+z_3^2-(z_1z_2+z_1z_3+z_2z_3) & = & \frac{3}{2}\left(\rho_0^2+\lambda_0^2\right).
\end{eqnarray}
The following linear combination suppress the center of mass dependence
\begin{eqnarray}
 \lefteqn{\sqrt{\frac{2}{3}}\psi_{(0s)^2(0d)} - \sqrt{\frac{1}{3}}\psi_{(0s)(0p)^2} =} \nonumber \\
 			& &  \sqrt{\frac{2}{3}}\frac{1}{3}\left(\frac{4\alpha^3}{\sqrt{\pi}}\right)^{3/2}\frac{1}{(4\pi)^{3/2}}\alpha^2\left[3(z_1^2+z_2^2+z_3^2)-(r_1^2+r_2^2+r_3^2)\right.\nonumber \\ & & \left. -3(z_1z_2+z_3z_2+z_1z_3)+(r_1r_2+r_3r_2+r_1r_2)\right] e^{-\alpha^2(r^2_1+r^2_2+r^2_3)/2}  \\
			& & = \sqrt{\frac{1}{6}}\left(\frac{4\alpha^3}{\sqrt{\pi}}\right)^{3/2}\frac{1}{(4\pi)^{3/2}}\alpha^2\left[3\left(\rho_0^2+\lambda_0^2\right)-\left(\rho^2+\lambda^2\right)\right] e^{-\alpha^2(R^2+\rho^2+\lambda^2)/2}.
\end{eqnarray}

Using  this procedure we obtain Table \ref{harmonicconfigurations} describing baryons with $N_c=3$. In a similar way one obtains the result of Table \ref{harmonicconfigurationsnc} valid for an arbitrary number of quarks in a baryon.

\chapter{Isoscalar factors of the permutation group S$_n$}
\label{Isoscalar factors of the permutation group S$_n$}

\thispagestyle{empty}

Here we shortly recall the definition of isoscalar factors of the
permutation group S$_n$ \cite{stancu96}. Let us denote a basis vector in the
invariant subspace of the irrep
$[f]$ of S$_n$ by $|[f] Y \rangle $, where $Y$ is the
corresponding Young tableau or Yamanouchi symbol. A basis vector obtained
from the inner product of two irreps $[f']$ and $[f'']$ is defined by
the sum over  products of basis vectors of $[f']$ and $[f'']$ as
\begin{equation}
|[f] Y \rangle = \sum_{Y'Y''}
S([f']Y' [f'']Y'' | [f]Y ) |[f']Y'\rangle |[f'']Y'' \rangle,
\end{equation}
where $S([f']Y' [f'']Y'' | [f]Y )$ are  Clebsch-Gordan (CG) coefficients
of S$_n$.
Any CG coefficient can be factorized into an isoscalar factor, here called
$K$ matrix \cite{stancu96}, and a CG coefficient of S$_{n-1}$. To apply the
factorization property it is necessary to specify the row $p$ of the 
$n$-th particle and the row $q$ of the $(n-1)$-th particle. The remaining
particles are distributed in a Young tableau with $n-2$ boxes, denoted by $y$.
Then the isoscalar factor $K$ associated to a given CG of S$_n$ is defined as
\begin{equation}
S([f']p'q'y' [f'']p''q''y'' | [f]pqy ) =
K([f']p'[f'']p''|[f]p) S([f'_{p'}]q'y' [f''_{p''}]q''y'' | [f_p]qy ),
\end{equation}
where  the right-hand side contains a CG coefficient
of S$_{n-1}$ depending on $[f_p]$, $[f'_{p'}]$ and $[f''_{p''}]$, which are the partitions obtained from $[f]$ after
the removal of the $n$-th particle.
The $K$ matrix obeys the following orthogonality relations
\begin{eqnarray}
\sum_{p'p''}  K([f']p'[f'']p''|[f]p) K([f']p'[f'']p''|[f_1]p_1) & = &\delta_{f f_1}
\delta{p p_1}, \label{K1}\\
\sum_{fp}  K([f']p'[f'']p''|[f]p) K([f']p'_1[f'']p''_1|[f]p) & = &
\delta_{p'p'_1} \delta_{p'' p''_1} \label{K2}.
\end{eqnarray}
The isoscalar factors used to construct the spin-flavor symmetric state
(\ref{FS}) are denoted by
\begin{eqnarray}\label{SYM}
c^{[N_c]}_{11} & = & K([f']1[f']1|[N_c]1),\nonumber \\
c^{[N_c]}_{22} & = & K([f']2[f']2|[N_c]1),
\end{eqnarray}
with $[f'] = [N_c/2+S,N_c/2-S]$. One has analog
 isoscalar factors, \emph{i.e.} with the last particle in the first or last row both in spin and flavor part, needed to construct a state of mixed-symmetry $[N_c-1,1]$ from the same inner product. There are
\begin{eqnarray}\label{MS}
c^{[N_c-1,1]}_{11}  & = & K([f']1[f']1|[N_c-1,1]2),\nonumber \\
c^{[N_c-1,1]}_{22}  & = & K([f']2[f']2|[N_c-1,1]2).
\end{eqnarray}
The above coefficients and the orthogonality
relation (\ref{K2}) give 
\begin{eqnarray}
c_{11}^{[N_c-1,1]} & = & -c_{22}^{[N_c]}, \nonumber \\
c_{22}^{[N_c-1,1]} & = & c_{11}^{[N_c]}.
\end{eqnarray} \\

When the last particle is located in different rows in the flavor and
spin parts the needed coefficients are
\begin{eqnarray}
c^{[N_c-1,1]}_{12}  & = & K([f']1[f']2|[N_c-1,1]2) = 1,\nonumber \\
c^{[N_c-1,1]}_{21}  & = & K([f']2[f']1|[N_c-1,1]2) = 1,
\end{eqnarray}
which are identical  to each other because of the symmetry properties of $K$.
 \\

 Now we show how to obtain the analytic form of the above coefficients for the states with $N_c$ quarks.
 To get (\ref{SU2}) we write the matrix elements of the generators $S_i$
in two different ways. One is to use the Wigner-Eckart theorem (\ref{SPIN}).
The other is  to calculate the matrix
elements of $S_i$ by using (\ref{FS}), (\ref{ONEB}) and (\ref{SI}). By comparing the two expressions we obtain the equality
\begin{eqnarray}\label{B1}
\sqrt{S(S+1)}  & = &  (-)^{2S} N_c \sqrt{\frac{3}{2}}\sqrt{2S+1}
\left[\left(c_{22}^{[N_c]}\right)^2 \left\{\begin{array}{ccc} 1 & S & S \\ S+1/2 & 1/2 & 1/2 \end{array}\right\}\right. \nonumber \\ & & - \left.\left(c_{11}^{[N_c]}\right)^2 \left\{\begin{array}{ccc} 1 & S & S \\ S-1/2 & 1/2 & 1/2 \end{array}\right\}\right],
\end{eqnarray}
which is an equation for the unknown quantities $c_{11}^{[N_c]}$ and $c_{22}^{[N_c]}$. 
The other equation is the normalization relation (\ref{K1})
\begin{equation}\label{normaliz}
\left(c_{11}^{[N_c]}\right)^2 +\left(c_{22}^{[N_c]}\right)^2 = 1.
\end{equation}
We found
\begin{eqnarray}
c^{[N_c]}_{11}(S) & = & \sqrt{\frac{S[N_c+2(S + 1)]}{N_c(2 S + 1)}}, \nonumber \\
c^{[N_c]}_{22}(S) & = & \sqrt{\frac{(S + 1)(N_c - 2 S)}{N_c(2 S + 1)}},
\end{eqnarray}
which are  the relations (\ref{SU2}). This phase convention is consistent
with Ref. \cite{stancu99}.
Similarly, to get (\ref{SU3}) we calculate the matrix elements of the generators
$T_a$ from (\ref{FS}), (\ref{ONEB}) and (\ref{TA}) and compare to the
Wigner-Eckart theorem (\ref{FLAVOR}).
This leads to the equation
\begin{eqnarray}\label{B4}
\sqrt{\frac{g_{\lambda\mu}}{3}} &  =  &
  \frac{2}{\sqrt{3}} N_c \left[\left(c_{22}^{[N_c]}\right)^2 U((\lambda+1,\mu-1)(10)(\lambda\mu)(11);(\lambda\mu)(10))_{\rho=1}\right. \nonumber \\
 & & + \left.\left(c_{11}^{[N_c]}\right)^2  U((\lambda-1,\mu)(10)(\lambda\mu)(11);(\lambda\mu)(10))_{\rho=1}\right],
\end{eqnarray}
which together with the normalization condition (\ref{normaliz}) give
\begin{eqnarray}
c^{[N_c]}_{11} (\lambda \mu) & = & \sqrt{\frac{2g_{\lambda\mu}-N_c(\mu-\lambda+3)}{3N_c(\lambda +1)}},      \nonumber \\
c^{[N_c]}_{22} (\lambda \mu) & = & \sqrt{\frac{N_c(6+2\lambda+\mu)-2g_{\lambda\mu}}{3N_c(\lambda+1)}},
\end{eqnarray}
\emph{i.e.} the relations (\ref{SU3}).
In addition, we found that the following identity holds for $\rho = 2$
\begin{eqnarray}\label{B5}
0 & = &
  \frac{2}{\sqrt{3}} N_c
  \left[\left(c_{22}^{[N_c]}\right)^2
  U((\lambda+1,\mu-1)(10)(\lambda\mu)(11);(\lambda\mu)(10))_{\rho=2}\right. \nonumber \\
 & & + \left.\left(c_{11}^{[N_c]}\right)^2 
 U((\lambda-1,\mu)(10)(\lambda\mu)(11);(\lambda\mu)(10))_{\rho=2}\right].
\end{eqnarray}
 This  cancellation is consistent with the definition of the matrix elements
 of the SU(3) generators Eqs. (\ref{FLAVOR}), (\ref{REDUCED}) and it is an 
 important check of our results. \\
 
 One can identify  the so called ``elements of orthogonal basis
rotation" of Ref. \cite{carlson98b} with the above isoscalar factors of S$_n$, as discussed in Sec. \ref{SU(6) symmetric  wave functions}. 
\chapter{Racah coefficients and some isoscalar factors of SU(3)}
\label{su3racahandisoscalarfactors}

\thispagestyle{empty}

This Appendix gives details about the SU(3) Racah coefficients. We present also tables containing some SU(3) isoscalar factors, needed to compute the SU(3) Racah coefficients used to derived the matrix elements of the SU(6) generators   presented in Chapter 4 and of the operators used in Chapter 5. \\

SU(3) Racah coefficients are defined in Ref. \cite{hecht65}. They can be expressed in terms of summations involving Racah coefficients of SU(2) and SU(3) isoscalar factors
\begin{eqnarray}
 \lefteqn{U((\lambda_1\mu_1)(\lambda_2\mu_2)(\lambda\mu)(\lambda_3\mu_3);(\lambda_{12}\mu_{12})\rho_{12}\rho_{12,3}(\lambda_{23}\mu_{23})\rho_{23}\rho_{1,23})=} \nonumber \\
 & & \sum_{Y_1Y_2(Y_3) \atop I_1I_2I_3I_{12}I_{23}} U(I_1I_2II_3;I_{12}I_{23})\langle (\lambda_1\mu_1)Y_1I_1;(\lambda_2\mu_2)Y_2I_2||(\lambda_{12}\mu_{12})Y_{12}I_{12}\rangle_{\rho_{12}}  \nonumber \\   & & \times \langle(\lambda_{12}\mu_{12}) Y_{12}I_{12};(\lambda_3\mu_3)Y_3I_3||(\lambda\mu)YI\rangle_{\rho_{12,3}}
 \langle (\lambda_2\mu_2)Y_2I_2;(\lambda_3\mu_3)Y_3I_3||(\lambda_{23}\mu_{23})Y_{23}I_{23}\rangle_{\rho_{23}} \nonumber \\   & & \times\langle (\lambda_1\mu_1)Y_1I_1;(\lambda_{23}\mu_{23})Y_{23}I_{23}||(\lambda\mu)YI\rangle_{\rho_{1,23}}, \label{su(6)racahhecht}
\end{eqnarray}
where the SU(2) $U$ coefficients can be related to the 6-j symbols \cite{rotenberg59}
\begin{equation}
 U(I_1I_2II_3;I_{12}I_{23}) = (-1)^{I_1+I_2+I+I_3}\sqrt{(2I_{12}+1)(2I_{23}+1)}\left\{\begin{array}{ccc}
                                                                                       I_1 & I_2 & I_{12} \\
										       I_3 & I  & I_{23}
                                                                                      \end{array}\right\}.
\end{equation}
To get the Racah coefficients used in chapter 4, we have the simplified expression 
\begin{eqnarray}
\lefteqn{U((\lambda_1\mu_1)(10)(\lambda\mu)(11);(\lambda\mu)(10))_\rho = } \nonumber \\
& & \sum_{Y_1Y_2(Y_a) \atop I_1I_2I_aI_{12}I_{23}} (-1)^{I_1+I_2+I+I_a}\sqrt{(2I_{12}+1)(2I_{23}+1)}\left\{\begin{array}{ccc}
                                                                                       I_1 & I_2 & I_{12} \\
										       I_a & I  & I_{23}
                                                                                      \end{array}\right\}\nonumber \\  
& & \times
\langle (\lambda_1\mu_1)Y_1I_1;(10)Y_2I_2||(\lambda_{12}\mu_{12})Y_{12}I_{12}\rangle \langle(\lambda_{12}\mu_{12}) Y_{12}I_{12};(11)Y_aI_a||(\lambda\mu)YI\rangle_{\rho} 
 \nonumber \\  
& & \times  \langle (\lambda_2\mu_2)Y_2I_2;(11)Y_aI_a||(10)Y_{23}I_{23}\rangle \langle (\lambda_1\mu_1)Y_1I_1;(10)Y_{23}I_{23}||(\lambda\mu)YI\rangle. \label{racahsu3}
\end{eqnarray}
This form is simpler than the first one as only one  $\rho$ label is needed. One can directly notice that the sum in Eq. (\ref{racahsu3}) is independent of $Y$ and $I$ by definition. For the calculation of the Racah, we need to choose a fixed set of $Y, I$. One classical choice is to take $I=0$ and $Y=2/3(\mu-\lambda)$. In Tables \ref{vergados}--\ref{hecht4} one can find the isoscalar factors used to derive the SU(3) Racah coefficients out of Eq. (\ref{racahsu3}).

\begin{sidewaystable}[h!]
\renewcommand{\arraystretch}{1.5}
\begin{center}
 \begin{tabular}{c|lll}
  \hline \hline
$Y_2\ I_2$ &   &  &  \\
$I_1$ &   \raisebox{2.45ex}[0cm][0cm]{$(\lambda'\mu') = (\lambda + 1,\mu)$}     &  \raisebox{2.45ex}[0cm][0cm]{$(\lambda'\mu') = (\lambda, \mu-1)$}   & \raisebox{2.45ex}[0cm][0cm]{$(\lambda'\mu')=(\lambda-1,\mu-1)$} \\
\hline
-2/3 0 &    &   & \\
$I_1=I$      & \raisebox{2.45ex}[0cm][0cm]{$\sqrt{\frac{(\lambda+1-p)(\lambda+\mu+2-q)}{(\lambda+1)(\lambda+\mu+2)}}$} & \raisebox{2.45ex}[0cm][0cm]{$\sqrt{\frac{(\mu+1+p)(q+1)}{(\mu+1)(\lambda+\mu+2)}}$} & \raisebox{2.45ex}[0cm][0cm]{$-\sqrt{\frac{(\mu+1-q)(p+1)}{(\lambda+1)(\mu+1)}}$} \\ \hline
1/3 1/2 & & & \\
$I_1 = I+1/2$ & \raisebox{2.45ex}[0cm][0cm]{$-\sqrt{\frac{(\lambda+1-p)(\mu+1-q)q}{(\lambda+1)(\lambda+\mu+2)(\mu+1+p-q)}}$} &\raisebox{2.45ex}[0cm][0cm]{$\sqrt{\frac{(\lambda+\mu+1-q)(\mu-q)(\mu+1+p)}{(\lambda+\mu+2)(\mu+1)(\mu+p-q)}}$} & \raisebox{2.45ex}[0cm][0cm]{$\sqrt{\frac{(\lambda+\mu+2-q)q(p+1)}{(\lambda+1)(\mu+1)(\mu+2+p-q)}}$} \\ \hline
1/3 1/2 & & & \\
$I_1 = I-1/2$ & \raisebox{2.45ex}[0cm][0cm]{$\sqrt{\frac{(\mu+1+p)(\lambda+\mu+2-q)p}{(\lambda+1)(\lambda+\mu+2)(\mu+1+p-q)}}$} & \raisebox{2.45ex}[0cm][0cm]{$-\sqrt{\frac{(\lambda+1-p)(q+1)p}{(\mu+1)(\lambda+\mu+2)(\mu+p-q)}}$} & \raisebox{2.45ex}[0cm][0cm]{$\sqrt{\frac{(\lambda-p)(\mu+2+p)(\mu+1-q)}{(\lambda+1)(\mu+1)(\mu+2+p-q)}}$} \\ \hline \hline
 \end{tabular}
 \caption{$\langle (\lambda\mu)Y_1I_1;(10)Y_2I_2||(\lambda'\mu')YI\rangle$, $Y=-1/3(2\lambda'+\mu'-3p-3q)$, $I=1/2\mu'+1/2(p-q)$ \cite{vergados68}}\label{vergados}
\end{center}
\end{sidewaystable}

\begin{sidewaystable}[h!]
\renewcommand{\arraystretch}{1.5}
\begin{center}
 \begin{tabular}{c|c|c}
  \hline \hline
  $Y_2\ I_2$ &  &   \\
  $I_1$      &  \raisebox{2.45ex}[0cm][0cm]{$(\lambda'\mu') = (\lambda+2, \mu-1)$} & \raisebox{2.45ex}[0cm][0cm]{$(\lambda'\mu') = (\lambda-2, \mu+1)$}  \\ \hline
  $-1\ 1/2$ &  & \\
  $I_1=I+1/2$ & \raisebox{2.45ex}[0cm][0cm]{$\sqrt{\frac{(\lambda+1-p)(\lambda +2-p)(\mu +1+p)(\mu -q)(\lambda+\mu+2-q)}{(\lambda+1)(\lambda+2)(\mu+1)(\lambda+\mu+2)(\mu+p-q)}}$} & \raisebox{2.45ex}[0cm][0cm]{$\sqrt{\frac{(p+1)(p+2)(\mu+p+3)(\lambda+\mu+1-q)(\mu+1-q)}{\lambda(\lambda+1)(\mu+1)(\lambda+\mu+2)(\mu+p-q+2)}}$} \\ \hline
  $-1\ 1/2$ & & \\
  $I_1=I-1/2$ & \raisebox{2.45ex}[0cm][0cm]{$-\sqrt{\frac{p(\lambda+2-p)(q+1)(\lambda+\mu+1-q)(\lambda+\mu+2-q)}{(\lambda+1)(\lambda+2)(\mu+1)(\lambda+\mu+2)(\mu+p-q)}}$} & \raisebox{2.45ex}[0cm][0cm]{$\sqrt{\frac{(p+1)(\lambda-1-p)(q+1)(\mu-q)(\mu+1-q)}{\lambda(\lambda+1)(\mu+1)(\lambda+\mu+2)(\mu+p-q+2)}}$} \\ \hline
  0 0 & & \\
  $I_1=I$ &  \raisebox{2.45ex}[0cm][0cm]{$\sqrt{\frac{3p(\lambda+2-p)(\mu-q)(\lambda+\mu+2-q)}{2(\lambda+1)(\lambda+2)(\mu+1)(\lambda+\mu+2)}}$} & \raisebox{2.45ex}[0cm][0cm]{$-\sqrt{\frac{3(p+1)(\lambda-1-p)(\mu+1-q)(\lambda+\mu+1-q)}{2\lambda(\lambda+1)(\mu+1)(\lambda+\mu+2)}}$} \\ \hline
  0 1 & & \\
  $I_1=I+1$ & \raisebox{2.45ex}[0cm][0cm]{$-\sqrt{\frac{(\lambda+1-p)(\lambda+2-p)(\mu+1+p)q(\mu-q)(\mu+1-q)}{(\lambda+1)(\lambda+2)(\mu+1)(\lambda+\mu+2)(\mu+p-q)(\mu+p-q+1)}}$} & \raisebox{2.45ex}[0cm][0cm]{$-\sqrt{\frac{(p+1)(p+2)(\mu+p+3)q(\lambda+\mu+1-q)(\lambda+\mu+2-q)}{\lambda(\lambda+1)(\mu+1)(\lambda+\mu+2)(\mu+p-q+2)(\mu+p-q+3)}}$} \\ \hline
  0 1 & & \\
  $I_1=I-1$ & \raisebox{2.45ex}[0cm][0cm]{$-\sqrt{\frac{p(p-1)(\mu+p)(q+1)(\lambda+\mu+1-q)(\lambda+\mu+2-q)}{(\lambda+1)(\lambda+2)(\mu+1)(\lambda+\mu+2)(\mu+p-q)(\mu+p-q-1)}}$} & \raisebox{2.45ex}[0cm][0cm]{$-\sqrt{\frac{(\lambda-p)(\lambda-1-p)(\mu+2+p)(q+1)(\mu-q)(\mu+1-q)}{\lambda(\lambda+1)(\mu+1)(\lambda+\mu+2)(\mu+p-q+2)(\mu+p-q+1)}}$} \\ \hline
  0 1 & & \\
  $I_1=I$ & \raisebox{2.45ex}[0cm][0cm]{$\frac{(\mu+1+p+q)\sqrt{p(\lambda+2-p)(\mu-q)(\lambda+\mu+2-q)}}{\sqrt{2(\lambda+1)(\lambda+2)(\mu+1)(\lambda+\mu+2)(\mu+p-q-1)(\mu+p-q+1)}}$} & \raisebox{2.45ex}[0cm][0cm]{$-\frac{(\mu+3+p+q)\sqrt{(p+1)(\lambda-1-p)(\mu+1-q)(\lambda+1+\mu-q)}}{\sqrt{2\lambda(\lambda+1)(\mu+1)(\lambda+\mu+2)(\mu+p-q+1)(\mu+p-q+3)}}$} \\ \hline 
  $1\ 1/2$ & & \\
  $I_1=I+1/2$ & \raisebox{2.45ex}[0cm][0cm]{$-\sqrt{\frac{p(\lambda+2-p)q(\mu-q)(\mu+1-q)}{(\lambda+1)(\lambda+2)(\mu+1)(\lambda+\mu+2)(\mu+p-q)}}$} & \raisebox{2.45ex}[0cm][0cm]{$\sqrt{\frac{(p+1)(\lambda-1-p)q(\lambda+\mu+1-q)(\lambda+\mu+2-q)}{\lambda(\lambda+1)(\mu+1)(\lambda+\mu+2)(\mu+p-q+2)}}$} \\ \hline
  $1\ 1/2$ & & \\
  $I_1=I-1/2$ & \raisebox{2.45ex}[0cm][0cm]{$\sqrt{\frac{p(p-1)(\mu+p)(\mu-q)(\lambda+\mu+2-q)}{(\lambda+1)(\lambda+2)(\mu+1)(\lambda+\mu+2)(\mu+p-q)}}$} & \raisebox{2.45ex}[0cm][0cm]{$\sqrt{\frac{(\lambda-p)(\lambda-1-p)(\mu+2+p)(\mu+1-q)(\lambda+\mu+1-q)}{\lambda(\lambda+1)(\mu+1)(\lambda+\mu+2)(\mu+p-q+2)}}$} \\ \hline \hline
     \end{tabular}
 \caption{$\langle (\lambda\mu)Y_1I_1;(11)Y_2I_2||(\lambda'\mu')YI\rangle$, $Y=-1/3(2\lambda'+\mu'-3p-3q)$, $I=1/2\mu'+1/2(p-q)$ \cite{hecht65}}\label{hecht1}
\end{center}
\end{sidewaystable}

\begin{sidewaystable}[h!]
\renewcommand{\arraystretch}{1.5}
\begin{center}
 \begin{tabular}{c|c|c}
  \hline \hline
  $Y_2\ I_2$ &  &   \\
  $I_1$      &  \raisebox{2.45ex}[0cm][0cm]{$(\lambda'\mu') = (\lambda+1, \mu+1)$} & \raisebox{2.45ex}[0cm][0cm]{$(\lambda'\mu') = (\lambda-1, \mu-1)$}  \\ \hline
  $-1\ 1/2$ &  & \\
  $I_1=I+1/2$ & \raisebox{2.45ex}[0cm][0cm]{$\sqrt{\frac{(p+1)(\lambda-p)(\lambda+1-p)q(\lambda+\mu+3-q)}{(\lambda+1)(\mu+1)(\lambda+\mu+2)(\lambda+\mu+3)(\mu+p-q+2)}}$} & \raisebox{2.45ex}[0cm][0cm]{$-\sqrt{\frac{(p+1)(\mu+p+1)(\mu+p+2)(q+1)(\lambda+\mu-q)}{(\lambda+1)(\mu+1)(\lambda+\mu+1)(\lambda+\mu+2)(\mu+p-q)}}$} \\ \hline 
  $-1\ 1/2$ & & \\
  $I_1=I-1/2$ &   \raisebox{2.45ex}[0cm][0cm]{$\sqrt{\frac{(\lambda+1-p)(\mu+2+p)(\mu+1-q)(\lambda+\mu+2-q)(\lambda+\mu+3-q)}{(\lambda+1)(\mu+1)(\lambda+\mu+2)(\lambda+\mu+3)(\mu+p-q+2)}}$} & \raisebox{2.45ex}[0cm][0cm]{$-\sqrt{\frac{(\lambda-p)(\mu+1+p)(q+1)(q+2)(\mu-1-q)}{(\lambda+1)(\mu+1)(\lambda+\mu+1)(\lambda+\mu+2)(\mu+p-q)}}$} \\ \hline
  0 0 & & \\
  $I_1=I$ & \raisebox{2.45ex}[0cm][0cm]{$\sqrt{\frac{3(\lambda+1-p)(\mu+2+p)q(\lambda+\mu+3-q)}{2(\lambda+1)(\mu+1)(\lambda+\mu+2)(\lambda+\mu+3)}}$} & \raisebox{2.45ex}[0cm][0cm]{$\sqrt{\frac{3(\lambda-p)(\mu+1+p)(q+1)(\lambda+\mu-q)}{2(\lambda+1)(\mu+1)(\lambda+\mu+1)(\lambda+\mu+2)}}$} \\ \hline
  0 1 & & \\
  $I_1=I+1$ & \raisebox{2.45ex}[0cm][0cm]{$-\sqrt{\frac{q(q-1)(\mu+2-q)(p+1)(\lambda-p)(\lambda+1-p)}{(\lambda+1)(\mu+1)(\lambda+\mu+2)(\lambda+\mu+3)(\mu+p-q+2)(\mu+p-q+3)}}$} & \raisebox{2.45ex}[0cm][0cm]{$-\sqrt{\frac{(p+1)(\mu+1+p)(\mu+2+p)(\mu-q)(\lambda+\mu-q)(\lambda+\mu+1-q)}{(\lambda+1)(\mu+1)(\lambda+\mu+1)(\lambda+\mu+2)(\mu+p-q)(\mu+p-q+1)}}$} \\ \hline
  0 1 &  &  \\
  $I_1=I-1$ & \raisebox{2.45ex}[0cm][0cm]{$\sqrt{\frac{p(\mu+1+p)(\mu+2+p)(\mu+1-q)(\lambda+\mu+2-q)(\lambda+\mu+3-q)}{(\lambda+1)(\mu+1)(\lambda+\mu+2)(\lambda+\mu+3)(\mu+p-q+2)(\mu+p-q+1)}}$} & \raisebox{2.45ex}[0cm][0cm]{$\sqrt{\frac{p(\lambda-p)(\lambda+1-p)(q+1)(q+2)(\mu-q-1)}{(\lambda+1)(\mu+1)(\lambda+\mu+1)(\lambda+\mu+2)(\mu+p-q)(\mu+p-q-1)}}$} \\ \hline
  0 1 & & \\
  $I_1=I$ & \raisebox{2.45ex}[0cm][0cm]{$-\frac{(\mu+1-p-q)\sqrt{(\lambda+1-p)(\mu+2+p)q(\lambda+\mu+3-q)}}{\sqrt{2(\lambda+1)(\mu+1)(\lambda+\mu+2)(\lambda+\mu+3)(\mu+p-q+1)(\mu+p-q+3)}}$} & \raisebox{2.45ex}[0cm][0cm]{$-\frac{(\mu-1-p-q)\sqrt{(\lambda-p)(\mu+1+p)(q+1)(\lambda+\mu-q)}}{\sqrt{2(\lambda+1)(\mu+1)(\lambda+\mu+1)(\lambda+\mu+2)(\mu+p-q+1)(\mu+p-q-1)}}$} \\ \hline
  $1\ 1/2$ & & \\
  $I_1=I+1/2$ & \raisebox{2.45ex}[0cm][0cm]{$-\sqrt{\frac{(\lambda+1-p)(\mu+2+p)q(q-1)(\mu+2-q)}{(\lambda+1)(\mu+1)(\lambda+\mu+2)(\lambda+\mu+3)(\mu+p-q+2)}}$} & \raisebox{2.45ex}[0cm][0cm]{$\sqrt{\frac{(\lambda-p)(\mu+1+p)(\mu-q)(\lambda+\mu-q)(\lambda+\mu+1-q)}{(\lambda+1)(\mu+1)(\lambda+\mu+1)(\lambda+\mu+2)(\mu+p-q)}}$} \\ \hline
  $1\ 1/2$ & & \\
  $I_1=I-1/2$ & \raisebox{2.45ex}[0cm][0cm]{$\sqrt{\frac{p(\mu+1+p)(\mu+2+p)q(\lambda+\mu+3-q)}{(\lambda+1)(\mu+1)(\lambda+\mu+2)(\lambda+\mu+3)(\mu+p-q+2)}}$} & \raisebox{2.45ex}[0cm][0cm]{$-\sqrt{\frac{p(\lambda-p)(\lambda+1-p)(q+1)(\lambda+\mu-q)}{(\lambda+1)(\mu+1)(\lambda+\mu+1)(\lambda+\mu+2)(\mu+p-q)}}$} \\ \hline \hline
     \end{tabular}
 \caption{$\langle (\lambda\mu)Y_1I_1;(11)Y_2I_2||(\lambda'\mu')YI\rangle$, $Y=-1/3(2\lambda'+\mu'-3p-3q)$, $I=1/2\mu'+1/2(p-q)$ \cite{hecht65}}\label{hecht2}
\end{center}
\end{sidewaystable}

\begin{sidewaystable}[h!]
\renewcommand{\arraystretch}{1.5}
\begin{center}
 \begin{tabular}{c|c|c}
  \hline \hline
  $Y_2\ I_2$ &  &   \\
  $I_1$      &  \raisebox{2.45ex}[0cm][0cm]{$(\lambda'\mu') = (\lambda+1, \mu-2)$} & \raisebox{2.45ex}[0cm][0cm]{$(\lambda'\mu') = (\lambda-1, \mu+2)$}  \\ \hline
  $-1\ 1/2$ &  & \\
  $I_1=I+1/2$ & \raisebox{2.45ex}[0cm][0cm]{$\sqrt{\frac{(\mu+p)(\mu+1+p)(\lambda+1-p)(q+1)(\mu-q-1)}{(\lambda+1)\mu(\mu+1)(\lambda+\mu+2)(\mu+p-q-1)}}$} & \raisebox{2.45ex}[0cm][0cm]{$-\sqrt{\frac{(p+1)(p+2)(\lambda-1-p)q(\mu+2-q)}{(\lambda+1)(\mu+1)(\mu+2)(\lambda+\mu+2)(\mu+p-q+3)}}$} \\ \hline 
  $-1\ 1/2$ & & \\
  $I_1=I-1/2$ & \raisebox{2.45ex}[0cm][0cm]{$-\sqrt{\frac{p(\mu+p)(q+2)(q+1)(\lambda+\mu-q)}{(\lambda+1)\mu(\mu+1)(\lambda+\mu+2)(\mu+p-q-1)}}$} & \raisebox{2.45ex}[0cm][0cm]{$-\sqrt{\frac{(p+1)(\mu+p+3)(\mu+1-q)(\mu+2-q)(\lambda+\mu+2-q)}{(\lambda+1)(\mu+1)(\mu+2)(\lambda+\mu+2)(\mu+p-q+3)}}$} \\ \hline
  0 0 & & \\
  $I_1=I$ & \raisebox{2.45ex}[0cm][0cm]{$\sqrt{\frac{3p(\mu+p)(q+1)(\mu-1-q)}{2(\lambda+1)\mu(\mu+1)(\lambda+\mu+2)}}$} & \raisebox{2.45ex}[0cm][0cm]{$-\sqrt{\frac{3(p+1)(\mu+3+p)q(\mu+2-q)}{2(\lambda+1)(\mu+1)(\mu+2)(\lambda+\mu+2)}}$} \\ \hline
  0 1 & & \\
  $I_1=I+1$ & \raisebox{2.45ex}[0cm][0cm]{$\sqrt{\frac{(\mu+p)(\mu+p+1)(\lambda+1-p)(\mu-q)(\mu-1-q)(\lambda+\mu+1-q)}{(\lambda+1)\mu(\mu+1)(\lambda+\mu+2)(\mu+p-q)(\mu+p-q-1)}}$} & \raisebox{2.45ex}[0cm][0cm]{$\sqrt{\frac{(p+1)(p+2)(\lambda-1-p)q(q-1)(\lambda+\mu+3-q)}{(\lambda+1)(\mu+1)(\mu+2)(\lambda+\mu+2)(\mu+p-q+3)(\mu+p-q+4)}}$} \\ \hline
  0 1  & & \\
  $I_1=I-1$ & \raisebox{2.45ex}[0cm][0cm]{$\sqrt{\frac{p(p-1)(\lambda+2-p)(q+1)(q+2)(\lambda+\mu-q)}{(\lambda+1)\mu(\mu+1)(\lambda+\mu+2)(\mu+p-q-2)(\mu+p-q-1)}}$} & \raisebox{2.45ex}[0cm][0cm]{$\sqrt{\frac{(\lambda-p)(\mu+2+p)(\mu+3+p)(\mu+1-q)(\mu+2-q)(\lambda+\mu+2-q)}{(\lambda+1)(\mu+1)(\mu+2)(\lambda+\mu+2)(\mu+p-q+3)(\mu+p-q+2)}}$} \\ \hline
  0 1 & & \\
  $I_1=I$ & \raisebox{2.45ex}[0cm][0cm]{$-\frac{(2\lambda+\mu+2-p-q)\sqrt{p(\mu+p)(q+1)(\mu-1-q)}}{\sqrt{2(\lambda+1)\mu(\mu+1)(\lambda+\mu+2)(\mu+p-q)(\mu+p-q-2)}}$} & \raisebox{2.45ex}[0cm][0cm]{$\frac{(2\lambda+\mu+2-p-q)\sqrt{(p+1)(\mu+p+3)q(\mu+2-q)}}{\sqrt{2(\lambda+1)(\mu+1)(\mu+2)(\lambda+\mu+2)(\mu+p-q+2)(\mu+p-q+4)}}$} \\ \hline
  $1\ 1/2$ & & \\
  $I_1=I+1/2$ & \raisebox{2.45ex}[0cm][0cm]{$\sqrt{\frac{p(\mu+p)(\mu-1-q)(\mu-q)(\lambda+\mu+1-q)}{(\lambda+1)\mu(\mu+1)(\lambda+\mu+2)(\mu+p-q-1)}}$} & \raisebox{2.45ex}[0cm][0cm]{$\sqrt{\frac{(p+1)(\mu+3+p)q(q-1)(\lambda+\mu+3-q)}{(\lambda+1)(\mu+1)(\mu+2)(\lambda+\mu+2)(\mu+p-q+3)}}$} \\ \hline
  $1\ 1/2$ & & \\
  $I_1=I-1/2$ & \raisebox{2.45ex}[0cm][0cm]{$-\sqrt{\frac{p(p-1)(\lambda+2-p)(q+1)(\mu-1-q)}{(\lambda+1)\mu(\mu+1)(\lambda+\mu+2)(\mu+p-q-1)}}$} & \raisebox{2.45ex}[0cm][0cm]{$\sqrt{\frac{(\lambda-p)(\mu+2+p)q(\mu+3+p)(\mu+2-q)}{(\lambda+1)(\mu+1)(\mu+2)(\lambda+\mu+2)(\mu+p-q+3)}}$} \\ \hline\hline
    \end{tabular}
 \caption{$\langle (\lambda\mu)Y_1I_1;(11)Y_2I_2||(\lambda'\mu')YI\rangle$, $Y=-1/3(2\lambda'+\mu'-3p-3q)$, $I=1/2\mu'+1/2(p-q)$ \cite{hecht65}}\label{hecht3}
\end{center}
\end{sidewaystable}

\begin{sidewaystable}[h!]
\renewcommand{\arraystretch}{1.5}
\begin{center}
 \begin{tabular}{c|c|c}
  \hline \hline
  $Y_2\ I_2$ & $(\lambda'\mu') = (\lambda\mu)$ &  $(\lambda'\mu') = (\lambda\mu)$ \\
  $I_1$      & $\rho=1$                           & $\rho=2$  \\ \hline
 $-1\ 1/2$ & & \\
 $I_1=I+1/2$ &  \raisebox{2.45ex}[0cm][0cm]{$\sqrt{\frac{3(p+1)(\lambda-p)(\mu+2+p)}{2g_{\lambda\mu}(\mu+p-q+1)}}$} & \raisebox{2.45ex}[0cm][0cm]{$\frac{2g_{\lambda\mu}q-\mu(\lambda+\mu+1)(\lambda+2\mu+6)\sqrt{(p+1)(\lambda-p)(\mu+2+p)}}{\sqrt{\lambda(\lambda+2)\mu(\mu+2)(\lambda+\mu+1)(\lambda+\mu+3)2g_{\lambda\mu}(\mu+p-q+1)}}$} \\
 \hline
$-1\ 1/2$ & & \\
 $I_1=I-1/2$ &  \raisebox{2.45ex}[0cm][0cm]{$\sqrt{\frac{3(q+1)(\mu-q)(\lambda+\mu+1-q)}{2g_{\lambda\mu}(\mu+p-q+1)}}$} & \raisebox{2.45ex}[0cm][0cm]{$\frac{2g_{\lambda\mu}p+\lambda(\mu+2)(\lambda-\mu+3)\sqrt{(q+1)(\mu-q)(\lambda+\mu+1-q)}}{\sqrt{\lambda(\lambda+2)\mu(\mu+2)(\lambda+\mu+1)(\lambda+\mu+3)2g_{\lambda\mu}(\mu+p-q+1)}}$} \\ \hline
 0 0 & & \\
 $I_1=I$ & \raisebox{2.45ex}[0cm][0cm]{$-\frac{2\lambda+\mu-3p-3q}{2\sqrt{g_{\lambda\mu}}}$} & \raisebox{2.45ex}[0cm][0cm]{$\frac{\sqrt{3}[\lambda\mu(\mu+2)(\lambda+\mu+1)-\mu(\lambda+\mu+1)(\lambda+2\mu+6)p+\lambda(\mu+2)(\lambda-\mu+3)q+2g_{\lambda\mu}pq]}{2\sqrt{\lambda(\lambda+2)\mu(\mu+2)(\lambda+\mu+1)(\lambda+\mu+3)g_{\lambda\mu}}}$} \\ \hline
 0 1 & &  \\
 $I_1=I+1$ & \raisebox{2.45ex}[0cm][0cm]{0} & \raisebox{2.45ex}[0cm][0cm]{$\sqrt{\frac{2(p+1)(\lambda-p)(\mu+2+p)q(\mu+1-q)(\lambda+\mu+2-q)g_{\lambda\mu}}{\lambda(\lambda+2)\mu(\mu+2)(\lambda+\mu+1)(\lambda+\mu+3)(\mu+p-q+1)(\mu+p-q+2)}}$} \\ \hline
 0 1 & &  \\
 $I_1=I-1$ & \raisebox{2.45ex}[0cm][0cm]{0} & \raisebox{2.45ex}[0cm][0cm]{$-\sqrt{\frac{2p(\lambda+1-p)(\mu+1+p)(q+1)(\mu-q)(\lambda+\mu+1-q)g_{\lambda\mu}}{\lambda(\lambda+2)\mu(\mu+2)(\lambda+\mu+1)(\lambda+\mu+3)(\mu+p-q+1)(\mu+p-q)}}$} \\ \hline
  0 1  &  & \\
  $I_1=I$ & \raisebox{2.45ex}[0cm][0cm]{$\frac{1}{2}\sqrt{\frac{3(\mu+p-q)(\mu+p-q+2)}{g_{\lambda\mu}}}$} &
  \raisebox{-3ex}[0cm][0cm]{$\frac{{\scriptsize \left\{\begin{array}{c} \lambda(\lambda+\mu+1)\mu(\mu+2)(2\lambda+\mu+6)+2(\lambda+\mu+1)\mu[\lambda(\lambda+2)-(\mu+2)(\mu+3)]p \\ -\mu(\lambda+\mu+1)(\lambda+2\mu+6)p^2-2\lambda[(\mu+1)(\lambda+\mu+1)(2\lambda+\mu+6)-\mu g_{\lambda\mu}]q \\ +\lambda(\mu+2)(\lambda-\mu+3)q^2 - 2[\lambda(\lambda+\mu+1)(2\lambda+\mu+6)-g_{\lambda\mu}]pq+2g_{\lambda\mu}(p^2q+pq^2) \end{array}\right\}}}{2\sqrt{\lambda(\lambda+2)\mu(\mu+2)(\lambda+\mu+1)(\lambda+\mu+3)g_{\lambda\mu}(\mu+p-q)(\mu+p-q+2)}}$} \\ & & \\
\hline
$1\ 1/2$ & & \\
$I_1=I+1/2$ & \raisebox{2.45ex}[0cm][0cm]{$\sqrt{\frac{3q(\mu+1-q)(\lambda+\mu+2-q)}{2g_{\lambda\mu}(\mu+p-q+1)}}$} & \raisebox{2.45ex}[0cm][0cm]{$\frac{2g_{\lambda\mu}p+\lambda(\mu+2)(\lambda-\mu+3)\sqrt{q(\mu+1-q)(\lambda+\mu+2-q)}}{\sqrt{\lambda(\lambda+2)\mu(\mu+2)(\lambda+\mu+1)(\lambda+\mu+3)2g_{\lambda\mu}(\mu+p-q+1)}}$} \\ \hline
   $1\ 1/2$ & & \\
$I_1=I-1/2$ & \raisebox{2.45ex}[0cm][0cm]{$-\sqrt{\frac{3p(\lambda+1-p)(\mu+1-p)}{2g_{\lambda\mu}(\mu+p-q+1)}}$} &\raisebox{2.45ex}[0cm][0cm]{$-\frac{2g_{\lambda\mu}q-\mu(\lambda+\mu+1)(\lambda+2\mu+6)\sqrt{p(\lambda+1-p)(\mu+1+p)}}{\sqrt{\lambda(\lambda+2)\mu(\mu+2)(\lambda+\mu+1)(\lambda+\mu+3)2g_{\lambda\mu}(\mu+p-q+1)}}$} \\ \hline \hline
   \end{tabular}
 \caption{$\langle (\lambda\mu)Y_1I_1;(11)Y_2I_2||(\lambda'\mu')YI\rangle$, $Y=-1/3(2\lambda'+\mu'-3p-3q)$, $I=1/2\mu'+1/2(p-q)$ \cite{hecht65}}\label{hecht4}
\end{center}
\end{sidewaystable}

\chapter{Some useful  matrix elements}
\label{generalformylasmatrix elements}

\thispagestyle{empty}

To derive the matrix elements of the operators $O_i$ and $B_i$ used in Chapter 5, one need first to calculate the matrix elements of the generators of the SU(4) or SU(6) group, depending if we work with non-strange or strange baryons (see Chapter 4), and of the O(3) group.
Concerning the O(3) group we have
\begin{equation}
 \langle \ell'm'_\ell|\ell^i|\ell m_\ell\rangle = \delta_{\ell\ell'} \sqrt{\ell(\ell+1)} \\
 \left(\begin{array}{cc|c}
        \ell  & 1 & \ell \\
	m_\ell & i & m'_\ell
       \end{array}\right).
\end{equation}
We also need the matrix elements of the  tensor operator $\ell^{(2)}$. This can be written as
\begin{equation}
 \langle \ell'm'_\ell|\ell^{(2)ij}|\ell m_\ell\rangle  =  \sum_{\mu} \left(\begin{array}{cc|c}
              1 & 1 & 2 \\
	      i & j & \mu 
	      \end{array}\right)
	      \langle \ell'm'_\ell|T^2_\mu|\ell m_\ell\rangle,
\end{equation}
where \cite{brink68}
\begin{equation}
 T^2_\mu = \sum_{ij} \left(\begin{array}{cc|c}
              1 & 1 & 2 \\
	      i & j & \mu 
	      \end{array}\right)  \ell^i \ell^j.
\end{equation}
The one can write
\begin{equation}
 \langle \ell'm'_\ell|\ell^{(2)ij}|\ell m_\ell\rangle =   \sum_{\mu} \left(\begin{array}{cc|c}
              1 & 1 & 2 \\
	      i & j & \mu 
	      \end{array}\right)
	   \left(\begin{array}{cc|c}
	             \ell & 2 & \ell' \\
		     m_\ell & \mu & m'_\ell
	            \end{array}\right)
\langle \ell'||T^2||\ell\rangle,	        
\end{equation}
where we have applied the Wigner-Eckart theorem. One needs now to derive the reduce matrix element $\langle \ell'||T^2||\ell\rangle$. As it is independent of $\mu$ one can choose $\mu=i+j=2$. This gives
\begin{equation}
 T^2_2  =  \ell_1\ell_1,
\end{equation}
where 
 \begin{equation}
  \ell_1=-\frac{1}{\sqrt{2}}\left(\ell_x+i\ell_y\right).
 \end{equation}
We have 
\begin{equation}
 \ell_1|\ell, m_\ell\rangle = -\sqrt{\frac{(\ell-m_\ell)(\ell+m_\ell+1)}{2}}|\ell, m_\ell +1\rangle.
\end{equation}
One can then write
\begin{eqnarray}
 \langle \ell,\ell|T_2^2 |\ell,\ell-2\rangle & = & \sqrt{\ell(2\ell-1)} \\
 					     & = &  \left(\begin{array}{cc|c}
	             \ell & 2 & \ell \\
		     \ell-2 & 2 & \ell
	            \end{array}\right)
\langle \ell||T^2||\ell\rangle		 \\
      & = & \sqrt{\frac{6}{(\ell+1)(2\ell+3)}} \langle \ell||T^2||\ell\rangle.
\end{eqnarray}
We then obtain the matrix elements of the tensor operator $\ell^{(2)ij}$ as
\begin{equation}
 \langle \ell'm'_\ell|\ell^{(2)ij}|\ell m_\ell\rangle = \delta_{\ell\ell'}\sqrt{\frac{\ell(\ell+1)(2\ell-1)(2\ell+3)}{6}} \sum_{\mu} \left(\begin{array}{cc|c}
              1 & 1 & 2 \\
	      i & j & \mu 
	      \end{array}\right)
	      \left(\begin{array}{cc|c}
	             \ell & 2 & \ell \\
		     m_\ell & \mu & m'_\ell
	            \end{array}\right). \label{O3tensor}
\end{equation}

After introduced in the matrix elements of the generators, to calculate the matrix elements of $O_i$ and $B_i$, we use the following relations \cite{brink68,rotenberg59}:
\begin{enumerate}
 \item 
\begin{equation}
 \left(\begin{array}{cc|c}
        a & b & c \\
	\alpha & \beta & -\gamma 
       \end{array}\right)
= (-1)^{a-b-\gamma}\sqrt{2c+1}\left(\begin{array}{ccc}
                                     a & b & c \\
				     \alpha & \beta & \gamma
                                    \end{array}\right).
\end{equation}
This formula relates an SU(2) Clebsch-Gordan coefficient in the left-hand side to the corresponding 3-j symbols.
\item The following two formulas relate  6-j symbols to  3-j symbols,
\begin{eqnarray}
\left\{\begin{array}{ccc}
                  j_1 & j_2  & j_3 \\
		  l_1 & l_2 &  l_3
                 \end{array}\right\}& = & 
\sum_{\mathrm{all}\ m,n} (-1)^S 
\left(\begin{array}{ccc}
       j_1 & j_2 & j_3 \\
       m_1 & m_2 & m_3
      \end{array}\right)
\left(\begin{array}{ccc}
       j_1 & l_2 & l_3 \\
       m_1 & n_2 & -n_3
      \end{array}\right) \nonumber \\ & & \times
\left(\begin{array}{ccc}
       l_1 & j_2 & l_3 \\
       -n_1 & m_2 & n_3
      \end{array}\right)
\left(\begin{array}{ccc}
       l_1 & l_2 & j_3 \\
       n_1 & -n_2 & m_3
      \end{array}\right),  
\end{eqnarray}
 and
\begin{eqnarray}
 \lefteqn{\left(\begin{array}{ccc}
       j_1 & j_2 & j_3 \\
       m_1 & m_2 & m_3
      \end{array}\right)\left\{\begin{array}{ccc}
                  j_1 & j_2  & j_3 \\
		  l_1 & l_2 &  l_3
                 \end{array}\right\}=} \nonumber \\ & &
\sum_{\mathrm{all}\ n} (-1)^S
\left(\begin{array}{ccc}
       j_1 & l_2 & l_3 \\
       m_1 & n_2 & -n_3
      \end{array}\right)
  \left(\begin{array}{ccc}
       l_1 & j_2 & l_3 \\
       -n_1 & m_2 & n_3
      \end{array}\right)
\left(\begin{array}{ccc}
       l_1 & l_2 & j_3 \\
       n_1 & -n_2 & m_3
      \end{array}\right),    
\end{eqnarray}
where $S=l_1+l_2+l_3+n_1+n_2+n_3$.
\item A 9-j symbol can be written in terms of  3-j symbols as
\begin{eqnarray}
 \lefteqn{\left\{\begin{array}{ccc}
                  j_{11} & j_{12} & j_{13} \\
		  j_{21} & j_{22} & j_{23} \\
		  j_{31} & j_{32} & j_{33} 
                 \end{array}\right\} =} \nonumber \\ & &
\sum_{\mathrm{all}\ m} 
\left(\begin{array}{ccc}
       j_{11} & j_{12} & j_{13} \\
       m_{11} & m_{12} & m_{13}
      \end{array}\right)
 \left(\begin{array}{ccc}
       j_{21} & j_{22} & j_{23} \\
       m_{21} & m_{22} & m_{23}
      \end{array}\right)
 \left(\begin{array}{ccc}
       j_{31} & j_{32} & j_{33} \\
       m_{31} & m_{32} & m_{33}
      \end{array}\right)     \nonumber \\
& & \times
\left(\begin{array}{ccc}
       j_{11} & j_{21} & j_{31} \\
       m_{11} & m_{21} & m_{31}
      \end{array}\right)
\left(\begin{array}{ccc}
       j_{12} & j_{22} & j_{32} \\
       m_{12} & m_{22} & m_{32}
      \end{array}\right)
\left(\begin{array}{ccc}
       j_{13} & j_{23} & j_{33} \\
       m_{13} & m_{23} & m_{33}
      \end{array}\right).
\end{eqnarray}
\end{enumerate}
The 3-j and 6-j coefficients are tabulated \cite{rotenberg59}. One can also use analytic calculators to derive 3-j, 6-j and 9-j coefficients \cite{stevenson02}. \\

Let us now write the matrix elements of the $O_i$ coefficients for the  multiplets studied in various chapters. 
\section{The $[{\bf 70},\ell^+]$ multiplet}

\subsection{Non-strange baryons}

Here we give  explicit matrix elements of the operators $O_i$ written in Table \ref{70,ell+nonstrangeoperators}. The notations used are the ones of Ref. \cite{carlson98b} where we have $\rho = S-I=\pm 1,0$ and $\eta/2=I_c-I=\pm 1/2$. Concerning the isoscalar factors the  identification between the notations used  in Eq. (\ref{isoscalarms}) and the ones introduced here is 
\begin{equation}
c^{[N_c-1,1]}_{0+}=c^{[N_c-1,1]}_{22},
\end{equation} and 
\begin{equation}
c^{[N_c-1,1]}_{0-}=c^{[N_c-1,1]}_{11}.
\end{equation} \\

Let us first write the matrix elements of the SU(4) generators for a symmetric spin-flavor wave function in these notations \cite{carlson98b}:
\begin{eqnarray}
\langle [N_c]S'=I';S_3',I_3'|S^i|[N_c]S=I;S_3,I_3\rangle = \delta_{SS'}\delta_{II'}\delta_{I_3I_3'} \sqrt{S(S+1)}  \left(\begin{array}{cc|c}
         S & 1 & S' \\
	 S_3 & i & S_3'
        \end{array}\right),
\end{eqnarray}
\begin{eqnarray}
\langle [N_c]S'=I';S_3',I_3'|T^a|[N_c]S=I;S_3,I_3\rangle = \delta_{SS'}\delta_{S_3S_3'}\delta_{II'} \sqrt{I(I+1)}  \left(\begin{array}{cc|c}
        I & 1 & I' \\
	I_3 & a & I_3'
       \end{array}\right),
       \end{eqnarray}
\begin{eqnarray}
 \lefteqn{\langle [N_c]S'=I';S_3',I_3'|G^{ia}|[N_c]S=I;S_3,I_3\rangle =} \nonumber \\  & & \frac{1}{4}\sqrt{(2I+1)(2I'+1)}\sqrt{(N_c+1)^2-(I'-I)^2(I'+I+1)^2} 
 \left(\begin{array}{cc|c}
         S & 1 & S' \\
	 S_3 & i & S_3'
        \end{array}\right)
\left(\begin{array}{cc|c}
        I & 1 & I' \\
	I_3 & a & I_3'
       \end{array}\right). \nonumber \\
\end{eqnarray}
They can be derived by using Eq. (\ref{GENsu4}) and Table \ref{genesu(4)sym}. To obtain the matrix elements of $S_c,\ T_c$  and $G_c$, one replaces each $N_c$ by $N_c-1$ and $S,\ I$ by $S_c,\ I_c$. For the matrix elements of $s,\ t$ and $g$, one has to replace $N_c$ by 1 and   $S$ and $I$ by 1/2. \\

Using the above formulas, one can derive the matrix elements of all the $O_i$ operators. The results are
\begin{eqnarray}
\label{70,ell+nonstrangelqs}
\langle \ell_q  s \rangle  & =  & \delta_{JJ'}\delta_{J_3J'_3}\delta_{II'}\delta_{I_3I'_3}(-1)^{J-I+\ell_q+\ell_c+1} \sqrt{\frac{3}{2}}\sqrt{(2\ell+1)(2\ell'+1)}\sqrt{\ell_q(\ell_q+1)(2\ell_q+1)} \nonumber \\ 
& & \times  \sqrt{(2S+1)(2S'+1)} 
 \left\{\begin{array}{ccc}
		\ell & 1 & \ell' \\
		\ell_q & \ell_c & \ell_q
	\end{array}\right\}
\left\{\begin{array}{ccc}
		1 & \ell & \ell' \\
		J & S'& S 
	\end{array}\right\} \nonumber \\
& & \times  \sum_{\eta=\pm 1} (-1)^{(1-\eta)/2} c^{[N_c-1,1]}_{\rho' \eta} c^{[N_c-1,1]}_{\rho \eta} 
\left\{\begin{array}{ccc}
		S & 1 & S' \\
		\frac{1}{2} & I+\frac{\eta}{2} & \frac{1}{2} 
	\end{array}\right\}, 
\end{eqnarray}

\begin{eqnarray}
\langle \ell^{(2)}_q g G_c \rangle & = & \delta_{JJ'}\delta_{J_3J_3'}\delta_{II'}\delta_{I_3I_3'}\delta_{\ell_c\ell_c'}\delta_{\ell_q\ell_q'} (-1)^{J-2I+\ell_q+\ell_c+S}\frac{1}{8}\sqrt{\frac{15}{2}}\sqrt{(2S+1)(2S'+1)} \nonumber \\
& & \times \sqrt{(2\ell+1)(2\ell'+1)}\sqrt{\ell_q(\ell_q+1)(2\ell_q-1)(2\ell_q+1)(2\ell_q+3)}\nonumber \\
& & \times \left\{\begin{array}{ccc}
            2 & \ell & \ell' \\
	    J & S'   &  S
           \end{array}\right\}
\left\{\begin{array}{ccc}
        \ell  & 2  & \ell' \\
	\ell_q & \ell_c & \ell_q 
       \end{array}\right\}
\sum_{\eta,\eta'=\pm 1}c^{[N_c-1,1]}_{\rho'\eta'}c^{[N_c-1,1]}_{\rho\eta}(-1)^{(1+\eta')/2} \nonumber \\
& & \times \sqrt{(2I_c+1)(2I_c'+1)}\sqrt{(N_c+1)^2-\left(\frac{\eta'-\eta}{2}\right)^2(2I+1)^2} \nonumber \\
& & \times \left\{\begin{array}{ccc}
        1/2 & 1 & 1/2 \\
	I_c & I & I_c'
       \end{array}\right\}
\left\{\begin{array}{ccc}
        I_c' & I_c & 1 \\
	S'   & S   & 2 \\
	1/2  & 1/2 & 1
	 \end{array}\right\}, 
\end{eqnarray}

\begin{eqnarray}
\langle S_c^2 \rangle  & = & \delta_{J' J} \delta_{J'_3 J_3}
\delta_{L' L} \delta_{I' I} \delta_{I'_3 I_3}
\delta_{S' S} \sum_{\eta = \pm 1} \left(c^{[N_c-1,1]}_{\rho \eta}\right)^2 \left(I+\frac{\eta}{2}\right)\left(I+\frac{\eta}{2}+1\right), 
\end{eqnarray}

\begin{eqnarray}
\langle s S_c \rangle & = & \delta_{J J'} \delta_{J_3 J'_3}
\delta_{\ell\ell'} \delta_{II'} \delta_{I_3 I'_3}
\delta_{SS'} (-1)^{S-I} \sqrt{\frac{3}{2}}\sum_{\eta = \pm 1} (-1)^{(1-\eta)/2} \left(c^{[N_c-1,1]}_{\rho \eta}\right)^2\nonumber \\ & & \times  \sqrt{\left(I+\frac{\eta}{2}\right)\left(I+\frac{\eta}{2}+1\right)\left(2I+\eta+1\right)} 
 \left\{\begin{array}{ccc}
		S & I+\frac{\eta}{2} & \frac{1}{2} \\
		1 & \frac{1}{2} & I+\frac{\eta}{2}
		\end{array}\right\}.
\end{eqnarray}

\subsection{Strange baryons}
The matrix elements of the SU(6) generators for a symmetric spin-flavor wave function have been presented in Chapter 4.
Let us write explicitly the matrix elements of the operators $t^{-a}$,
$g^{ia}$ and $G_c^{ja}$. As explained in Chapter 4, we have

\begin{eqnarray}\label{70,ell+strangeTA}
\langle (10)y'i'i'_3
|t^{-a}| (10)yii_3\rangle
= \sqrt{\frac{4}{3}}
  \left(\begin{array}{cc|c}
	i  &    I^a    &   i'   \\
        i_3 &    -I^a_3  &   i'_3
      \end{array}\right)
\left(\begin{array}{cc||c}
(10)  &  (11)   &  (10)  \\
yi & -Y^aI^a & y'i' \\
\end{array}\right),
\end{eqnarray}

\begin{equation}
\langle \frac{1}{2} m_2;(10) y' i' i'_3 |g^{ia}| \frac{1}{2} m_2; (10) y i i_3 \rangle =
 \left(\begin{array}{cc|c}
	\frac{1}{2} & 1 & \frac{1}{2}   \\
	     m_2     & i & m'_2
	 \end{array}\right)
 \left(\begin{array}{cc|c}
	(10)      &   (11)  & (10) \\
	 y i i_3 &     y^ai^ai^a_3 &  y' i' i'_{3}
      \end{array}\right),\label{gia1bodysu6}
     \end{equation}
and
\begin{eqnarray}\label{Gcia1bodysu6}
\lefteqn{\langle [N_c-1] S_c'm_1';(\lambda'_c\mu'_c) Y'_cI'_cI'_{c_3}|G^{ja}_c|[N_c-1] S_cm_1;(\lambda_c\mu_c) Y_cI_cI_{c_3}\rangle = } \nonumber \\
& & \frac{1}{\sqrt{2}} \sqrt{\frac{5}{12}(N_c-1)(N_c+5)} \left(\begin{array}{cc|c}
	S_c & 1 & S_c'   \\
	   m_1  & j & m'_1
	 \end{array}\right)
\left(\begin{array}{cc|c}
	I_c & I^a & I_c'   \\
        I_{c_3}  & I^a_3 & I'_{c_3}
	 \end{array}\right)\nonumber \\
& &\times \sum_{\rho = 1,2}
 \left(\begin{array}{cc||c}
	(\lambda_c \mu_c)    &  (11)   &   (\lambda'_c \mu'_c)\\
	Y_c I_c   &  Y^a I^a  &  Y'_c I'_c
      \end{array}\right)_{\rho}
\left(\begin{array}{cc||c}
	[N_c-1]    &  [21^4]   & [N_c-1]   \\
	(\lambda_c \mu_c) S_c  &  (11)1  &  (\lambda'_c \mu'_c) S'_c
      \end{array}\right)_{\rho},
\end{eqnarray}
where the SU(3) isoscalar factors are from Tables \ref{hecht1}--\ref{hecht4} and the SU(6) isoscalar factors can be found in Table \ref{su6isoscalar}. \\

Let us first write  the matrix elements of the operator $O_2$, using the notations used in Section \ref{excitedwavefunctions},
\begin{eqnarray}
\label{70,ell+strangelqs}
\langle \ell_q  s \rangle   =   \delta_{J'J}\delta_{J'_3J_3}
\delta_{\lambda' \lambda} \delta_{\mu' \mu} \delta_{Y' Y}
\delta_{I'I}\delta_{I'_3I_3}
(-1)^{J-1/2+\ell_q+\ell_c}
\sqrt{\frac{3}{2} (2S+1)(2S'+1)} \nonumber \\ 
\times 
\sqrt{(2\ell+1)(2\ell'+1)
\ell_q(\ell_q+1)(2\ell_q+1)} 
 \left\{\begin{array}{ccc}
		\ell & 1 & \ell' \\
		\ell_q & \ell_c & \ell_q
	\end{array}\right\}
\left\{\begin{array}{ccc}
		1 & \ell & \ell' \\
		J & S'& S 
	\end{array}\right\} \nonumber \\
 \times
 \sum_{p p' p''} (-1)^{-S_c} c_{p' p}(S) c_{p'' p}(S')
\left\{\begin{array}{ccc}
		S & 1 & S' \\
		\frac{1}{2} & S_c & \frac{1}{2}
	\end{array}\right\}, 
\end{eqnarray}
where $S_c = S - 1/2$ for $p' = 1$ and $S_c = S + 1/2$ for  $p' = 2$
and similarly $S_c = S' - 1/2$ for $p'' = 1$ and $S_c = S' + 1/2$ for $p'' = 2$. Eq. (\ref{70,ell+strangelqs}) is equivalent to Eq. (\ref{70,ell+nonstrangelqs}). \\

 The compact form of $O_3$  given in Table 
\ref{70,ell+strangeoperators} is
\begin{equation}\label{OPO3}
O_3 = \frac{3}{N_c}\ell^{(2),ij}_{q}g^{ia}G_c^{ja}.
\end{equation}
Writing the scalar products in an explicit form we have
\begin{equation}
O_3 = \frac{3}{N_c}\sum_{ij}(-1)^{i+j}\ell^{(2),-i,-j}_{q}
\sum_{Y^a I_3^a}(-1)^{I_3^a+Y^a/2}g^{ia}G_c^{ja},
\end{equation}
with $i,j = 1,2,3$ and $a = 1,2, ...,8$. 
The final formula for the matrix elements of $O_3$ between states of
mixed orbital symmetry  $[N_c-1,1]$ is
\begin{eqnarray}
\label{70,ell+strangeO3}
\lefteqn{\langle  \ell S'; JJ_3; (\lambda \mu) Y I I_3
| O_3 | \ell S; JJ_3;
(\lambda \mu) Y I I_3 \rangle = } \nonumber \\
& & (-1)^{\ell_q+\ell_c+S'+J+1} \frac{5}{4N_c}(2 \ell+1)
 \sqrt{\ell_q (\ell_q+1)(2 \ell_q-1)(2 \ell_q+1)(2 \ell_q+3)}
 \left\{\begin{array}{ccc}
	\ell     &  2       &  \ell  \\
	\ell_q &  \ell_c  & \ell_q 
      \end{array}\right\} 
 \nonumber \\
 & & \times
 \sqrt{2(N_c-1)(N_c+5)(2S+1)(2S'+1)}
    \left\{\begin{array}{ccc}
	S    &   2   & S'   \\
	\ell    &   J   & \ell
      \end{array}\right\} 
    \nonumber \\
& &\times
       \sum_{p,p',q,q'} c^{[N_c-1,1]}_{pp'}(S)    c^{[N_c-1,1]}_{qq'}(S')\sqrt{2S'_c+1}
       \left\{\begin{array}{ccc}
	S'_c    &  S'  &  1/2  \\
	1       &  2   &  1  \\
	S_c     &  S   &  1/2
      \end{array}\right\}
      \nonumber \\
& &\times\sum_{\rho=1,2}
      U((\lambda_c \mu_c)(11)(\lambda \mu)(10);(\lambda'_c \mu'_c)(10))_{\rho}
       \left(\begin{array}{cc||c}
	[N_c-1]    &   [21^4]   &  [N_c-1]   \\
	(\lambda_c \mu_c) S_c   &   (11) 1   &  (\lambda'_c \mu'_c) S'_c
      \end{array}\right)_{\rho},
\end{eqnarray} 
where the coefficients $c^{[N_c-1,1]}_{pp'}(S)$ are given by Eqs. (\ref{isoscalarms}).
We recall that $S_c=S-1/2$ for $p=1$ and $S_c=S+1/2$ for $p=2$
and by analogy $S'_c=S'-1/2$ for $q=1$ and $S'_c=S'+1/2$ for $q=2$.
Also $(\lambda_c \mu_c) = (\lambda-1,\mu)$ for $p'=1$,
$(\lambda_c \mu_c) = (\lambda+1,\mu-1)$ for $p'=2$ and $(\lambda_c \mu_c) = (\lambda,\mu+1)$ for $p'=3$ and an analogous
situation for $(\lambda'_c \mu'_c)= (\lambda-1,\mu)$ if $q'=1$,
 $(\lambda'_c \mu'_c)=(\lambda+1,\mu-1)$ if $q'=2$ and $(\lambda'_c \mu'_c)= (\lambda,\mu+1)$ if $q'=3$. \\

When applied on the excited quark the operator $O_4$ reads 
\begin{equation}\label{70,ell+OPO4}
O_4 = \frac{4}{N_c+1}\ell^i_{q}t^{a}G_c^{ia}.
\end{equation}
Writing the scalar products explicitly we have
\begin{equation}
O_4 = \frac{4}{N_c+1}\sum_{i}(-1)^{i}\ell^{i}_{q}
\sum_{Y^a I_3^a}(-1)^{I_3^a+Y^a/2}t^{-a}G_c^{-ia}~.
\end{equation}
Inserting the above expression and the matrix elements of $G^{ia}_c$,
 Eq. (\ref{Gcia1bodysu6}), into (\ref{70,ell+OPO4}) one obtains
\begin{eqnarray}
\label{70,ell+strangeO4}
\lefteqn{\langle  \ell S'; JJ_3; (\lambda \mu) Y I I_3
| O_4 | \ell S; JJ_3;
(\lambda \mu) Y I I_3 \rangle = (-1)^{\ell_q+\ell_c+S'-S+J+1/2} \frac{4}{N_c+1}
} \nonumber \\
& & \times
(2 \ell+1)
 \sqrt{\ell_q (\ell_q+1)(2 \ell_q+1)}
 \left\{\begin{array}{ccc}
	\ell     &  1       &  \ell  \\
	\ell_q &  \ell_c  & \ell_q 
      \end{array}\right\} 
 \nonumber \\
 & & \times
 \sqrt{\frac{5}{18} (N_c-1)(N_c+5)(2S+1)(2S'+1)}
    \left\{\begin{array}{ccc}
	J   &   \ell   & S   \\
	1   &   S'   & \ell
      \end{array}\right\} 
    \nonumber \\
& &\times
       \sum_{p,p',q,q'} c^{[N_c-1,1]}_{pp'}(S) c^{[N_c-1,1]}_{qq'}(S')\sqrt{2S'_c+1}(-1)^{-S_c'}
       \left\{\begin{array}{ccc}
	S'  &  1/2  & S'_c   \\
	S_c &  1    & S
      \end{array}\right\}
      \nonumber \\
& &\times\sum_{\rho=1,2}
      U((\lambda_c \mu_c)(11)(\lambda \mu)(10);(\lambda'_c \mu'_c)(10))_{\rho}
       \left(\begin{array}{cc||c}
	[N_c-1]    &   [21^4]   &  [N_c-1]   \\
	(\lambda_c \mu_c) S_c   &   (11) 1   &  (\lambda'_c \mu'_c) S'_c
      \end{array}\right)_{\rho}. \nonumber \\
\end{eqnarray} 
The unitary Racah coefficients $U$, defined according to Eq. (\ref{su(6)racahhecht}), which are needed to calculate (\ref{70,ell+strangeO3}) and (\ref{70,ell+strangeO4}) can be obtain by using the definition
\begin{eqnarray}
 \lefteqn{U((\lambda_1\mu_1)(11)(\lambda\mu)(10);(\lambda_{12}\mu_{12})(10))_\rho = } \nonumber \\
& & \sum_{Y_1Y_a(Y_3) \atop I_1I_aI_3I_{12}I_{23}} (-1)^{I_1+I_a+I+I_3}\sqrt{(2I_{12}+1)(2I_{23}+1)}\left\{\begin{array}{ccc}
                                                                                       I_1 & I_a & I_{12} \\
										       I_3 & I  & I_{23}
                                                                                      \end{array}\right\}\nonumber \\  
& & \times
\langle (\lambda_1\mu_1)Y_1I_1;(11)Y_aI_a||(\lambda_{12}\mu_{12})Y_{12}I_{12}\rangle_{\rho} \langle(\lambda_{12}\mu_{12}) Y_{12}I_{12};(10)Y_3I_3||(\lambda\mu)YI\rangle 
 \nonumber \\  
& & \times  \langle (11)Y_aI_a;(10)Y_3I_3||(10)Y_{23}I_{23}\rangle \langle (\lambda_1\mu_1)Y_1I_1;(10)Y_{23}I_{23}||(\lambda\mu)YI\rangle, \label{racahsu32forme}
\end{eqnarray}
where the  isoscalar factors can be found in Tables \ref{vergados}--\ref{hecht4}. The Racah coefficients are
\begin{equation}
U((\lambda-1,\mu)(11)(\lambda\mu)(10);(\lambda+1,\mu-1)(10)) = 
-\frac{1}{2}\sqrt{\frac{3(\lambda+2)\mu}{2(\lambda+1)(\mu+1)}},
\end{equation}
\begin{equation}
U((\lambda+1,\mu-1)(11)(\lambda\mu)(10);(\lambda-1,\mu)(10)) = 
\frac{1}{2}\sqrt{\frac{3\lambda(\lambda+\mu+1)}{2(\lambda+1)(\lambda+\mu+2)}}, 
\end{equation}
\begin{equation}
U((\lambda,\mu +1)(11)(\lambda\mu)(10);(\lambda,\mu +1)(10))_{\rho=1}  =  \frac{\lambda+2\mu+8}{4\sqrt{g_{\lambda,\mu+1}}}, 
\end{equation}
\begin{equation}
U((\lambda,\mu +1)(11)(\lambda\mu)(10);(\lambda,\mu +1)(10))_{\rho=2}  =  \frac{1}{4} \sqrt{\frac{3\lambda(\lambda+2)(\mu+3)(\lambda+\mu+4)}{(\mu+1)(\lambda+\mu+2)g_{\lambda,\mu+1}}}, 
\end{equation}
\begin{equation}
U((\lambda-1,\mu)(11)(\lambda \mu)(10);(\lambda-1, \mu)(10))_{\rho=1}
  =  - \frac{2 \lambda + \mu - 2}
 { 4 \sqrt{g_{\lambda-1,\mu}}},
\end{equation}
\begin{equation}
 U((\lambda-1,\mu)(11)(\lambda \mu)(10);(\lambda-1, \mu)(10))_{\rho=2}
 =  \frac{1}{4} 
 \sqrt{\frac{3(\lambda + \mu)(\lambda - 1) \mu (\mu + 2)}
{(\lambda + 1)(\lambda + \mu + 2) g_{\lambda-1,\mu}}},
\end{equation}
\begin{equation}
U((\lambda+1,\mu-1)(11)(\lambda \mu)(10);(\lambda+1, \mu-1)(10))_{\rho=1}
 =  \frac{\lambda - \mu + 5}{4 \sqrt{g_{\lambda+1,\mu-1}}},
\end{equation}
\begin{eqnarray}
\lefteqn{U((\lambda+1,\mu-1)(11)(\lambda \mu)(10);(\lambda+1, \mu-1)(10))_{\rho=2}
 =} \nonumber \\ & &   - \frac{1}{4}
\sqrt{\frac{3(\lambda + \mu + 1)(\lambda + \mu + 3)(\lambda + 3)(\mu - 1)}
{(\lambda + 1)(\mu + 1) g_{\lambda+1,\mu-1}}}.
  \end{eqnarray} 
All $U$ coefficients, but the 4th one, are of order $\mathcal{O}(N_c^0)$ which can 
be seen by inserting $\lambda=2S$ and $\mu=N_c/2-S$. This helps in finding the 
order of the matrix elements of $O_4$. \\

The matrix elements of $O_6$ are given by the product of $1/N_c$ and
\begin{eqnarray}\label{70,ell+strangeHARD}
\langle \ell S JJ_3;(\lambda'\mu') Y' I' I'_3|t^aT^a_c|\ell S JJ_3;(\lambda \mu) Y I I_3\rangle = \delta_{\lambda\lambda'}\delta{\mu\mu'}\delta_{YY'}\delta_{II'}\delta_{I_3I_3'} \nonumber \\
\times (-1) \sum_{pp'} \left[c^{[N_c-1,1]}_{pp'}(S)\right]^2 \frac{2\sqrt{g_{\lambda_c\mu_c}}}{3} U((\lambda_c\mu_c)(11)(\lambda\mu)(10);(\lambda_c\mu_c)(10))_{\rho=1},
\end{eqnarray}
where the $(-1)$ sign results from a phase entering the symmetry property of
SU(3) Clebsch-Gordan coefficients \cite{deswart63}. This is
\begin{equation}
 \left(\begin{array}{cc||c}
	(10)    &  (11)   &   (10)\\
	Y I   &  Y^a I^a  &  Y'I'
      \end{array}\right)
= \xi_1 (-1)^{I+I^a-I'}
\left(\begin{array}{cc||c}
	(11)    &  (10)   &   (10)\\
	Y^aI^a   &  Y I  &  Y'I'
      \end{array}\right),
\end{equation}
where $\xi_1=-1$ in this case.  The same property has also been used
in the calculation of the matrix elements of 
$O_3$ and $O_4$. A simpler alternative is to calculate the matrix elements 
of $O_6$ by using the identity
\begin{equation}
t\cdot T_c = \frac{1}{2}\left(T^2-T^2_c-t^2\right),
\end{equation}
which gives
\begin{equation}\label{70,ell+strangeEASY}
\langle \ell S JJ_3;(\lambda\mu) Y I I_3|t^aT^a_c|\ell S JJ_3;(\lambda \mu) Y I I_3\rangle 
= \frac{1}{6} \left\{ g_{\lambda\mu}
- \sum_{pp'} \left[c^{[N_c-1,1]}_{pp'}(S)\right]^2 g_{\lambda_c\mu_c}- 4 \right\}.
\end{equation}
The formulas (\ref{70,ell+strangeHARD}) and (\ref{70,ell+strangeEASY}) give identical results.\\

Let us write the matrix elements 
of the flavor breaking operators $t_8$, $T^c_8$ and $\ell_q^ig^{i8}$ which have been used 
to generate Table \ref{70,ell+strangeT8} and \ref{70,ell+strangeb3}. These are
\begin{eqnarray}
\langle \ell S JJ_3;(\lambda'\mu') Y' I' I'_3|T^8_c|\ell S JJ_3;(\lambda \mu) Y I I_3\rangle =
 \delta_{YY'}\delta_{II'}\delta_{I_3I_3'}\sum_{p,p',p''}c_{pp'}(S)c_{p p''}(S) \nonumber \\
\times \sum_{Y_c,I_c,y,i} \frac{3Y_c}{2\sqrt{3}}
\left(
\begin{array}{cc||c}
(\lambda_c\mu_c) & (10) & (\lambda\mu) \\
Y_cI_c  & yi  & YI \\
\end{array}
\right)
\left(
\begin{array}{cc||c}
(\lambda_c\mu_c) & (10) & (\lambda'\mu') \\
Y_cI_c  & yi  & YI
\end{array}
\right),
\end{eqnarray}

\begin{eqnarray}
\langle \ell S JJ_3;(\lambda'\mu') Y' I' I'_3|t^8|\ell S JJ_3(\lambda \mu) Y I I_3\rangle =
\delta_{YY'}\delta_{II'}\delta_{I_3I_3'}\sum_{p,p',p''}c_{pp'}(S)c_{p p''}(S) \nonumber \\
\times \sum_{Y_c,I_c,y,i}
\frac{3y}{2\sqrt{3}}
\left(
\begin{array}{cc||c}
(\lambda_c\mu_c) & (10) & (\lambda\mu) \\
Y_cI_c  & yi  & YI \\
\end{array}
\right)
\left(
\begin{array}{cc||c}
(\lambda_c\mu_c) & (10) & (\lambda'\mu') \\
Y_cI_c  & yi  & YI
\end{array}
\right),
\end{eqnarray}
and
\begin{eqnarray}\label{70,ell+strangeB4}
 \langle \ell S' JJ_3;(\lambda'\mu') Y' I' I'_3|\ell_q^ig^{i8}|\ell S JJ_3(\lambda \mu) Y I I_3\rangle =\delta_{YY'}\delta_{II'}\delta_{I_3I_3'} (-1)^{J+\ell_q+\ell_c-1/2} \nonumber \\
 \times  (2\ell+1) \sqrt{\ell_q(\ell_q+1)(2\ell_q+1)} \sqrt{(2S+1)(2S'+1)} 
 \left\{
\begin{array}{ccc}
 \ell & 1 & \ell \\
 \ell_q & \ell_c  & \ell_q 
\end{array}\right\} 
\left\{
\begin{array}{ccc}
 J & \ell & S \\
 1 & S'  & \ell 
\end{array}\right\} \nonumber \\
\times \sum_{p,p',q,q'} (-1)^{S_c} c^{[N_c-1,1]}_{pp'}(S)c^{[N_c-1,1]}_{q q'}(S')
\left\{
\begin{array}{ccc}
 S' & 1 & S \\
 1/2 & S_c  & 1/2 
\end{array}\right\} \nonumber \\
\times \sum_{Y_c,I_c,y,i}
\frac{3y}{2\sqrt{2}}
\left(
\begin{array}{cc||c}
(\lambda_c\mu_c) & (10) & (\lambda\mu) \\
Y_cI_c  & yi  & YI \\
\end{array}
\right)
\left(
\begin{array}{cc||c}
(\lambda_c\mu_c) & (10) & (\lambda'\mu') \\
Y_cI_c  & yi  & YI
\end{array}
\right).
\end{eqnarray}
To obtain Table \ref{70,ell+strangeb3} and \ref{70,ell+strangeT8} we have used the Eqs. (\ref{su6isoscalar}) for the coefficients
$c^{[N_c-1,1]}_{pp'}$ and Table \ref{vergados} 
for the isoscalar factors of SU(3).
From their expressions one can find that all these coefficients and
isoscalar factors 
are of order $N^0_c$.  Then it follows that for states with 
spin and strangeness of order $N^0_c$, the matrix elements of $T^c_8$ are 
of order $N_c$ because $Y_c = Y - y$, $Y= N_c/3 + \mathcal{S}$ so
that $Y_c \sim N_c$.

 \section{The $[{\bf 56},4^+]$ multiplet}
 \label{appendixd564+}
 For this multiplet, another method has been used to derived the matrix elements of $O_i$ and $B_i$. One just needs  the spin states $\chi$ and the flavor states $\phi$ explicitly. \\
 
 Consider first the octet. One can write
\begin{equation}
|SS_z; (11) YII_z\rangle_{\mathrm{Sym}}= \frac{1}{\sqrt{2}}\left( \chi^\rho \phi^\rho +\chi^\lambda \phi^\lambda\right),
\end{equation}
with the flavor states
$\phi^\rho$ and $\phi^\lambda$ defined in Table \ref{msymflstates}.
The states
$\chi^\rho$ and $\chi^\lambda$ have spin $S=1/2$ and permutation 
symmetry $\rho$ and $\lambda$ respectively. They can be obtained from 
$\phi^\rho_{p(n)}$ and $\phi^\lambda_{p(n)}$ by making the replacement
$$u\rightarrow \uparrow, \ d\rightarrow \downarrow,$$
where $\uparrow$ and $\downarrow$ are spin 1/2 single-particle states of projection $S_z=+1/2$ and $S_z=-1/2$, respectively. The three particle states of mixed symmetry and $S=1/2$, $S_z=+1/2$ are
\begin{equation}
\chi^\lambda_+=-\frac{1}{\sqrt{6}}\left(\uparrow\downarrow\uparrow+\downarrow\uparrow\uparrow-2\uparrow\uparrow\downarrow\right),
\end{equation}
\begin{equation}
\chi^\rho_+=\frac{1}{\sqrt{2}}\left( \uparrow\downarrow\uparrow-\downarrow\uparrow\uparrow\right),
\end{equation}
and those having $S=1/2$, $S_z=-1/2$ are
\begin{equation}
\chi^\lambda_-=\frac{1}{\sqrt{6}}\left(\uparrow\downarrow\downarrow+\downarrow\uparrow\downarrow-2\downarrow\downarrow\uparrow\right),
\end{equation}
\begin{equation}
\chi^\rho_-=\frac{1}{\sqrt{2}}\left(\uparrow\downarrow\downarrow-\downarrow\uparrow\downarrow\right).
\end{equation}\\

For the decuplet, we have
\begin{equation}
|SS_z; (30) Y II_z\rangle_{\mathrm{Sym}}=\chi\phi^S,
\end{equation}
with $\phi^S$ defined in Table \ref{symflstates}. $\chi$ must be an $S=3/2$ state. With the previous notation, the $\chi_{\frac{3}{2}m}$ states take the form
\begin{equation}
\chi_{\frac{3}{2}\frac{3}{2}}=\uparrow\uparrow\uparrow,
\end{equation}
\begin{equation}
\chi_{\frac{3}{2}\frac{1}{2}}=\frac{1}{\sqrt{3}}\left( \uparrow\uparrow\downarrow+\uparrow\downarrow\uparrow+\downarrow\uparrow\uparrow\right),
\end{equation}
\begin{equation}
\chi_{\frac{3}{2}-\frac{1}{2}}=\frac{1}{\sqrt{3}}\left(\uparrow\downarrow\downarrow+\downarrow\uparrow\downarrow+\downarrow\downarrow\uparrow\right),
\end{equation}
\begin{equation}
\chi_{\frac{3}{2}-\frac{3}{2}}=\downarrow\downarrow\downarrow.
\end{equation}
\chapter{Matrix elements of the spin-orbit operator for  $[{\bf 70},1^-]$ baryons}
\thispagestyle{empty}

In this Appendix, we  derive explicitly two matrix elements of the operators $\ell^i(j)s^i(j)$ and $\ell^i(j)s^i(k),\ j\neq k$. One can decompose the two operators \cite{brink68}
\begin{eqnarray}
 \ell^i(j)s^i(j) & =  & \ell_0(j)s_0(j) + \frac{1}{2}\left(\ell_+(j)s_{-}(j)+\ell_{-}(j)s_+(j)\right), \label{decomp1body} \\
 \ell^i(j)s^i(k) & = & \ell_0(j)s_0(k) + \frac{1}{2}\left(\ell_+(j)s_{-}(k)+\ell_{-}(j)s_+(k)\right), \label{decomp2body}
\end{eqnarray}
with
\begin{eqnarray}
\begin{array}{ll}
 \ell_0  =  \ell_z & s_0  =  s_z, \\
 \ell_+  =  \ell_x+i\ell_y, & s_+  =  s_x + is_y,\\
 \ell_{-}  =  \ell_x-i\ell_y, & s_{-}  =  s_x-is_y.\\
 \end{array}
 \label{decompositionappende}
 \end{eqnarray}
One has the following matrix elements 
\begin{eqnarray}
 \begin{array}{ll}
  \ell_0|\ell m_\ell\rangle  = m_l|\ell m_\ell\rangle, & s_0|ss_3\rangle = s_3|ss_3\rangle, \\
  \ell_+|\ell m_\ell\rangle  = \sqrt{(\ell-m_\ell)(\ell+m_\ell+1)}|\ell m_\ell+1\rangle, & s_+  = \sqrt{(s-s_3)(s+s_3+1)}s_3|ss_3+1\rangle, \\
 \ell_{-} |\ell m_\ell\rangle  = \sqrt{(\ell+m_\ell)(\ell-m_\ell+1)}|\ell m_\ell-1\rangle, & s_-  = \sqrt{(s+s_3)(s-s_3+1)}s_3|ss_3-1\rangle. \\
 \end{array}
 \label{decompmatrixelemappende}
\end{eqnarray}

Let us first calculate the matrix element of the operators Eqs. (\ref{decomp1body}) and  (\ref{decomp2body}) for the state $^4N[{\bf 70},1^-]1/2^-1/2$ with $N_c=3$. The wave function is given by \cite{stancu96} 
{\small\begin{eqnarray}
\lefteqn{|^4N[{\bf 70},1^-] \frac{1}{2}^-\frac{1}{2}\rangle =} \nonumber \\ & & \sqrt{\frac{1}{12}} \left(\psi^{\rho}_{11}\phi^{\rho}+ \psi^{\lambda}_{11}\phi^{\lambda}\right)\chi_{\frac{3}{2}-\frac{1}{2}}-\sqrt{\frac{2}{12}}\left(\psi^{\rho}_{10}\phi^{\rho}+ \psi^{\lambda}_{10}\phi^{\lambda}\right)\chi_{\frac{3}{2}\frac{1}{2}}+\sqrt{\frac{3}{12}}\left(\psi^{\rho}_{1-1}\phi^{\rho}+ \psi^{\lambda}_{1-1}\phi^{\lambda}\right)\chi_{\frac{3}{2}\frac{3}{2}}. \nonumber \\  \label{newlook4n1/2}
\end{eqnarray}}
where 
\begin{eqnarray}
\psi^{\lambda}_{1m} & = & \frac{1}{\sqrt{6}}\left(2ssp-sps-pss\right), \\
\psi^{\rho}_{1m} & = & \frac{1}{\sqrt{2}}\left(sps-pss\right),
\end{eqnarray}
where $m=-1,0,1$ because $|p\rangle \sim Y_{1m}$ (see Eq (\ref{0pwavefunction})). The spin part 
$\chi$ is given explicitly in Section \ref{appendixd564+} and the flavor part $\phi$ in Tables \ref{symflstates}--\ref{antisymstate}.
Using Eqs. (\ref{decomp1body}), (\ref{decompmatrixelemappende}) and  (\ref{newlook4n1/2}) one obtains 
\begin{eqnarray}
 \langle ^4N[{\bf 70},1^-] \frac{1}{2}^-\frac{1}{2}|\ell^i(3)s^i(3)|^4N[{\bf 70},1^-] \frac{1}{2}^-\frac{1}{2}\rangle & =  & -\frac{5}{18}, \\
  \langle ^4N[{\bf 70},1^-] \frac{1}{2}^-\frac{1}{2}|\ell^i(2)s^i(3)|^4N[{\bf 70},1^-] \frac{1}{2}^-\frac{1}{2}\rangle & =  & -\frac{5}{18}.
\end{eqnarray}
Taking $N_c=3$ in Eq. (\ref{spinorbitonebody})  one obtains 
\begin{equation}
 \sum_{j=1}^{3} \langle ^4N[{\bf 70},1^-] \frac{1}{2}^-\frac{1}{2}|\ell^i(j)s^i(j)|^4N[{\bf 70},1^-] \frac{1}{2}^-\frac{1}{2}\rangle  =  -\frac{5}{6},
\end{equation}
 which is identical to $\langle N'_{1/2} | \ell s|  N'_{1/2}\rangle$ of Ref. \cite{carlson98b}. Taking $N_c=3$ in  Eq. (\ref{spinorbittwobody}), one obtains
 \begin{eqnarray}
  \sum_{j=1}^{3} \sum_{k\neq j =1}^{3}  \langle ^4N[{\bf 70},1^-] \frac{1}{2}^-\frac{1}{2}|\ell^i(j)s^i(k)|^4N[{\bf 70},1^-] \frac{1}{2}^-\frac{1}{2}\rangle = -\frac{5}{3},
 \end{eqnarray}
which is identical to $\langle N'_{1/2} | \ell S_c|  N'_{1/2}\rangle$ of Ref. \cite{carlson98b}. \\

Let us now consider the state \cite{stancu96}
\begin{equation}
 |^2\Delta[{\bf 70},1^-] \frac{1}{2}^-\frac{1}{2}\rangle = \left[\sqrt\frac{2}{6} \left(\psi^{\rho}_{11}\chi^{\rho}_{-}+\psi^{\lambda}_{11}\chi^{\lambda}_{-}\right) - \sqrt\frac{1}{6} \left(\psi^{\rho}_{10}\chi^{\rho}_{+}+\psi^{\lambda}_{10}\chi^{\lambda}_{+}\right)\right]\phi^S.
\end{equation}
Using the same method as above, one has
\begin{eqnarray} 
 \langle ^2\Delta[{\bf 70},1^-] \frac{1}{2}^-\frac{1}{2}|\ell^i(3)s^i(3)|^2\Delta[{\bf 70},1^-] \frac{1}{2}^-\frac{1}{2}\rangle & =  & \frac{1}{9}, \\
  \langle ^2\Delta[{\bf 70},1^-] \frac{1}{2}^-\frac{1}{2}|\ell^i(2)s^i(3)|^2\Delta[{\bf 70},1^-] \frac{1}{2}^-\frac{1}{2}\rangle & =  & -\frac{2}{9}.
\end{eqnarray}
Then 
\begin{eqnarray}
 \sum_{j=1}^{3} \langle ^2\Delta[{\bf 70},1^-] \frac{1}{2}^-\frac{1}{2}|\ell^i(j)s^i(j)|^2\Delta[{\bf 70},1^-] \frac{1}{2}^-\frac{1}{2}\rangle & =  & \frac{1}{3}, \\
 \sum_{j=1}^{3} \sum_{k\neq j =1}^{3}  \langle ^2\Delta[{\bf 70},1^-] \frac{1}{2}^-\frac{1}{2}|\ell^i(j)s^i(k)|^2\Delta[{\bf 70},1^-] \frac{1}{2}^-\frac{1}{2}\rangle & =  & -\frac{4}{3}.
\end{eqnarray}
which correspond  to $\langle \Delta_{1/2}|\ell s| \Delta_{1/2}\rangle$ and $\langle \Delta_{1/2}|\ell S_c| \Delta_{1/2}\rangle$ of Ref. \cite{carlson98b} respectively.

\end{document}